 \documentclass[preprint,aps,amsmath,nofootinbib,tightenlines,floatfix]{revtex4}
\usepackage[dvips]{graphics}
\usepackage{epsfig}
\usepackage{latexsym}
\usepackage{amssymb}

\newcommand{\lsim}{\buildrel < \over {_\sim}}
\newcommand{\gsim}{\buildrel > \over {_\sim}}

\newcommand{\be}{\begin{equation}}
\newcommand{\ee}{\end{equation}} 
\newcommand{\bea}{\begin{eqnarray}}
\newcommand{\eea}{\end{eqnarray}} 
\newcommand{\ba}{\begin{array}}
\newcommand{\ea}{\end{array}}

\newcommand{\msusy}{{\tilde m}}

\def\sinhat{\hat{s}^2}

 \def\sstw{{\sin^2\theta_W}}

\newcommand{\dslash}[1]{#1\!\!\!/}

\newcommand{\ket}[1]{\left\lvert #1\right\rangle}
\newcommand{\bra}[1]{\left\langle #1\right\rvert}

\newcommand{\diracslash}[1]{#1\!\!\!\!\!/}

\newcommand{\CPV}{CP\!\!\!\!\!\!\!\raisebox{1pt}{\scriptsize$\diagup$}}

\newcommand{\beq}{\begin{equation}}
\newcommand{\eeq}{\end{equation}}
\newcommand{\beqa}{\begin{eqnarray}}
\newcommand{\eeqa}{\end{eqnarray}}

\newcommand{\apv}{{A_{PV}}}

\def\rnu{{R_\nu}}
\def\rnubar{{R_{\bar\nu}}}
\def\sst#1{{\scriptscriptstyle #1}}
\def\sstw{{\sin^2\theta_\sst{W}}}
\def\mzs{{M^2_{Z}}}

\def\sinhat{\sin^2\hat\theta_W}
\def\mz{{M_{Z}}}
\def\mw{{M_{W}}}

\def\Journal#1#2#3#4{{#1} {#2} (#4) #3 }
\def\AIP{{\em AIP Conf. Proc.}} 
\def\PT{{\em Phys. Today}}

\def\NPA{{\em Nucl. Phys.} A}

\def\NPB{{\em Nucl. Phys.} B}

\def\PLB{{\em Phys. Lett.} B}

\def\PRL{\em Phys. Rev. Lett.}
\def\PREV{\em Phys. Rev.}

\def\PRA{{\em Phys. Rev.} A}
\def\PRD{{\em Phys. Rev.} D}
\def\PRC{{\em Phys. Rev.} C}

\def\RMP{{\em Rev. Mod. Phys.}}

\def\INT{{\em Int. J. Mod. Phys.} A}

\def\JETP{\em J. Exp. Theor. Phys.} 
\def\SJNP{\em Sov. J. Nucl. Phys.}

\newcommand{\etal}{{\it et al.}}

\def\sinhat{\hat{s}^2}

\def\sstw{{\sin^2\theta_W}}

\def\alred{{A_{\sst PV}^{eD,\ \rm DIS}}}

\newcommand{\mbold}[1]{\mbox{\boldmath$ #1$}}

\begin{document}

\title{Low Energy Precision Test of Supersymmetry}

\author{M.J. Ramsey-Musolf}
\affiliation{California Institute of Technology,
Pasadena, CA 91125\ USA}
\affiliation{University of Wisconsin-Madison,
Madison, WI 53706\ USA}

\author{S.~Su}
\affiliation{University of Arizona,
Tucson, AZ 85721\ USA}

\begin{abstract}

Supersymmetry (SUSY) remains one of the leading candidates for physics
beyond the Standard Model, and the search for SUSY will be a central focus of
future collider experiments. Complementary information on the viability and
character of SUSY can be obtained via the analysis of precision electroweak
measurements. 
In this review, we discuss the prospective implications for SUSY of
present and future precision studies at low energy.

\end{abstract}


\maketitle

\tableofcontents

\pagenumbering{arabic}


\section{Introduction}
\label{sec:intro}

With the advent of the Large Hadron Collider (LHC), the search for physics beyond the Standard Model (SM) will enter a new era. While the establishment of the Standard Model constitutes one of the major achievements in 20th century physics, particle physicists have long been aware of its limitations and have looked for clues as to the larger framework in which it is embedded. These limitations include the number of {\em a priori} unknown parameters (nineteen), the absence of any requirement of electric charge quantization, the lack of coupling unification at high scales, and the instability of the electroweak scale under radiative corrections. From the standpoint of cosmology, the SM also falls short, as it provides no explanation for the abundance of either the visible, baryonic matter of the universe or the cold dark matter. The recent observation of neutrino oscillations and the corresponding implication that neutrinos have small, non-vanishing masses also points to physics beyond the SM as originally written down by Glashow, Weinberg, and Salam \cite{SM}.

One of the leading candidates for the larger framework in which the SM lies is supersymmetry (SUSY). For over two decades, the attractive features of SUSY have inspired particle physicists to explore its theoretical basis and phenomenological implications with great vigor. These attractive features include its elegant mechanism for producing electroweak symmetry-breaking and stabilizing the electroweak scale; its generation of coupling unification at the grand unification scale; its introduction of a candidate for the cold dark matter (the lightest supersymmetric particle, or LSP); and its possibilities for explaining the abundance of baryonic matter. In addition, SUSY is a rather generic feature of superstring theories, so one might expect a low-energy remnant of string theory to exhibit features of supersymmmetry. The  presence of these elements that could resolve many (but not all) of the shortcomings of the SM have outweighed the costs of introducing SUSY, such as the additional large number of {\em a priori} unknown parameters, and have inspired a vast literature in particle physics during the past two decades.

It is hoped among SUSY enthusiasts that experiments at the LHC will finally uncover direct evidence for low-energy SUSY, and the variety of corresponding high-energy signatures have been discussed extensively elsewhere \cite{wang}. In this review, we focus on another frontier in the search for SUSY: the high-precision, low-energy frontier. This low-energy frontier, which lies at the intersection of particle physics with nuclear and atomic physics, has seen substantial recent advances in both experimental techniques and theoretical analysis, making precision low-energy studies a powerful probe of SUSY. Roughly speaking, the sensitivity required for these studies to probe supersymmetric effects is given by $\delta_{\rm SUSY}\sim (\alpha/\pi)(M/{\tilde m})^2$, where $M$ is an appropriate standard model mass scale and ${\tilde m}$ is a typical superpartner mass\footnote{As we discuss in later in this review, exceptions occur.}. The prefactor of $\alpha/\pi$ appears because SUSY contributions typically first arise at one loop order and are of electroweak strength\footnote{The SUSY loop corrections scale as $\alpha/\pi$ rather than $\alpha/4\pi$       
since the SU(2)$_L$ coupling $g_2$  that enters one loop amplitudes is          
related to the electric charge via the weak mixing angle as                     
$g_2=e/\sin\theta_W$. The value of $\sin^2\theta_W\approx 1/4$ compensates      
for the factor of four that would ordinarily appear in the denominator. }. A well-known illustration occurs in the case of the muon anomalous magnetic moment, $(g_\mu-2)$, where $M\sim m_\mu$ and where, for ${\tilde m}\sim M_W$, one has $\delta_{\rm SUSY}\sim 10^{-9}$. Both the precision of the latest experimental measurement of $(g_\mu-2)$ \cite{Bennett:2006fi} as well as the uncertainty in the theoretical, SM prediction, are at this level, making this observable an important means of accessing SUSY loop effects. Indeed, the $\sim 2\sigma$ deviation from the SM prediction reported by the E821 collaboration, gives the first tantalizing hints of low-energy SUSY. 

While there has been considerable recent attention paid to $(g_\mu-2)$ for these reasons, the high energy physics community may have less appreciation for the analogous power of other precision, low-energy observables to provide a window on supersymmetry. In the case of weak decays or electroweak scattering, for example, one has $M\sim M_W$, making $\delta_{\rm SUSY}\sim 10^{-3}$ rather than the $10^{-9}$ in the case of $(g_\mu-2)$. As we discuss in detail throughout the remainder of this review, both the precision of low-energy weak decay and lepton scattering studies -- as well as that of the corresponding theoretical, SM predictions -- is now at the $10^{-3}$ level or better, making an analysis of their implications for SUSY a timely endeavor. As with the study of precision electroweak observables at the end of the last decade, the study of low-energy precision electroweak processes can provide important information that complements what we may learn from the LHC or a future, $e^+ e^-$ linear collider. Indeed, comparisons of the value of the top quark mass implied by precision data with the value determined by direct top quark observation at the Tevatron provided a significant test of the self-consistency of the SM at the level of one-loop radiative corrections and stands as a major success in the history of SM. Given the level of experimental and theoretical precision now available for the low-energy studies discussed here, one can anticipate a similarly useful comparison of indirect and direct search information in the LHC era. Moreover, there exist special cases -- such as searches for lepton flavor violation or permanent electric dipole moments -- where the SM predictions lie well below what will be achievable in the next generation of precision studies, where the expectations of SUSY effects lie well within the reach of these experiments, and where the physics reach of the low-energy measurements can significantly exceed what is accessible at the LHC or a linear collider. 

In the remainder of this article, we describe the recent experimental and theoretical advances that have led to this situation and discuss the corresponding present and prospective implications of precision, low-energy measurements for SUSY. In doing so, we attempt to provide sufficient background material for readers from both the low-energy and collider communities to be able to appreciate these developments. We begin with a brief review of low-energy SUSY, and refer readers to the excellent, more extensive ``primer" by S. Martin\cite{Martin:1997ns} for additional details. Because many of the SUSY effects discussed here arise at loop level, we also provide a brief review of renormalization as it is applied to the observables of interest here. The bulk of our subsequent discussion involves a review of low-energy charged current and neutral current experiments, searches for lepton flavor and number violation, tests of CP-violation and the corresponding implications for cosmology. Because members of the high energy community may not be so familiar with the phenomenology of these studies, we provide some background material while referring readers to recent, comprehensive studies of low-energy precision tests\cite{Erler:2004cx}. In addition, the information on SUSY obtained from high energy studies is important in analyzing the low-energy sector, so we also give brief summaries of the present implications of  high energy experiments (for recent reviews, see {\em e.g.}, Ref.~\cite{wang,Heinemeyer:2004gx}). Finally, the reader will notice one significant, but intentional, omission from this review: a discussion of the present situation regarding $(g_\mu-2)$. Because the recent literature on this topic is so vast and because there exist useful, recent reviews (see, {\em e.g.}, Refs.~\cite{Hertzog:2006sc,Erler:2004cx,Czarnecki:2001pv}), we believe that a truncated discussion in this article would be  redundant and would not do justice to this important measurement. Thus, we refer the reader to  the literature for a proper review of the muon anomalous magnetic moment, and concentrate on the other precision tests in the remainder of this article.

\section{Minimal Supersymmetric Extension of Standard Model}
\label{sec:susy}

\subsection{Introduction}
The Standard Model of elementary particle physics has been
confirmed to high precision by a wide array of experiments. The
strong, weak and electromagnetic interactions  are described by ${\rm
SU(3)}_C\times{\rm SU(2)}_L\times{\rm U(1)}_Y$ gauge interactions.  At
low energies,  ${\rm SU(2)}_L\times{\rm U(1)}_Y$ is broken to ${\rm
U(1)}_{EM}$ symmetry by the Higgs mechanism, which generates masses
for the $W^{\pm}$ and $Z$ bosons, while keeping the photon  massless.
For this purpose,  a complex scalar ${\rm SU(2)}_L$ Higgs doublet
$H=(H^+, H^0)$ is introduced.  The electroweak symmetry is broken
spontaneously when the  neutral component $H^0$ gets a vacuum
expectation value (VEV):  $\langle H^0 \rangle = v/\sqrt{2}$.  While
three degrees of freedom for the Higgs doublet are eaten by $W^{\pm}$
and $Z$ (corresponding to their longitudinal degrees of freedom) , one
physical Higgs boson, $h$, which is the real part of $H^0$, 
remains in the spectrum.   
For the Higgs potential $V(H)=m_H^2H^\dagger H + \lambda |H^\dagger H|^2$, 
the mass of the physical Higgs
is related to the Higgs quartic coupling $\lambda$ and Higgs VEV
$v=246$ GeV: $m_{h}^2=2 \lambda v^2$.  For $\lambda$ of order unity,
the Higgs mass is around the electroweak scale of a few hundred GeV.



Being a fundamental scalar particle, the Higgs boson can receive large
corrections  to its mass from quantum loop effects. Assuming the SM is
an effective theory valid below a cut-off scale $\Lambda_{\rm UV}$, the
Higgs mass, $m_h$, depends strongly on  physics at the scale
$\Lambda_{\rm UV}$.   For example, any SM fermion $f$ with Yukawa
interaction $(\lambda_f/\sqrt{2}) h \bar{f} f$ 
induces a one loop correction to
the squared mass of  the Higgs.  
The leading contribution to the mass of the physical Higgs 
depends quadratically on $\Lambda_{\rm UV}$\cite{Drees:1996ca}:
\begin{equation}
\Delta {m_{h}^2}=  \frac{|\lambda_f|^2}{16\pi^2}\left[ -2 \Lambda_{\rm
UV}^2+ 6 m_f^2\ln (\Lambda_{\rm UV}^2/m_f^2) + \ldots \right],
\label{eq:mh_f}
\end{equation}
where $\Lambda_{\rm UV}$  is the cutoff scale, which   could be as
large as the Planck scale  $M_{\rm pl}=(8\pi G_{\rm
Newton})^{-1/2}=2.4 \times 10^{18}$ GeV.   A precise cancellation of
32 orders of magnitude between  the tree level bare Higgs mass and the
radiative corrections is needed to obtain a physical Higgs mass around
electroweak scale.  Such a high level of fine tuning is usually
refered to as the  ``hierarchy problem'' \footnote{The hierarchy problem has two elements: the 
large scale difference between
$M_{\rm pl}$ and $M_{\rm weak}$, and the need to cancel the radiative corrections to maintain a light Higgs scalar. The need for fine tuning to achieve this cancellation might be more accurately termed a \lq\lq naturalness problem".}.  Finding a solution to the hierarchy problem 
points to  new physics beyond the SM, such as supersymmetry
\cite{Martin:1997ns,susy},  extra dimensions \cite{exD}, little  Higgs \cite{lh},
composite Higgs \cite{ch}, Higgsless models\cite{hless}, etc.

Supersymmetry  -- a symmetry under interchange of  bosonic and
fermionic degrees of freedom  --  is one  of the most  promising new
physics scenarios among various proposals.  For each   particle in a
supersymmetric theory, there exists a superpartner with spin differing
by a half unit.   When SUSY is exact, the masses and the gauge quantum numbers
of  superpartners are the same, and the couplings are related by the
symmetry.  These features protect the Higgs mass from receiving the
problematic quadratic dependence on $\Lambda_{\rm UV}$ as these
contributions from fermionic and bosonic superpartners cancel.  


For
example, the $\Lambda_{\rm UV}^2$ term in  Eq.~(\ref{eq:mh_f}) from
fermion $f$  is cancelled precisely by the contribution from its
scalar partners\footnote{One 
Dirac fermion $f$ has two complex scalar superpartners.
Eq.~(\ref{eq:mh_s}) assumes the both scalar superpartners have the 
same mass $m_S$.} $S$ with mass $m_S$ and Higgs coupling $\lambda_S
|H|^2|S|^2$\cite{Drees:1996ca}: 
\begin{equation}
\Delta {m_{h}^2}=  \frac{\lambda_S}{16\pi^2}\left[ 2\Lambda_{\rm
UV}^2-2m_S^2 \ln (\Lambda_{\rm UV}^2/m_S^2) + \ldots \right]\ \ \ ,
\label{eq:mh_s}
\end{equation}
A cancellation of the quadratic divergence occurs when we employ the supersymmetric relation $\lambda_S=|\lambda_f|^2$ and add Eqs.~(\ref{eq:mh_f}) and (\ref{eq:mh_s}). The remaining $\Lambda_{\rm UV}$-dependence is only  logarithmic level and
fine tuning is no longer necessary. In Eq.~(\ref{eq:mh_s}) the logarithmic term proportional to $m_S^2$ arises from the tadpole graph containing the full quartic scalar interaction. 
An additional logarithmic contribution to the Higgs mass
arises from the diagram containing two insertions of the triscalar interaction
$\sqrt{2}\lambda_S v h |S|^2$  induced by the quartic interaction after electroweak symmetry breaking (EWSB). 
\begin{equation}
\Delta {m_{h}^2}=  \frac{\lambda_S}{16\pi^2}\left[
-4m_f^2\ln (\Lambda_{\rm UV}^2/m_S^2) + \ldots \right],
\label{eq:mh_s2}
\end{equation}
The explicit dependence on $m_f$ appears because
\begin{eqnarray}
\lambda_S |H|^2 |S|^2 & =& |\lambda_f|^2  |H|^2 |S|^2 = |\lambda_f|^2\left(\frac{v^2}{2}+\sqrt{2} v h +\frac{h^2}{2}+\cdots\right) |S|^2 \\
\lambda_f \sqrt{2}v & = & 2 m_f \ \ \ 
\end{eqnarray}
with  the \lq\lq $+\cdots$"  denoting the other scalar doublet degrees of freedom.
Adding Eq.~(\ref{eq:mh_s}) and Eq.~(\ref{eq:mh_s2}), the total contribution from 
a pair of complex scalars to the physical Higgs mass are 
\begin{equation}
\Delta {m_{h}^2}=  \frac{\lambda_S}{16\pi^2}\left[ 2\Lambda_{\rm
UV}^2-(2m_S^2+4 m_f^2)\ln (\Lambda_{\rm UV}^2/m_S^2) + \ldots \right].
\label{eq:mh_s3}
\end{equation}
It is obvious from Eqs.~(\ref{eq:mh_f}) and (\ref{eq:mh_s3}) 
that in the supersymmetric limit, when $m_f=m_S$, the logarithmic 
contributions from scalars and fermion also cancel each other.

Besides providing an elegant solution to the hierarchy problem, SUSY
has a variety of other  attractive features.  In the SM, the ${\rm
SU}(3)_C$, ${\rm SU}(2)_L$  and ${\rm U}(1)_Y$ gauge couplings come
close to unifying at high scales, providing a tantalizing hint of
grand unification.  With TeV scale mass for superpartners, the SUSY
$\beta$ functions lead to coupling unification at a scale $M_{\rm
GUT}\sim 10^{16}$ GeV --  close to the Planck scale \cite{unification}.   In addition,
electroweak symmetry breaking can be generated radiatively with SUSY,
due to the ${\cal O}(1)$  Yukawa coupling of top quarks and their
scalar superpartners. Finally, SUSY provides viable
particle physics solutions to problems in cosmology. In the minimal
supersymmetric extension of the SM, for example, the lightest
supersymmetric particle is a natural candidate for cold dark
matter (CDM) if it is protected from decays into SM particles by a
symmetry known as $R$-parity. Similarly, SUSY contributions to the
Higgs potential and the introduction of new CP-violating interactions
involving superpartners can make supersymmetric electroweak
baryogenesis a viable mechanism for explaining the baryon asymmetry of
the universe. In short, the theoretical and cosmological motivation
for considering low energy SUSY is strong.

\subsection{The Minimal Supersymmetric Extension of Standard Model}
In the simplest supersymmetric extension of the SM -- the Minimal Supersymmetric Standard Model (MSSM) --
provides a useful framework for discussing the phenomenology of low
energy SUSY. Although there is considerable interest in amplifications
of the MSSM -- particularly in the neutrino and Higgs sectors -- we
will concentrate on the MSSM throughout this article. In the MSSM,
each SM particle is accompanied by a superpartner  with the same gauge
quantum numbers as given  in Table~\ref{table:MSSMmatter} for the
matter fields and in  Table~\ref{table:MSSMgauge} for the gauge
sector.  The symbols for the  SM superpartners are the same as the
corresponding SM particles, but with a  tilde on the top.  For each
quark and lepton, its spin 0 superpartner is  called a {\it squark}
and {\it slepton}, respectively.  The fermionic superpartner of each
Higgs boson is called a {\it Higgsino}.  Note that  in MSSM, introduction
of two Higgs doublets with opposite hypercharge --  $H_u$ and $H_d$ --
is dictated  by the requirement of anomaly cancellation among
fermionic Higgsinos.  In addition, the Yukawa interactions in
supersymmetric models are derived  from the superpotential, which must
be holomorphic in order to be supersymmtric.  In contrast to the SM,
where the same Higgs  gives mass both to the up and down type quark,
the latter receive mass in the MSSM from the VEVs of the neutral
$H_u$ and $H_d$, respectively.  Finally, the spin-$1/2$ superpartners of
the  ${\rm SU}(3)_C$, ${\rm SU}(2)_L$  and ${\rm U}(1)_Y$ gauge bosons
are called  the {\it gluino, Wino} and {\it Bino}, respectively.
\begin{table}
\begin{tabular}{c|ccc|cc|cc}
\hline &\multicolumn{3}{c|}{quark sector}& \multicolumn{2}{c|}{lepton
sector}& \multicolumn{2}{c}{Higgs sector} \\ \hline
&$Q$&$\bar{u}$&$\bar{d}$&$L$&$\bar{e}$&$H_u$&$H_d$\\  ${\rm SU(3)}_c$,
${\rm SU(2)}_L$, ${\rm U(1)}_Y$
&(3,2,$\frac{1}{6})$&($\bar{3}$,1,$-\frac{2}{3}$)&
($\bar{3}$,1,$\frac{1}{3}$)&(1,2,$-\frac{1}{2}$)&
(1,1,$1$)&(1,2,$\frac{1}{2}$)&(1,2,$-\frac{1}{2}$) \\ \hline spin 0&
$(\tilde{u}_L, \tilde{d}_L)$&$\tilde{u}_R^*$&$\tilde{d}_R^*$
&$(\tilde{\nu}, \tilde{e}_L)$&$\tilde{e}_R^*$& $(H_u^+,
H_u^0)$&$(H_d^0, H_d^-)$ \\ spin $1/2$ &$(u_L,
d_L)$&$u_R^\dagger$&$d_R^\dagger$ &$(\nu,
e_L)$&$e_R^\dagger$&$(\tilde{H}_u^+, \tilde{H}_u^0)$& $(\tilde{H}_d^0,
\tilde{H}_d^-)$ \\ \hline
\end{tabular}
\caption{Field content for the quark, lepton and Higgs sectors of the MSSM.}
\label{table:MSSMmatter}
\end{table}

\begin{table}
\begin{tabular}{c|ccc}
\hline &\multicolumn{3}{c}{gauge sector}\\ \hline ${\rm SU(3)}_c$,
${\rm SU(2)}_L$, ${\rm U(1)}_Y$ &(8,1,0)&(1,3,0)&(1,1,0)\\ \hline spin
$1/2$ & $\tilde{g}$&$\tilde{W}^{\pm},\tilde{W}^0$& $\tilde{B}^0$\\
spin 1&$g$&$W^{\pm}, W^0$&$B^0$\\ \hline
\end{tabular}
\caption{Field content for the gauge sector of the  MSSM.}
\label{table:MSSMgauge}
\end{table}

The Lagrangian of MSSM can be written as
\begin{equation}
{\cal L}={\cal L}_{\rm gauge} + {\cal L}_{\rm chiral}-
\sqrt{2}g[(\phi^*T^a\psi)\lambda^a +  \lambda^{\dagger
a}(\psi^{\dagger}T^a\phi)] -\frac{1}{2}\sum_{i}
g^2_i(\phi^*T^a\phi)^2,
\end{equation}
where $i$ runs over the  ${\rm SU(3)}_C$, ${\rm SU(2)}_L$ and $ {\rm
U(1)}_Y$ gauge groups; $\phi$ denotes a spin-$0$ complex scalar field
and   $\psi$ is the corresponding fermionic superpartner;  $\lambda^a$
is the gaugino field for ${\rm SU(3)}_C$,  ${\rm SU(2)}_L$ and ${\rm
U(1)}_Y$, with $g_i$ being the corresponding  gauge coupling and $T^a$
is the hermitian matrix  for the gauge group in the fundamental
representation.  The Lagrangian for the gauge fields ${\cal L}_{\rm
gauge}$ contains the  kinetic term for gauge bosons and two-component
gaugino spinors  $\lambda^a$:
\begin{equation}
{\cal L}_{\rm gauge}=-\frac{1}{4}F_{\mu\nu}^aF^{\mu \nu a}-
i\lambda^{a\dagger}\bar{\sigma}^{\mu}D_{\mu}\lambda^a\ \ \ ,
\end{equation}
where the metric is $\eta^{\mu\nu}={\rm diag}(-1,1,1,1)$,  
${\bar\sigma}^\mu = (-1, {\vec\sigma})$, and
$D_\mu$ is the gauge covariant derivative\footnote{Here, we have followed the conventions of Ref.~\cite{Martin:1997ns}.}.  The Lagrangian for the
matter fields ${\cal L}_{\rm chiral}$ contains  kinetic term and
interactions:
\begin{equation}
{\cal L}_{\rm chiral}=-D^{\mu}\phi^*D_{\mu}\phi-
i\psi^{\dagger}\bar{\sigma}^\mu D_{\mu}\psi +{\cal L}_{\rm int},
\end{equation}
where $\psi$ is a two component spinor for either left- or
right-handed fermions and ${\cal L}_{\rm int}$ can be obtained from
the  superpotential $W$
\begin{equation}
W_{\rm MSSM}=\bar{u}{\bf y_u}QH_u - \bar{d}{\bf y_d}QH_d -
\bar{e}{\bf y_e}LH_d + \mu H_u H_d.
\label{eq:MSSMsuperpotential}
\end{equation}
using
\begin{equation} 
{\cal L}_{\rm int}=(\partial W / \partial \phi_i \phi_j)\psi_i \psi_j
+(\partial W / \partial \phi_i)(\partial W / \partial \phi_i)^*.
\label{eq:lag}
\end{equation}
The first term in Eq.~(\ref{eq:lag}) gives rise to the usual Yukawa
coupling [from the first three terms in
Eq.~(\ref{eq:MSSMsuperpotential})], and the Higgsino mass [from the
last term in  Eq.~(\ref{eq:MSSMsuperpotential})].  The second term in
Eq.~(\ref{eq:lag})  gives rise to all the cubic and quartic scalar
interactions.

The general MSSM superpotential also  includes baryon and  lepton
 number violating interactions:
\begin{eqnarray}
W_{\Delta L=1}&=&\frac{1}{2}\lambda_{ijk} L_i L_j\bar{e}_k  +
\lambda^{\prime}_{ijk}L_iQ_j\bar{d}_k + \mu^{\prime}_i L_i H_u,
\label{eq:RPVL}\\
W_{\Delta B=1}&=&\frac{1}{2}\lambda_{ijk}^{\prime\prime} \bar{u}_i \bar{d}_j
\bar{d}_k,
\label{eq:RPVB}
\end{eqnarray}
The simultaneous presence of non-vanishing $\lambda^\prime$ and
$\lambda^{\prime\prime}$ couplings allows for rapid proton decay that
conflicts with present bounds on the proton lifetime.  One way to
eliminate  such terms is to introduce  a new symmetry called
$R$-parity, defined by conservation of the quantum number
\begin{equation}
P_R=(-1)^{3(B-L)+2s},
\end{equation}
where $s$ is the spin of the particle.  All SM particles have
$P_R=+1$ while all the superpartners  have $P_R=-1$.  If $R$-parity is
an exact symmetry, then   all the terms appearing in
Eq.~(\ref{eq:RPVL}) and ~(\ref{eq:RPVB})  are forbidden and no
dangerous proton decay can occur via these interactions.

There are two important phenomenological consequence if $R$-parity is
exactly conserved:
\begin{itemize}
\item{The lightest supersymmetric particle is absolutely stable.}
\item{SM particles are coupled to even numbers of superpartners
(usually two).}
\end{itemize}
If LSP is colorless and charge neutral, it can be a viable candidate 
for the cold dark matter. 
$R$-parity conservation also implies that sparticles are produced in
pairs in collider experiments and that each sparticle other than LSP
eventually  decays into final states containing odd numbers of LSPs.
Moreover, for low-energy processes involving only SM particles in the
initial and final states -- such as those of interest in this article
--  supersymmetric contributions appear only at loop-level ({\em
e.g.}, virtual superpartners are pair produced).  However, one may
relax the constrains of $R$-parity conservation while preserving
proton stability via, {\em e.g.}, forbidding  baryon number violating
terms in Eq.~(\ref{eq:RPVB}). In this case, the  LSP is no longer
stable and tree level SUSY contributions to low energy  processes
appear through $R$-parity violating interactions. In what follows, we
will consider the implications of both $R$-parity conserving and
$R$-parity violating (RPV) supersymmetry.

\subsection{Soft SUSY Breaking}
If supersymmetry is exact, superpartners have  the same mass as the
corresponding SM particles.  However, supersymmetry must be broken in
nature because superpartners have not been experimentally observed at
energies where they could be pair produced if they are degenerate with
SM particles.  In order  to retain the exact cancellation of quadratic
$\Lambda_{\rm UV}$ dependence of the  Higgs mass corrections,  all the
SUSY breaking couplings must be \lq\lq soft" (of positive mass
dimension).  After adding the fermion and scalar contributions of Eqs. (\ref{eq:mh_f},\ref{eq:mh_s}), the remaining logarithamic correction to the Higgs mass
is  proportional to the soft SUSY breaking masses\footnote{There 
are additional logarithmic contributions proportional to the square of the
 triscalar coupling, $a_f$, defined below\cite{Drees:1996ca}.}:
\begin{equation}
\Delta {m_{h}^2}=  -\frac{\lambda_S}{8\pi^2}\left[\ \delta m_S^2\ln
  (\Lambda_{\rm UV}^2/m_S^2) + \ldots \right]\ \ \ ,
\label{eq:mhsoft}
\end{equation}
where we have taken $m_S^2=m_f^2+\delta m_S^2$.  Therefore, the soft
SUSY breaking mass parameters ({\em e.g.} , $\delta m_S^2$) are
should be below a few TeV to avoid reintroduction of the
naturalness problem.  Throughout this work, we will refer to this scale of SUSY-breaking mass parameters as $\msusy$.
A brief description of soft SUSY breaking
parameters, SUSY particle  mass spectra and interactions is given
below.  For a more detailed  review of MSSM and related phenomenology,
see Refs.~\cite{Martin:1997ns, susy}.

In MSSM, the Lagrangian for soft SUSY breaking terms are
\begin{eqnarray}
{\cal L}_{\rm
soft}&=&-\frac{1}{2}(M_3 {\tilde{g}}\tilde{g}+M_2 {\tilde{W}}\tilde{W}
+M_1 {\tilde{B}}\tilde{B})+c.c. \nonumber \\
&&-(\tilde{\bar{u}}{\bf a_u}\tilde{Q}H_u-\tilde{\bar{d}}{\bf
a_d}\tilde{Q}H_d -\tilde{\bar{e}}{\bf a_e}\tilde{L}H_d)+c.c. \nonumber
\\ &&-\tilde{Q}^\dagger{\bf m_Q^2}\tilde{Q} -\tilde{L}^\dagger{\bf
m_L^2}\tilde{L} -\tilde{\bar{u}}{\bf
m_{\bar{u}}^2}\tilde{\bar{u}}^\dagger -\tilde{\bar{d}}{\bf
m_{\bar{d}}^2}\tilde{\bar{d}}^\dagger -\tilde{\bar{e}}{\bf
m_{\bar{e}}^2}\tilde{\bar{e}}^\dagger
-m_{H_u}^2H_u^*H_u-m_{H_d}^2H_d^*H_d
\nonumber \\
&&-(bH_uH_d+c.c.)
\label{eq:soft}
\end{eqnarray}
The first line gives the gaugino mass $M_i$, $i=1,2,3$ for ${\rm
U}(1)_Y$, ${\rm SU}(2)_L$ and ${\rm SU}(3)_C$ gauginos, respectively, and where the boldfaced quantities indicate matrices in flavor space. 
The second line gives the trilinear ``$A$-term'' that couples Higgs
scalars with left- and right- squarks and sleptons.  The third line
gives the  scalar mass $m_{\tilde{q}_{L,R}}^2$,
$m_{\tilde{l}_{L,R}}^2$,  and $m_{H_{u,d}}^2$ for squarks, sleptons
and Higgs scalars, respectively.  Finally, the last line is the
bilinear $b$-term, which couples up- and down-type Higgses.  In
principle, one may also include RPV soft interactions that correspond
to the terms in the superpotentials $W_{\Delta L=1}$ and $W_{\Delta
B=1}$. However, pure scalar RPV interactions are generally not
relevant to the low-energy observables discussed here, so we will not
include them.

The trilinear $A$-terms and the soft SUSY breaking squark and slepton
masses are in general non-diagonal in the flavor basis, a feature that
introduces  flavor-changing-neutral-current (FCNC) effects beyond
those that are GIM-suppressed in the SM.   Moreover, after performing
an appropriate set of field redefinitions, ${\cal L}_{\rm soft}$ --
together with the $\mu$-term in the superpotential -- includes 40
CP-violating phases beyond those of the SM (for a useful discussion, see, {\em e.g.}, 
Ref.~\cite{Dimopoulos:1995ju}). In contrast to the effects
of the CP-violating phase in the Cabibbo-Kobayashi-Maskawa (CKM)
matrix, the effects of these new phases are not suppressed by the
Jarlskog invariant\cite{Jarlskog:1985ht} and light quark Yukawa couplings. Thus, the
interactions in ${\cal L}_{\rm soft}$ can lead to unsuppressed FCNC
and CP-violating effects at low energy. On the other hand, both FCNC
and CP violation have been tightly constrained by  experiment.
Attemps to reconcile these two phenomenological implications of
${\cal L}_{\rm soft}$ with experimental bounds on FCNCs and
CP-violation are known as the \lq\lq SUSY flavor" and \lq\lq SUSY CP"
problem, respectively.

A detailed discussion of the SUSY flavor and CP problems appears in
Section \ref{sec:cpv}. However, for purposes of illustration, we consider one
approach to the flavor problem in which it is assumed that  and  ${\bf
m_Q^2}$,  ${\bf m_{\bar{u}}^2}$, ${\bf m_{\bar{d}}^2}$, ${\bf m_L^2}$
and ${\bf m_{\bar{e}}^2}$  are diagonal in flavor basis and that ${\bf
a_u}$, ${\bf a_d}$, and ${\bf a_e}$  are proportional to the
corresponding Yukawa matrices, ${\bf y_{u,d,e}}$.  The conventional
parameterization thus gives -- after diagonalization of the Yukawa
matrices --
\begin{equation}
\label{eq:triscalaryukawa}
{\bf a_u}=A_u\left(
\begin{array}{ccc}
y_u&&\\ &y_c&\\ &&y_t
\end{array}
\right),\ \ \  {\bf a_d}=A_d\left(
\begin{array}{ccc}
y_d&&\\ &y_s&\\ &&y_b
\end{array}
\right),\ \ \  {\bf a_e}=A_e\left(
\begin{array}{ccc}
y_e&&\\ &y_\mu&\\ &&y_\tau
\end{array}
\right).
\end{equation}

Specific SUSY breaking scenarios have been studied in the literature, which could
solve the FCNC and CP-violation problems introduced by the soft SUSY breaking terms.
If SUSY breaking is mediated from an unseen, high energy ``hidden sector'' to the visible weak scale sector (superparticles
in MSSM) via gravity, it is called Gravity Mediated SUSY Breaking (SUGRA) \cite{sugra}.
If SUSY breaking is mediated from the hidden sector to the visible sector via gauge interactions,
it is called Gauge Mediated SUSY Breaking (GMSB) \cite{gmsb}.  Recently,  Anomaly Mediated SUSY breaking (AMSB) \cite{amsb} and gaugino Mediated SUSY Breaking scenarios~\cite{gauginomsb} have also been considered.
These models relate the large number of parameters  in ${\cal          
L}_{\rm soft}$ to a few parameters associated with SUSY-breaking physics at     
high scales. We will occasionally refer to these model-dependent relations      
throughout this review.

\subsection{Superparticle Spectrum}

The superpartner mass spectrum emerges after diagonalization of the
relevant mass matrices.  For squarks and sleptons, the mass matrices
contain three components.  In the  $(\tilde{f}_L, \tilde{f}_R)$ basis,
there are mass matrices ${\bf M^2_{LL}}$ for the  LH fermion
superpartners, $\tilde{f}_L$; ${\bf M^2_{RR}}$ for the $\tilde{f}_R$;
and matrices ${\bf M^2_{LR}}$ that mix the two. The $LR$ mixing
matrices arise only after electroweak symmetry breaking, and to the
extent that the triscalar couplings ${\bf a_f}$ are proportional to
the Yukawa couplings as in Eq. (\ref{eq:triscalaryukawa}), one expects
the effects of this mixing to be relatively small except for the third
generation sfermions. In Section \ref{sec:cc}  we discuss low-energy
tests of this expectation. In flavor space, each of these matrices is
$6\times 6$ ($3\times 3$ in the case of the sneutrino). Since an extensive
discussion of the flavor problem appears in Section 
\ref{sec:cpv}, we will assume
momentarily that the ${\bf M_{AB}^2}$ ($A,B=L,R$) are flavor diagonal
for purposes of illustration. In general, one has
\begin{equation}
{\bf{M_{LL}^2}}  =    {\bf m_Q^2}+ {\bf m_q^2 }+{\bf \Delta_f}
\end{equation}
\begin{equation}
{\bf{M_{RR}^2}}  =    {\bf m_{\bar f}^2}+ {\bf m_q^2 }+{\bf
\bar\Delta_f}
\end{equation}
with
\begin{equation}
{\bf \Delta_f} = \left(I^f_3-Q_f\sin^2\theta_W\right)\ \cos 2\beta
M_Z^2
\end{equation}
\begin{equation}
{\bf \bar\Delta_f} = Q_f\sin^2\theta_W \ \cos 2\beta M_Z^2
\end{equation}
and
\begin{equation}
{\bf M_{LR}^2}={\bf M_{RL}^2} =
\begin{cases}
v\left[{\bf a_f} \sin\beta -\mu {\bf y_f} \cos\beta\right]\ , &
{\tilde u}-{\rm type\ sfermion}\\ v\left[{\bf a_f} \cos\beta -\mu {\bf
y_f} \sin\beta\right]\ , & {\tilde d}-{\rm type\ sfermion}
\end{cases}\ \ \ .
\end{equation}
Here ${\bf m_q^2}$ is the mass matrix for the corresponding fermion specie,
$I_3^f$ and $Q_f$ are the third component of isospin, and the charge of the 
fermion, respectively, and $\tan\beta$ is the ratio of the neutral Higgs vevs
$\tan\beta=\langle H_u^0 \rangle / \langle H_d^0 \rangle$.

The diagonal elements depend on the unknown soft SUSY breaking
parameters ${\bf m_Q^2}$, ${\bf m_{\bar{u}}^2}$, {\em etc.} while the
off-diagonal elements depend on the supersymmetric parameter $\mu$,
the soft-triscalar coupling ${\bf a_f}$, $v$  and $\tan\beta$.
Assuming no flavor mixing among different sfermion generations, the
sfermion mass matrix reduces to a set of $2\times 2$ matrices for each
flavor. The corresponding mass eigenstates ${\tilde F}_{1,2}$ are
mixtures of the ${\tilde f}_{L,R}$, with the mixing angle $\theta_{\tilde{f}}$:
\beq
\left(
\begin{tabular}{c}
${\tilde F}_{1}$\\
${\tilde F}_{2}$
\end{tabular}
\right)
=\left(
\begin{tabular}{cc}
$\cos\theta_{\tilde{f}}$&$\ \ \ -\sin\theta_{\tilde{f}}$\\
$\sin\theta_{\tilde{f}}$&$\ \ \ \cos\theta_{\tilde{f}}$\\
\end{tabular}
\right)
\left(
\begin{tabular}{c}
${\tilde f}_{L}$\\
${\tilde f}_{R}$
\end{tabular}
\right).
\eeq

In the more general situation where one allows for flavor mixing, the
$6\times 6$ diagonal mass matrix is given by
\begin{equation}
\left( {\bf M_f^2}\right)_{ \rm diag} = {\bf Z_f}^\dag\ {\bf M_f^2}\ {\bf
Z_f}
\end{equation}
where
\begin{equation}
{\bf M_f^2} =\left(
\begin{array}{cc}
{\bf M_{LL}^2} & {\bf M_{LR}^2}\\ {\bf M_{LR}^2} & {\bf M_{RR}^2}
\end{array}\right)
\end{equation}
for each species of sfermion. Hence, a given sfermion mass eigenstate
${\tilde F}_j$ is given in terms of the flavor eigenstates ${\tilde
f}_I$ as\footnote{Here, we follow the notation and conventions of Ref.~\cite{Rosiek:1995kg}.}
\begin{equation}
{\tilde f}_I = Z_f^{Ij}\, {\tilde F}_j
\end{equation}
where $I=1,2,3$ indicate the left-handed (LH) flavor states\footnote{For 
scalars, the handness simply indicates that they are the superpartners of 
the left-handed or the right-handed fermions.} ${\tilde f}_{L_I}$ and
$I=4,5,6$ refer to the right-handed (RH) flavor states  ${\tilde
f}_{R_{I-3}}$. Hence, the simultaneous presence of non-vanishing
$Z_f^{1j}$ and $Z_f^{2j}$ would indicate flavor mixing among first and
second generation LH sfermions, while having both  $Z_f^{1j}\not= 0$
and $Z_f^{4j}\not=0$ would imply mixing among the LH and RH sfermions
of the first generation.  Note that for sneutrinos, the indices $I,j$
run over only $1,2,3$ and we have no left-right mixing in this case.




The gauginos and higgsinos mix with each other  since both are charged
under the electroweak gauge group.  The mass matrix for  the neutral
states $\psi^0=(\tilde{B}, \tilde{W}^0, \tilde{H}_d^0, \tilde{H}_u^0)$
is
\begin{equation}
{\bf M}_{\tilde{N}}=\left(
\begin{array}{cccc}
M_1&0&-c_{\beta}s_WM_Z&s_{\beta}s_WM_Z\\ 0&M_2&c_\beta c_WM_Z&-s_\beta
c_WM_Z\\ -c_\beta s_WM_Z&c_\beta c_W M_Z&0&-\mu\\ s_\beta
s_WM_Z&-s_\beta c_W M_Z&-\mu&0
\end{array}
\right).
\end{equation}
Here we have introduced the abbreviation  $s_\beta=\sin\beta$,
$c_{\beta}=\cos\beta$, $s_W=\sin\theta_W$,  and $c_W=\cos\theta_W$.
The mass matrix can be diagonalized by a $4\times 4$ unitary matrix
$N$:
\begin{equation}
{\bf M_{\chi^0}}^{\rm diag}={\bf N}^* {\bf M_{\tilde{N}}} {\bf N}^{-1},
\end{equation}
The mass eigenstates are called neutralinos
$\chi_i^0=N_{ij}\psi_j^0$, $i=1 \ldots 4$, with
$m_{\chi_1^0}<m_{\chi_2^0}<m_{\chi_3^0}<m_{\chi_4^0}$.    In the limit
that $M_Z \ll M_1, M_2, |\mu|$, each neutralino is a pure gaugino or
Higgsino state,  while in general, the neutralino is a mixture  of
gauginos and Higgsinos.

Similarly, the mass matrix for charged gaugino and Higgsino
$\psi^{\pm}=(\tilde{W}^+, \tilde{H}_u^+, \tilde{W}^-, \tilde{H}_d^-)$
is
\begin{equation}
{\bf M_{\tilde{C}}}=\left(
\begin{array}{cc}
{\bf 0}&{\bf X^T}\\ {\bf X}&{\bf 0}
\end{array}
\right); \ \ \ {\bf X}=\left(
\begin{array}{cc}
M_2&\sqrt{2}s_{\beta}M_W\\ \sqrt{2}c_{\beta}M_W&\mu
\end{array}\right).
\end{equation}
The mass eigenstates are called charginos $\chi_i^{\pm}$, $i=1,2$
($m_{\chi_1^\pm}<m_{\chi_2^\pm}$), which are related to the gauge
eigenstates by two unitary $2\times2$ matrices  $U$ and $V$ that
diagonalize the chargino mass matrix:
\begin{equation}
\left(\begin{array}{c} \chi_1^+\\ \chi_2^+ \end{array}\right) ={\bf
V}\left(\begin{array}{c} \tilde{W}^+\\ \tilde{H}_u^+
\end{array}\right);\ \ \  \left(\begin{array}{c} \chi_1^-\\ \chi_2^-
\end{array}\right) ={\bf U}\left(\begin{array}{c} \tilde{W}^-\\
\tilde{H}_d^- \end{array}\right);\ \ \  {\bf U^*XV^{-1}}=\left(
\begin{array}{cc}m_{\chi_1^{\pm}}&0\\0&m_{\chi_2^{\pm}}\end{array}\right)
\end{equation}
The ${\rm SU}(3)_C$ gluino is a color octet fermion, and  does not mix
with other particles in MSSM.  Its mass is parametrized by $M_3$ as
defined in Eq.~(\ref{eq:soft}).

In gravity-mediated and gauge-mediated SUSY breaking
models,  there is a unification relation for ${\rm SU}(3)_C$,  ${\rm
SU}(2)_L$ and ${\rm U}(1)_Y$ gaugino masses $M_{3,2,1}$: 
\beq
\frac{M_3}{\alpha_s}=\frac{M_2}{\alpha_2}=\frac{M_1}{\alpha_1}, 
\eeq
where $\alpha_s$, $\alpha_2$ and $\alpha_1$ are related to the
couplings of ${\rm SU}(3)_C$, ${\rm SU}(2)_L$ and ${\rm U}(1)_Y$ via
\beq \alpha_s=\frac{{g_s}^2}{4\pi},\ \ \
\alpha_2=\frac{\alpha}{\sin^2\theta_W}=\frac{{g_2}^2}{4\pi},\ \ \
\alpha_1=\frac{5}{3}\frac{\alpha}{\cos^2\theta_W}
=\frac{5}{3}\frac{{g_Y}^2}{4\pi}.  \eeq  
This relation holds at any
energy scale to one-loop order\footnote{We do not discuss possible
threshold effects at the GUT or Plank scale.} .  In particular, at
electroweak scale, if we take $\alpha_s=0.118$, $\alpha=1/128$ and
$\sin^2\theta_W=0.23$, the ratio betweem gaugino masses are \beq
M_3:M_2:M_1\approx 7:2:1.  \eeq However, such a unification relation
need not hold in the most general  MSSM, and, $M_3$, $M_2$ and $M_1$ could
be completely independent of  each other. 

The MSSM has two complex Higgs doublets,  $H_u$ and $H_d$,  which give
mass to up and down type fermions, respectively.   The potential for
the neutral Higgs fields is \beq
V=(|\mu|^2+m_{H_u}^2)|H_u^0|^2+(|\mu|^2+m_{H_d}^2)|H_d^0|^2
-(bH_u^0H_d^0+c.c)+\frac{1}{8}(g^2+g^{\prime
2})(|H_u^0|^2-|H_d^0|^2)^2.  \eeq As in the Standard Model, the
minimum of the potential corresponds to a non-zero VEV for the neutral
Higgs fields, thereby breaking  electroweak symmetry.  Let us write
$\langle H_u^0 \rangle = v_u/\sqrt{2}$,  $\langle H_d^0 \rangle =
v_d/\sqrt{2}$.  The sum of $v_u^2$ and $v_d^2$ is related to the $Z$
boson mass and the gauge couplings as   $v_u^2+v_d^2=v^2=4
M_Z^2/(g^2+g^{\prime 2})\approx(246\ {\rm GeV})^2$.   It is convenient
to write the ratio of $v_u$ and $v_d$  as $\tan\beta$:
$\tan\beta\equiv v_u/v_d$, where $v_u=v \sin\beta$ and
$v_d= v \cos\beta$.

The two complex Higgs doublets contain eight real scalar degrees of
freedom.  After EWSB, two charged and one neutral degree  of freedom
are the would-be Nambu-Goldstone bosons, $G^{\pm}$ and $G^0$, that are
eaten by $W^{\pm}$ and $Z$ to become their longitudinal modes.   We
are, thus,  left with five physical Higgs bosons: two neutral CP-even
Higgses, $h^0$ and $H^0$;  one  neutral CP-odd Higgs, $A^0$; and a
pair of charged Higgses, $H^{\pm}$.  When $m_{A^0}\gg M_W$, $h^0$ is 
the SM-like Higgs.  The tree level Higgs masses can
be obtained via  expanding the potential around the Higgs VEVs and
diagonalizing the $2\times 2$ mass matrices. One finds \beqa
m_{A^0}^2&=&2b/\sin 2 \beta,\\  m_{H^{\pm}}^2&=&m_{A^0}^2 + M_W^2,\\
m_{h^0,H^0}^2&=&\frac{1}{2}\left(
m_{A^0}^2+M_Z^2\mp\sqrt{(m_{A^0}^2+M_Z^2)^2 -4 M_Z^2 m_{A^0}^2\cos^2 2
\beta} \right).  \eeqa The mass eigenstates can be written in terms of
the gauge eigenstates as 
\beqa
A^0&=&\sqrt{2}(\cos\beta\ {\rm Im}[H_u^0]+\sin\beta\ {\rm
Im}[H_d^0]),\\ H^{+}&=&\cos\beta\ H_u^+ +\sin\beta\ H_d^{-*},\ \ \
H^{-}\ =\ \cos\beta\ H_u^{+*} +\sin\beta\ H_d^{-},\\
\left(\begin{array}{c} h^0\\ H^0
\end{array}
\right)&=& \sqrt{2}\left(\begin{array}{cc} \cos\alpha&-\sin\alpha\\
\sin\alpha&\cos\alpha
\end{array}
\right) \left(\begin{array}{c} {\rm Re}[H_u^0]-v_u\\ {\rm
Re}[H_d^0]-v_d\\
\end{array}
\right), 
\label{eq:alpha}
\eeqa 
where \beq \frac{\sin 2 \alpha}{\sin 2 \beta}=-
\frac{m_{A^0}^2+M_Z^2}{m_{H^0}^2-m_{h^0}^2};\ \ \  \frac{\cos 2
\alpha}{\cos 2 \beta}=- \frac{m_{A^0}^2-M_Z^2}{m_{H^0}^2-m_{h^0}^2}.
\eeq The tree level Higgs masses are determined by only two
parameters: $b$ and  $\tan\beta$ (or $m_{A^0}$ and $\tan\beta$). For
large $m_{A^0}$, $A^0$,  $H^0$ and $H^{\pm}$ are heavy and  decouple
from low-energy observables. On the  other hand, the 
tree level mass of the light
CP-even Higgs $h^0$ is bounded from  above: $m_{h^0} < M_Z$, which has
already been excluded by the current  LEP Higgs searches
\cite{LEPHiggs}.  However, the mass of the   $h^0$ receives large
radiative corrections from third generation quarks and their
superpartners due to the large top Yukawa couplings,
allowing for a mass large enough to be consistent with present direct
search limits. The  dominant contribution is from the stop loop: 
\beq
\Delta(m_{h^0}^2)=\frac{3}{4\pi^2}v^2 y_t^4 \sin^4\beta \ln  \left(
\frac{m_{\tilde{t}_1}m_{\tilde{t}_2}}{m_t^2}\right) \ \ \ .  
\eeq 
For
stop masses around 1 TeV, this correction can push the  $m_{h^0}$  above
the current  experimental bound.   Detailed two loop calculations for
the light CP-even Higgs mass indicate that the upper bound is about
135 GeV. Uncertainties of a few GeV arise from neglected  higher order
effects and the experimental error in the top quark mass \cite{mh2loop}.

The couplings of the Higgses to the (s)quarks and (s)leptons are
proportional to the Yukawa couplings and are,   therefore,  non-negligible
only for the third generation.   For the low energy precision
measurements where only the light quark and  leptons are involved, the
contribution from the Higgs sector can almost  always be
neglected. However, the details of the Higgs sector do affect the
phenomenology of SUSY CP-violation, so we will provide a brief summary
of the status of the experimental searches of MSSM
neutral Higgs bosons in Sec~\ref{sec:higgs}.

\subsection{SUSY Interactions}
The gauge interactions in MSSM can be obtained 
from the usual SM gauge interactions by  replacing two of
the SM particles with their superpartners.  For example, the coupling
of the SU(2)$_L$ gauge boson to quarks is (before EWSB and
diagonalization of the quark mass matrices) 
\be
\label{eq:gauge-quark}
-g{Q}^{\dagger}\bar\sigma^\mu 
\frac{\vec\tau}{2}\cdot \vec{W}_{\mu} Q 
\ee while
the corresponding gauge-squark-squark interaction is 
\be
i g \partial^{\mu} \tilde{Q}^\dagger \frac{\vec\tau}{2} \cdot \vec{W}_{\mu} \tilde{Q}
+c.c.
\label{eq:gauge-squark}
\ee 

\begin{figure}
\resizebox{5 in}{!}{
\includegraphics*[0,570][290,660]{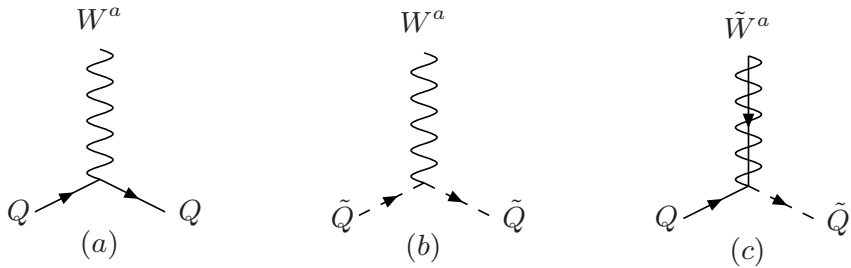}}
\caption{Feynman diagrams for supersymmetric (a) gauge-quark-quark, 
(b) gauge-squark-squark and (c) gaugino-quark-squark vertices.}
\label{fig:Feyn_gqq}
\end{figure}

Similarly, supersymmetry leads to a
squark-quark-SU(2)$_L$ gaugino interaction 
\be
-\sqrt{2}g (\tilde{Q}^\dagger \frac{\vec\tau}{2} Q) 
\cdot 
\vec{\tilde{W}} + c.c.
\label{eq:gauge-quark-squark}
\ee
The corresponding Feynman rules are illustrated in the diagrams of
Fig.~\ref{fig:Feyn_gqq}.  Additional 
gauge boson-gauge boson-squark-squark interactions
appear via the $D^{\mu}\tilde{Q}^*D_{\mu}\tilde{Q}$ term.

\begin{figure}
\resizebox{4 in}{!}{
\includegraphics*[0,570][200,650]{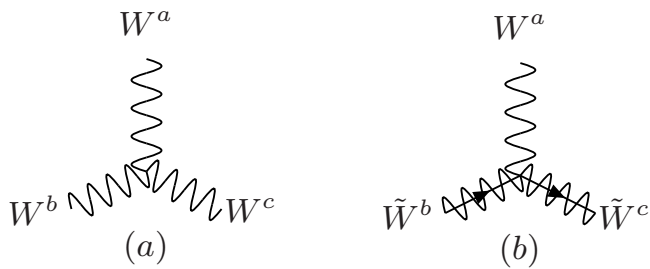}}
\caption{Feynman diagrams for supersymmetric (a) tri-gauge boson coupling, 
(b) gauge boson-gaugino-gaugino coupling.}
\label{fig:Feyn_ggg}
\end{figure}

The other fermion-fermion-gauge boson,
sfermion-sfermion-gauge boson, and fermion-sfermion-gaugino
interactions follow a similar pattern.  
Similarly, the supersymmetrized gauge boson self-interation leads to 
gauge boson-gaugino-gaugino coupling, as shown in Fig.~\ref{fig:Feyn_ggg}.

The Higgs Yukawa interactions are obtained from the superpotential
$W_{\rm MSSM}$ in Eq.~(\ref{eq:MSSMsuperpotential}).  It also give rise
to Higgsino-quark-squark interactions via the 
first term in Eq.~(\ref{eq:lag}) and Higgs-squark-squark interactions via
the second term in Eq.~(\ref{eq:lag}).  The soft SUSY breaking trilinear 
$A$-term give rise to additional Higgs-squark-squark couplings.  
The coupling between the Higgs and the lepton sector can be obtained similarly.

Including the effects of EWSB leads to
modifications of these expressions, due to (a) diagonalization of the
quark mass matrices, leading to the presence of the CKM matrix in
Eqs. (\ref{eq:gauge-quark},\ref{eq:gauge-quark-squark}); left-right
mixing (as well as possible flavor mixing) among sfermions, leading to
the presence of mixing matrices $Z_f^{Ij}$ in
Eqs.~(\ref{eq:gauge-squark},\ref{eq:gauge-quark-squark}); and mixing
of gauginos and Higgsinos into the charginos and neutralinos, leading
to factors of the matrices $N_{ij}$, $V_{ij}$, and $U_{ij}$ {\em etc.}
in Eq.~(\ref{eq:gauge-quark-squark}). The
Feynman rules for these interactions appear several places in the
literature  and we do not reproduce a complete list here. Throughout
this article, we generally follow the conventions given in
Refs.~\cite{Rosiek:1995kg,Rosiek:1989rs}.

\subsection{$R$-parity Violating Interactions}
\label{sec:rpv}

Additional $B$- and $L$- violating interactions may appear in the MSSM
if $R$-parity violation is allowed.  Rapid proton decay can still be
avoided if we only turn on $B$ or $L$ violating terms, but not both
simultaneously.   The RPV terms in the Lagrangian can be obtained from
the superpotentials Eqs.~(\ref{eq:RPVL}) and (\ref{eq:RPVB}) via
Eq.~(\ref{eq:lag}).  For low energy process where light quarks are
present  in the initial and final states, the RPV terms that  are of
interests are Yukawa-type interactions:
\begin{eqnarray}
\label{eq:rpv-super}
{\cal L}_{RPV, \ \Delta{L}=1}&\!\!\!= &\!\!\!\lambda_{ijk} (\frac{1}{2} L_i
L_j\tilde{\bar{e}}^\dagger_k  +\tilde{L}_i L_j\bar{e}^{\dagger}_k)
+\lambda_{ijk}^{\prime}( L_i Q_j\tilde{\bar{d}}^\dagger_k  +\tilde{L}_i
Q_j\bar{d}^{\dagger}_k +{L_i} \tilde{Q}_j\bar{d}^{\dagger}_k)+\mu_i^\prime L_i\tilde{H}_u; \\ 
{\cal L}_{RPV, \ \Delta{B}=1}&\!\!\!=&\!\!\!\lambda_{ijk}^{\prime\prime} (\bar{u}^{\dagger}_i
\bar{d}^{\dagger}_j \tilde{\bar{d}}^\dagger_k +\tilde{\bar{u}}^\dagger_i \bar{d}^{\dagger}_j
\bar{d}^{\dagger}_k).
\end{eqnarray}
Note that here we follow the notation of Ref.~\cite{Martin:1997ns} in which
fermion fields are denoted by two-compoments Weyl spinors.

\begin{figure}[ht]
\begin{center}
\resizebox{3 in}{!}
{\includegraphics*[0,640][220,740]{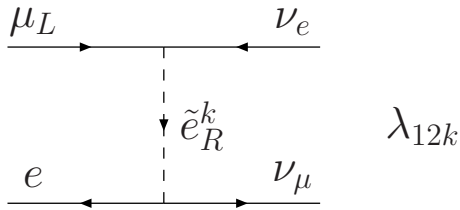}}
\caption{
Tree-level $P_R$-violating contributions to  muon decay.
}
\label{fig:muondecay}
\end{center}
\end{figure}

Such terms will contribute to the low energy SM process via the
exchange of  heavy scalar quarks or scalar leptons.  For example,
Fig.~\ref{fig:muondecay} shows the RPV contribution
from the purely leptonic  term  proportional to
$\lambda_{12k}$ to the muon
decay amplitude that determines the Fermi constant, $G_\mu$.
After a Fierz
reordering, the resulting four-fermion amplitude has the same
structure as the tree-level $(V-A)\times(V-A)$ amplitude of the SM,
but with a normalization determined by the ratio of
$|\lambda_{12k}|^2$ to the square of the exchanged slepton mass. More
generally,  for momentum transfer $q^2 \ll {\tilde m}^2$,  the
correction to low-energy SM amplitudes from RPV interactions can be
parametrized in terms of the following quantities:
\beq
\label{eq:deltas}
\Delta_{ijk}(\tilde f)={|\lambda_{ijk}|^2\over 4\sqrt{2}G_\mu
m_{\tilde f}^2}\ge 0 \eeq with a similar definition for the primed
quantities and double  primed quantities.

The quantities $\Delta_{ijk}$ {\it etc.} are constrained by other
precision measurements and rare decay searches.  A summary of the current experimental bounds
can be found in \cite{RPV}.  Here, we up-date our earlier global
analyses of  low-energy constraints on RPV obtained from the
low-energy observables in Table~\ref{tab:rpv-constrain}. 
For each observable, we indicate
the sensitivity to the various $\Delta_{ijk}^{(\prime)}(\tilde f)$ along with a
reference to the chapter in this article where a more detailed
discussion appears.

\begin{table}
\begin{tabular}{|c|ccccc|c|c|}
\hline Quantity & $\Delta_{11k}^{\prime}(\tilde{d}_R^k)$ &
$\Delta_{1k1}^{\prime}(\tilde{q}_L^k)$ & $\Delta_{12k}(\tilde{e}_R^k)$
& $\Delta_{21k}^{\prime}(\tilde{d}_R^k)$ 
& $\Delta_{2k1}^{\prime}(\tilde{d}_L^k)$
& Value 
& Discussion
\\ \hline\hline
$\delta |V_{ud}|^2/|V_{ud}|^2$ &2&0&-2&0&0&$-0.0032\pm 0.0014$(a)
&Sec.~\ref{sec:cc}
\\
 &&&&&&$-0.0002\pm 0.0015$(b)
&Sec.~\ref{sec:cc}
\\
$\delta Q_W^{\rm Cs}/Q_W^{\rm Cs}$ &-4.82&5.41&0.05&0&0&$-0.0040\pm
0.0066$
&Sec.~\ref{sec:nc}
\\ 
$\delta R_{e/\mu}$ &2&0&0&-2&0&$-0.0042 \pm 0.0033$ 
&Sec.~\ref{sec:cc}
\\
$\delta G_\mu/G_\mu$ &0&0&1&0&0&$0.00025\pm 0.001875$ 
&Sec.~\ref{sec:nc}
\\ 
$\delta Q_W^e/Q_W^e$
&0&0&-29.8&0&0&$0.14\pm0.11$
&Sec.~\ref{sec:nc}
\\ 
$\delta R_{\nu}$
&0&0
&-0.21& 0.22& 0.08&$-0.0033\pm 0.0007$
&Sec.~\ref{sec:nc}
\\
$\delta R_{\bar\nu}$
&0&0&-0.077& 0.132 &0.32
&$-0.0019\pm 0.0016$
&Sec.~\ref{sec:nc}
\\
\hline
\end{tabular}
\caption{$P_R$-violating contributions to $\delta
|V_{ud}|^2/|V_{ud}|^2$, $\delta Q_W^{\rm Cs}/Q_W^{\rm Cs}$, $\delta
R_{e/\mu}$, $\delta G_\mu/G_\mu$, $\delta Q_W^p/Q_W^p$, $\delta
Q_W^e/Q_W^e$, $\rnu$ and $\rnubar$.
Here $\delta |V_{ud}|^2/|V_{ud}|^2$ means the possible correction to the value of       
$|V_{ud}|^2$ extracted from beta-decay that is allowed by first row CKM unitarity     
tests.  See text for description of scenario (a) and (b) for  $\delta |V_{ud}|^2/|V_{ud}|^2$.
Columns give the coefficients of the various
corrections from $\Delta_{ijk}^{\prime}$ and $\Delta_{12k}$ to the
different quantities.  Next to last column gives the 
 value of the corresponding quantity extracted from experiment assuming only Standard Model contributions to the relevant process. Final column gives section of this review containing relevant discussion..} 
\label{tab:rpv-constrain}
\end{table}

The results of our fit are particularly sensitive to tests of the
unitarity of the first row of the CKM matrix, discussed in Section
\ref{sec:cc}. As we discuss in there, the status of
first row CKM unitarity is presently unsettled, so we provide a fit
for two scenarios: (a) using a value of the kaon decay form factor
$f_+(0)$ obtained from large $N_C$ QCD studies, leading to a deviation
from CKM unitarity by roughly two standard deviations; (b)  a value
for $f_+(0)$ obtained from recent lattice QCD simulations that implies
agreement with unitarity. The resulting 95\% C.L. ranges for the
$\Delta_{ijk}$ and $\Delta_{ijk}^\prime$ under these two scenarios are given in 
Table~\ref{tab:rpvrange}.

\begin{table}
\begin{tabular}{|c|c|c|}\hline 
&(a) large $N_c$ QCD&(b)lattice QCD \\ \hline 
$\Delta_{11k}^{\prime}(\tilde{d}_R^k)$ & $0-0.0020$
& $0-0.0024$\\
$\Delta_{1k1}^{\prime}(\tilde{q}_L^k)$ &$0-0.0017$ &$0-0.0019$\\
$\Delta_{12k}(\tilde{e}_R^k)$ & $0.0013-0.0039$&$0.0006-0.0031$\\
$\Delta_{21k}^{\prime}(\tilde{d}_R^k)$ &$0-0.0015$&$0-0.0014$\\
$\Delta_{2k1}^{\prime}(\tilde{d}_L^k)$ &$0-0.0013$&$0-0.0009$\\ \hline
\end{tabular}
\caption{95\% C.L. ranges for  the $\Delta_{ijk}^{(\prime)}$ 
obtained from fitting
to the low energy observables listed in Table~\ref{tab:rpv-constrain}.}
\label{tab:rpvrange}
\end{table}

\subsection{SUSY Searches}
Both direct and indirect searches for superparticles have been
carried out in various experiments \cite{pdg}.   Sparticles can be
pair produced at $e^+e^-$, $p\bar{p}$ and $pp$ colliders via the
intermediate $\gamma^*$, $Z^*$, gluon or sparticles.   Each sparticle
subsequently decays into energetic lepton, jets plus an LSP for
$R$-parity conserving scenarios.  In most cases, the LSP is a neutral
weakly interacting particle ({\em e.g.}, a neutralino) which travels
through the detector without  depositing significant energy.
Therefore, typical signatures consist of  some combination of jets,
leptons, possibly photons, and  large missing energy.

The large electron-position (LEP) collider  at CERN has completed its
running at center of mass energy more than twice the  mass of $Z$
boson with a few hundred ${\rm pb}^{-1}$ integrated luminosity.   No
events inconsistent with the SM have been reported and all visible
sparticles with mass up to  half the $Z$ mass have been excluded.
Data taken at center of mass energy  above $M_Z$ can set stronger
limits on the  neutralino, chargino, squark and slepton masses,
although limits would  depend on the interplay between the sparticle
masses, cross sections  and decay branching ratios.  Charginos are
excluded up to 103 GeV except  in the case of low  acceptance or low
cross section.  The limits on the  slepton masses are based on the
pair production of $\tilde{l}\tilde{l}$ with $\tilde{l}\rightarrow l
\chi_1^0$, which gives a lower  mass bound of about 80$-$100 GeV.
Limits on the stop and sbottom masses  are about 90 GeV, which varies
with the left-right squark mixing angle.

The Tevatron Run I has accumulated about $110\ {\rm pb}^{-1}$ of  data for $p\bar{p}$ collisions 
at $\sqrt{s}=1.8$  TeV.  Pairs of squarks
and gluinos  can be produced because of the large strong interaction
cross sections.  Signals of energetic multijets plus missing
transverse energy have been  searched for,  with null results being
reported.  The exclusion region in  $(m_{\tilde{q}}, m_{\tilde{g}})$
has been derived,  and mass larger than 300 GeV  has been excluded if
$m_{\tilde{g}}=m_{\tilde{q}}$.  The gluino mass has been excluded up
to 195 GeV for any squark mass.  Charginos and neutralinos can be
produced via $q\bar{q}^{\prime}\rightarrow \chi_1^{\pm}\chi_2^0$.
Leptonic decay of both $\chi_1^\pm$ and $\chi_2^0$ leads to trilepton
signals which reduces the background significantly.  The same sign
dilepton signal is possible for charginos produced  in the squark and
gluino cascade decay. Bounds on charginos and neutralinos, however,
are not as strong as the LEP results.  The upgraded Tevatron Run II
with $\sqrt{s}=2$ TeV and designed integrated luminosity $L=2\ {\rm
fb}^{-1}$ could cover large regions of SUSY parameter space. The
Large Hadron Collider (LHC) at CERN of $pp$ collision with
$\sqrt{s}=14$ TeV and  $L=100\ {\rm fb}^{-1}$ would cover the region
of squarks and gluinos  with mass up to a few TeV.

Precision measurements of $Z$-pole observables, $M_W$, $m_t$ and several low-energy
observables could constrain the MSSM mass spectrum. The bounds from the current precision $Z$-pole measurements on the SUSY parameters
are discussed in Sec.~\ref{sec:zpole}. In addition, FCNC and CP
violation experiments impose strong limits on the possible
flavor structure and CP-violating phases of MSSM, which heavily
constrain the  structure of soft-SUSY breaking parameters (see Section \ref{sec:cpv}).  In the $R$-parity conserving MSSM, sparticles could affect the low
energy  measurements {\em via} radiative corrections. In general, loop-induced SUSY
effects in low-energy observables are proportional to $\alpha/{\pi} (M/\msusy)^2$, where $M$ is the relevant SM particle mass. For $\msusy\sim M_W$ and $M\sim M_W$, probing these effects, therefore, requires precision of better than one percent. The analsyis and implications of these precision measurements constitutes the bulk of the remainder of this review.

\section{Renormalization}
\label{sec:renorm}

In this review, we focus on possible manifestations of SUSY in low energy processes that may complement the information obtained from high energy collider searches. The low energy processes of interest here further break down into two broad classes: (a) those allowed by the Standard Model and for which there exist precise SM predictions, and (b) processes that are either forbidden by SM symmetries or that are highly suppressed. In order to discern the effects of SUSY in the first category of observables, one generally requires knowledge of SM predictions at the level of one-loop (or higher) radiative corrections. In the case of SUSY models that conserve R-parity, SUSY contributions to precision electroweak observables will also appear solely via loop effects. 

In order to make meaningful comparisons between experimental results and predictions based on supersymmetric extensions of the SM, one must compute and renormalize superpartner loop effects using the same framework as adopted for the SM predictions. Doing so requires care, however, as one must employ a regulator that preserves supersymmetry. Here, we use dimensional reduction (DR), wherein one works in $d=4-2\varepsilon$ spacetime dimensions while retaining the Clifford algebra appropriate to fermion field operators in $d=4$ dimensions. Renormalized quantities are obtained by introducing counterterms  that remove the factors of $1/\varepsilon-\gamma+\ln 4\pi$ that arise in divergent one-loop graphs -- a subtraction scheme known as modified dimensional reduction, or ${\overline {\rm DR}}$. This scheme represents a variation on the more familiar modified minimal subtraction (${\overline {\rm MS}}$) commonly used for the SM. In ${\overline {\rm MS}}$ renormalization, one regularizes divergent graphs using dimensional regularization, which differs form dimensional reduction by continuing both the number of spacetime dimensions as well as the Clifford algebra into $d=4-2\varepsilon$ dimensions. Doing so entails an explicit breaking of supersymmetry, however, as dimensional regularization effectively changes the number of fermionic degrees of freedom relative to those of their bosonic superpartners
in $d\neq 4$. In contrast, ${\overline {\rm DR}}$ retains the correspondence between the number of bosonic and fermionic degrees of freedom.

Most low energy precision electroweak observables are mediated at lowest order by the exchange of a virtual gauge boson (GB), so we consider first the renormalization of GB propagators and GB-fermion interactions. Low energy processes generally involve the lightest quarks and leptons, and since the corresponding Higgs-fermion interactions are suppressed by small Yukawa couplings, we will not discuss renormalization of the Higgs sector in detail here\footnote{A brief discussion 
of MSSM neutral Higgs searches is given in Section \ref{sec:higgs}; and more extensive review can be found in Ref.~\cite{Heinemeyer:2004gx}.}. To set our notation, we first discuss renormalization relevant to charged current (CC) processes and subsequently discuss the neutral current (NC) sector.

\subsection{Charged Current Processes}

\begin{figure}[ht]
\begin{center}
\resizebox{6 in}{!}{
\includegraphics*[20,160][580,620]{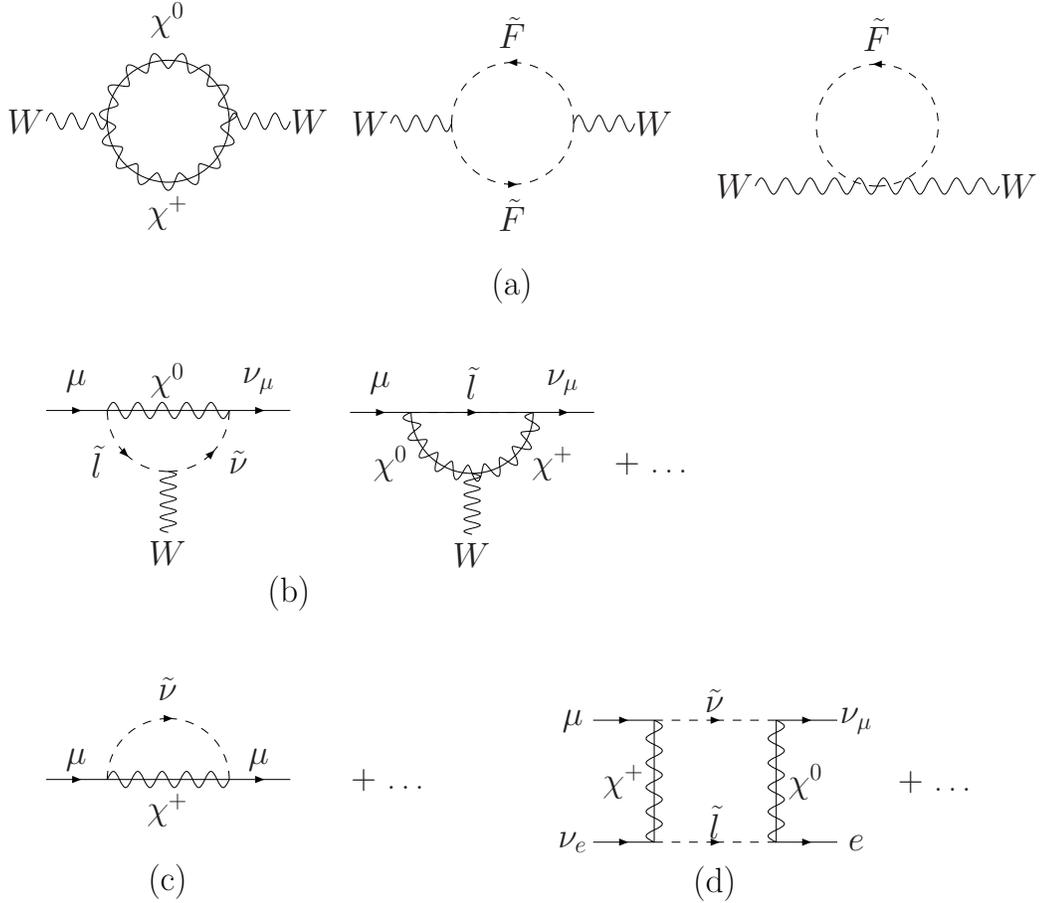}}
\caption{Representative supersymmetric corrections to charged current observables: (a) $W$-boson propagator corrections; (b) vertex corrections; (c) external leg corrections; and (d) box graph contributions.}
\label{fig:cccorr}
\end{center}
\end{figure}

Radiative corrections to CC amplitudes naturally divide into four topologies: (a) $W$-boson propagator corrections; (b) corrections to the $W$-fermion vertices; (c) fermion propagator corrections; and (d) box graphs. These corrections have been well-studied in the SM and we refer the reader to the extensive literature on the subject for further details (see, {\em e.g.}, Refs.~\cite{Sirlin:1977sv,Marciano:1985pd,Marciano:1993sh,Marciano:2005ec}). As our emphasis in this review lies with the effects of supersymmetric particles, we show in Fig. \ref{fig:cccorr} illustrative superpartner contributions to each type of radiative correction. 

One loop corrections to the $W$-boson propagator, fermion propator, and $W$-fermion vertices are divergent, so one must carry out the appropriate renormalizaton. After such renormalization, the W-boson propagator, $iD_{\mu\nu}(k)$ takes the general form in the Feynman gauge
\be
iD_{\mu\nu}(k) = -i\left[T_{\mu\nu}{\hat D}_{WW}^T(k^2)+L_{\mu\nu}{\hat D}_{WW}^L(k^2)\right]
\ee
where the transverse and longitudinal projection operators are given by
\bea
T_{\mu\nu} & = & -g_{\mu\nu}+ k_\mu k_\nu/k^2 \\
L_{\mu\nu} & = & k_\mu k_\nu/k^2
\eea
and ${\hat D}_{WW}^{T,L}(k^2)$ are finite scalar functions and the hat  indicates quantities renormalized in the ${\overline {\rm DR}}$ scheme. In low-energy processes, effects associated with the longitudinal term are suppressed by light fermion masses, so we will not discuss the component further. The renormalized transverse component is given by
\be
\left[ {\hat D}_{WW}^T(k^2)\right]^{-1} = k^2-{\hat M}_W^2 +{\hat\Pi}_{WW}^T(k^2)\ \ \ ,
\ee
where ${\hat M_W}$ is the finite part of the bare W-boson mass parameter appearing in the renormalized Lagrangian after electroweak symmetry breaking and ${\hat\Pi}_{WW}^T(k^2)$ gives the finite loop contribution after ${\overline {\rm DR}}$ subtraction is performed. Both ${\hat M_W}$ and ${\hat\Pi}_{WW}^T(k^2)$ depend on the t'Hooft renormalization scale $\mu$. However, the physical W-boson mass -- defined by the value of $k^2$ giving $[{\hat D}_{WW}^T(k^2
=M_W^2)]^{-1}=0$ -- is $\mu$-independent. The finite residue ${\hat Z}_W$ of the pole in ${\hat D}_{WW}^T$ is given by 
\be
{\hat Z}_W = \left[1+{\hat\Pi}_{WW}^{T\ \prime}(M_W^2)\right]^{-1}\ \ \ .
\ee

The corresponding expression for the renormalized, inverse fermion propagator is 
\be
{\hat S}_f^{-1}(k) = \dslash{k}-{\hat m}_f +\left[ {\hat A}_L(k^2)\dslash{k} +{\hat B}_L(k^2)\right] P_L 
+ \left[ {\hat A}_R(k^2)\dslash{k} +{\hat B}_R(k^2)\right]P_R
\ee
where $P_{L,R}$ are the left- and right-handed projectors and the ${\hat A}_{L,R}$ and ${\hat B}_{L,R}$ contain the finite loop contributions.  The physical fermion mass is given by 
\be
m_f=\left[ {\hat m}_f -\frac{1}{2}{\hat B}_L(m_f^2)
-\frac{1}{2}{\hat B}_R(m_f^2)\right]\,  \left[1+\frac{1}{2}{\hat A}_L(m_f^2)  + \frac{1}{2}{\hat A}_R(m_f^2) \right]^{-1}\ \ \ ,
\ee
while the residue of the pole is
\be
{\hat Z}_\psi = \left[1+{\hat A}_L(m_f^2) P_L + {\hat A}_R(m_f^2) P_R\right]^{-1} \ \ \ .
\ee
Note that for CC interactions in the SM,  the left-handed (LH) components given the dominant contribution to physical amplitudes, as the presence of right-handed (RH) components will be suppressed by factors of the fermion masses\footnote{For example, the weak magnetic moment operator in the SM is chirality-odd and  is generated by one-loop vertex corrections that contain single insertions of the Yukawa interaction.}.  

The renormalized vertex functions for CC amplitudes are relatively straightforward. We illustrate using the muon decay process  $\mu^-\to \nu_\mu W^-$, for which the tree-level amplitude is 
\be
i{\cal M}_{0}^{\rm CC} = i \frac{g}{\sqrt{2}}{\bar \nu }_\mu \diracslash{W}^{\ +} P_L \mu
\ee
After one-loop renormalization, one has
\be
i{\cal M}_{0}^{\rm CC}+i{\cal M}_{\rm vertex}^{\rm CC} =i \frac{{\hat g}(\mu)}{\sqrt{2}}\left[1+{\hat F}_V(k^2)\right] {\bar \nu }_\mu \diracslash{W}^{\ +} P_L \mu
\ee
where ${\hat g}(\mu)$ is the running SU(2)$_L$ gauge coupling and ${\hat F}_V(k^2)$ is the finite part of the one-loop vertex correction. 

The vertex and propagator corrections outlined above will contribute to the four-fermion amplitudes that describe low energy CC processes of interest to us, such as $\mu$- and $\beta$-decay. Additional, but finite, SM one-loop contributions are generated by box graphs involving the exchange of two vector bosons. To the extent that the external masses and  momenta are small compared to the weak scale, the box contributions will have the form of a product of two left-handed currents, $(V-A)\otimes(V-A)$.  In the case of $\mu^-\to \nu_\mu e^-{\bar\nu}_e$ 
\be
i{\cal M}_{\rm box}^{\rm CC} = -i \frac{{\hat g}^2}{2{\hat M}_W^2}{\hat \delta}_{\rm box} \ {\bar \nu}_\mu  \gamma^\lambda P_L \mu \ {\bar {e } }\gamma_\lambda P_L {\nu_{\bar e}} + \cdots\ \ \ \ ,
\ee
where the $+\cdots$ indicate terms whose structure differs from the $(V-A)\otimes(V-A)$  structure of the tree-level CC amplitude\footnote{Here, $\nu_{\bar e}$ is the v-spinor for the electron antineutrino.}. In the SM, such terms will be suppressed by factors of $m_\mu^2/M_W^2$. Superpartner loops can lead to relatively unsuppressed non-$(V-A)\otimes(V-A)$ contributions in the presence of mixing between left- and right-handed sfermions (see Section \ref{sec:cc}).

Including the box contribution along with the other renormalized one-loop contributions, taking into account the factors of $1/\hat Z_\psi^{1/2}$ that arise in the standard reduction formulae, and working in the $k^2 << M_W^2$ limit, one has
\bea
i{\cal M}_{\rm tree}^{\rm CC}&+&i{\cal M}_{\rm vertex}^{\rm CC}+i{\cal M}_{\rm propagator}^{\rm CC}+i{\cal M}_{\rm box}^{\rm CC} = 
-i\frac{{\hat g}^2}{2{\hat M}_W^2}\Bigl[1+\frac{{\hat\Pi}_{WW}^T(0)}{{\hat M}_W^2}  \\
\nonumber
& -& \frac{1}{2}\left\{
{\hat A}_L^\mu(m_\mu^2)+{\hat A}_L^e(m_e^2)+{\hat A}_L^{\nu_e}(0)+{\hat A}_L^{\nu_\mu}(0)\right\}
+{\hat F}_V^e(0)+{\hat F}_V^\mu(0)+{\hat \delta}_{\rm box}\Bigr] \\
\nonumber
&&\times \ {\bar \nu }_\mu \gamma^\lambda P_L \mu \ {\bar {e } }\gamma_\lambda P_L {\nu_{\bar e}} +\cdots \ \ \ ,
\eea
or
\be
i{\cal M}_{\rm one-loop}^{\rm CC}=-i\frac{{\hat g}^2}{2{\hat M}_W^2}\left[1+\frac{{\hat\Pi}_{WW}^T(0)}{{\hat M}_W^2}+{\hat\delta}_{VB}\right]{\bar \nu }_\mu \gamma^\lambda P_L \mu \ {\bar {e } }\gamma_\lambda P_L {\nu_{\bar e}} +\cdots \ \ \ ,
\ee
where ${\hat\delta}_{VB}$ denotes the fermion propagator, vertex, and box graph contributions. 

The resulting rate for muon decay -- including the bremstraahlung contribution -- is then given by
\bea
\label{eq:taumu}
\frac{1}{\tau_\mu} & = & \frac{m_\mu^5}{96\pi^3}\left(\frac{{\hat g}^2}{8{\hat M}_W^2}\right)^2  \left[1+\frac{{\hat\Pi}_{WW}^T(0)}{{\hat M}_W^2}+{\hat\delta}_{VB}\right]^2 \  + \ {\rm brem} \\
\nonumber
& = & \frac{m_\mu^5}{192\pi^3} G_\mu^2\left[1+\delta_{\rm QED}\right] \ \ \ ,
\eea
where $\tau_\mu$ is the muon lifetime and the second equality defines the $\mu$-decay Fermi constant, $G_\mu$, and where 
\be
\delta_{\rm QED} = \frac{\alpha}{2\pi} \left( \frac{25}{4}-\pi^2 \right)+\cdots 
\ee
denotes the contributions from real and virtual photons computed in the Fermi theory of the decay.
Thus, one has
\be
\frac{G_\mu}{\sqrt{2} }= \frac{{\hat g}^2}{8{\hat M}_W^2}\left[1+\frac{{\hat\Pi}_{WW}^T(0)}{{\hat M}_W^2}+{\hat\delta}_{VB}^{(\mu)}\right] \equiv \frac{{\hat g}^2}{8{\hat M}_W^2}\left(1+{\Delta \hat r}_\mu\right) \ \ \ ,
\ee
where ${\hat\delta}_{VB}^{(\mu)}$ is given by ${\hat\delta}_{VB}$ but with the Fermi theory QED contributions subtracted out. 

Along with the fine structure constant and the Z-boson mass, $M_Z$, the value of $G_\mu$  is one of the three most precisely determined parameters in the gauge sector of the SM. Thus, for purposes of computing other electroweak observables, it is conventional to express ${\hat g}^2$ in terms of $G_\mu$, ${\hat M}_W$, and the correction $\Delta \hat r_\mu$:
\be
\label{eq:ghat}
{\hat g}^2 = \frac{8{\hat M}_W^2 G_\mu}{\sqrt{2}}\ \frac{1}{1+{\Delta \hat r}_\mu}\ \ \ .
\ee
As a particular application, we consider the corresponding amplitude for the $\beta$-decay $d\to u e^-{\bar\nu}_e$:
\bea
\label{eq:betaampl}
i{\cal M}_{\beta-{\rm decay} } &=& i\frac{{\hat g}^2}{2{\hat M}_W^2}\, V_{ud}\, \left(1+{ \Delta \hat r}_\beta\right) \, {\bar u}
\gamma^\lambda P_L \, d\  {\bar e }\gamma_\lambda P_L {\nu_{\bar e}} \\
\nonumber
&=& i \frac{G_\mu}{\sqrt{2}}\, V_{ud}\, \left(1+{ \Delta \hat r}_\beta-{ \Delta \hat r}_\mu\right)
{\bar u} \gamma^\lambda (1-\gamma_5)\, d\  {\bar e} \gamma_\lambda (1-\gamma_5)  {\nu_{\bar{e}}}\ \ \ 
\eea
where we have omitted terms that are suppressed by quark masses and where ${ \Delta \hat r}_\beta$ includes virtual photon contributions, in contrast to ${\Delta \hat r}_\mu$.

\subsection{Neutral Current Processes}

\begin{figure}[ht]
\begin{center}
\resizebox{6 in}{!}{
\includegraphics*[20,160][620,620]{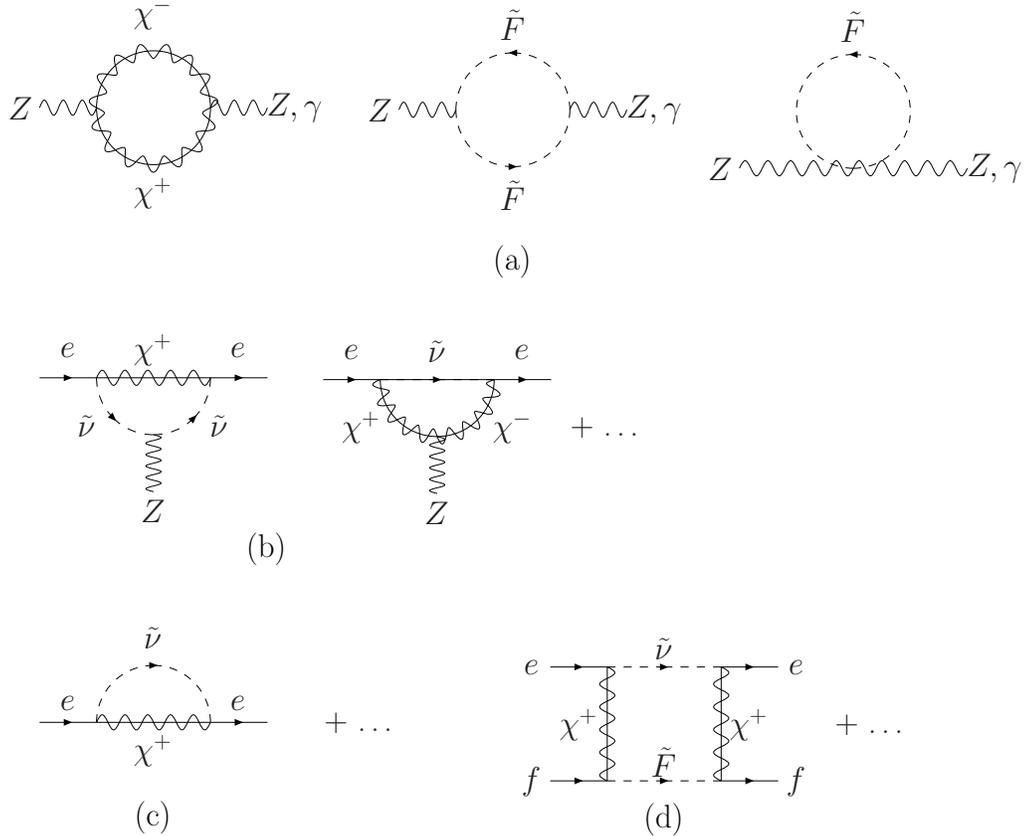}}
\caption{Representative supersymmetric corrections to neutral current observables: (a) $Z$-boson propagator and $Z$-$\gamma$ mixing conributions; (b) vertex corrections; (c) external leg corrections; and (d) box graph contributions}
\label{fig:nccorr}
\end{center}
\end{figure}

The renormalization of neutral current (NC) amplitudes follows similar lines, though additional features arise due to mixing between the SU(2)$_L$ and U(1)$_Y$ sectors (for early computations in the SM, see, {\em e.g.}, Refs.~\cite{Marciano:1980pb,Marciano:1982mm,Sarantakos:1982bp,Marciano:1983ss}). The general structure of the renormalized amplitude for the neutral current process $\ell+f\to\ell +f$ is
\be
i{\cal M}_{\rm one - loop}^{\rm NC} = -i\frac{G_\mu}{2\sqrt{2}} {\hat \rho}_{\rm NC}(k^2) {\bar\ell}\,  \gamma^\lambda({\hat g}_V^\ell+{\hat g}_A^\ell \gamma_5)\, \ell \ {\bar f}\, \gamma_\lambda({\hat g}_V^f+{\hat g}_A^f \gamma_5)\,  f + {\rm box} \ \ \ ,
\ee
where $\ell$ and $f$ denote the lepton and fermion spinors, respectively, and 
``+box" denotes the box diagram contributions.  The quantity  $\hat\rho_{\rm NC}$ is a normalization factor common to all four-fermion NC processes that can be expressed in terms of gauge boson masses, the ${\hat \Pi}^T_{VV}(k^2)$, and ${\Delta \hat{r}}_\mu$:
\bea
\label{eq:rhonc1}
\hat\rho(k^2)_{\rm NC} & = & \frac{M_Z^2}{k^2-M_Z^2+i M_Z\Gamma_Z}\Bigl\{1+
\frac{{\rm Re}\ {\hat\Pi}_{ZZ}^T(M_Z^2)}{M_Z^2}-\frac{{\hat\Pi}_{WW}^T(0)}{M_W^2}\\
\nonumber
&&-\frac{\left[{\hat\Pi}_{ZZ}^T(k^2)-{\hat\Pi}_{ZZ}^T(M_Z^2)\right]}{k^2-M_Z^2} -{\hat\delta}_{VB}^{(\mu)}\Bigr\}\ \ \ ,
\eea
where
\be
\label{eq:mzhat}
M_Z^2={\hat M}_Z^2-{\hat\Pi}_{ZZ}^T(M_Z^2)
\ee
and $M_Z\Gamma_Z={\rm Im}\, {\hat\Pi}_{ZZ}^T(k^2)$. For $k^2<M_Z^2$, $\Gamma_Z=0$. Representative superpartner contributions to ${\hat\Pi}_{ZZ}$ are shown in Fig. \ref{fig:nccorr}.

The renormalized vector and axial vector couplings of the $Z^0$ to fermion $f$ -- ${\hat g}_V^f$ and ${\hat g}_A^f$ -- can be expressed in terms of the weak mixing angle, $\theta_W$, and an associated universal renormalization factor, along with process-specific vector and axial vector radiative corrections: 
\bea
{\hat g}_V^f & = & 2 I_3^f- 4 {\hat\kappa}(k^2,\mu) \sin^2{\hat\theta}_W(\mu)Q_f + {\hat\lambda}_V^f\\
{\hat g}_A^f & = & -2 I_3^f +{\hat\lambda}_A^f \ \ \ .
\eea
where $I_3^f$ and $Q_f$ are the fermion isospin and charge, respectively; $\sin^2{\hat\theta}_W(\mu)\equiv{\hat s}^2(\mu)$ defines the weak mixing angle in the ${\overline {\rm MS}}$ scheme:
\be
\sin^2{\hat\theta}_W(\mu) = \frac{{\hat g}^\prime(\mu)^2}{{\hat g}(\mu)^2 +{\hat g}^\prime(\mu)^2} \ \ \ ;
\ee
and ${\hat\lambda}_{V,A}^f$  are process-dependent corrections that vanish at tree-level.
Here $\hat{g}$ and $\hat{g}^\prime$ are the ${\rm SU}(2)_L$ and ${\rm U}(1)_Y$ coupling, respectively.

As with the running QED and QCD couplings, ${\hat\alpha}(\mu)$ and ${\hat\alpha}_s(\mu)$, respectively, the running of the weak mixing angle is a prediction of the SM and provides a useful benchmark for precision studies in the NC sector\cite{Czarnecki:1995fw} . A renormalization group-improved SM prediction for  ${\hat s}^2(\mu)$ in the ${\overline {\rm MS}}$ scheme has recently been carried out in Ref.~\cite{Erler:2004in}, where logarithmic contributions of the form $\alpha^n \ln^n(\mu/\mu_0)$ (with $\mu_0$ being a reference scale) have been summed to all orders. Additional subleading contributions of the form $\alpha^{n+1}\ln^n(\mu/\mu_0)$ and $\alpha\alpha_s^{n+k}\ln^n(\mu/\mu_0)$ with $k=0,1,2$ were also included in that analysis, and a refined estimate of the hadronic physics uncertainty associated with light-quark loops at low scales performed (see below). The results are shown in Fig.~\ref{fig:sin2theta}, where the scale $\mu$ has been chosen to be $Q=\sqrt{|k^2|}$ for a process occurring at squared momentum transfer $k^2$. The reference scale has been chose to be $\mu_0=M_Z$ and the running of ${\hat s}^2(Q)$ normalized to reproduce its value at the $Z^0$-pole : 
$\sin^2{\hat\theta}_W(M_Z)=0.23122(15)$. The discontinuities in the curve of Fig.~\ref{fig:sin2theta} correspond to particle thresholds, below which a particle of the corresponding mass decouples from the running. The change in sign of the slope at $Q=M_W$ arises from the difference in sign of the gauge boson and fermion contributions to the $\beta$ function for ${\hat s}^2(\mu)$. Note that threshold matching conditions in the ${\overline {\rm DR}}$-scheme will differ from those in the ${\overline {\rm MS}}$ framework due to the differences in continuation of the Clifford algebra into $d=4-2\varepsilon$ dimensions\cite{Antoniadis:1982vr, Langacker:1992rq}. 

For purposes of this review, it is particularly interesting to quote the value of the the running weak mixing angle at $Q=0$\cite{Erler:2004in}:
\be
\sin^2{\hat\theta}_W(0)= 0.23867\pm 0.00016
\ee
where the error is dominated by the experimental error in $\sin^2{\hat\theta}_W(M_Z)$ and where the value of ${\hat s}^2(\mu)$ at the two scales differs by roughly three percent. In Section \ref{sec:nc} we describe a variety of low-energy NC experiments designed to test this running.

\begin{figure}
\begin{center}
\includegraphics[width=4in, angle=-90]{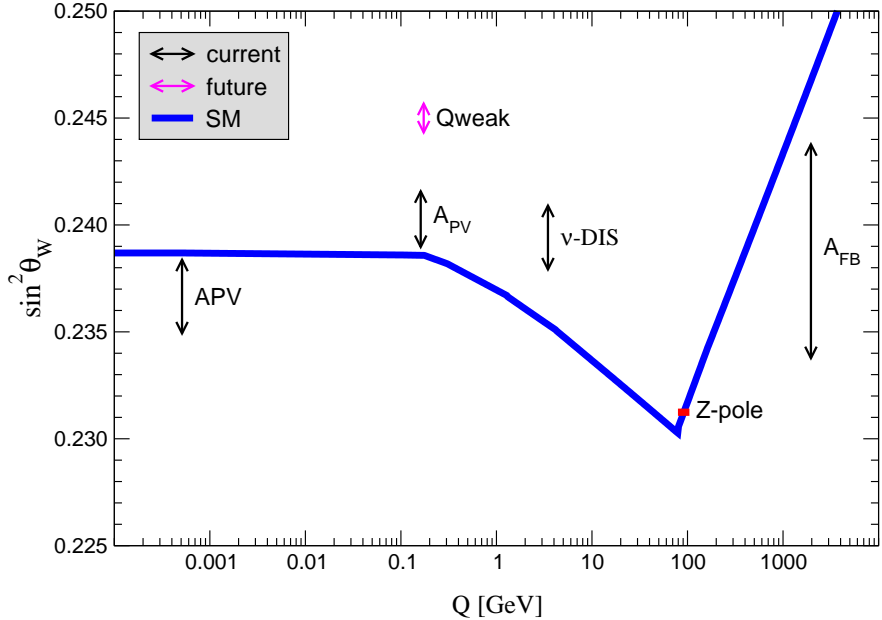}
\caption{Calculated running of the weak mixing angle in the SM, defined in the
$\overline{\rm MS}$ renormalization scheme.
Also shown are the experimental results from APV,
neutrino DIS ($\nu$-DIS),   
parity violating asymmetry measurement at E158 ($A_{PV}$), 
the expected precision of Qweak 
and the lepton forward-backward asymmetry measurement
at CDF ($A_{FB}$).  This plot
is taken from Ref.\cite{sin2theta}.}
\label{fig:sin2theta}
\end{center}
\end{figure}


By itself, ${\hat s}^2(\mu)$ is not an observable since it depends on the renormalization scale. One may, however, define an effective weak mixing angle that is $\mu$-independent and may in principle be isolated experimentally by comparing experiments with different species of fermions:
\be
\label{eq:sweff}
\sin^2{\hat\theta}_W(k^2)^{\rm eff} \equiv {\hat\kappa}(k^2,\mu) \sin^2{\hat\theta}_W(\mu)
\ee
where the quantity ${\hat\kappa}(k^2,\mu)$ describes a class of electroweak radiative corrections that is independent of the species of fermion involved in the NC interaction. Contributions to ${\hat\kappa}(k^2,\mu)$  arise primarily from the $Z$-$\gamma$ mixing tensor:
\be
{\hat\Pi}^{\mu\nu}_{Z\gamma}(k^2) = {\hat\Pi}^T_{Z\gamma}(k^2) T^{\mu\nu} +{\hat\Pi}^{L}_{Z\gamma}(k^2) L^{\mu\nu} \ \ \ .
\ee
Note that in general, the functions ${\hat\Pi}_{Z\gamma}^T(k^2)$ depend on the choice of electroweak gauge parameter, so to arrive at a gauge-independent ${\hat\kappa}(k^2,\mu)$, a prescription for removing the gauge-dependent components of ${\hat\Pi}_{Z\gamma}^T(k^2)$ must be employed\cite{Ferroglia:2003wa}. 

For processes involving $|k^2| \ll M_Z^2$, contributions from light fermions to ${\hat\kappa}(k^2,\mu)$ can lead to the presence of large logarithms when one chooses $\mu=M_Z$. The presence of these logarithms can spoil the expected behavior of the perturbation series unless they are summed to all orders. To illustrate, consider the amplitude for low-energy, parity-violating M\o ller scattering:
\be
\label{eq:moller1}
{\cal M}_{PV}^{ee} = \frac{G_\mu}{2\sqrt{2}} {\hat \rho}_{\rm NC}(0) {\hat g}_V^e{\hat g_A}^e\ {\bar e}\gamma_\mu e\ {\bar e} \gamma^\mu\gamma_5 e \ \ \ 
\ee
with
\be
\label{eq:moller2}
Q_W^e \equiv {\hat \rho}_{\rm NC}(0)\, {\hat g}_V^e {\hat g}_A^e = {\hat \rho}_{\rm NC}(0)\left[-1+4{\hat\kappa}(0,\mu){\hat s}^2(\mu)+{\hat\lambda}_V^f+ {\hat\lambda}_A^f(-1+4{\hat s}^2)\right]+\cdots
\ee
being the \lq\lq weak charge" of the electron and with the $+\cdots$ indicating box diagram contribution and terms of order $(\alpha/4\pi)^2$. At tree-level (${\hat\kappa}\to 1$, ${\hat\lambda}_{V,A}^e\to 0$), the weak charge is suppressed, since ${\hat s}^2$ is numerically close to $1/4$: $Q_W^{e,\ \rm tree}\sim -0.1$. Inclusion of one-loop SM radiative corrections reduce the magnitude of $Q_W^e$ by nearly 40 \%, owing largely to the near cancellation between the first two terms in Eq. (\ref{eq:moller2}) and the presence of large logarithms in ${\hat\kappa}(0,\mu)$ when $\mu$ is chosen to be $M_Z$ as is conventional\cite{Czarnecki:1995fw}. Given these two considerations, one would expect the relative size of two-loop corrections to $Q_W^e$ to be considerably larger than the nominal $\alpha/4\pi$ scale. 

In order to improve the convergence of the SM prediction for $Q_W^e$, one would like to sum the large logarithms to all orders. The use of the running $ \sin^2{\hat\theta}_W(\mu)$ provides a means for doing so. By choosing $\mu\sim Q$ in both ${\hat\kappa}(k^2,\mu)$  and $ \sin^2{\hat\theta}_W(\mu)$, using the requirement that their product is $\mu$-independent as per Eq.~(\ref{eq:sweff}), and solving the RG equations for $ \sin^2{\hat\theta}_W(\mu)$ as in Ref.~\cite{Erler:2004in}, one effectively moves all the large logarithms from ${\hat\kappa}(k^2,\mu)$ into $\sin^2{\hat\theta}_W(\mu)$ and sums them to all orders. The result is a SM prediction for $\sin^2{\hat\theta}_W(k^2)^{\rm eff}$ with substantially smaller truncation error than would be obtained by the naive application of perturbation theory to one-loop order. 

In the case of superpartner loop contributions to low-energy observables, it is sufficient to include their effects solely in the form factor ${\hat\kappa}(k^2,\mu)$ while choosing $\mu=M_Z$ (illustrative contributions to ${\hat\Pi}_{Z\gamma}$ are shown in Fig. \ref{fig:nccorr}). In addition, one should include their effects in the value of $\sin^2{\hat\theta}_W(M_Z)$. One may adopt two different strategies for doing so: 

\begin{itemize}

\item[(1)] Include their effects implicitly in the value of $\sin^2{\hat\theta}_W(M_Z)$ that is obtained from fits to precision $Z$-pole observables. To be consistent, such fits must include the effects of superpartner contributions to ${\cal O}(\alpha)$ electroweak radiative corrections, and to our knowledge, such an extraction has not been carried out using LEP and SLD data in a way that does not rely on a model for SUSY-breaking mediation (see, {\em e.g.}, Ref.~\cite{Erler:1998ur,Cho:1999km} and references therein).

\item[(2)] Use the requirements of electroweak symmetry to compute superpartner contributions to $\sin^2{\hat\theta}_W(M_Z)$ explicitly. Specifically, using 
\bea
{\hat e}^2(\mu) & =&  {\hat g}^2(\mu) {\hat s}^2(\mu) \\
{\hat M}_W^2 & = & {\hat M}_Z^2 {\hat c}^2 \ \ \ ,
\eea
writing 
\be
{\hat \alpha}(\mu) = \alpha + \Delta{\hat \alpha}(\mu)
\ee
where $\alpha$ is the fine structure constant, employing Eqs.~(\ref{eq:ghat},\ref{eq:mzhat}), and choosing
$\mu=M_Z$ we obtain
\be
\label{eq:Gfswmz}
{\hat s}^2 (M_Z) {\hat c}^2 (M_Z) = \frac{\pi\alpha}{\sqrt{2} M_Z^2 G_\mu\left[1-\Delta{\hat r}(M_Z)\right]}
\ee
where
\be
\label{eq:deltarhat}
\Delta{\hat r}(\mu) = \Delta{\hat r}_\mu+\frac{\Delta{\hat\alpha}}{\alpha} -\frac{{\hat\Pi}_{ZZ}^T(M_Z^2, \mu)}{M_Z^2}\ \ \ .
\ee
Thus, by computing the superpartner loop corrections to the various terms in Eq.~(\ref{eq:deltarhat}) and employing Eq.~(\ref{eq:Gfswmz}), one may determine the predicted shift in ${\hat s}^2(M_Z)$ explicitly for a given set of SUSY parameters. In the remainder of this article, we follow this second strategy.

\end{itemize}

The corrections contained in the ${\hat\lambda}_{V,A}^f$  are fermion-specific, containing the $Zff$ vertex and external leg contributions. Representative superpartner contributions are shown in Fig.  \ref{fig:nccorr}. In the case of low-energy interactions involving one or more charged fermions, an additional contribution to the ${\hat\lambda}_V^f$ is generated by  $\gamma$ exchange and involves the so-called anapole coupling of the fermion\cite{zeldovich,Musolf:1990sa}
\be
\label{eq:anapole1}
{\cal L}_{\rm anapole} = \frac{eF_A}{M^2} {\bar\psi} \gamma_\mu\gamma_5 \psi \ \partial_\nu F^{\mu\nu} \ \ \ ,
\ee
where $F_A$ is the dimensionless anapole moment and $M$ is an appropriate mass scale. The interaction ${\cal L}_{\rm anapole}$ generates a contribution to the fermion martrix element of the electromagnetic current:
\bea
\label{eq:anapole2}
\bra{p'} J_\mu^{EM}(0)\ket{p} & = & {\bar U}(p') \Bigl[ F_1 \gamma_\mu + \frac{iF_2}{2 M} \sigma_{\mu\nu} k^\nu \\
\nonumber
&& + \frac{F_A}{M^2} \left(k^2\gamma_\mu -\dslash{k}k_\mu\right)\gamma_5 +\frac{iF_E}{2 M}\sigma_{\mu\nu} k^\nu\gamma_5\Bigr] U(p) \ \ \ ,
\eea
where $k=p'-p$ and where $F_1$, $F_2$, $F_A$, and $F_E$ give the Dirac, Pauli, anapole, and electric dipole form factors, respectively\footnote{Note that the overall sign of the anapole term in Eq.~(\ref{eq:anapole2}) differs from the convention used in Ref.~\cite{Kurylov:2003zh}.}.  Both the anapole and electric dipole couplings to the photon are parity-odd, while the electric dipole coupling is also odd under time-reversal. Since only weak interactions can give rise to a parity-odd photon-fermion coupling, we choose $M=M_Z$ in Eq.~(\ref{eq:anapole1}).

\begin{figure}[ht]
\begin{center}
\resizebox{5 in}{!}{
\includegraphics*[60,520][470,640]{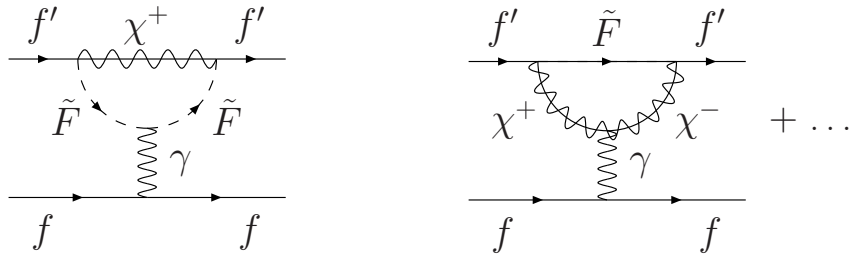}}
\caption{Anapole contributions to the NC interaction between two fermions.}
\label{fig:anapole}
\end{center}
\end{figure}

From  Eq. (\ref{eq:anapole1}) one sees that the anapole coupling gives rise to a contact interaction in co-ordinate space, since $\partial_\nu F^{\mu\nu} = j_\mu$ with $j_\mu$ being the current of the other fermion involved in the low-energy interaction. This feature is illustrated in momentum space in Fig.~\ref{fig:anapole}, where the contribution of the anapole coupling of fermion $f^\prime$ to the scattering from fermion $f$ is shown. The $\dslash{k}k_\mu$ term in Eq.~(\ref{eq:anapole2}) gives a vanishing contribution due to current conservation, while the $k^2$ term cancels the $1/k^2$ from the photon propagator to yield a four-fermion contact interaction proportional to $F_A^{f^\prime} (k^2)Q_{f}$. Since this interaction involves a coupling to the vector current of fermion $f$, it corresponds to a contribution to $g_V^f$. 

Note that at low energies for which $|k^2| \ll M_Z^2$, the prefactor in $\hat\rho(k^2)_{\rm NC}$ becomes $k^2$-independent constant, signaling that the $Z$-exchange contribution is also a contact interaction. In this regime, one has no experimental, kinematic handle with which to separate the anapole and $Z$-exchange contributions\footnote{In contrast, at $k^2\sim M_Z^2$, the anapole contribution becomes negligible in contrast to the resonating $Z$-exchange amplitude.}. Indeed, the coupling $F_A$ itself depends on the choice of electroweak gauge, and only the complete one-loop scattering amplitude that includes all ${\cal O}(\alpha)$ electroweak radiative corrections (including $F_A$) is gauge-independent (see Ref.~\cite{Musolf:1990sa} and references therein). Nonetheless, when classifying various topologies of the one-loop corrections, it is useful to separate out the anapole contributions, and in doing so, we note that superpartner loop contributions to $F_A$ are gauge-independent. In what follows, we focus on the low-energy $ff^\prime$, in which the $F_A^{f^\prime}$-contribution to the product of vector coupling ${\hat g}_V^f$ and the axial vector coupling ${\hat g}_A^{f^\prime}$ is given by
\be
\left( {\hat g}_A^{f^\prime} {\hat g}_V^f \right)_{\rm anapole} = -16{\hat c}^2{\hat s}^2 Q_f F_A^{f^\prime}\ \ \ .
\ee

In studies of precision $Z$-pole observables, it has been useful to characterize possible corrections to the gauge boson propagators from new heavy particles 
in terms of the so-called oblique parameters, $S$, $T$, $U$ \cite{Peskin:1990zt,Golden:1990ig,Marciano:1990dp, Kennedy:1990ib,Kennedy:1991sn,Altarelli:1990zd,Holdom:1990tc,Hagiwara:1994pw}: 

\begin{eqnarray}
\label{eq:stu-sirlin}
S&=&\frac{4{\hat s}^2{\hat c}^2}{{\hat \alpha}M_Z^2}{\rm Re}\Biggl\{
{\hat \Pi}_{ZZ}(0)-{\hat \Pi}_{ZZ}(M_Z^2)+\frac{{\hat c}^2-{\hat
s}^2}{{\hat c}{\hat s}} \left[{\hat \Pi}_{Z\gamma}(M_Z^2)-{\hat
\Pi}_{Z\gamma}(0)\right] +{\hat \Pi}_{\gamma\gamma}(M_Z^2)
\Biggr\}^{\rm New} ~,\nonumber \\ 
T&=&\frac{1}{{\hat \alpha}M_W^2}
\Biggl\{ 
{\hat c}^2\left( {\hat \Pi}_{ZZ}(0)+\frac{2{\hat s}}{\hat c}
{\hat \Pi}_{Z\gamma}(0) \right) -{\hat \Pi}_{WW}(0) \Biggr\}^{\rm
New} ~,\nonumber \\ 
U&=&\frac{4{\hat s}^2}{\hat \alpha} \Biggl\{
\frac{{\hat \Pi}_{WW}(0)-{\hat \Pi}_{WW}(M_W^2)}{M_W^2} +{\hat
c}^2\frac{{\hat \Pi}_{ZZ}(M_Z^2)-{\hat \Pi}_{ZZ}(0)}{M_Z^2} \nonumber
\\ 
&+&2{\hat c}{\hat s}
\frac{ {\hat \Pi}_{Z\gamma}(M_Z^2)-{\hat
\Pi}_{Z\gamma}(0)}{M_Z^2} +{\hat s}^2 \frac{{\hat
\Pi}_{\gamma\gamma}(M_Z^2)}{M_Z^2} \Biggr\}^{\rm New}
~,\end{eqnarray}
where the superscript \lq\lq New" indicates that only the new physics
contributions to the self-energies are included.  Contributions to
gauge-boson self energies  can be expressed entirely in terms of the
oblique parameters $S$, $T$, and $U$ in the limit that $M_{\rm NEW}\gg
\mz$. 

However, since present collider limits allow for fairly light
superpartners, we do not work in this limit\footnote{It is possible to extend the oblique parameterization in this case with three additional parameters\cite{Maksymyk:1993zm}. For the low-energy observables of interest here, this extended oblique approximation is not especially useful.}. Consequently, the
corrections arising from the photon self-energy ($\Pi_{\gamma\gamma}$)
and $\gamma$-$Z$ mixing tensor ($\Pi_{Z\gamma}$) contain a residual
$k^2$-dependence not embodied by the oblique parameters.  Expressing the contributions to 
${\hat\rho}$ and $\sin^2{\hat\theta}_W(k^2)^{\rm eff} = {\hat\kappa}(k^2,\mu) \sin^2{\hat\theta}_W(\mu)$ in terms of $S$,$T$, and $U$ we obtain:
\begin{eqnarray}
\delta{\hat\rho}^{\rm SUSY} & = & {\hat\alpha} T-{\hat\delta}_{VB}^\mu
~,\nonumber \\ 
\nonumber \\
\left(\frac{\delta\sin^2{\hat\theta}_W^{\rm eff}}{\sin^2{\hat\theta}_W^{\rm eff}}\right)^{\rm SUSY} & = & \left(
\frac{{\hat c}^2}{{\hat c}^2-{\hat s}^2} \right) 
\left(\frac{{\hat\alpha}}{4{\hat s}^2
{\hat c}^2} S-{\hat \alpha} T +{\hat\delta}_{VB}^\mu \right) +
\frac{{\hat c}}{{\hat s}}\Bigl[  \frac{{\hat\Pi}_{Z\gamma}(k^2)}{k^2}-
\frac{{\hat\Pi}_{Z\gamma}(M_Z^2)}{M_Z^2}\Bigr] \nonumber \\
&&+\Bigl(\frac{{\hat c}^2}{{\hat c}^2-{\hat s}^2}
\Bigr)\Bigl[-\frac{{\hat\Pi}_{\gamma\gamma}(M_Z^2)}{M_Z^2}
+\frac{\Delta{\hat\alpha}}{\alpha} \Bigr] ~,
\label{eq:rho-kappa-stu}
\end{eqnarray}
where
$k^2$ is the typical momentum
transfer for a given process.  For low energy interactions, 
$k^2\rightarrow 0$. Note that we have included in  $\delta\sin^2{\hat\theta}_W^{\rm eff}$ both the the contribution from ${\hat\Pi}_{Z\gamma}(k^2)/k^2$ that enters ${\hat\kappa}(k^2,\mu)$ as well as the shift in ${\hat s}^2(M_Z^2)$ obtained from Eq.~(\ref{eq:Gfswmz}) as discussed above. 

In analyzing SUSY radiative corrections to low-energy observables, Eqs.~(\ref{eq:rho-kappa-stu}) provide a useful means of incorporating constraints on new physics from precision $Z^0$-pole observables. For example, $\delta{\hat\rho}^{\rm SUSY}$ is highly constrained by bounds on $T$ obtained from such observables. In contrast, $[\delta \sin^2{\hat\theta}_W(k^2)^{\rm eff}]^{\rm SUSY}$ is less stringently constrained. As we discuss in Section \ref{sec:nc} below, the  unconstrained contributions to the effective weak mixing angle can lead to relatively large effects in some low-energy NC processes.

\subsection{Theoretical Uncertainties in Electroweak Radiative Corrections}

An important consideration in exploiting low-energy, precision electroweak observables as a probe  of SUSY is to ensure that the theoretical uncertainties associated with SM contributions are well-below the level of possible SUSY effects. The SM uncertainties generally involve one of two considerations: (i) neglect of higher order electroweak contributions, and (ii) contributions from strong interactions. While an extensive discussion of these considerations goes beyond the scope of the present article, we give here a brief overview of the strategies employed to address them. 

Nominally, one expects the one-loop contributions to quantities such as ${\Delta \hat{r}}_\mu$, $\hat\kappa$, {\em etc.} to be of order $\alpha/\pi\sim 10^{-3}$, so that neglect of two- and higher-loop effects is well justified for the present level of experimental sensitivity. Moreover, since SUSY loop contributions must generally decouple in the ${\tilde m}\to\infty$ limit, one expects the relative magnitude of their contributions to be
\be
\delta_{\rm SUSY\ loop} = \frac{{\delta\cal O}^{\rm SUSY loop}}{{\cal O}^{\rm SM}} \sim \frac{\alpha}{\pi}\left(\frac{M}{\tilde m}\right)^2 \  \  \ ,
\ee
where $M$ is the relevant SM mass and $\tilde m$ is a generic superpartner mass. For weak processes, such as $\mu$- and $\beta$-decay, one has $M\to M_W$, and to the extent that ${\tilde m}$ is not too different from the weak scale, one would expect $\delta_{\rm SUSY\ loop}$ to be comparable in magnitude, or slighter smaller than, the scale of one-loop, SM electroweak corrections. Thus, one would expect neglect of two-loop SM contributions to be a justifiable approximation. As discussed above, however,  exceptions may occur when the one-loop SM contributions contain large logarithms, when the tree-level SM amplitudes are suppressed, or both. In such situations, the summing terms of the form $\alpha^n \ln^n(\mu/\mu_0)$ is essential, and the RG equations can be employed for this purpose. 

Reduction of theoretical uncertainties associated with QCD corrections is generally more challenging. Short-distance QCD contributions can  be treated using the operator product expansion (OPE), and the resulting correction to a given order in $\alpha_s$ achieved with sufficient effort. In the case of PV electron-proton scattering, for example, the one-loop $WW$ box contribution is anomalously -- but not logarithmically -- enhanced, and its effect on the proton weak charge, $Q_W^p$, nearly cancels that of the large logarithms appearing in ${\hat\kappa}$. Since the semileptonic, $WW$ box graphs involve hadronic intermediate states one might expect relatively important QCD corrections to the one-loop amplitude. In this case, the loop integral is dominated by  dominated by high momentum scales ($k^2\sim M_W^2$), so the corrections can be computed using the OPE, leading to\cite{Erler:2003yk}
\be
\delta Q_W^p(WW\ {\rm box}) = \frac{\hat\alpha}{4\pi{\hat s}^2}\left[-2+4\left(1-\frac{\alpha_s(M_W)}{\pi}\right)\right]\ \ \ 
\ee
for a total QCD correction of $\approx -0.7\%$.

A more problematic situation arises for one-loop corrections that sample momenta of order the hadronic scale.  To illustrate, we again consider PV electron scattering. For both PV M\o ller and elastic $ep$ scattering, light quark loop contributions to ${\hat\Pi}_{Z\gamma}^T$ lead to hadronic uncertainties in ${\hat\kappa}(0,\mu)$. Traditionally, light quark contributions have been computed by relating ${\hat\Pi}_{Z\gamma}^T$ to the $\sigma(e^+ e^-\to{\rm hadrons})$ via dispersion relations\cite{wjmkappa}, much as one does in computing hadronic vacuum polarization contributions to the muon anomalous magnetic moment. In the case of ${\hat\Pi}_{Z\gamma}^T$, however, additional assumptions regarding flavor symmetry in the current-current correlator are needed in order to make use of $e^+ e^-$ data. Recently, these assumptions have been examined and more stringent bounds on the hadronic uncertainty in ${\hat\kappa}(0,\mu)$ obtained\cite{Erler:2004in}. 

For semileptonic processes, additional hadronic uncertainties appear in box graphs that contain one $\gamma$ and one weak gauge boson. In contrast to the situation for the $WW$-box graphs, the $\gamma Z$ loop integral samples all mometum from the hadronic scale to the weak scale. Neglecting the short-distance, perturbative QCD corrections, one finds
\be
\label{eq:zgbox}
\delta Q_W^p(\gamma Z\ {\rm box}) = \frac{5\hat\alpha}{2\pi}\left(1-4{\hat s}^2\right)\left[\ln\left(\frac{M_Z^2}{\Lambda^2}\right)+C_{\gamma Z}(\Lambda)\right]\ \ \ ,
\ee
where $\Lambda$ is a scale characterizing the transition between the perturbative and non-perturbative domains and $C_{\gamma Z}(\Lambda)$ parameterizes contributions to the loop integral from momenta
$\sqrt{|k^2|} \lsim \Lambda$. The coefficient of  logarithm in Eq. (\ref{eq:zgbox}) is determined by short distance dynamics and can be calculated reliably in perturbation theory. However, the values of both $\Lambda$ and $C_{\gamma Z}(\Lambda)$ are sensitive to long-distance scales and have not, as yet, been computed from first principles in QCD. A similar contribution arises in neutron, nuclear, and pion $\beta$-decay. An estimate of the theoretical uncertainty associated with these contributions had been made by varying $\Lambda$ over the range $ 400 \leq \Lambda \leq 1600$ MeV. Recently, Marciano and Sirlin observed that for the $\gamma W$ box, both the pQCD corrections to the logarithmic term as well as the value of $\Lambda$ could be obtained by comparison with the theoretical expression for the Bjorken Sum Rule using isospin symmetry\cite{Marciano:2005ec}. As a result, these authors have reduced the previously-quoted theoretical error by a factor of two. The analogous treatment of the $\gamma Z$ box is more complex, since one cannot obtain the isoscalar contribution from isospin arguments. In both cases, the more refined estimates of the  uncertainty associated with the low-energy constants $C_{\gamma Z}$ and $C_{\gamma W}$ remain to be performed. Fortunately, the impact of the  uncertainty in $\delta Q_W^p(\gamma Z\ {\rm box})$ due to $C_{\gamma Z}$ is suppressed by the overall factor of $1-4{\hat s}^2\sim 0.1$.

When discussing the implications of various low-energy observables for SUSY, we will also summarize the current situation regarding hadronic uncertainties in the SM predictions.

\section{Charged Current Processes}
\label{sec:cc}

Historically, the study of low energy charged current (CC) processes have played an important role in  developing  the SM, in  determining its parameters, and in testing its self-consistency at the level of one-loop radiative corrections. Indeed, the observation of a parity-violating asymmetry in the $\beta$-decay of polarized $^{60}$Co\cite{Wu:1957my} and $\mu^+$-decay\cite{Garwin:1957hc} confirmed Lee and Yang's hypothesis of parity violation in the weak interaction\cite{Lee:1956qn} and  pointed the way toward the $V-A$ structure of the weak charged currents. Measurements of the muon lifetime yield the parameter $G_\mu$ that is one of the three independent, experimental inputs needed for the gauge sector of the theory. Studies of nuclear $\beta$-decay  give the most precisely-known element of the CKM matrix -- $V_{ud}$, while measurements of branching ratios for kaon leptonic decays yield a precise value for $V_{us}$ that allows for stringent tests of the unitarity property of the CKM matrix\footnote{The value of $V_{ub}$ is also required, but its magnitude is too small to be relevant.}(for recent discussions, see Refs.~\cite{Hardy:2004id,Severijns:2006dr,Blucher:2005dc}). Comparisons of the widths $\Gamma[\pi\to \mu \nu_\mu (\gamma)]$ and $\Gamma(\pi\to e\nu_e (\gamma)]$ have provided tests of the universality of CC leptonic interactions at the few parts per thousand level\cite{Britton:1992pg,Czapek:kc}. The theoretical interpretation of these precise measurements in terms of the SM has required comprehensive calculations of one-loop radiative corrections to the tree-level amplitudes. The vast majority of this work has been carried out by Sirlin and Marciano, dating back to the classic treatment within the current algebra framework by Sirlin\cite{Sirlin:1977sv}. The implications for various SM extensions have been analyzed extensively by Herczeg and others\cite{Herczeg:2001vk,Deutsch}.

Interest in precise studies of low energy processes remains high, as reviewed recently in Ref.~\cite{Erler:2004cx}. While an extensive survey of the field can be found in that work, we highlight recent developments that motivate the study of CC processes from the standpoint of SUSY. Experimentally one has seen:

\begin{itemize}

\item[i)] New measurements of the Michel parameters that characterize the spectral shape, angular distribution, and polarization properties in polarized $\mu$-decay\cite{Gaponenko:2004mi,Jamieson:2006cf,Danneberg:2005xv} (for a recent global analysis, see Ref.~\cite{Gagliardi:2005fg})

\item[ii)] New efforts to measure $\tau_\mu$ with an order-of-magnitude improvement in precision\cite{fast,mulan}

\item[iii)] Recent Penning trap measurements of \lq\lq superallowed" nuclear $\beta$-decay Q-values \cite{Hardy:2005qv,Eronen:2006if} with significant implications for tests of first-row CKM unitarity 

\item[iv)] New measurements of the neutron lifetime, $\tau_n$\cite{Serebrov:2004zf}, and decay correlation coefficients that determine $V_{ud}$ in a manner free from possible  nuclear structure ambiguities\cite{Abele02,ucnA,Wietfeldt:2005wz,bowman06} 

\item[v)] Extensive new measurements of kaon leptonic decay branching ratios that could imply significant changes in the decades-old value of $V_{us}$ (recently reviewed in Ref.~\cite{Blucher:2005dc}; also, see below)

\item[vi)] Improved precision in the pion $\beta$-decay branching ratio\cite{Pocanic:2002av}

\item[vii)] New efforts to measure the ratio $R_{e/\mu} = \Gamma[\pi\to e\nu_e (\gamma)]/\Gamma[\pi\to \mu \nu_\mu (\gamma)]$ \cite{TRIUMFnew,PSInew}

\end{itemize}

The theoretical interpretation of the semileptonic decays has also been sharpened through 

\begin{itemize}

\item[i)] A new analysis of strong interaction uncertainties in $\Delta \hat r^V_\beta$ that associated with $W\gamma$ box graphs that reduces the theoretical uncertainty in the extraction of $V_{ud}$ from $\beta$-decay rates by a factor of two\cite{Marciano:2005ec}

\item[ii)] Computations of the ${\cal O}(p^6)$ loop corrections to the kaon decay form factor $f_{+}^K(t)$ whose value at $t=0$ is needed in order to extract $V_{us}$ from kaon decay branching ratios\cite{Post:2001si,Bijnens:2003uy}

\item[iii)] New analyses of the ${\cal O}(p^6)$ counterterm contributions to $f_{+}^K(0)$ using large $N_C$ QCD\cite{Cirigliano:2001mk} and lattice QCD computations (see below)

\end{itemize}

Given the level of activity in this area, consideration of the implications for SUSY is a timely activity. In reviewing these implications, we also provide additional background on the theoretical and experimental issues for the benefit of readers who may not be familiar with the field. We begin with the purely leptonic CC interaction that gives rise to muon decay and follow with an extensive discussion of low-energy semileptonic CC processes. We divide the latter discussion into several parts: (1) general considerations for semileptonic CC processes; (2) pion leptonic decays and the related implications for SUSY; (3) neutron and nuclear $\beta$-decay; (4) pion and (5) kaon $\beta$-decay; (5) implications of first row CKM unitarity tests  for SUSY.

\subsection{Muon Decay}

As discussed in Section \ref{sec:renorm}, the measurements of the muon lifetime generally do not, by themselves, provide information on non-SM physics. Rather, the value of $G_\mu$ as extracted from $\tau_\mu$ using  Eq. (\ref{eq:taumu}) provides  a key input into SM predictions for other observables, and the presence or absence of deviations from these predictions provides information about various SM extensions. For example, one may use of Eq.~(\ref{eq:Gfswmz}) for this purpose, treating $\alpha$, $G_\mu$, $M_Z$, and $\sin^2{\hat\theta}_W(M_Z)$ as independent, experimentally determined quantities and computing the correction $ \Delta{\hat r}(M_Z)$ in the SM\cite{Marciano:1999ih}. The degree of self-consistency among these quantities in Eq.~(\ref{eq:Gfswmz}) will lead to constraints on any deviation of $ \Delta{\hat r}(M_Z)$ from its SM value associated with possible new physics, such as SUSY radiative corrections to the $\mu$-decay amplitude or the $Z$-boson self-energy [see Eq.~(\ref{eq:deltarhat})].

Studies of the muon spectral shape and angular distribution, along with the electron polarization, can provide additional handles on non-SM contributions. The spectrum and polarization are typically described by the eleven  Michel parameters\cite{michel1,michel2}.  Four of them ($\rho$, $\eta$, $\delta$, $\xi$) characterize  the spectral shape and angular distribution:
\bea
\nonumber
   d\Gamma& = & {G_\mu^2 m_\mu^5\over 192\pi^3} {d\Omega\over 4\pi} x^2\ dx
  \times \Biggl\{ {1+h(x)\over 1 + 4\eta(m_e/m_\mu)}\left[
  12(1-x)+\frac{4}{3}\rho(8x-6)+ 24\frac{m_e}{m_\mu}{(1-x)\over x}\eta\right]\\
\label{eq:michel1}
   && \pm P_\mu\; \xi\cos\theta \left[ 4 (1-x) + \frac{4}{3}\delta(8x - 6) +
   {\alpha\over 2\pi}{g(x)\over x^2}\right]\Biggl\},
\eea
where $x=|{\vec p}_e|/|{\vec p}_e|_{\rm max}$, 
$\theta=\cos^{-1}({\hat p}_e\cdot{\hat s}_\mu)$, $P_\mu$ is the $\mu^{\pm}$ 
polarization, and $h(x)$ and $g(x)$ are momentum dependent radiative 
corrections. Five additional parameters ($\xi^\prime$, $\xi^{\prime\prime}$, $\eta^{\prime\prime}$, $\alpha/A$, $\beta/A$) are needed to describe the electron transverse and longitudinal polarization and two more  ($\alpha^\prime/A$, $\beta^\prime/A$)  parameterize the T-odd correlation of the  final state lepton spin and momenta with the muon polarization. The parameter $\eta$ also characterizes deviations of the muon lifetime from its value in the pure $V-A$ Fermi theory, as can be seen from the corresponding modification of the total decay rate:
\be
\label{eq:Gftaumu}
\frac{1}{\tau_\mu}=\frac{m_\mu^5}{192\pi^3} G_\mu^2\left[1+\delta_{\rm QED}\right]\left[1+4\eta\frac{m_e}{m_\mu}-8\left(\frac{m_e}{m_\mu}\right)^2\right]\left[1+\frac{3}{5}\left(
\frac{m_\mu}{M_W}\right)^2\right]\ \ \ .
\ee

In the Standard Model, one has $\rho=\delta=3/4$, $P_\mu\xi=1$, and $\eta=0$, so that from measurements of $\tau_\mu$ one obtains a fractional uncertainty in the Fermi constant of
$\Delta G_\mu/G_\mu = 9\times 10^{-6}$.  Allowing for $\eta\not=0$ due to possible non-SM contributions to the decay amplitude and constraining such effects with experimental determinations of the Michel parameters  can result in a larger uncertainty in $G_\mu$. A recent measurement of transverse positron polarization 
from $\mu^+$ decay yields $\eta=(71\pm 37 \pm 5)\times 10^{-3}$, leading to an increase in the relative error on $G_\mu$ by a factor of 40\cite{Danneberg:2005xv}. The results of this experiment also lead to new values for the parameters $\eta^{\prime\prime}$, $\alpha^\prime$ and $\beta^\prime$. 

New measurements of $\rho$ and $\delta$ have also been completed, and a new global analysis of the Michel parameters has been carried out in Ref.~\cite{Gagliardi:2005fg}. It is conventional to analyze the results in terms of the effective, four-fermion Lagrangian
\be
\label{eq:leff0}
{\cal L}^{\mu-\rm decay} = -\frac{4 G_\mu}{\sqrt{2}}\, \sum_\gamma \ g^\gamma_{\epsilon\mu}\ 
\ {\bar e}_\epsilon \Gamma^\gamma \nu_e\,  {\bar\nu}_\mu \Gamma_\gamma \mu_\mu
\ee
where the sum runs over Dirac matrices $\Gamma^\gamma= 1$ (S), $\gamma^\alpha$ (V), and $\sigma^{\alpha\beta}$ (T) and the subscripts $\epsilon$ and $\mu$ denote the chirality ($R$,$L$) of the final state lepton and muon, respectively\footnote{The use of the subscript \lq\lq $\mu$" to denote both the chirality of the muon and the flavor of the corresponding neutrino is an unfortunate historical convention.}. The SM has $g^V_{LL}=1$ with all other couplings vanishing, leading to the SM values for $\rho$, $\delta$, $P_\mu\xi$, and $\eta$ noted above. The sensitivity of the Michel parameters to non-SM interactions is obtained by expressing them  in terms of the general set of couplings $g^\gamma_{\epsilon\mu}$. For example, one has
\begin{equation}
\rho-\frac{3}{4}  =  -\frac{3}{4}\Bigl\{ |g^V_{LR}|^2+|g^V_{RL}|^2+2\left(|g^T_{LR}|^2+|g^T_{RL}|^2\right)+{\rm Re}\, \left(g^S_{LR} g^{T\,\ast}_{LR}+ g^S_{RL} g^{T\,\ast}_{RL}\right)\Bigr\}
\end{equation}

\begin{figure}[ht]
\begin{center}
\resizebox{6 in}{!}{
\includegraphics*[50,510][540,630]{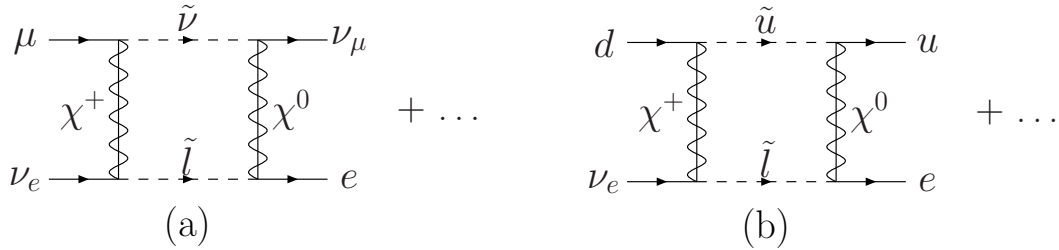}}
\caption{Box graphs contributing to (a) the muon decay parameter $g^S_{RR}$ and 
(b) $\beta$ decay parameters $a^S_{RR}$, $a^S_{RL}$, and $a^T_{RL}$. }
\label{fig:susybox}
\end{center}
\end{figure}

In order for SUSY to affect the Michel spectrum or lepton polarization in a  discernible way, it must generate contributions to the effective Lagrangian (\ref{eq:leff0}) other than those associated with the $g^V_{LL}$ term. If one assumes conservation of R-parity, then such contributions can be generated at the one-loop level {\em via} the box graphs of Fig. \ref{fig:susybox} (a). The corresponding amplitudes contribute to both $g^V_{LL}$ and $g^S_{RR}$. One has\cite{Profumo:2006yu}
\begin{eqnarray}
\label{eq:grrloop}
g^S_{RR,\, \rm loop} & = & \frac{\alpha M_Z^2}{2\pi}\Biggl\{ 2 |U_{j'1}|^2  Z_L^{2i\ast} Z_L^{5i} Z_L^{1i'} Z_L^{4i'\ast} |N_{j1}|^2\, {\cal F}_1\left(M_{\chi^0_j}^2, M_{\chi_{j'}}^2, M_{\tilde L_i}^2,
M_{\tilde L_{i'}}^2\right) \\
\nonumber
&&- Z_\nu^{1j\ast} Z_\nu^{2j} Z_L^{5i} Z_L^{4i\ast} \left(N_{j2}^\ast-\tan\theta_W N_{j1}^\ast\right)
N_{j1}  \left(N_{j'2}-\tan\theta_W N_{j'1}\right) N_{j'1}^\ast\\
\nonumber
&&\qquad \times\, {\cal F}_1\left(M_{\chi^0_j}^2, M_{\chi^0_{j'}}^2, M_{\tilde\nu_j}^2, M_{\tilde L_{i}}^2\right)\\
\nonumber
&&-Z_\nu^{1j\ast} Z_\nu^{2j} Z_L^{5i} Z_L^{4i\ast} \left(N_{j2}^\ast-\tan\theta_W N_{j1}^\ast\right)
N_{j1}  \left(N_{j'2}-\tan\theta_W N_{j'1}\right) N_{j'1}^\ast\\
\nonumber
&&\qquad\times\, M_{\chi^0_j} M_{\chi^0_{j'}}\, {\cal F}_2\left(M_{\chi^0_j}^2, M_{\chi^0_{j'}}^2, M_{\tilde\nu_j}^2, M_{\tilde L_{i}}^2\right)\Biggr\}
\end{eqnarray}
where the $Z_\nu^{Ij}$, $Z_L^{Ij}$, $U_{ij}$, and $N_{ij}$ are the sneutrino, slepton, chargino, and neutralino mixing matrices, respectively, defined in Section \ref{sec:susy} and 
\begin{equation}
{\cal F}_n\left(a,b,c,d\right) = \int_0^1 dx\,  \int_0^{1-x} dy\, \int_0^{1-x-y}dz\, \left[ax+by+cz+d(1-x-y-z)\right]^{-n}
\ \ \ .
\end{equation}
The $g^S_{RR}$ term in Eq.~(\ref{eq:leff0}) involves scalar couplings between the right-handed (RH) charged leptons and left-handed (LH) neutrinos. Although CC interactions are associated with the SU(2)$_L$ sector of the MSSM, couplings to the RH charged leptons can arise via either lepton Yukawa interactions or mixing of the LH and RH slepton weak eigenstates into mass eigenstates. In obtaining 
Eq.~(\ref{eq:grrloop}) we have retained only contributions associated with the latter, as reflected in the presence of the matrices $Z_L^{(I+3)i}$, $I=1,2$ in each of the terms. 


As discussed in Ref.~\cite{Profumo:2006yu}, the products $Z_\nu^{1j\ast} Z_\nu^{2j}$ and $Z_L^{5i} Z_L^{4i\ast}$ lead to lepton flavor changing amplitudes at one-loop order and are, thus, highly constrained by searches for lepton flavor violating processes such as $\mu\to e\gamma$. Thus, for practical purposes, one may neglect the last two terms in Eq.~(\ref{eq:grrloop}). In contrast, the first term in $g^S_{RR,\, \rm loop}$ is flavor diagonal but requires the presence of left-right mixing among first and second generation sleptons. The product $Z_L^{2i\ast} Z_L^{5i}$ also enters the SUSY contribution to $a_\mu=(g_\mu-2)/2$, the muon anomalous magnetic moment. Since the magnetic moment operator is  chirality odd, a non-vanishing SUSY contribution  requires the presence of left-right mixing either through the muon Yukawa coupling or smuon left-right mixing. Consequently, 
the degree of left-right mixing in the first term in Eq.~(\ref{eq:grrloop}) is constrained to some degree by the experimental results for $a_\mu$. Taking these considerations into account, the authors of Ref.~\cite{Profumo:2006yu} find that contributions to $g^S_{RR,\, \rm loop}$ as large as a few $\times 10^{-4}$ are possible, with the largest effects occuring when one of the two smuon mass eigenstates becomes light. 
Contributions of this magnitude would imply that either        
$|\mu|$ lies well above the electroweak scale or that all but the               
SM-like, lightest Higgs boson would decouple in order to avoid the presence of charge and color-breaking minima in the scalar potential. 

From the standpoint of the Michel parameters, $g^S_{RR}$ contributes quadratically to the combination of parameters that governs the spectral shape and spatial asymmetry
\begin{equation}
1-\xi\frac{\delta}{\rho} = 2|g^V_{RR}|^2+\frac{1}{2}|g^S_{RR}|^2 +\frac{1}{2} |g^S_{LR}-2 g^T_{LR}|^2
\end{equation}
as well as the parameter $\xi^\prime$ that enters the energy-dependence of the outgoing lepton longitudinal polarization. Experimentally, one has\cite{Stoker:1985sj,Jodidio:1986mz,Eidelman:2004wy} 
\begin{eqnarray}
P_\mu\xi\frac{\delta}{\rho} &=& 0.99787\pm0.00082 \\
\xi^\prime &=& 1.00\pm 0.04
\end{eqnarray}
leading to $|g^S_{RR}| < 0.067$ at 90\% confidence  from the recent global fit of Ref.~\cite{Gagliardi:2005fg} under the assumption that $P_\mu=1$. Given the maximum magnitude of $g^S_{RR,\, \rm loop}$ it appears unlikely that one will probe SUSY contributions using these quantities. In contrast, the parameters $\eta$, $\eta^{\prime\prime}$, and $\beta^\prime/A$ carries a linear dependence on $g^S_{RR,\, \rm loop}$. The impact on the parameter $\eta$  that characterizes the energy-dependence of the isotropic outgoing lepton spectrum is particularly interesting:
\begin{equation}
\label{eq:eta}
\eta= \frac{{\rm Re}\, g^V_{LL} g^{S,\, \ast}_{RR}}{2|g^V_{LL}|^2}+\cdots\ \ \ ,
\end{equation}
where the \lq\lq $+\cdots$" indicate contributions from the other $g^\gamma_{\epsilon\mu}$ that are not generated in the MSSM. The present limit on $\eta$ obtained in Ref.~\cite{Danneberg:2005xv} is about two orders of magnitude above the SUSY expectations, and a substantial improvement in precision would be needed to probe this parameter at an interesting level from the standpoint of SUSY. 

While a direct probe of $g^S_{RR,\, \rm loop}$ via measurements of the Michel parameters may be challenging, its contribution to $\eta$ could be large enough to affect the extraction of $G_\mu$ from $\tau_\mu$. As indicated by Eqs.~(\ref{eq:Gftaumu},\ref{eq:eta}), the fractional shift in $G_\mu$ due to this parameter is 
\begin{equation}
\label{eq:Gmushift}
\frac{\Delta G_\mu}{G_\mu} = -\frac{m_e}{m_\mu}\,  \frac{{\rm Re}\, g^V_{LL} g^{S,\, \ast}_{RR,\, \rm loop}}{|g^V_{LL}|^2}+\cdots\ \ \ ,
\end{equation}
so that contributions to $g^S_{RR,\, \rm loop}$ of order a few $\times 10^{-4}$ would lead to ppm effects in the value of the Fermi constant. Given the objectives of the new PSI experiments\cite{fast,mulan}, this correction may have to be considered if future collider experiments discover superpartners. In principle, one might also probe this SUSY-induced non-$(V-A)\otimes(V-A)$ operator using Eq.~(\ref{eq:Gfswmz}) if the uncertainty in $M_Z$ and the weak mixing angle can be improved by one and three orders of magnitude, respectively. While the latter appears to be an especially daunting task, obtaining a commensurate reduction in the theoretical uncertainty in $\Delta\hat r$ would likely be even more difficult.

\subsection{Semileptonic Decays of Light Quark Systems: General Considerations}
\label{sec:semi}

As with the analysis of $\mu$-decay, the theoretical interpretation of semileptonic decays requires careful attention to electroweak radiative corrections. Moreover, the presence of low-energy strong interactions introduces additional complications not present for purely leptonic decays. Nonetheless, considerable progress has been made in reducing the theoretical uncertainties associated with non-perturbative QCD effects, allowing one to derive information on SUSY and other possible new physics scenarios from precise studies of these decays.

In analyzing both the ${\cal O}(\alpha)$ electroweak corrections and non-perturbative QCD effects, it is useful to return to Eq.~(\ref{eq:betaampl}), replacing the quark spinors by the corresponding field operators:
\be
\label{eq:semi1}
{\cal L}^{\rm CC}_{\rm semileptonic} = -\frac{G_\mu}{\sqrt{2}}\, V_{ud}\, \left(1+{ \Delta \hat r}_\beta-{ \Delta \hat r}_\mu\right)\,  {\bar e}\gamma^\lambda(1-\gamma_5) \nu_e\, {\bar u}\gamma_\lambda(1-\gamma_5)d\, +{\rm h.c.} \ \ \ .
\ee
The computation of decay observables requires taking matrix elements of the effective operator between initial and final hadronic states. Note, however, that the ${\cal O}(\alpha)$ correction ${
 \Delta \hat r}_\beta$ can only be computed reliably in perturbation theory down to a scale $\Lambda_{\rm had}$ at which strong interactions between quarks become non-perturbative. 
Non-perturbative effects arising from QCD dynamics below this scale lead to a dependence of
the radiative corrections on the structure of the initial and final hadronic states 
as well as on the spacetime properties of the hadronic charged current. It is conventional to include these process-dependent QCD contributions to the ${\cal O}(\alpha)$ radiative corrections in process-dependent corrections. To this end, we write the decay amplitudes for various semileptonic processes of interest here as 
\begin{eqnarray}
\nonumber
{\cal M}_{\ell_2}^\pi & = & -\frac{G_\mu}{\sqrt{2}}\, V_{ud}\, \left(1+{\Delta\hat r^A_\pi}-{\Delta\hat r}_\mu\right)\,  {\bar \ell}\gamma^\lambda(1-\gamma_5) \nu_\ell\, \bra{0} {\bar u}\gamma_\lambda \gamma_5 d\ket{\pi^-}\\
\nonumber
{\cal M}_F^\beta & = & -\frac{G_\mu}{\sqrt{2}}\, V_{ud}\, \left(1+{ \Delta \hat r^V_\beta}-{ \Delta\hat  r}_\mu\right)\,  {\bar e}\gamma^\lambda(1-\gamma_5) \nu_e\, \bra{f} {\bar u}\gamma_\lambda d\ket{i} \\
\label{eq:semi2}
{\cal M}_{GT}^\beta & = & -\frac{G_\mu}{\sqrt{2}}\, V_{ud}\, \left(1+{ \Delta\hat  r^A_\beta}-{\Delta\hat r}_\mu\right)\,  {\bar e}\gamma^\lambda(1-\gamma_5) \nu_e\, \bra{f} {\bar u}\gamma_\lambda \gamma_5 d\ket{i} \ \ \ ,
\end{eqnarray}
where ${\cal M}_F^\beta$ (${\cal M}_{GT}^\beta$) denote the Fermi (Gamow-Teller) amplitudes for nuclear, neutron, or pion $\beta$-decay  that involve hadronic matrix elements of the charged vector (axial vector) current and ${\cal M}_{\ell_2}^\pi$ denotes the amplitude for pion leptonic decay. The ${\Delta\hat r^V_\beta}$, ${\Delta\hat r^A_\beta}$, and ${\Delta\hat r^A_\pi}$ denote the corresponding process-dependent ${\cal O}(\alpha)$ radiative corrections\footnote{The value of the correction ${ \Delta \hat r^V_\beta}$ for the neutron differs from the corresponding correction in pion $\beta$-decay.}. 
Since we have factored out the ${\cal O}(\alpha)$ contributions to the amplitudes explicitly, the hadronic matrix elements appearing in Eqs.~(\ref{eq:semi2}) involve purely strong interaction dynamics. 

For economy of notation, it is also useful to define a set of process-dependent Fermi constants that encode the ${\cal O}(\alpha)$ corrections and information on the hadronic matrix elements:
\begin{eqnarray}
\nonumber
G_A^\pi & \equiv & G_\mu V_{ud}\, \left(1+{\Delta\hat r^A_\pi}-{\Delta\hat r}_\mu\right) \\
\nonumber
G_V^\beta & \equiv & G_\mu V_{ud}\, \left(1+{\Delta\hat r^V_\beta}-{\Delta\hat r}_\mu\right)g_V(0) \\
\label{eq:semi3}
G_A^\beta & \equiv & G_\mu V_{ud}\, \left(1+{\Delta\hat r^A_\beta}-{\Delta\hat r}_\mu\right)g_A(0) \ \ \  ,
\end{eqnarray}
where $g_V(q^2)$ and $g_A(q^2)$ are the nucleon vector and axial vector form factors, respectively, defined in Eqs.~(\ref{eq:ncurrent}) below. The pion decay matrix element appearing in ${\cal M}_{\ell 2}^\pi$ also contains a dimensional form factor, $F_\pi(q^2)$, but it is conventional to keep the dependence of the decay rate on $F_\pi\equiv F_\pi(m_\pi^2)=92.4$ MeV explicit rather than absorbing it in $G_A^\pi$. 

\vskip 0.2in

\noindent{{\bf \ref{sec:semi}.1 Pion Decay Leptonic Decay}}

\vskip 0.2in

The purely leptonic channel $\pi^+\to \mu^+\nu (\gamma)$ yields a value for the pion decay constant $F_\pi$ that provides a key input for the analysis of chiral dynamics. Theoretically the SM prediction for $\Gamma[\pi \to \mu \nu (\gamma)]$ has been computed to one-loop order\cite{Marciano:1993sh}. In contrast to the situation for $\mu$-decay, where QED corrections have been factored out before extracting $G_\mu$ from $\tau_\mu$, the correction ${\Delta\hat r_A^\pi}$ contains QED corrections and the associated infrared divergences. Consequently, one considers the SM prediction for the total, infrared finite decay rate 
\begin{eqnarray}
\label{eq:pion0}
\Gamma[\pi^+ \to \ell^+ \bar\nu_\ell (\gamma)]& = & \Gamma[\pi^+ \to \ell^+ \bar\nu_\ell ]+\Gamma[\pi^+ \to \ell^+ \bar\nu_\ell  \gamma] \\
\nonumber
&=&\frac{ (G_A^\pi)^2}{4\pi}F_\pi^2 m_\pi m_\ell^2 \left[ 1 - {m_\ell^2\over m_\pi^2} \right]\ +\, {\rm brem} \ \ \ ,
\end{eqnarray}
where the \lq\lq $+\ {\rm brem}$" indicates ${\cal O}(\alpha)$ bremstraahlung contributions. 

The most recent analysis of $\Gamma[\pi^- \to \ell^- \bar\nu_\ell (\gamma)]$ has been carried out by Marciano and Sirlin, yielding the result\cite{Marciano:1993sh}
\begin{equation}
\label{eq:piona}
  \Gamma[\pi^+ \to \ell^+ \bar\nu_\ell (\gamma)]  =  {G_\mu^2 |V_{ud}|^2 \over 
  4\pi} F_\pi^2 m_\pi m_\ell^2 \left[ 1 - {m_\ell^2\over m_\pi^2} \right]
  \left[ 1  + {2\alpha\over\pi}\ln\frac{M_Z}{\mu} \right] 
\end{equation}
\begin{equation}
\nonumber
  \times \left[ 1 - {\alpha\over\pi} \left \{ \frac{3}{2} \ln\frac{\mu}{m_\pi}
   +  \bar{C}_1(\mu) + \bar{C}_2(\mu) \frac{m_\ell^2}{\Lambda_\chi^2}
  \ln\frac{\mu^2}{m_\ell^2} + \bar{C}_3(\mu) \frac{m_\ell^2}{\Lambda_\chi^2}
  + \cdots \right\} \right] \left[ 1 + \frac{\alpha}{\pi} F(x) \right]\ \ \ ,
\end{equation}
where the quantities proportional to $\alpha$ arise from both the electroweak and QED radiative corrections entering $G_A^\pi$ and contributions from real photon radiation. Note that the corrections are manifestly process- and hadron structure-dependent. The quantity
 $\Lambda_\chi=4\pi F_\pi$ is the scale of chiral symmetry breaking associated with the onset of nonperturbative dynamics; the ${\bar C}_i(\mu)$ denote low energy constants that parameterize presently incalculable non-perturbative QCD effects and that depend on the renormalization scale $\mu$; $F$ is a calculable function of $x=m_\ell^2/m_\pi^2$; and the $+\cdots$ indicate additional terms that are suppressed by $m_\ell^2/\Lambda_\chi^2$.  Both the function $F(x)$ and the terms containing the ${\bar C}_i(\mu)$ arise from QED corrections to the decay rate for a point-like pion\footnote{We have adopted a slightly different normalization convention from the one used in Ref.~\cite{Marciano:1993sh}.}. 

The large logarithms $\ln (M_Z/\mu)$ in Eq. (\ref{eq:pion0}) have been resummed to all orders in powers of $[(\alpha/\pi)\ln (M_Z/\mu)]^2$ using the renormalization group, yielding the electroweak correction factor $S_{EW}(\mu, M_Z)$ that replaces the factor $1+2(\alpha/\pi)\ln (M_Z/\mu)$\cite{Marciano:1993sh}. For $\mu=m_\rho$ one has $S_{EW}(m_\rho, M_Z)=1.0232$. The authors of Ref.~\cite{Marciano:1993sh} estimate that the uncertainty in $\Gamma[\pi \to \mu \nu (\gamma)]$ associated with the ${\bar C}_i(\mu)$ is $\pm 0.56\%$. This uncertainty dominates the error in $F_\pi$ since the pion lifetime and leptonic branching fraction are known to $\pm 0.02\%$ and $\pm 0.00004\%$ uncertainty, respectively. One thus obtains
\begin{equation}
\label{eq:fpiexp}
F_\pi = 92.4\pm 0.025 \pm 0.25\quad {\rm MeV}
\end{equation}
where the first error is associated with the value of $V_{ud}$ and the second with the effects of the ${\bar C}_i(m_\rho)$.

Since one cannot presently compute $F_\pi$ from first principles with a precision comparable to the uncertainties quoted in Eq.~(\ref{eq:fpiexp}), a measurement of $\Gamma[\pi \to \mu \nu (\gamma)]$ by itself does not provide a useful low-energy probe of SUSY effects. On the other hand, inclusion of SUSY corrections to the rate could alter the extracted value of $F_\pi$.  To illustrate this possibility, it is convenient to write Eq.~(\ref{eq:piona}) as
\begin{eqnarray}
\label{eq:pionc}
\Gamma[\pi^+ \to \ell^+ \bar\nu_\ell (\gamma)] &=& {G_\mu^2 |V_{ud}|^2 \over 
  4\pi} F_\pi^2 m_\pi m_\ell^2 \left[ 1 - {m_\ell^2\over m_\pi^2} \right]
\\
\nonumber
&\times & \left\{1+\left(2\left[{\Delta\hat r^A_\pi}-{\Delta\hat r}_\mu\right] +{\rm brem}\, \right)_{\rm SM} +2\left({\Delta\hat r^A_\pi}-{\Delta\hat r}_\mu\right)_{\rm new}\right\}
\end{eqnarray}
where $(2[{\Delta\hat r^A_\pi}-{\Delta\hat r}_\mu] +\, {\rm brem})_{\rm SM}$  denotes the ${\cal O}(\alpha)$ SM corrections to the rate appearing on the RHS of Eq.~(\ref{eq:piona})  and $2({\Delta\hat r^A_\pi}-{\Delta\hat r}_\mu)_{\rm new}$ indicate the corrections from new physics. Since the latter generally involves the exchange of heavy particles, one does not encounter infrared singularities in the new physics corrections and need not consider the corresponding bremstraahlung contributions. 
  
To illustrate the impact of these corrections in SUSY, we first consider the case of RPV interactions, neglecting left-right mixing among sfermions. In terms of the $\Delta_{ijk}({\tilde f})$ and $\Delta_{ijk}^\prime({\tilde f})$ defined in Eq.~(\ref{eq:deltas}), we have \cite{Ramsey-Musolf:2000qn,Barger:1989rk}
\be
\label{eq:pione}
\left({\Delta\hat r^A_\pi}-{\Delta\hat r}_\mu\right)_{\rm new}^{\rm RPV} =\left[\Delta_{\ell 1k}^\prime({\tilde d}_R^k)-\Delta_{12k}({\tilde e}_R^k)\right] \ \ \ ,
\ee
where the subscript \lq\lq $\ell$" denotes the generation of the final state leptons. 
From a fit to other low energy observables discussed in Section \ref{sec:rpv}, we obtain 
\begin{eqnarray}
\label{eq:pionf}
(a)&& -0.004 \leq ({\Delta\hat r^A_\pi}-{\Delta\hat r}_\mu)_{\rm new} \leq -0.001\ \ \ 
 {\rm 95 \%\ C.L.}
\nonumber \\
 (b)&&-0.003 \leq ({\Delta\hat r^A_\pi}-{\Delta\hat r}_\mu)_{\rm new} \leq -0.0004\ \ \ 
 {\rm 95 \%\ C.L.}
\end{eqnarray}
for the partial rate involving final state muons, 
where we have required the $\Delta_{ijk}$, $\Delta_{ijk}^\prime$ to be non-negative according to the definition of Eq.~(\ref{eq:deltas}). 
Here (a) or (b) refers to the different value of $\delta |V_{ud}|^2/|V_{ud}|^2$
in Table.~\ref{tab:rpv-constrain} that was used in the fit.
As a result, 
we obtain a small shift in the central value for the rate  (\ref{eq:piona}) and an             
additional uncertainty of roughly $\pm 0.5\%$, with a corresponding $\sim$  quarter percent increase in the uncertainty of $F_\pi$. It is interesting that the estimated hadronic structure uncertainty associated with SM radiative corrections is comparable to this RPV effect.

In order to obtain a probe of new physics using $\pi_{\ell 2}$ decays, one must attempt to circumvent the uncertainties associated with the hadronic matrix element that is parameterized by $F_\pi$. To that end, it is useful to consider the ratio of partial rates for final state electrons and muons
\be
\label{eq:remua}
R_{e/\mu} = {\Gamma[\pi^+ \to e^+ \bar\nu_e (\gamma)]\over \Gamma[\pi^+ \to \mu^+ \bar\nu_\mu (\gamma)]} \ \ \ ,
\ee
whose measurement provides a useful test of lepton universality of the CC weak interaction and its breakdown due to ${\cal O}(\alpha)$ corrections or new physics. From Eqs.~(\ref{eq:piona}) and (\ref{eq:pionc}) it follows that
\begin{eqnarray}
\label{eq:remub}
R_{e/\mu} &=&  \frac{m_e^2}{m_\mu^2} \left[ {m_\pi^2-m_e^2 \over m_\pi^2-m_\mu^2} \right]^2\Bigl\{1+\left(2\left[{\Delta\hat r^A_\pi}(e)-{\Delta\hat r^A_\pi}(\mu)\right] +\Delta_{\rm brem}\, \right)_{\rm SM}\\
\nonumber
&& \qquad\qquad\qquad\qquad\qquad+2\left[{\Delta\hat r^A_\pi}(e)-{\Delta\hat r_\pi^A(\mu)}\right]_{\rm new}\Bigr\}\\
\nonumber
&&\\
\nonumber
\nonumber
&\equiv& R_{e/\mu}^{\rm SM}\left\{ 1+2\left[{\Delta\hat r^A_\pi}(e)-{\Delta\hat r_\pi^A(\mu)}\right]_{\rm new}\right\}
\end{eqnarray}
where the  ${\Delta\hat r^A_\pi}(\ell)$ indicate the corrections for final state lepton $\ell$, the
\lq\lq $\Delta_{\rm brem}$" indicate the difference in the bremstraahlung contributions to the rate for final state electrons and muons, and
$R_{e/\mu}^{\rm SM}$ is the SM value for the ratio. The SM contributions have been computed in Ref.~\cite{Marciano:1993sh}
\be
\label{eq:remuc}
   R_{e/\mu}^{\rm SM} =  \frac{m_e^2}{m_\mu^2} \left[ {m_\pi^2-m_e^2 \over m_\pi^2-m_\mu^2} \right]^2\\
   \left\{ 1 +\frac{\alpha}{\pi} \left[ F({m_e\over m_\pi}) - 
   F({m_\mu\over m_\pi}) + \frac{m_\mu^2}{\Lambda_\chi^2} ( \bar{C}_2 
   \ln {m_\mu^2\over \Lambda_\chi^2} + \bar{C}_3) \right] \right\}\ \ \ .
\ee
The terms proportional to $\alpha m_e^2$ are numerically insignificant and have been omitted from this expression. To obtain a precise numerical prediction, the authors of Ref.~\cite{Marciano:1993sh} included structure-dependent bremsstrahlung corrections and performed a renormalization group resummation of all $[(\alpha/\pi) \ln(m_e/m_\mu)]^n$ corrections that enter
$F(m_e/m_\pi)-F(m_\mu/m_\pi)$, leading to
\begin{equation}
\label{eq:remud}
   R_{e/\mu}^{\rm SM} = (1.2352\pm 0.0005)\times 10^{-4}\ \ \ .
\end{equation}
where the error is dominated by theoretical uncertainties in the structure dependent 
bremsstrahlung contributions. 

The ratio $R_{e/\mu}$ has been measured by groups at TRIUMF~\cite{Britton:1992pg} and PSI~\cite{Czapek:kc}, yielding the world average
\begin{eqnarray}
\label{eq:remuresult}
   {R_{e/\mu}^{\rm exp}\over R_{e/\mu}^{\rm SM}}=0.9966\pm 0.0030\pm 0.0004,
\end{eqnarray}
where the first error is experimental and the second is the estimated 
theoretical uncertainty. A new measurement that has been approved to run at TRIUMF aims to reduce the experimental uncertainty to a level comparable with the theoretical error\cite{TRIUMFnew}. A  measurement at PSI with a similar improvement in precision has also recently been approved\cite{PSInew}.

The new measurements could provide significant tests of possible SUSY contributions to the breaking of CC lepton universality. In the case of RPV contributions, for example, one has
\be
\label{eq:remue}
\left[{\Delta\hat r_\pi^A}(e)-{\Delta\hat r_\pi^A}(\mu)\right]_{\rm RPV} = \Delta_{11k}^\prime({\tilde d}_R^k)-\Delta_{21k}^\prime({\tilde d}_R^k) \ \ \ .
\ee
Thus, the precise measurement of $R_{e/\mu}$ provides an important input into the global analysis of the RPV corrections $\Delta_{ijk}^\prime$, {\em etc.} (see Table~\ref{tab:rpv-constrain}). The global analysis leading to Eq.~(\ref{eq:pionf}) uses the present average (\ref{eq:remuresult}), and the new measurements should reduce the range on the RPV-related uncertainty in the value of $F_\pi$.

When $P_R$ is conserved, lepton universality can be broken by superpartner loop corrections when the first and second generation sleptons have unequal masses. SUSY loop corrections to the $W{\bar u} d$ vertices, light quark propagators, and $W$-boson  propagator are identical for the $\pi^+\to e^+\nu_e(\gamma)$ and $\pi^+\to \mu^+\nu_\mu(\gamma)$ amplitudes, thus canceling out of the ratio $R_{e/\mu}$ at ${\cal O}(\alpha)$. In contrast, corrections to the $W{\ell}\nu_\ell$ vertices, lepton propagators, and box graphs can differ in the two cases, so that $R_{e/\mu}$ provides a probe of the non-universality of the first and second generation sleptons. 
An analysis of these corrections has recently been performed in Ref.~\cite{tulin06}. In the limit that the charginos and neutralinos are approximately pure gaugino and higgsino states (a limit achieved in the absence of electroweak symmetry-breaking), the SUSY correction to $R_{e/\mu}$ is dominated by box graphs. The vertex and external leg corrections for the $W^+\ell^+\nu_{\ell}$ vertex sum to zero as required by the Ward Identity\footnote{When $\overline{DR}$ is used, the SM and superpartner corrections individually contribute $\mu$-independent constants of equal magnitude and opposite sign.}. The resulting correction is
\be
\label{eq:remuloop1}
\delta R_{e/\mu}^{SUSY}=\frac{\Delta R_{e/\mu}^{SUSY}}{R_{e/\mu}^{SM}} =\frac{\alpha V_{ud}}{6\pi {\hat s}^2}\, \left(\frac{M_W}{M_2}\right)^2 \left\{F\left(\frac{m_{\tilde L_1}^2}{M_2^2},\frac{m_{\tilde Q_1}^2}{M_2^2}\right)-F\left(\frac{m_{\tilde L_2}^2}{M_2^2},\frac{m_{\tilde Q_1}^2}{M_2^2}\right)\right\}\ \ \ ,
\ee
where $F(x,y)$ is a loop function associated with the box graph with $F(1,1)=1$ (corresponding to  $m_{\tilde L_i}=m_{\tilde Q_1} =M_2$). When the two sleptons are equal in mass, $\Delta R_{e/\mu}^{SUSY} =0$, whereas when $m_{\tilde L_i} >> m_{\tilde L_j}$, one has 
\be
\label{eq:remuloop2}
\vert \delta R_{e/\mu}^{SUSY}\vert =1.7 \times 10^{-3} \left(\frac{M_W}{M_2}\right)^2 \, F\left(\frac{m_{\tilde L_j}^2}{M_2^2},\frac{m_{\tilde Q_1}^2}{M_2^2}\right)\ \ \ ,
\ee
so that the correction can be as large as a few $\times 10^{-3}$ in this case. 

Allowing for significant gaugino-higgsino mixing can lead to can lead to logarithmic enhancements of $\delta R_{e/\mu}^{SUSY}$ as the vertex and external leg corrections no longer sum to zero in this case. The resulting contribution is
\begin{align}
\left. \delta R_{e/\mu}^{\textrm{SUSY}} \right|_{V+L} = & \; \frac{\alpha}{4 \pi s_W^2} \; \ln \left(\frac{m_{\widetilde{e}_L}^2}{m_{\widetilde{\mu}_L}^2} \right)
\times \left[ 2 - 2\: V^{*}_{j1}U^{*}_{j1} N_{i2} N_{i2} \frac{}{} \right. \nonumber \\ 
&\left. \quad\quad - V^{*}_{j1} U^{*}_{j2} N_{i2} N_{i3}/\sqrt{2} 
+ U^{*}_{j1} V^{*}_{j2} N_{i2}^{*} N_{i3}^{*}/\sqrt{2} \frac{}{} \right]  + \; ...  ,
\end{align}
where the pre-factor in the square brackets that contains the chargino and neutralino mixing matrices vanishes in the limit that $M_{1,2}$ and $\mu$ are much larger than $M_W$ and and can be as large as $0.5$ for gaugino and higgsino mass parameters of order 100 GeV. The resulting correction can, thus, be as large as 
\be
\vert \delta R_{e/\mu}^{SUSY}\vert \lsim 1.3 \times 10^{-3}\, \ln \left(\frac{m_{\tilde L_1}^2}{m_{\tilde L_2}^2}\right)\ \ \ .
\ee
A numerical scan over MSSM parameters performed by the authors of Ref.~\cite{tulin06} is consistent with this bound. Given that  the new TRIUMF and PSI experiments hope to reduce the experimental uncertainty to the level of the present SM theory uncertainty ($\sim 5\times 10^{-4}$), one could expect to see a departure from the SM expectations of several standard deviations if the masses of the selectron and smuon  differ by more than a factor of $\sim 2$ and the mass of the  lightest chargino is less than 300 GeV. Moreover, in the case of such non-degeneracy among sleptons, the sign of $\delta R_{e/\mu}^{SUSY}$ provides an indication of whether the lightest first or second generation slepton is heavier.
 
\vskip 0.2in

\noindent{{\bf \ref{sec:semi}.2 Neutron and Nuclear $\beta$ Decay: General Features}}

\vskip 0.2in

Studies of nuclear and neutron $\beta$-decay have yielded both important information about parameters of the SM as well as constraints on physics beyond it (for comprehensive reviews, see Refs.~\cite{Herczeg:2001vk,Deutsch,Severijns:2006dr}). In particular, the most precisely-known element of the Cabibbo-Kobayashi-Maskawa (CKM) matrix that characterizes the misalignment of quark weak interaction and mass eigenstates in CC interactions has been determined from \lq\lq superallowed" nuclear $\beta$-decays. The latter involve transitions between spin-parity $J^\pi=0^+$ nuclear states that are mediated solely by the vector charge component of the CC weak current. To the extent that the initial and final $0^+$ nuclear states are states of pure isospin and the energy transfer to the outgoing lepton pair is negligible compared to typical nuclear  scales, the transition matrix element is independent of nuclear structure. In contrast, the rates for nuclear decays involving initial and/or final states with non-zero spin -- including the decay of the neutron -- depend more strongly on the details of hadronic and nuclear structure. In these cases, the extraction of precise information on the electroweak sector of the SM or on new physics generally requires measurement of both the decay rate and one or more decay correlation coefficients. 

In all cases, the overall decay rate in the SM is characterized by the  the reduced half life or \lq\lq $ft$" value, given by
\begin{eqnarray}
\label{eq:ftsuper}
ft&=&\frac{K}{(G_V^\beta)^2  M_F^2 + (G_A^\beta)^2 M_{GT}^2}\\
\nonumber 
\\
\nonumber
K&=&\hbar (2\pi^3\ln2 )(\hbar c)^6 /(m_e c^2)^5 \ \ \ ,
\end{eqnarray}
where $t$ is the half-life; $f$ is a factor that takes into account the outgoing $\beta$-particle wavefunction in the presence of the nuclear Coulomb field; the Fermi matrix element $M_F$ is the nuclear transition matrix element of the vector current charge operator $J_0^\dag(x)=u^\dag(x) d(x)+{\rm h.c.}$, while the Gamow-Teller matrix element $ M_{GT}$ involves the spatial component of the axial vector current operator\footnote{For neutron decay, $M_F$ and $M_{GT}$ simply indicate the matrix elements of the vector and axial vector CC operators without respect to their spacetime components.}. 

It is often useful to consider the differential decay rate for the decay of a polarized nucleus, for which one has\cite{Jackson} (see also Refs.~\cite{Herczeg:2001vk,Deutsch,Severijns:2006dr})
\begin{eqnarray}
\label{eq:betacor}
  d\Gamma& \propto & {\cal N}(E_e)\Biggl\{ 1+a {{\vec p}_e\cdot{\vec p}_\nu\over E_e E_\nu}
   + b{\Gamma m_e\over E_e} + \langle {\vec J}\rangle\cdot \left[A{{\vec p}_e\over E_e} 
   + B{{\vec p}_\nu \over E_\nu} + D{{\vec p}_e\times {\vec p}_\nu \over E_e E_\nu}\right] \\
 \nonumber
&&+ {\vec\sigma}\cdot\left[N \langle{\vec J}\rangle + G\frac{{\vec p}_e}{E_e}+Q^\prime {\hat p}_e {\hat p}_e\cdot \langle{\vec J}\rangle+R \langle {\vec J}\rangle\times\frac{{\vec p}_e}{E_e}\right]
 \Biggr\}
   d\Omega_e d\Omega_\nu d E_e,
\end{eqnarray}
where $N(E_e)=p_e E_e(E_0-E_e)^2$, $E_e$ ($E_\nu$) and ${\vec p}_e$ 
(${\vec p}_\nu$) are the $\beta$ (neutrino) energy and momentum, respectively; $E_0$ is the endpoint energy; 
and ${\vec J}$ is the  polarization of the decaying nucleus, ${\vec \sigma}$ is the $\beta$ polarization, and $\Gamma=\sqrt{1-(Z\alpha)^2}$. The coefficients of the various correlations involving lepton momenta and nuclear spin depend on the structure of the underlying lepton-quark weak interaction. The correlations parameterized by $A$, $B$, and $G$ are odd under parity (P) and even under time-reversal (T); the $D$-term is T-odd but P-even; the $R$ correlation is both T- and P-odd; and all others are P and T-even. 

The various correlation coefficients in Eq.~(\ref{eq:betacor}) carry complementary dependences on the ratio of Fermi constants
\be
\label{eq:gvga}
\lambda = \frac{G_A^\beta}{G_V^\beta} = \frac{g_A(0)}{g_V(0)}\left(1+{\Delta\hat r}_\beta^A-{\Delta\hat r}_\beta^V \right)
\ee
as well as on operators that depart from the $(V-A)\otimes(V-A)$ form of the SM, CC low-energy current-current interaction. In the SM, the correlation coefficients can be expressed in terms of  $\lambda$ alone.
In the case of neutron decay, for           
example, one has:
\be
\label{eq:corcoeff}
a = {1-\lambda^2\over 1+3\lambda^2}, \hspace{80pt}
A = -2{\lambda(1+\lambda)\over 1+3\lambda^2}, \hspace{80pt}
B =  2{\lambda(\lambda-1)\over 1+3\lambda^2},
\ee
with analogous expressions for the coefficients $N$ and $G$.
The quantity $b$ appearing in the so-called Fierz interference term is zero for
purely vector and axial vector interactions, 
while $N$ vanishes for pure vector             
transitions. The T-odd correlations are zero in the SM\footnote{Final state QED interactions can induce non-vanishing contributions to $D$ that mimic the effects of {\em bona fide} T-violation.}. 

As with $\mu$-decay, it is useful to describe non-$(V-A)\otimes(V-A)$ possible departures using an effective, low-energy  Lagrangian. In analogy with Eq.~(\ref{eq:leff0}) we write
\begin{equation}
\label{eq:leffbeta}
{\cal L}^{\beta-\rm decay} = - \frac{4 G_\mu}{\sqrt{2}}\ \sum_{\gamma,\, \epsilon,\, \delta} \ a^\gamma_{\epsilon\delta}\, 
\ {\bar e}_\epsilon \Gamma^\gamma \nu_e\, {\bar u} \Gamma_\gamma d_\delta\ \ \ .
\end{equation}
The SM gives
\be
\label{eq:avll}
a^V_{LL}=V_{ud}\, \left(1+{\Delta\hat r}_\beta-{ \Delta\hat r}_\mu\right)\ \ \ ,
\ee
with all other $a^\gamma_{\epsilon,\, \delta}=0$. Note that in  hadronic matrix elements of the SM, $(V-A)\otimes(V-A)$ operator, non-perturbative QCD effects will lead to a renormalization of the axial vector current and to differences in hadronic contributions to the radiative corrections $\Delta{\hat r}_\beta^V$ and $\Delta{\hat r}_\beta^A$ as indicated in Eqs.~(\ref{eq:semi2},\ref{eq:semi3}). 

Tree-level, supersymmetric contributions to $a^V_{LL}$ can be generated by RPV interactions. One has\cite{Ramsey-Musolf:2000qn,Barger:1989rk}.
\be
\label{eq:ckm2}
\left[{ \Delta\hat r}_\beta^{V,A}-{\Delta\hat r}_\mu\right]_{\rm RPV} =  2\left[\Delta_{11k}^\prime({\tilde d}_R^k)-\Delta_{12k}({\tilde e}_R^k)\right] \ \ \ .
\ee
Note that because the RPV interactions preserve the $(V-A)\otimes(V-A)$ structure of the low-energy CC interaction, the RPV contributions to ${ \Delta\hat r}_\beta^{V}$ and ${ \Delta\hat r}_\beta^{A}$ are identical and, thus,  do not affect the value of $\lambda$.
As we discuss below, probes of these contributions can be obtained using a combination of $\beta$-decay, kaon semileptonic ($K_{\ell 3}$) decays, $B$-meson decays, and the unitarity property of the CKM matrix. The results of these unitarity tests have been used as input for the global analysis of RPV corrections $\Delta_{ijk}$, $\Delta_{ijk}^\prime$, {\em etc.} summarized in Table~\ref{tab:rpv-constrain}.

Supersymmetric loop corrections can also contribute to $[{ \Delta\hat r}_\beta^{V,A}-{ \Delta \hat r}_\mu]$. As with the RPV corrections, loop SUSY loop corrections to ${ \Delta\hat r}_\beta^{V}$ and ${ \Delta\hat r}_\beta^{A}$ are identical in the MSSM, since the superpartner mass scale lies well above the hadronic scale. Hence, the presence of these corrections leaves the value of $\lambda$ unchanged from its SM value.  
An analysis of these corrections has been carried out in Ref.~\cite{Kurylov:2001zx} using the MSSM with R-parity conservation. Doing so requires computation of the corrections illustrated in Fig.~\ref{fig:cccorr} (see Section \ref{sec:renorm}). In general, carrying out an analysis of these corrections on the vast number of MSSM parameters is a formidable task. Typical analyses resort to one of two strategies to contend with the large number of parameters: either one carries out a model-independent analysis by randomly generating a large number of MSSM parameter sets and computing the corrections for each set, or one reduces the number of independent parameters by resorting to a model for SUSY-breaking mediation in which the soft SUSY parameters are determined from a small number of parameters at a high scale and their RG running to the weak scale.

In the case of the corrections to $G_V^\beta$, however, cancellations between corrections to ${ \Delta\hat r}_\beta^{V,A}$ and ${ \Delta\hat r}_\mu$ allow for an alternate approach. Specifically, contributions to the $W$-boson propagators -- ${\hat\Pi}_{WW}^T$ -- cancel entirely between the two terms\footnote{Up to negligible corrections arising for the different kinematics of each process.}, as do corrections to the first generation lepton propagators and $We{\bar\nu}_e$ vertices. As a result, 
$[{ \Delta \hat r}_\beta^{V,A}-{ \Delta \hat r}_\mu]_{\rm SUSY-loop}$ depends only on the differences in the $W\, \mu\, \nu_\mu$ and $W\, u\, d$ and external leg corrections as well as the box graphs. The authors of Ref.~\cite{Kurylov:2001zx} found that these simplifications allowed completion of a model-independent analysis without resorting to randomly generated parameter sets. We review the results of this analysis below. 

%
%

One-loop radiative corrections in the MSSM may also induce non-zero scalar and tensor interactions via the box graphs of Fig.~\ref{fig:susybox} (b). These loop-induced non-$(V-A)\otimes(V-A)$ interactions have recently been analyzed by the authors of Ref.~\cite{Profumo:2006yu}, who obtained 
\bea
\label{eq:deltabetaa}
\tilde\delta_\beta^{\mathbf (a)}&=&\frac{\alpha M_Z^2 V_{ud}}{3\pi} \left|U_{k1}\right|^2Z_D^{1i*}Z_D^{4i}Z_L^{1m}Z_L^{4m*}\left|N_{j1}\right|^2{\mathcal F}_1\left(M_{\chi_j^0},M_{\chi_{k}^+},M_{\tilde d_i},M_{\tilde l_{m}}\right)\\
\nonumber
\tilde\delta_\beta^{\mathbf (b)}&=&\frac{-\alpha M_Z^2 V_{ud}}{3\pi} U_{j1}V_{j1}^* Z_U^{1i*}Z_U^{4i}Z_L^{1m}Z_L^{4m *}\left|N_{k1}\right|^2M_{\chi_j^+}M_{\chi_{k}^0}{\mathcal F}_2\left(M_{\chi_j^+},M_{\chi_{k}^0},M_{\tilde u_i},M_{\tilde l_{m}}\right)
\eea
where the notation is similar to that of Eq.~(\ref{eq:grrloop}). The corresponding operator coefficients are 
\bea
\label{eq:box1}
a^S_{RR} & = & \tilde\delta_\beta^{\mathbf (a)} \\
\nonumber
a^S_{RL} = -2 a^T_{RL} & = & \tilde\delta_\beta^{\mathbf (b)}\ \ \ .
\eea
Note that, in contrast to the situation with $\mu$-decay, the presence of SUSY loop-induced scalar and tensor interactions relevant to $\beta$-decay requires only flavor-diagonal L-R mixing among scalar fermions. These loop-induced scalar and tensor interactions generate contributions to the  Fierz interference parameter, $b$, of Eq.~(\ref{eq:betacor}); the energy-dependent components of the neutrino asymmetry parameter $B$ and spin-polarization coefficient $Q^\prime$; and the energy-independent component of the spin-polarization coefficient $N$. We discuss the prospects for future $\beta$-decay experimental probes of these loop-induced interactions below. 

%
%

\subsection{Superallowed Nuclear Decays}




For the superallowed Fermi transitions one has $M_{GT}=0$, leaving only the dependence on $M_F$ in the $ft$-value. To high degree of accuracy, $M_F$ is independent of the details of nuclear structure, making these transitions an ideal venue for determining $G_V^\beta$ and, thus, $V_{ud}$.  As Eqs.~(\ref{eq:semi3}) and (\ref{eq:ftsuper}) indicate, the  determination of $V_{ud}$ from these transitions requires both  experimental and theoretical input. Experimentally, the $ft$ values for twelve different transitions have been measured with uncertainties ranging from $\sim (3-25)\times 10^{-4}$\cite{Hardy:2004dm,Hardy:2004id}, while the value of $G_\mu$ is known to ten ppm accuracy (assuming $\eta=0$ as in the SM). As discussed in 
Refs.~\cite{Hardy:2004dm,Hardy:2004id}, a determination of the experimental half lives requires three distinct experimental measurements: the total decay half life, the branching ratio for the decay to the $0^+$ ground state of the daughter nucleus, and the energy release in the decay, or $Q$-value. New measurements of the $Q$-values for several superallowed decays have recently been completed, leading to shifts in the $ft$ values in some, but not all, cases\cite{Savard:2005cv,Eronen:2006if}. The impact of these new measurements on the overall fit to the twelve decays awaits completed analyses of Penning trap measurements of the $Q$-values of $^{26m}$Al and $^{42}$Sc.

Theoretically, one requires computations of the matrix element $M_F$ as well as the SM radiative correction factors ${\Delta\hat r}^V_\beta$ and ${\Delta\hat r}_\mu$. In the limit of zero energy transfer and exact isospin, the Fermi matrix element is 
\be
\label{eq:mv}
M_F = \langle I,I_Z\pm 1| J_0 |I, I_Z \rangle = [(I\mp I_Z)(I\pm I_Z+1)]^{1/2}\ \ \ .
\ee
For the most precisely-known cases, one has $I=1, I_Z=0$ so that $  M_F =\sqrt{2}$. For realistic nuclei and finite energy transfer, one must apply small nuclear structure-dependent corrections
\be
\label{eq:mvcorr}
 M_F^2 (1+\delta_R)(1-\delta_C) = [(I\mp I_Z)(I\pm I_Z+1)]
 \ee
 where $\delta_C$ is a correction that accounts for isospin-breaking and $\delta_R$ is a nucleus-dependent contribution to the ${\cal O}(\alpha)$ electroweak radiative corrections. After applying these one obtains a corrected $ft$ value
 \be
 \label{eq:ftcorr}
 {\cal F}t=ft(1+\delta_R)(1-\delta_C) \ \ \ .
\ee
It follows from Eqs.~(\ref{eq:ftsuper}) and (\ref{eq:ftcorr}) that $ {\cal F}t$ should be the same for each nucleus, since this quantity depends only on the nucleus-independent Fermi constant $G_V^\beta$ and the universal matrix element in Eq.~(\ref{eq:mv}) determined solely by the isospin quantum numbers. For historical reasons, this prediction is described as a consequence of the conserved vector current (CVC) property of the semileptonic, CC weak interaction. 
The corrections $\delta_{R,C}$ have been computed by the authors of 
Refs.~\cite{Hardy:2004dm,Hardy:2004id} using the nuclear shell model (NSM) and applied to the  twelve best-known superallowed transitions, leading to an average ${\cal F}t$ value
 \be
 \label{eq:ftcorrave}
{\overline{ {\cal F}t} }= 3072.7\pm 0.8\ s
 \ee
where the error includes the estimated theoretical uncertainty associated with the corrections $\delta_R$ and $\delta_C$\footnote{The $\chi^2$ per degree of freedom for the average is 0.42.}.   
This result for ${\overline{ {\cal F}t} }$ implies consistency with CVC at the $0.026\% $ level.

The result in Eq.~(\ref{eq:ftcorrave}) provides a test of the inter-nuclear consistency implied by CVC but does not account for the possibility of a nucleus-independent systematic theoretical error associated with the calculated nuclear corrections. To allow for this possibility, the authors of Refs.~\cite{Hardy:2004dm,Hardy:2004id} compared the NSM  ${\cal F}t$ values with those obtained by using the  $\delta_C$  Hartree-Fock calculations of Ormand and Brown\cite{ormandbrown}. Since the latter exist for nine out of the 12 most accurately measured superallowed transitions, Towner and Hardy (TH)averaged their values with those of Ormand and Brown (OB) for these cases, yielding the average
\be
 \label{eq:ftcorravenine}
{\overline{ {\cal F}t} }= 3073.5 \pm 1.2\ s
\ee
where an additional theoretical uncertainty given by half the difference of the TH and OB values for ${\overline {\cal F}t }$. In what follows, we use the average (\ref{eq:ftcorravenine}) in the discussion of $V_{ud}$.

A value for $G_V^\beta$ can be extracted from ${\overline{ {\cal F}t}}$ by employing Eqs.~(\ref{eq:ftsuper},\ref{eq:ftcorr}) and subsequently used to determine $V_{ud}$ by applying the SM radiative correction difference ${ \Delta \hat r}_\beta^V-{ \Delta \hat r}_\mu$ appearing in $G_V^\beta$. The dominant uncertainty entering the latter is the theoretical uncertainty associated with hadronic contributions to the correction ${ \Delta \hat r}_\beta^V$ arising from the $W\gamma$ box graph. 
To ${\cal O}(\alpha)$ one has \cite{Sirlin:1977sv}
\be
\label{eq:wgbox}
   {\Delta\hat r}_\beta^V(W\gamma\, {\rm box}) = {{\hat\alpha}\over 8\pi} \left[
   \ln\left( {M_W^2\over\Lambda^2} \right) + C_{\gamma W}(\Lambda)\right] \ \ \ .
\ee
Here,  the leading logarithmic term is generated by short-distance contributions to the loop integral and can be computed reliably in the SM. The 
the constant $C_{\gamma W}(\Lambda)$ parameterizes contributions to the loop 
integral below a momentum scale $\Lambda$. An estimate of $C_{\gamma W}(\Lambda)$ was
given by Marciano and Sirlin~\cite{Marciano:1985pd} using nucleon intermediate states in 
the box diagram, and this estimate had been retained by subsequent authors for many years. An estimate of the uncertainty in this contribution had been obtained by varying $\Lambda$ over a reasonable range, yielding an uncertainty of  $\sim\pm 0.00038$ in ${\Delta r}_\beta^V$. 

Recently, Marciano and Sirlin have reduced this uncertainty by a factor of two over previous estimates by relating the asymptotic part of the $W\gamma$ box integral to the Bjorken sum rule to include perturbative QCD corrections through ${\cal O}(\alpha_s^3)$ and by employing large $N_C$-based current-current correlators to treat the resonance region\cite{Marciano:2005ec}. The result is an uncertainty of $\sim\pm 0.00019$ in ${ \Delta\hat  r}_\beta^V$. 
Using the results of this new analysis and an up-date of the fit in 
Refs.~\cite{Hardy:2004dm,Hardy:2004id} by Savard {\em et al} \cite{Savard:2005cv} one has
\be
\label{eq:vudnuc}
V_{ud}=0.97377(11)(15)(19)
\ee
where the first error arises from combining the experimental error in $\overline{{\cal F}t}$ and nuclear structure theory uncertainty; the second error is associated with nuclear Coulomb distortion effects; and the final error is the theoretical hadronic structure error associated with $C_{\gamma W}(\Lambda)$.

In light of this new result, the uncertainty associated with  ${ \Delta \hat r}_\beta^V$ and with the nuclear corrections $\delta_{R,C}$ are comparable. In order to test theoretical, nuclear structure computations of these corrections, new measurements of additional superallowed decays are underway with nuclei in which the magnitude of the calculated nuclear corrections are larger than the corrections for the nine most precisely measured transitions used to obtain the result (\ref{eq:vudnuc}). These studies involve even-Z, $I_Z=-1/2$ parent nuclei in the mass range $18 \leq A\leq 42$ and four odd-Z, $I_Z\geq 62$ $0^+$ parent nuclei. Obtaining refined measurements of half lives, branching ratios, and $Q$-values for both series of decays is challenging, though use of Penning trap techniques have allowed for new, highly precise mass measurements for the light cases. A more extensive review of these challenges and prospects for meeting them can be found in Ref.~\cite{Hardy:2005kc}.

In addition to providing a determination of $G_V^\beta$, studies of superallowed transitions also provide a probe of the Fierz interference coefficient $b_F$ (the subscript \lq\lq $F$" denotes those contributions allowed for Fermi decays).  In terms of the effective operator coefficients $a^\gamma_{\epsilon\, \delta}$ one has
\begin{equation}
b_F = \pm \frac{2\, g_S}{g_V}\, {\rm Re}\, \left(\frac{a^S_{RL}+a^S_{RR}}{a^V_{LL}}\right)
\end{equation}
independent of the details of the nuclear matrix elements (the upper and lower signs correspond to $\beta^-$ and $\beta^+$ decay, respectively). Here, $g_V$ and $g_S$ are the nucleon vector and scalar current form factors, respectively [see Eq.~(\ref{eq:ncurrent}) below]. An global analysis of superallowed decays leads to the $b_F=0.0026(26)$\cite{Hardy:2004dm,Hardy:2004id}, implying stringent bounds on scalar interactions. In order to probe the  non-$(V-A)\otimes(V-A)$ interactions generated by superpartner loops, roughly an order-of-magnitude improvement in sensitivity would be required. Specifically, from Eqs.~(\ref{eq:deltabetaa},\ref{eq:box1}) we have
\begin{eqnarray}
\nonumber
b_F&=&\pm\frac{2\alpha}{3\pi}\, \left(\frac{g_S}{g_V}\right)\, {\rm Re}\, Z_L^{1m}Z_L^{4m*} \Bigl[ \left|U_{k1}\right|^2Z_D^{1i*}Z_D^{4i}\left|N_{j1}\right|^2\, M_Z^2{\mathcal F}_1\left(M_{\chi_j^0},M_{\chi_{k}^+},M_{\tilde d_i},M_{\tilde l_{m}}\right)\\
\label{eq:bFsusy}
&&-U_{j1}V_{j1}^* Z_U^{1i*}Z_U^{4i}\left|N_{k1}\right|^2\, M_Z^2M_{\chi_j^+}M_{\chi_{k}^0}{\mathcal F}_2\left(M_{\chi_j^+},M_{\chi_{k}^0},M_{\tilde u_i},M_{\tilde l_{m}}\right)\Bigr]\ \ \ .
\end{eqnarray}
The prefactor $2\alpha/3\pi$ is ${\cal O}(10^{-2})$ while the product of $M_Z^2$ and the loop functions can be as large as $10^{-1}$ for ${\tilde m}\sim M_Z$. For nearly maximal L-R mixing among sfermions, the product of rotation matrices $Z_F^{1k} Z_F^{4k\ast}$ {\em etc.} is ${\cal O}(1)$. Thus, contributions to $b_F$ as large as $\sim 10^{-3}$ can occur in the regime of maximal mixing and superpartner masses of order the weak scale. Measurements designed to observe $b_F$ at the few $\times 10^{-4}$ level would provide interesting probes of L-R mixing among first generation scalar fermions. 
Note that non-vanishing results at this scale would         
disfavor the alignment hypothesis of Eq.~(\ref{eq:triscalaryukawa}), which, for    
$|\mu|$ of order the electroweak scale, implies 
that L-R mixing for the first two generations is Yukawa suppressed. As in the case of the muon decay parameter $g^S_{RR}$, large L-R mixing for the first generation also implies that the masses of the Higgs bosons $H^0$, $A^0$, and $H^\pm$ are super heavy, leaving only the light SM-like Higgs $h^0$ as an experimentally accessible degree of freedom. On the other hand, null results would provide added experimental plausibility to the alignment idea.

\subsection{$\beta$-Decay Correlations}




Experimental studies of the $\beta$ spectral shape, angular distribution, and polarization can provide information on both non-$(V-A)\otimes(V-A)$ interactions as well as an alternate means of obtaining $V_{ud}$. As a specific illustration, we consider polarized neutron decay, for which 
both the vector and axial vector components of the CC contribute to the $ft$ value in Eq. (\ref{eq:ftsuper}) with
\begin{eqnarray}
\label{eq:ngt}
M_F^2 & = & 1 \\
\nonumber
M_{GT}^2 & = & 3\, g_A^2 
\end{eqnarray}
and where $g_A$ is the hadronic axial vector coupling that  characterizes the strong interaction renormalization of the axial vector quark current that enters the neutron decay matrix elements. The latter are given by 
\begin{eqnarray}
\label{eq:ncurrent}
\bra{p} {\bar u}(0) \gamma_\mu d(0)\ket{n}& = & {\bar U}_p (P') \left[ g_V(q^2)  \gamma_\mu + \frac{ i\,g_M(q^2)}{2 m_N} \sigma_{\mu\nu}\, q^\nu \right] U_n(P)\\
\nonumber
\bra{p} {\bar u}(0) \gamma_\mu\gamma_5 d(0)\ket{n} & = & {\bar U}_p (P') \left[ g_A(q^2) \gamma_\mu\gamma_5 + \frac{ g_P(q^2)}{m_N}\, q_\mu\gamma_5\right] U_n(P)\ \ \ .
\end{eqnarray}
At $q^2=0$ one has $g_V(0)=1$ and $g_M(0)=\kappa_P-\kappa_n$ according to the CVC property of the vector CC, whereas $g_A\equiv g_A(0) \approx 1.26$ and $g_P\approx 8.5$ (Here, we quote a value for $g_P$ taken from chiral perturbation theory that is in agreement with the results of ordinary muon capture experiments but that differs from the result obtained from radiative muon capture. For a recent discussion, see, {\em e.g.}, Ref.~\cite{Kammel:2002sd}.). Neglecting the small $q^2$-dependent corrections associated with the $g_M$ and $g_P$ terms, Eq.~(\ref{eq:ncurrent}) leads to the matrix elements in Eq. (\ref{eq:ngt}). Nucleon matrix elements of scalar and tensor operators, ${\bar u} d$ and ${\bar u}\sigma_{\mu\nu} d$, associated with non-SM interactions are parameterized by analogous form factors, $g_S$ and $g_T$, respectively.

At present, it is not possible to compute $g_A$ with the $0.1 \%$ precision needed for a $ 0.1\%$ determination of $V_{ud}$ from the neutron lifetime, $\tau_n$. Consequently, the axial vector contribution to $ft$ in Eq.~(\ref{eq:ftsuper}) must be separated experimentally from the vector contribution. Doing so requires measurement of a neutron decay correlation coefficient appearing the partial rate Eq.~(\ref{eq:betacor}). The coefficients $A$, $a$, and $B$ carry a dependence on $\lambda$ [see Eq.~(\ref{eq:corcoeff})], so that their measurement can yield the requisite separation. The most precise value of $\lambda$ has been obtained with a 0.6\% measurement of the $A$ parameter by the PERKEO collaboration, leading to $\lambda=-1.2739\pm0.0019$\cite{Abele:2002wc}. Since the publication of that result, a new value for  $\tau_n$ has been obtained at ILL that differs from the previous world average by more  than six standard deviations\cite{Serebrov:2004zf}. Including that result and performing a one-parameter fit to neutron decay measurements yields $\lambda=-1.27293(46)$. The resulting value for $V_{ud}$ is
\be
\label{eq:vudnnew}
V_{ud}=0.97757(65)\ \ \ 
\ee
compared with the value $V_{ud}=0.97192(65)$ obtained using the previous world average for $\tau_n$ and the new average for $\lambda$.

Several efforts are underway in order to obtain a value of $V_{ud}$ using a combination of the neutron lifetime, $\tau_n$, and correlation coefficient measurements, and a review of these studies can be found in Ref.~\cite{Erler:2004cx}. In light of the new $\tau_n$ result, additional precise determinations of the neutron lifetime are either underway or are being planned at ILL, NIST, and LANSCE.  

Precise measurements of neutron decay correlation coefficients may also probe the SUSY loop-induced non-$(V-A)\otimes(V-A)$ interactions. For example, the $\beta$ energy-dependent component of the neutrino asymmetry parameter $B$ depends linearly on both the scalar and tensor interactions of Eq.~(\ref{eq:box1}):
\bea
B_{\rm SUSY\,  box} & = & -2\left(\frac{\Gamma m}{E}\right)\, \frac{\lambda}{1+3\lambda^2}\,
{\rm Re}\, \Biggl\{ 4\lambda \left(\frac{g_T}{g_A}\right)\, \left(\frac{a^{T}_{RL}}{a^{V}_{LL}}\right)^\ast\\
\nonumber
&& +\left[2 \left(\frac{g_T}{g_A}\right)\,  \left(\frac{a^{T}_{RL}}{a^{V}_{LL}}\right)^\ast - \left(\frac{g_S}{g_V}\right)\, 
\left(\frac{a^{S}_{RL}+a^S_{RR}}{a^{V}_{LL}}\right)^\ast\right]\Biggr\}
\eea
As with the Fierz interference term, the SUSY contributions to $B$ can approach the $10^{-3}$ level for nearly maximal L-R mixing. Experimentally, future measurements using cold or ultracold neutrons may be able to determine the energy-dependent component of $B$ with a sensitivity of a few $\times 10^{-4}$\cite{brad}. At present, the results of nuclear $\beta$-decay correlation measurements do not appear to be sensitive to the linear interference of scalar and tensor interactions with the SM amplitude; improvements in superallowed sensitivity to $b_F$ and in neutron decay correlation measurements appear to hold the brightest prospects for probing the SUSY loop-induced non-$(V-A)\otimes(V-A)$ interactions.

\subsection{Pion $\beta$-decay}




In addition to the use of nuclear and neutron decay, measurements of the rate for pion $\beta$-decay ($\pi_\beta$) also yield a determination of $G_V^\beta$. In this case, the non-universal hadronic contributions to the radiative correction ${ \Delta \hat r}_\beta$ for $\pi_\beta$ differs from that for neutron and nuclear decays, and in the past it has been argued that the corresponding theoretical uncertainties for $\pi_\beta$ are smaller. However, a recent analysis using $\chi$PT  reported in Ref.~\cite{Cirigliano:2002ng} quotes a theoretical uncertainty in $V_{ud}$ of $\pm 0.0005$ -- a value that is larger than the new theoretical uncertainty associated with neutron and nuclear decays.

The rate for pion $\beta$-decay is given by
\be
\label{eq:pionbeta1}
\Gamma(\pi_\beta)= \frac{(G_V^\beta)^2 m_\pi^5 |f_{+}^\pi(0)|^2 I(\lambda_{+}^\pi)}{64\pi^3}\ \ \ ,
\ee
where $f_{\pm}^{\pi}(t)$ are the two pion form factors and $I(\lambda_{+}^\pi)$ is a phase space integral that results from inclusion of real photon emission and that is a function of the slope $\lambda_{+}^\pi$ of $f_{+}^\pi(t)$  at the photon point. The non-universal, long-distance ${\cal O}(\alpha)$ corrections can be included by 
replacing $f_{+}^\pi(t)$ by
\be
F_{+}^\pi(t,u) =  \left[1+\frac{\alpha}{4\pi}\Gamma(u, m_e^2, m_\pi^2, \lambda_\gamma)\right]\ \ \ ,
\ee
where $\lambda_\gamma$ is an infrared regulator whose effect on the total rate is cancelled by the corresponding $\lambda_\gamma$-dependence of $I(\lambda^\pi_{+})$. 

Contributions to $\Gamma(u, m_e^2, m_\pi^2, \lambda_\gamma)$ that are non-analytic in the various masses and momenta can be computed unambiguously at one-loop order and carry no dependence on {\em a priori} unknown parameters. However, there exist analytic contributions that arise at the same chiral order that are parameterized by three constants in the effective Lagrangian: $K_{12}^r(\mu)$,
$X_1$, and $X_6^r(\mu)$. Here, $K_{12}^r$ and $X_6^r$ depend on the renormalization scale $\mu$ since one-loop graphs generate divergences having the same structure as the corresponding terms in the Lagrangian.  A theoretical prediction for $K_{12}^r(m_\rho)$ with $\sim 10\%$ uncertainty has been given in Ref.~\cite{Moussallam:1997xx}, while bounds on $X_1$ and $X_6^r(m_\rho)$ were obtained by the authors of Ref.~\cite{Cirigliano:2002ng} using dimensional analysis. The uncertainty associated with these constants dominates the theoretical error in the extraction of $V_{ud}$ from $\Gamma(\pi_\beta)$.

Experimentally, the PIBETA collaboration has recently obtained the most precise determination of the ratio of branching ratios for the $\pi_{\beta(\gamma)}$ and $\pi_{e2(\gamma)}$ decays\cite{Pocanic:2003pf}. Multiplying by the current world average for the $\pi_{e2(\gamma)}$ branching ratio leads to the $\pi_\beta$ branching 
ratio\cite{Blucher:2005dc}
\be
\label{eq:pionbeta2}
B_{\pi_\beta(\gamma)} = \left[1.036 \pm 0.004 ({\rm stat}) \pm 0.004 ({\rm sys}) \pm 0.003 (\pi_{e2(\gamma)})\right]\, \times10^{-8} 
\ee
for a total experimental error of $\pm 0.006\times 10^{-8}$. Note that the fractional uncertainty is roughly a factor of ten times larger than the theoretical error and fifteen times larger than the combined experimental and nuclear structure theory uncertainty in the superallowed $\overline{{\cal F}t}$ value. Thus, considerable experimental and theoretical progress is necessary before a value for $V_{ud}$ can be obtained from $\pi_\beta$ with precision competitive with the superallowed value.

\subsection{Kaon decays and $V_{us}$}




In order to use the value of $V_{ud}$ (\ref{eq:vudnuc}) as a probe of SUSY, one must compare it with the SM expectation. In this case, the SM implies that the CKM matrix is unitary, so that
\be
\label{eq:ckm1}
|V_{ud}|^2+|V_{us}|^2+|V_{ub}|^2= 1  \qquad {\rm SM}\ \ \ .
\ee
The value of $V_{ub}=0.0032\pm 0.0009$ is obtained from the decays of $B$-mesons. Given the level of uncertainty in $V_{ud}$, both the magnitude of $V_{ub}$ and its error are too small to affect a test of Eq.~(\ref{eq:ckm1}). On the other hand, the value of $V_{us}$ as well its uncertainty are critically important. The theoretically cleanest determination of $V_{us}$ is obtained from the branching ratios for $K_{\ell 3}$ decays, $K\to\pi \ell \nu$. The partial rate for this decay mode is given by\cite{Blucher:2005dc}
\be
\label{eq:ke3partial}
   d\Gamma(K^+_{\ell 3}) = \frac{G_\mu^2 m_K^5}{128\pi^3} S_{\rm EW} C(t) |V_{us}|^2 |f_{+}^K(0)|^2 \left[ 1 +
   \frac{\lambda_{+}^K\, t}{m_\pi^2}\right]^2\left[1+2\Delta^K_{SU(2)}+2\Delta^{K\ell}_{EM}\right]\ \ \ ,
\ee  
where $f_{+}^K(t)$ is the $K$-to-$\pi$ transition form factor with $t=(p_K - p_\pi)^2$, $\lambda_{+}^K/m_\pi^2$ is the slope of the form factor at $t=0$, and $C(t)$ is a kinematic function that depends on the kaon form factor, $S_{\rm EW}$ contains the short-distance electroweak radiative corrections, and $\Delta^K_{SU(2)}$ and $\Delta^{K\ell}_{EM}$ indicate corrections generated by the breaking of flavor SU(2) and long-distance electromagnetic corrections, respectively\cite{Cirigliano:2001mk}. To carry out a test of CKM unitarity at the $0.1\%$ level, one must include the  $\Delta^K_{SU(2)}$ and $\Delta^{K\ell}_{EM}$ corrections and determine the value of $f^K_{+}(0)$ with one percent uncertainty or better.

As with the determination of $V_{ud}$ using the $\beta$-decay half lives, arriving at a value for $V_{us}$ from the $K_{\ell 3}$ partial rates requires both experimental and theoretical input. Experimentally, new determinations of the $K_{\ell 3}$ branching ratios\cite{Alexopoulos:2004sx,Ambrosino:2005ec, Lai:2004bt,Sher:2003fb,ambrosino}  -- combined with experimental values for $\lambda_{+}^K$  and $C(t)$ \cite{Alexopoulos:2004sy,Lai:2004kb,Yushchenko:2004zs}-- have yield a shift in the world average for the product $V_{us}\times f_{+}^K(0)$. The 2005 Particle Data Group value is
\be
V_{us}\, \left[ f_{+}^K(0)/0.961\right ] = 0.2257(9)  
\ee
Here the value of $f_{+}^K(0)$ has been normalized to the combined chiral perturbation theory ($\chi$PT)-quark model prediction of Leutwyler and Roos\cite{Leutwyler:1984je}  that had been used for many years.  That prediction includes contributions from non-analytic, one-loop terms through chiral order $p^4$, analytic ${\cal O}(p^4)$ terms that can be obtained from fits to other pseudoscalar meson observables, and a quark model estimate of the ${\cal O}(p^6)$ contribution. 

Recently, the ${\cal O}(p^6)$ loop contributions have been computed in Refs.~\cite{Post:2001si,Bijnens:2003uy}. In this context, the dominant, remaining uncertainty is associated with the ${\cal O}(p^6)$ analytic contributions that depend on the square of the ${\cal O}(p^4)$ constant $L^r_5(\mu)$ and  two of the 94 unknown constants appearing in the ${\cal O}(p^6)$ chiral Lagrangian, $C^r_{12}(\mu)$ and $C^r_{34}(\mu)$:
\be
\label{eq:fplus6}
   f_{+}^{K\, (6),\, {\rm analytic}} = 
   8{(m_\pi^2-m_K^2)^2\over F_\pi^4}\left[\frac{L_5^r(\mu)^2}{F_\pi^2}-C_{12}^r(\mu)-C_{34}^r(\mu)\right]+\cdots \ , 
\ee
where $\mu$ is the renormalization scale, usually taken to be $\sim \Lambda_\chi$. The constant $L^r_5(\mu)$ is presently well known from experiment, whereas the determination of $C^r_{12,34}(\mu)$ require additional  input. 
In principle, new measurements of the pion and kaon scalar form factors $f^{\pi,\, K}_0(t)$ could allow a determination of these unknown constants, removing this remaining uncertainty. In particular, a 5\% measurement of the slope of $f_0^K(t)$ and a 20\% determination of its curvature -- coupled with a 1\% theoretical determination of the ratio of decay constants $F_K/F_\pi$ from theory -- would be sufficient to determine $C^r_{12,34}(\mu)$ at a level needed for the first row CKM unitarity test\cite{Blucher:2005dc,Bijnens:2003uy}. 

Alternately, values can be determined theoretically. Recent work using large N$_C$ QCD has yield a value for $C^r_{12, 34}$ leading to the prediction\cite{Cirigliano:2005xn} 
\be
\label{eq:fplusres}
f_{+}^K(0)_{{\rm large}\, N_C} = 0.984 \pm 0.012\ \ \ .
\ee
A number of lattice QCD computations of $f_{+}^K(0)$ have been carried out that yield, in effect, the sum of the nonanalytic and analytic terms in the $\chi$PT analysis.  The results tend to favor a smaller value for $f_{+}^K(0)$ that is consistent with the Leutwyler and Roos estimate\cite{Leutwyler:1984je}:
\be
f_{+}^K(0)_{\rm lattice} =
\left\{
\begin{array}{ll}
0.960\pm 0.005_{\rm stat} \pm 0.007_{\rm sys}\, &  {\rm quenched,\ Wilson}
[119]
\\
0.962(6)(9)\,&  {\rm unquenched,\ staggered}
[120]
\\
0.952(6)\,&  {\rm unquenched,\ Wilson} 
[121]
\\
0.955(12)\,& {\rm unquenched,\ doman\, wall} 
[122]
\end{array}
\right. , 
\ee
where recent quenched and unquenched results are shown for different lattice fermion actions (Wilson, staggered, domain wall). Note that the quenched results of Ref.~ \cite{Becirevic:2004ya} do not include any systematic uncertainty associated with the quenched approximation. 

An alternative determination of $V_{us}$ can be made by comparing the rates for the leptonic decays of the charged kaon and pion\cite{Marciano:2004uf}. From Eq.~(\ref{eq:piona}) and the analogous expression for the charged kaon decay rate one has
\be
\frac{  \Gamma[K^+ \to \mu^+ \nu (\gamma)]}{  \Gamma[\pi^+ \to \mu^+ \nu (\gamma)]} = \frac{V_{us}^2}{V_{ud}^2}\, \frac{F_K^2}{F_\pi^2}\, \frac{m_\pi^3}{m_K^3}\, \frac{(m_K^2-m_\mu^2)^2}{(m_\pi^2-m_\mu^2)^2}\, \left[1-\frac{\alpha}{\pi}\Delta_{K\pi}\right]\ \ \ ,
\ee
where $\Delta_{K\pi}$ gives the difference in the radiative corrections entering the RHS of Eq.~(\ref{eq:piona}). The  uncertainty from the hadron structure-dependent contributions to this difference has been estimated to be $\pm 0.75$, corresponding to an uncertainty of $\sim 0.1\%$ in the ratio $(V_{us}/V_{ud})^2$. Preliminary  lattice QCD results for the pseudoscalar decay constants obtained by the MILC collaboration give $F_K/F_\pi=1.198\pm 0.003^{+0.016}_{-0.005}$\cite{Bernard:2005ei}. The resulting theoretical uncertainty in $V_{us}$ obtained by this technique is comparable with that entering the analysis of $K_{\ell 3}$ branching ratios. Using a new result for the $K_{\mu 2}$ branching ratio obtained by KLOE one obtains $V_{us}=0.2245^{+0.0011}_{-0.0031}$.

Using these results and those for $V_{ud}$ from the superallowed decays, one obtains for the first row of the CKM matrix
\be
|V_{ud}|^2+|V_{us}|^2+|V_{ub}|^2=
\begin{cases}
0.9968\pm 0.0014, & {\rm large}\   N_C
[118]
\\
0.9998\pm0.0015, & {\rm unquenched\, lattice,\ domain\, wall }
[122]
\end{cases}\ \ \ .
\ee

\subsection{CKM Unitarity Tests: Implications for SUSY}




The implications of the CKM unitarity test for SUSY can be significant. In the presence of RPV interactions, a CKM unitarity deficit could be remedied by having $[{\Delta\hat r}^V_\beta-{\Delta\hat r}_\mu]_{\rm RPV} < 0$, thereby implying that $\Delta_{12k}({\tilde e}_R^k) > \Delta_{11k}^\prime({\tilde d}_R^k)>0$ [see Eq.~(\ref{eq:ckm2})]. On the other hand, consistency with CKM unitarity would allow both $\Delta_{12k}({\tilde e}_R^k)$ and $\Delta_{11k}^\prime({\tilde d}_R^k)$ to be nonzero, but would imply a strong correlation on their magnitudes. 
RPV corrections of order $\sim 0.1\%$ are not unreasonable from the standpoint of expected magnitudes of superpartner masses or couplings. Indeed, having
\be
\Delta_{ijk}({\tilde e}_R^k)\sim 0.1
\ee 
implies that
\be
\frac{m_{\tilde e_R^k}}{100\, {\rm GeV}} \sim 40 \lambda_{ijk}
\ee
or $M_{\tilde e_R^k}\sim 1$ TeV for $\lambda_{ijk} \sim\sqrt{4\pi\alpha}$. As discussed in Ref.~\cite{Ramsey-Musolf:2000qn}, RPV corrections of this magnitude are not inconsistent with analogous bounds on RPV corrections obtained from other precise measurements, such as the studies of rare decays or flavor-changing neutral current processes. 

The implications for superpartner loop corrections for CKM unitarity have been studied in Ref.~\cite{Kurylov:2001zx}. The contributions from $W$ propagator  $We{\bar\nu}_e$ vertex and external leg corrections cancel from the difference $[{\Delta\hat r}^V_\beta-{\Delta\hat r}_\mu]$, leaving only the dependence on the $W\mu{\bar\nu}_\mu$ and $Wd{\bar u}$ vertex and external leg corrections as well as the box graphs entering the $\beta$- and $\mu$-decay amplitudes. As with the corrections to $R_{e/\mu}$, these box graphs are subdominant in the presence of gaugino-higgsino mixing, as the vertex and and external leg corrections receive logarithmic enhancements. As a result, the dependence of the correction $[{\Delta\hat r}^V_\beta-{\Delta\hat r}_\mu]$ on the parameters in ${\cal L}_{\rm soft}$ simplifies considerably, and the authors of Ref.~\cite{Kurylov:2001zx} were able to perform an analytic study of the corresponding parameter dependence. 

To illustrate the dependence of $[{\Delta\hat r}^V_\beta-{\Delta\hat r}_\mu]$ on the SUSY parameters, it is useful to consider several representative cases:

\begin{itemize}

\item[(i)] For situations in which the scalar superpartners of left- and right-handed fermions mix via the $\mu$-term and tri-scalar SUSY-breaking interactions into mass eigenstates ${\tilde f_{1,2}}$, one has 
\be
\left[{\Delta\hat r}^V_\beta-{\Delta\hat r}_\mu\right]_{\rm SUSY\ loop}\sim \frac{\alpha({\hat c}^2-{\hat s}^2)}{32\pi^2 {\hat c}^2 {\hat s}^2}\ \ln\left(\frac{m^2_{\tilde q_2}}{m^2_{\tilde q_1}} \frac{m^4_{\tilde \mu_1}}{m^4_{\tilde \mu_2}}\right)+\cdots\ \ \ ,
\ee
where the $m_{\tilde f_i}$ are the corresponding physical masses and where this expression holds in the limit $|m_{\tilde f_2}-m_{\tilde f_1}| >> m_{\tilde f_{1}}$. Note that the overall sign of $[{\Delta\hat r}^V_\beta-{\Delta\hat r}_\mu]_{\rm SUSY\ loop}$  depends on the relative degree of mass splitting among the sleptons and squarks. At the time when the analysis of Ref.~\cite{Kurylov:2001zx} was carried out, the CKM unitarity deviation favored a negative sign, leading to the requirement
\be
\label{eq:susybeta1}
\frac{m_{\tilde \mu_2}}{m_{\tilde \mu_1}} >\left( \frac{m_{\tilde q_2}}{m_{\tilde q_1}} \right)^{1/2}
\ee
For nearly degenerate squarks, one would need $m_{\tilde\mu_2}\gsim 3 m_{\tilde\mu_1}$. For values of $m_{\tilde\mu_1}$ close to the current direct search bounds, a mass splitting of this magnitude is ruled out by the recent muon $(g-2)_\mu$ results\cite{Bennett:2004pv}. 

\item[(ii)] The constraints from $(g-2)_\mu$ may be evaded by taking $M_{LR}^2\approx  0$ for the scalar fermions\footnote{From Eq. (\ref{eq:susybeta1}) we observe that the degree of mixing among squarks must always be less than that for smuons. Hence, setting $M_{LR}^2=0$ for the smuons implies a similar condition for the squarks.}. In this case, only the superparters of the left-handed fermions contribute to the CC weak interaction, and the requirements on the scalar fermion spectrum are rather different. To illustrate, we consider the limit of large sfermion masses, yielding the asymptotic expression
\be
\left[{\Delta\hat r}^V_\beta-{\Delta\hat r}_\mu\right]_{\rm SUSY-loop}\sim\frac{\alpha}{2\pi}\, \cos{2\beta}\, \left[\frac{1}{3}\frac{M_Z^2}{m_{\tilde q}^2}\ln\frac{m_{\tilde q}^2}{\langle M_{\tilde\chi}^2\rangle}-\frac{M_Z^2}{m_{\tilde \mu}^2}\ln\frac{m_{\tilde \mu}^2}{\langle M_{\tilde\chi}^2\rangle}\right]+\cdots\  \ \ ,
\ee
where $\langle M_{\tilde\chi}^2\rangle^{1/2} $ is the mass scale associated with the charginos and neutralinos. For $\tan\beta>1$ as currently favored, one requires $m_{\tilde\mu}^2\gsim 3
m_{\tilde q}^2$ in order to obtain a negative sign for $[{\Delta\hat r}^V_\beta-{\Delta\hat r}_\mu]_{\rm SUSY-loop}$ (up to small logarithmic corrections).  Note that such a sfermion spectrum would conflict with models that assume a universal sfermion mass at high scales, since in this case SU(3)$_C$ contributions to the renormalization group evolution of the first generation squark masses increases their magnitude at the weak scale relative to that of the first and second generation slepton masses. 

\item[(iii)] A final possibility arises when there exists significant mixing among either the $u$ or $d$-type squarks, but in a way that is not identical for both. In this case, gluino loop effects dominate the SUSY corrections to the $Wud$ vertex, and one can accommodate a negative sign for $[{\Delta\hat r}^V_\beta-{\Delta\hat r}_\mu]_{\rm SUSY-loop}$ without requiring significant mixing among the smuons. In order to evade the $(g-2)_\mu$ constraints, the dominant contribution to $M_{LR}^2$ for the $d$ squarks must arise from a large value for the triscalar coupling $A_d$. In this case, avoiding color or charge-breaking minima in the scalar potential implies that all by the CP-even Higgs $h^0$ must be quite heavy. One can suppress these supersymmetric SU(3)$_C$  loop corrections for $M_3\gsim 500$ GeV. In this case, however, one returns effectively to the situation characterized by item (i), leaving item (ii) as the only viable option. 

\end{itemize}

At present, the CKM first row unitarity situation is unclear, so one cannot draw strong conclusions about either PRV or R-parity conserving scenarios. However, the prospective implications for the superpartner spectrum or presence of RPV interactions underlines the importance of resolving the various theoretical and experimental uncertainties germane to CKM unitarity outlined above.

\section{Neutral Current Experiments}
\label{sec:nc}

\subsection{Introduction}
Historically, the study of parity-violating (PV) neutral current interactions has played an
important role in elucidating the structure of the electroweak
interaction. In the 1970's,  PV deep inelastic scattering (DIS)
measurements performed at 
SLAC confirmed the SM prediction for the structure of weak neutral
current interactions \cite{SLAC}. These results were consistent with a
value for the weak mixing angle given by $\sstw\approx 1/4$,
implying a tiny $V$(electron)$\times A$(quark) neutral current
interaction.  Subsequent PV measurements -- performed at both very low
scales using atoms as well as at the $Z$-pole in $e^+e^-$ annihilation
-- have  been remarkably consistent with the results of the SLAC DIS
measurement.

The value of $\sinhat$ at scale $Q=\sqrt{|q^2|}=M_Z$ 
has been determined precisely via $Z$-pole
precision measurements at LEP and SLD.   The global fit to all the 
precision observables gives \cite{pdg}
\beq
\sinhat(M_Z)=0.23118 \pm 0.00017
\eeq
The SM predicted scale-dependence of $\sinhat$, on the other hand,
has never been established experimentally to a high precision.  
The solid curve in  Fig.~\ref{fig:sin2theta} (see Section \ref{sec:renorm}) shows the running of weak
mixing angle  $\sstw$ [in the modified minimal subtraction 
($\overline{\rm MS}$) scheme] 
as a function of the scale $Q$. 
The dip around $\mw$ is due to the
decoupling of $W$ boson when $Q<\mw$.  
When we approach lower and lower energy, quarks with mass $m_q > Q$ 
decouple and the slope of the $\sinhat$ running changes.
For $Q<1$ GeV, all the heavy
quarks  decouple and $\sinhat$ is roughly a constant.  The
difference between $\sinhat$ at $Z$-pole and $\sinhat$
at low energy is: $\sinhat(0)-\sinhat(\mz)=0.00749\pm 0.00015 \pm 0.00007$, where the first error is the experimental error\footnote{Here, we have taken the value of  $\sinhat(\mz)$ from a fit to precision data rather than the value computed from Eq.~(\ref{eq:Gfswmz}).}  $\sinhat(\mz)$ and the second is the theoretical error associated with the running to $Q=0$.

Recently, determinations of $\sinhat$ at various low energy
scales have been performed, although the experimental error 
is still relatively large.
The cesium atomic parity-violation (APV) experiment measured the parity-forbidden transition between
atomic states by exploiting Stark interference effects \cite{APV}.  The cesium  weak charge,
which depends on $\sinhat$ at $Q\approx 0$, 
appears to be consistent with the SM prediction.
At higher energies, the NuTeV collaboration measured $\nu$- ($\bar\nu$-) nucleus 
deep inelastic scattering~\cite{NuTeV}, and studied the
ratio between the cross section of the neutral current and 
that of the charged current.  The result can be interpreted as
a determination of $\sinhat$ at $Q\approx 3$ ${\rm GeV}$.
The observed deviation of cross section ratios from the SM predictions 
implies a $+3\sigma$ deviation in $\sinhat$ at that scale.

More recently, the SLAC E158 collaboration measured the electron weak charge via 
the  PV  M{\o}ller ($ee$) scattering \cite{E158}. 
It determined 
$\sinhat$ at $Q^2\approx 0.026\ {\rm GeV}^2$ and obtained a result that  is consistent with the 
SM prediction at the 1.1 $\sigma$ level.
The Qweak experiment, which plans to measure the proton weak charge 
via elastic PV $ep$ scattering using polarized electron beam at the Jefferson Laboratory (JLab),
is currently under construction.  The expected $4\%$ measurement of 
proton weak charges at $Q^2\approx 0.03 {\rm GeV}^2$ 
corresponds to $0.3\%$ determination of $\sstw$:
$\delta \sstw =0.0007$ \cite{QWEAK}, better than any of the 
current measurements.

Several future PV experiments are being considered.
One proposal involves a more precise version of PV M{\o}ller ($ee$) scattering at 
JLab  using the planned 12 GeV upgrade of the accelerator.  With a $2.5\%$ precision in electron weak charge 
measurement at $Q^2\approx 0.008\ {\rm GeV}^2$, $\sinhat$ could
be determined with an error of $\delta \sinhat =0.00025$ \cite{jlabmoller},
comparable to the $Z$-pole measurements.
Another proposal is to study the PV electron-Deuterium DIS at JLab 
with current 6 GeV electron beam, or at future 12 GeV upgrade. 
Although the expected precision is worse than the Qweak experiments, 
such measurements have unique opportunity to probe the $V(e)\times A(q)$
interactions,  QCD higher twist effects, possible charge symmetry violation in the parton distribution functions, and the ratio $d(x)/u(x)$ at $x\to 1$. An DIS electron-Deuterium experiment has been approved for running with the present 6 GeV beam\cite{pvdis6GeV}, and several options for experiments using the future 12 GeV beam are under consideration\cite{eDDIS}.

\begin{table}
\begin{tabular}{lcc}
\hline
Measurements&$\delta\sstw/\sstw$&$\delta\sstw$\\ \hline
Z-pole&0.07\%&0.00017\\
0.6\% APV $Q_W({\rm Cs})$&0.7\%&0.0016 \\
NuTeV $\nu$-DIS &0.7\%&0.0016\\
13.1\% SLAC E158 $Q_W(e)$&0.5\%&0.0013\\
$*$2.5\% JLab  M{\o}ller  $Q_W(e)$&0.1\%&0.00025\\
4\% JLab Qweak $Q_W(p)$&0.3\%&0.00072\\
$*$0.8\% JLab eD DIS-parity &0.45\%&0.0011\\ \hline
\end{tabular}
\caption{Precision of various experiments which are sensitive to the value 
of $\sstw$ at low energies.  Entries with $*$ are ideas for future 
experiments.}
\label{tab:sin2thetaprecision}
\end{table}

In Table~\ref{tab:sin2thetaprecision}, we list the precision of 
various current measurements and possible future experiments, along with the 
sensitivity of the measurements of $\sinhat$.  Precise determinations of 
the value of $\sinhat$ at different scales, obtained from different types of experiments,
will provide a consistency check of the SM at loop level.
Any significant
deviation from the SM prediction 
would constitute striking evidence for new physics.  
These high precision,  
low energy measurements will be sensitive 
to the new physics up to TeV scale.  For example, we can write down the 
effective four fermion operators that contribute to 
the parity-violation $eq$ scattering as \cite{MRM99}
\begin{equation}
{\cal{L}}_{eq}^{PV}={\cal{L}}_{\rm SM}^{PV}+{\cal{L}}_{\rm new}^{PV}
=-\frac{G_{\mu}}{2 \sqrt{2}}\bar{e}\gamma_{\mu}\gamma_5e
\sum_{q} Q_W^q\bar{q}\gamma^{\mu}q+\frac{g^2}{4 \Lambda^2}
\bar{e}\gamma_{\mu}\gamma_5e\sum_{q}h_V^q\bar{q}\gamma^{\mu}q,
\end{equation}
where the first term is the SM contribution, and the second term
gives new physics effects. 
$g$ is the typical new physics coupling, $\Lambda$ is the new 
physics scale, and $h_V^q$ is an ${\cal O}(1)$ coefficient that 
parameterizes the new physics contributions for different quarks.  A $4\%$ measurement of 
proton weak charge $Q_W^p$ corresponds to a probe 
of the new physics scale of 
\begin{equation}
\frac{\Lambda}{g}\sim \frac{1}{\sqrt{\sqrt{2}G_{\mu}|\delta Q_W^p|}}
\sim 4.6\  {\rm TeV}.
\end{equation}

Table~\ref{tab:newphysicsscale} summarizes the sensitivity to various 
new physics scale from current and future parity-violation 
experiments \cite{jlabmoller}.
Also shown are the direct collider search limits for new physics
from current colliders (LEP2, CDF and HERA) and indirect search limit from current 
electroweak precision fit.
The scales that the low energy precision measurements
can probe is close to --  and in some cases  even exceeds --  the ones accessible in high energy, direct searches.  Even after LHC begins running,  the low energy measurements will be able to probe new physics scale 
comparable to the LHC reach.  Once any new physics is discovered at LHC, low energy
precision measurements can be used to probe details of the new physics, such as 
the couplings and charges of new particles.

\begin{table}
\begin{tabular}{c|cc|cc|cc}
\hline
&\multicolumn{2}{c|}{$Z^\prime$ models}&
\multicolumn{2}{c|}{leptoquark}&\multicolumn{2}{c}{compositeness}\\
&$m(Z_X)$&$m(Z_{LR})$&$m_{LQ}$(up)&$m_{LQ}$(down)&$e-q$&$e-e$\\ \hline
Current direct search limits&0.69&0.63&0.3&0.3&--&--\\
Current electroweak fit&0.78&0.86&1.5&1.5&11$-$26&8$-$10\\
0.6\% $Q_W({\rm Cs})$&1.2&1.3&4.0&3.8&28&-- \\
13.1\% $Q_W(e)$&0.66&0.34&--&--&--&13\\
$*$ 2.5\% $Q_W(e)$&1.5&0.77&--&--&--&29\\
4\% $Q_W(p)$&0.95&0.45&3.1&4.3&28&--\\
\hline
\end{tabular}
\caption{Sensitivity of new physics scale from various current and future
low energy precision measurements.  Entry with $*$ is an idea for future 
experiment.  Also shown are direct search limits from current colliders
(LEP, CDF and Hera) and indirect search limit from current electroweak precision fit.  The various new physics scale presented here 
are mass of $Z^\prime$ with extra U(1) [$m({Z_X})$], or 
in left-right models [$m(Z_{LR})$],  mass of 
leptoquark in up quark sector [$m_{LQ}$(up)], or down quark sector
[$m_{LQ}$(down)], composite scale for $e-q$ composite, or $e-e$
composite.   
Entries with ``--'' either do not exist or do not apply.  This Table is updated from 
Ref.~ \cite{jlabmoller}.}
\label{tab:newphysicsscale}
\end{table}

In the following sections, 
we discuss in detail three different types of neutral current
experiments.   Sec.\ref{sec:pves} is devoted to 
parity violating electron scattering (PVES),
which includes $ee$ M{\o}ller scattering,
$ep$ elastic scattering, and $eD$ deep inelastic scattering.
Atomic parity violation is discussed in Sec.~\ref{sec:apv}, and
neutrino-nucleus DIS is discussed in  Sec.~\ref{sec:nutev}. 

\subsection{Parity Violating Electron Scattering: 
M{\o}ller and Qweak}
\label{sec:pves}

The SLAC E158 experiment measured PV $ee$ M{\o}ller scattering at 
$Q^2 \sim 0.026 {\rm GeV}^2$ \cite{E158}, while the 
Qweak experiment at JLab will measure 
PV $ep$ scattering at $Q^2 \sim 0.03 {\rm GeV}^2$  \cite{QWEAK}.  
In both cases, polarized electron beams are used to measure the PV asymmetry  
\beq
A_{PV}=\frac{N_R-N_L}{N_R+N_L} \ \ \ ,
\eeq 
where $N_R$ ($N_L$) is the number of detected events for incident electrons with positive (negative) helicity.  Although the dominant contribution to 
the scattering is via  parity  conserving photon exchange,  the 
interference between the photon exchange and the  parity violating $Z$ exchange
generates the PV asymmetry and filters out the much larger electromagnetic scattering effects.  
At leading order in $Q^2$, the contributions to
$\apv$ are governed by $A(e)\times V(f)$ operator, with the coefficient being
$Q_W^f$, the \lq\lq weak charge" of the target
fermion, $f$ ($f=e$ for $ee$ scattering, and $f=p$ for $ep$
scattering, where $Q_W^p=2 Q_W^u+Q_W^d$).  
%
\begin{eqnarray}
\label{eq:Leff}
{\cal L}_{EFF}^{ef}=-\frac{G_\mu}{2\sqrt 2}Q_W^f {\bar e}
\gamma_\mu\gamma_5 e {\bar f}\gamma_\mu f~.
\label{eq:qw}
\end{eqnarray}
At tree-level in the SM the weak charges of both the electron and the
proton are suppressed: $Q_W^p=-Q_W^e=1-4\sstw\approx
0.1$. One-loop SM electroweak radiative corrections further reduce
this small number, leading to the predictions
$Q_W^e=-0.0449$ \cite{Mar96,Erl-MJRM-Kur02} and
$Q_W^p=0.0716$ \cite{Erl-MJRM-Kur02}. The   factor of $\gsim$10
suppression of these couplings in the SM renders them more transparent
to the possible effects of new physics. Consequently, experimental
precision of order a few percent, rather than a few tenths of a
percent, is needed to probe SUSY loop corrections.  
The advantages of  measuring $\sstw$
in these two PVES experiments are that hydrogen target  is used in
both experiments, which is a relatively clean environment  since the
QCD contamination of heavy nuclei can be avoid.  It is also
theoretically clean since the hadronic uncertainties are relatively
small and  under control\cite{Mar96,Erl-MJRM-Kur02} .

The $ee$ M{\o}ller scattering  experiment measured a 
parity violating asymmetry \cite{E158}: $A_{LR}=
(-131 \pm 14 \ ({\rm stat.}) \pm 10 \ 
({\rm syst.})) \times 10^{-9}$, leading to
the determination of the weak mixing angle 
$\sstw=0.2397 \pm  0.0010 \ ({\rm stat.}) \pm 0.0008 \ ({\rm syst.})$,
evaluated at $Q^2=0.026 {\rm GeV}^2$.  Comparing to the SM prediction of 
$\sstw = 0.2381 \pm 0.0006$ at this energy scale, the E158 results 
agreed with the SM value at 1.1 $\sigma$.  When expressed in terms of 
the electron weak change, the difference between the measured value
and the theoretical predicted one is 
$\delta Q_W^e=(Q_W^e)_{\rm exp}-(Q_W^e)_{\rm SM}=0.0064 \pm 0.0051$.

In the Qweak experiment, the SM prediction for the left-right asymmetry
is $290\times10^{-9}$.  
The expected experimental precision for $ep$
scattering at Qweak experiment is $4\%$.  When translated
to a precision in the $\sstw$ measurement, it corresponds to
$\delta\sinhat$ of 0.0007, 
which is two times smaller than the precision in Cs APV
and in NuTeV measurement.

In order to interpret these  high precision  electron scattering experiments in terms of possible new physics, 
it is crucial to have the hadronic uncertainties under control.  There are two
kinds of QCD uncertainties that one must consider: (a)  QCD corrections to the weak charge 
itself, and (b) QCD effects that impact the extraction of 
the weak charge from the experimentally measured 
PV asymmetry.  
The QCD corrections to weak charges have been  discussed in Sec.~\ref{sec:renorm}.
In  the case of Qweak, the latter set of uncertainties  is brought under control by a judicious choice of kinematics and by extrapolating the results of  other experimental measurements.
At forward angle $\theta$, the PV asymmetry can be written 
as~\cite{musolf1995}
\beq
A_{PV}=\frac{G_{\mu} Q^2}{4 \sqrt{2}\pi\alpha}\left[
Q_W^p+F(\theta, Q^2)
\right],
\eeq
where $F(\theta, Q^2)$ is the unknown form factor,  
that is proportional to $Q^2$ at low energy. 
The form factor  $F(\theta, Q^2)$ depends on a linear combination of isovector and isoscalar electromagnetic (EM) form factors as well as those associated explicitly with strange quarks. While the results of parity conserving electron scattering experiments provide the needed isovector and isocalar EM contributions to  $F(\theta, Q^2)$, the strange quark contributions can only be determined by additional PV electron scattering measurements. These contributions to $F(\theta, Q^2)$ -- which vanish with $Q^2$ as $Q^2\to 0$ -- have been studied with an extensive program of measurements by the SAMPLE~\cite{sample}, 
HAPPEX~\cite{HAPPEX}, PV A4 ~\cite{A4}, and G0~\cite{G0} Collaborations.  The results yield tight constraints on the strange quark contributions to $F(\theta, Q^2)$, which one can then extrapolate 
to the much smaller $Q^2$ relevant to the  Qweak experiments.   On the other hand, one cannot take
$Q^2$ to be too small since $\apv$  itself is proportional to $Q^2$  and
the statistical error increases for smaller $Q^2$.  Experimentally, these
two effects are optimized and $Q^2$ is
chosen to be about 0.03 ${\rm GeV}^2$.  The resulting hadronic uncertainty 
from the form factor is about 2\%, half of the total 
experimental uncertainties. 

The precise measurements of the weak charges could probe both supersymmetric loop effects as well as tree-level, RPV contributions. For the $R$-parity conserving MSSM, contributions to the weak charge $Q_W^e$
and $Q_W^p$ appear at loop level.   With higher-order corrections
included, the weak charge of a fermion $f$ can be written in terms of the parameters $\hat\rho_{NC}(0)$, $\hat\kappa$, and $\hat\lambda^f_{V,A}$ introduced in Section \ref{sec:renorm}:
\begin{equation}
\label{eq:C1f-radcorr}
Q_W^f = \hat\rho_{NC}(0) \left[2 I_3^f -4
Q_f\hat\kappa(0,\mu)\sinhat(\mu)\right]+\hat\lambda^f_V +\left(2I_3^f-4Q_f {\hat s}^2\right)\hat\lambda^f_A+{\rm box}\ \ \ ,
\end{equation}
where $I_3^f$ and $Q_f$ are, respectively, the weak isospin and the
electric charge of the fermion $f$, and 
$\hat{s}^2\equiv \sstw(M_Z^2)$.
The quantities $\hat\rho_{NC}(0)$ and
$\hat\kappa(0,\mu)$ are universal in that they do not depend on the fermion
$f$ under consideration. 
Detailed expressions for $\hat\rho$ and $\hat\kappa$ can 
be found in Sec.~\ref{sec:renorm}. 
The corrections $\hat\lambda^f_{V,A}$, on the other
hand, depend on the fermion species. They include
the vertex and  external leg corrections to the weak charge.
The relevant expressions for supersymmetric contributions to the $\hat\lambda^f_{V,A}$  can be found in 
Ref.~\cite{sumichaelpves}\footnote{In these studies, the electron anapole contribution was included in the parameter $\hat\kappa$ and we denoted $\kappa_{PV}$; the quantity $\hat\rho_{NC}(0)$ was denoted $\rho_{PV}$; and the sum of the terms containing the $\hat\lambda^f_{V,A}$ in Eq.~(\ref{eq:C1f-radcorr} were denoted by $\lambda_f$, which did not include the electron anapole moment contribution.} .
Note that at tree-level, one has
$\hat\rho_{NC}=1=\hat\kappa$ and $\hat\lambda^f_{V,A}=0$, while 
both  SM and new physics contributions enter at loop level. Consequently, in what follows we will refer to $\delta\hat\rho=\hat\rho_{NC}-1$ and $\delta\hat\kappa=\hat\kappa-1$.

\begin{figure}[ht]
\begin{center}
\includegraphics[width=4in]{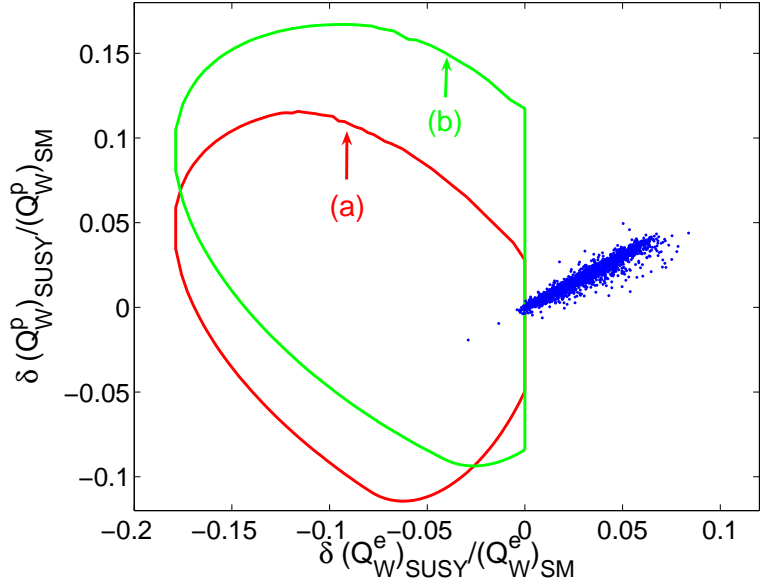}
\caption{Relative shifts in electron and proton weak charges
due to SUSY effects (updated plot from Ref. \cite{sumichaelpves}).  
Dots indicate MSSM loop corrections for $\sim 3000$
randomly-generated SUSY-breaking parameters. Interior of truncated
elliptical regions (a) and (b) give possible shifts
due to R-parity non-conserving SUSY interactions (95\% confidence),
using value of 
$\delta|V_{ud}|^2/|V_{ud}|^2$ (a) and (b) in Table.~\ref{tab:rpv-constrain}, respectively. 
}
\label{fig:mssm-vs-parity}
\end{center}
\end{figure}

The MSSM loop contributions to the electron and proton weak charges
have been analyzed in detail  in \cite{sumichaelpves}, which 
is reproduced in Fig.~\ref{fig:mssm-vs-parity}. Here, we plot 
the MSSM loop contributions
to the shift in the weak charge of the proton,  $\delta Q_W^p =
2\delta Q_W^u+ \delta Q_W^d$, versus the corresponding shift in the
electron's weak charge, $\delta Q_W^e$, normalized to the respective
SM values.  The dots show the results of a random scan over a range of MSSM parameters.
The loop corrections in the $R$-parity conserving MSSM can be as
large as $\sim 4\%$ ($Q_W^p$) and $\sim 8\%$ ($Q_W^e$) -- roughly the
size of the experimental errors for the two PVES
measurements.  Given the current results on E158, the SUSY loop contributions 
is consistent with the measurement at about 2 $\sigma$ level.
In general, the MSSM effects are larger for large
$\tan\beta$, light SUSY particles, and large splitting between
sfermions, although the latter is an isospin breaking effect and
therefore constrained by oblique 
$T$ parameter.  

The shifts $\delta Q_W^{e,p}$ are dominated by $\delta\hat\kappa^{\rm
SUSY}$ since  the  corrections to $Q_W^{e,p}$ due to shifts in the
$\rho_{PV}$ parameter are suppressed by $1-4\sinhat$.  In addition,
the non-universal corrections involving vertex corrections and
wavefunction renormalization experience significant cancellations.
Since $\delta\hat\kappa^{\rm SUSY}$ is universal, and thus identical for both
$Q_W^e$ and $Q_W^p$, the $\delta\hat\kappa$ dominance produces a
linear correlation between the two weak charges. 
The correction to  $\delta\kappa^{\rm SUSY}$ is nearly always
negative, corresponding to a reduction in the  value of
$\sstw^{eff}(Q^2)=\hat\kappa(Q^2,\mu)\sinhat({\mu})$ for the
parity-violating electron scattering experiments [see
Eq.~(\ref{eq:C1f-radcorr})].

\begin{figure}[ht]
\resizebox{5 in}{!}{
\includegraphics*[30,480][430,750]{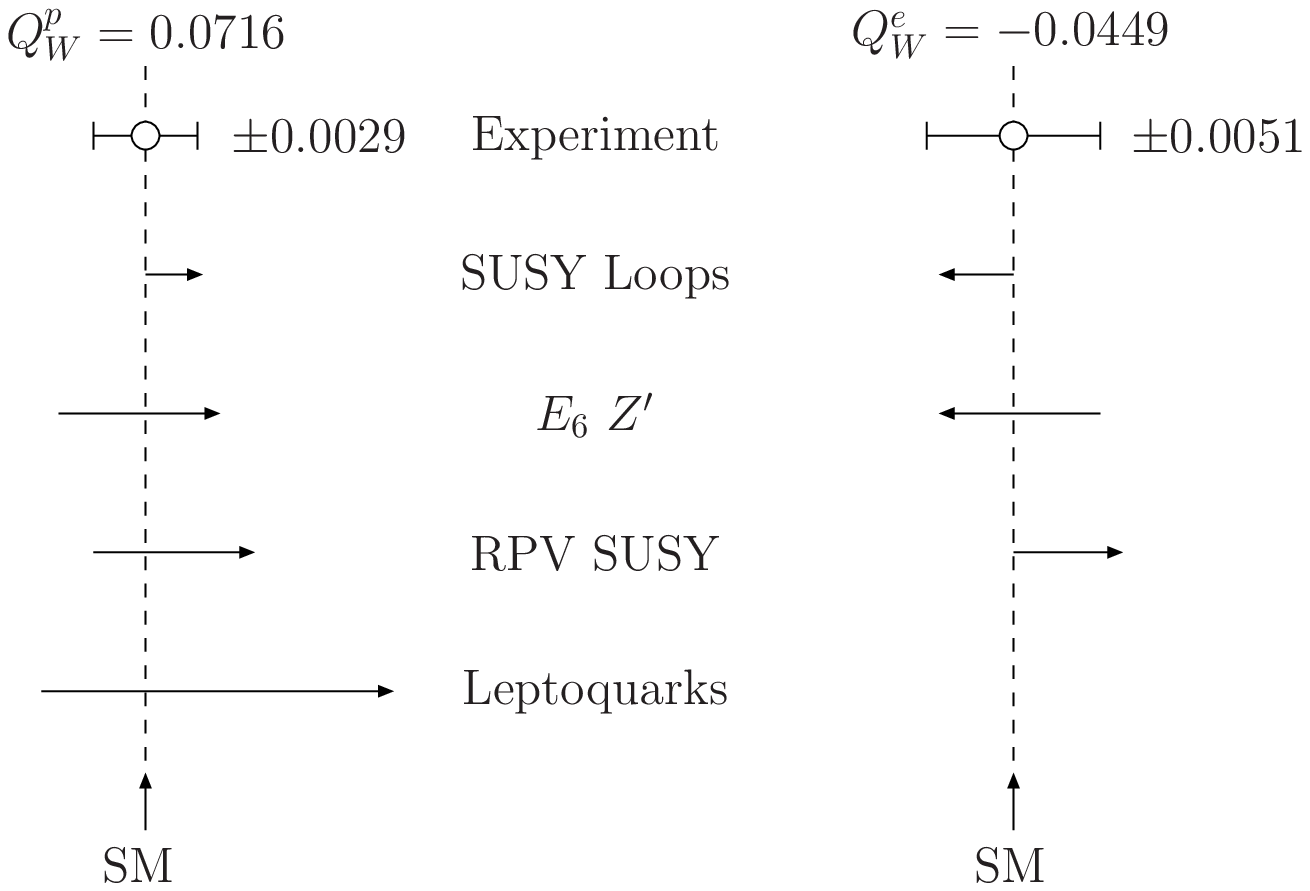}}
\caption{Comparison of anticipated errors 
for $Q_W^p$ and $Q_W^e$ with deviation from the SM expected from 
various extensions and allowed range (at 95\% CL) by fits to existing 
data \cite{Erl-MJRM-Kur02-update}.}
\label{fig:newphy}
\end{figure}

As evident from Fig.~\ref{fig:mssm-vs-parity}, the relative sign of
the corrections to both $Q_W^p$ and $Q_W^e$ -- normalized to the
corresponding SM values -- is nearly always the same and nearly always
positive. Since $Q_W^p>0$ ($ Q_W^e<0$) in the SM, SUSY loop
corrections give $\delta Q_W^p>0$ ($\delta Q_W^e<0$).  This
correlation is significant, since the effects of other new physics
scenarios can display different signatures. 
The combined measurements of the electron and proton weak charge can 
be a probe to distinguish other new physics, as illustrated in 
Fig.~\ref{fig:newphy} \cite{Erl-MJRM-Kur02-update}.  The arrows
indicate  correlated effects.  For example, while both the
superpartner loops and leptoquark exchange give a positive contribution to the proton weak
charge, only MSSM give rise to a sizable effect on the electron weak 
charge \cite{Erl-MJRM-Kur02,MRM99}.
for the
general class of $Z^{\prime}$ theories based on $E_6$ gauge group, with neutral
gauge bosons having mass $\lsim$ 1000 GeV, the effects on $Q_W^p$ and
$Q_W^e$ also correlate, but $\delta Q_W^{e,p}/Q_W^{e,p}$  can have
either sign in this case \cite{Erl-MJRM-Kur02, MRM99}. 
In the case when $E_6~Z^\prime$ models and MSSM have similar effects on the 
electron and proton weak charge, measurement of the cesium weak charge using  atomic parity violation can further tell these two apart, as explained
in Sec.~\ref{sec:apv} below.

If we relax the assumption of $R$-parity conservation, tree level
corrections to the weak charges are generated  RPV interactions.  In this manner
one obtains the following effective four-fermion Lagrangian, using the 
notation of $\Delta_{ijk}^{({\prime})}$ as defined in Eq.~(\ref{eq:deltas}):%
\begin{eqnarray}
\label{eq:rpveffectivepves}
{\cal L}_{{RPV}}^{{EFF}} &=& -{\Delta^{\prime}_{1k1}}(\tilde{q}_L^k){\bar d}_R
\gamma^\mu d_R {\bar e}_{L}\gamma_\mu
e_{L} + {\Delta^{\prime}_{11k}({\tilde d}_R^k)}{\bar
u}_L\gamma^\mu u_L {\bar e}_{L}\gamma_\mu e_{L} \nonumber   \\
&&-{\Delta_{12k}({\tilde e}_R^k)}\left[{\bar \nu}_{\mu
L}\gamma^\mu \mu_L {\bar e}_{ L}\gamma_\mu \nu_{e L}+{\rm
h.c.}\right],
\end{eqnarray}
where we have taken $|q^2|\ll m_{\tilde f}^2$ and have retained only
the terms relevant for the PVES scattering. 
The last term contributes to the muon decay, which affects the extraction
of the fermi constant from the muon decay lifetime.  
Note the absence from Eq.~(\ref{eq:rpveffectivepves}) 
of the parity-violating contact four-electron
interaction. This is because the superpotential in
Eq.~(\ref{eq:RPVL}) can only produce parity-conserving contact
interactions between identical leptons.

The relative
shifts in the weak charges are \cite{MRM00}:
\begin{eqnarray}
\label{eq:rpv-weak}
\frac{\delta Q_W^{e}}{Q_W^{e}}&\approx&-\left[1+ \left(\frac{4}{
1-4\hat{s}^2}\right)\lambda_x \right]\Delta_{12k}({\tilde e}_R^k)
=-29.8 \Delta_{12k}({\tilde e}_R^k)~,\nonumber \\
\frac{\delta Q_W^{p}}{Q_W^{p}}&\approx &\left(\frac{2}{
1-4\hat{s}^2}\right) \left[ -2\lambda_x \Delta_{12k}({\tilde
e}_R^k) +2\Delta_{11k}^\prime({\tilde
d}_R^k)-\Delta_{1k1}^\prime({\tilde
q}_L^k)\right]-\Delta_{12k}({\tilde e}_R^k)~,\nonumber \\
&=&-18.7\Delta_{12k}({\tilde
e}_R^k) +55.9\Delta_{11k}^\prime({\tilde
d}_R^k)-27.9\Delta_{1k1}^\prime({\tilde
q}_L^k)~,\nonumber \\
\lambda_x&=&\frac{{\hat s}^2(1-{\hat s}^2)}{1-2{\hat s}^2} \frac{1}{
1-\Delta {\hat r^{\rm SM}}} \approx 0.35 ~.
\end{eqnarray}
Since  the $\Delta_{ijk}^{(\prime)}$  
are non-negative, Eq.~(\ref{eq:rpv-weak}) indicates
that the relative shift in $Q_W^e$ is negative semidefinite. On the
other hand, the relative shift in $Q_W^p$ can have either sign
depending on the relative magnitudes of $\Delta_{12k}$,
$\Delta_{11k}^\prime$, and $\Delta_{1k1}^\prime$.

The quantities $\Delta_{ijk}$, {\em etc.} in 
Eq.~(\ref{eq:rpv-weak}) are constrained from the existing precision
data~\cite{MRM00}. A summary of the existing constraints 
is given in Table~\ref{tab:rpv-constrain} in Sec.~\ref{sec:susy}, which 
includes superallowed  nuclear
$\beta$-decay that constrains $|V_{ud}|$ \cite{towner-super}, atomic
PV measurements of the cesium weak charge $Q_W^{\rm Cs}$ \cite{Ben99},
the ratio $R_{e/\mu}$ of $\pi_{l2}$ decays \cite{pil2}, and a
comparison of the Fermi constant $G_\mu$ with the appropriate
combination of $\alpha$, $M_Z$, and $\sstw$ \cite{marciano99}. 

The 95\% CL region allowed by this fit in the $\delta Q_W^p/Q_W^p$
vs. $\delta Q_W^e/Q_W^e$ plane is shown by the closed curves (a) and (b)  in
Fig.~\ref{fig:mssm-vs-parity}, corresponding to the RPV fit with the value of 
$\delta|V_{ud}|^2/|V_{ud}|^2$ of case (a) and (b) in Table.~\ref{tab:rpv-constrain}, respectively.
Note that the truncation of the initially
elliptical curves is due to the sign requirements
$\Delta_{ijk}(\tilde f),~\Delta_{ijk}^\prime(\tilde f)\ge 0$ [see
Eq.~(\ref{eq:deltas})].  The correction to $Q_W^e$ and $Q_W^p$ from RPV 
SUSY could be two to three times larger than the SUSY loop effects. 
In addition, the prospective effects of $P_R$ non-conservation are
quite distinct from SUSY loops. The value of $\delta Q_W^e/Q_W^e$ is
never positive in contrast to the situation for SUSY loop effects,
whereas $\delta Q_W^p/Q_W^p$ can have either sign. 
Thus, a comparison of results for the two
parity-violating electron scattering experiments could help determine
whether this extension of the MSSM is to be favored over other new
physics scenarios (see also Ref.~\cite{Erl-MJRM-Kur02}).
If SUSY is the new physics beyond the SM, it is in particular
important to know whether $R$-parity is conserved or not.  Indeed, if $R$-parity is
conserved, then the neutral LSP (for example, 
the lightest ${\chi}^0$) would be a suitable candidate for
dark matter. If any deviation of the electron and proton weak charge is observed,
the correlation between these two would help us to identify whether or not
there is $R$-parity conservation and, thus, shed light on the feasibility of SUSY dark matter. 

Ideas for measuring $\sstw$ at low energy with higher precision have been
explored recently.  
There is a proposal of a similar M{\o}ller ($ee$) scattering 
measurement at JLab 12 GeV upgrade \cite{jlabmoller}.
The estimated precision for electron weak charge is 2.5\%. 
It could be used to determine the value of $\sstw$ at 
$Q^2\sim 0.008\ {\rm GeV}^2$, with a 0.1\% 
precision: $\delta\sstw=0.00025$, which is comparable to the precision of 
$\sstw$
determined from $Z$-pole precision measurements \cite{jlabmoller}. 
Such high precision enables us to constrain the SUSY parameter space, 
whether or not a deviation of $Q_W^e$ is observed.

\subsection{Electron-Deuterium Parity Violating Deep Inelastic Scattering}
\label{sec:PV-DIS}
In light of the recent developments in parity-violating $ee$ and $ep$ scattering
discussed in Sec.~\ref{sec:pves}, 
a new generation of PV DIS measurements with deuterium targets 
has been considered at JLab 6 GeV beam \cite{pvdis6GeV} and  12 GeV upgrade \cite{eDDIS}.  
The DIS-parity experiments seek to study the 
deep inelastic scattering of 
a longitudinally polarized electron beam on unpolarized deuterium target.
Neglecting target mass and higher-twist corrections as well as
contributions from sea quarks, 
the PV asymmetry for eD DIS has the simple form: 
\begin{equation}
\label{eq:pvasym}
\alred={3 G_\mu Q^2\over
2\sqrt{2}\pi\alpha}\frac{2 C_{1u}-C_{1d}+ Y(2 C_{2u}-C_{2d})}{5}\ \ \ ,
\end{equation}
where 
\begin{equation}
Y=\frac{1-(1-y)^2}{1+(1-y)^2-y^2R/(1+R)},\ \ \ {\rm and}\ \ \ 
R(x,Q^2)=\frac{\sigma_L}{\sigma_R}\approx 0.2,
\end{equation}
and $y\in [0,1]$ is the fractional energy transfer to the target in the lab
frame.
The quantities $C_{iq}$ parameterize the low-energy, PV
electron-quark interaction
\begin{equation}
{\cal L}^{eq}_{\rm PV} = {G_\mu\over \sqrt{2}}\sum_q\ \left[ C_{1q}
{\bar e}\gamma^\mu\gamma_5 e {\bar q}\gamma_ \mu q \ + \ C_{2q} {\bar
e}\gamma^\mu e {\bar q} \gamma_\mu\gamma_5 q\right].
\end{equation}
Using the SM values for $C_{iq}$ at tree level, one obtains
\begin{equation}
\alred\approx 10^{-4}Q^2\left[
\frac{3}{2}(1+Y)-\left(\frac{10}{3}+6Y\right)\sstw
\right].
\end{equation}

\begin{figure}
\begin{center}
\includegraphics[width=4in, angle=0]{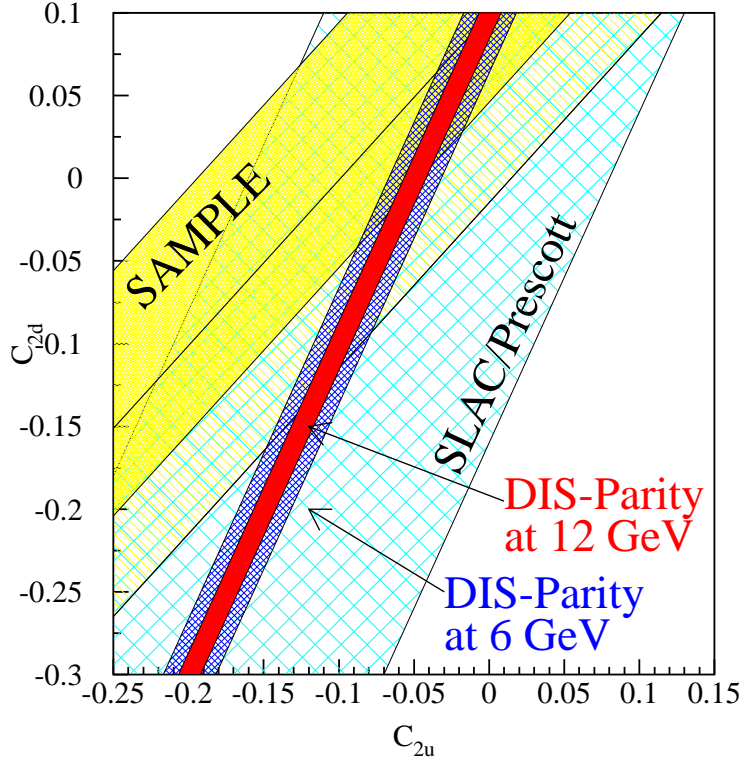}
\caption{The limits on $C_{2u}$ and $C_{2d}$ listed by the particle 
data group \cite{pdg}, by the SAMPLE experiment \cite{sample}, 
and by DIS-parity \cite{eDDIS}.  The plot is taken 
from Ref.~\cite{eDDISplot}.}
\label{fig:eDDIS}
\end{center}
\end{figure}

The DIS asymmetry is much larger than $A_{PV}$ in the M{\o}ller scattering
and Qweak experiments: for $Q^2$=3.7 ${\rm GeV}^2$, $\alred=0.0003$.
An expected 0.8\% measurement in $\alred$ corresponds to 0.45\%
precision in $\sinhat$: $\delta\sinhat=0.0011$.
While the sensitivity to $\sinhat$ in eD DIS is not as good as in 
the SLAC E158 and Qweak experiments, it has the unique opportunity to 
constrain the combination of 
$2 C_{2u}-C_{2d}$.
Assuming the successful completion of the Qweak experiment, an 
absolute uncertainty of $\delta C_{1u(d)}=0.005$ 
will be  obtained.  With this prospective limit, DIS-parity 
experiment places an absolute uncertainty of 
$\delta(2 C_{2u}-C_{2d})=0.026$.  When taken together with the 
results from the SAMPLE experiment~\cite{sample}, much tighter bounds
are placed on $C_{2u}$ and $C_{2d}$ than were previously 
available \cite{pdg}, as illustrated in 
Fig.~\ref{fig:eDDIS} \cite{eDDISplot}.

For the R-parity conserving MSSM, loop corrections to $C_{iq}$ appear.
The $C_{iq}$ are conveniently computed using the
expressions
\begin{eqnarray}
C_{1q} & = & 2\hat\rho_{NC} I_3^e(I_3^q-2 Q_q {\hat\kappa}
\sinhat)-\frac{1}{2}{\hat\lambda}_{1}^q 
\\ C_{2q} & = &
2{\hat\rho}_{NC} I_3^q(I_3^e-2 Q_e{\hat\kappa} \sinhat)
-\frac{1}{2}{\hat\lambda}_{2}^q\ \ \ ,
\end{eqnarray}
where the ${\hat\lambda}^q_{1,2}$ contain the appropriate combinations of the $\hat\lambda^q_{V,A}$ that are process dependent. Detailed expressions for expressions for the
$\hat\lambda_{i}^q$ can be found in Ref.~\cite{eDDISsu-musolf}.

As before, tree level contributions to $C_{iq}$ arise for RPV SUSY. 
In terms of the $\Delta_{ijk}({\tilde f})$ and
$\Delta_{ijk}^\prime({\tilde f})$, one has the following shifts in the
$C_{iq}$:
\begin{eqnarray}
\Delta C_{1u}^{\rm RPV} & = & -[C_{1u}-\frac{4}{3}\lambda_x
]\Delta_{12k}({\tilde e}^k_R)-\Delta^\prime_{11k}({\tilde d}^k_R), \\
\Delta C_{1d}^{\rm RPV} & = & -[C_{1d}+\frac{2}{3}\lambda_x
]\Delta_{12k}({\tilde e}^k_R)+\Delta^\prime_{1k1}({\tilde q}^k_L),\\
\Delta C_{2u}^{\rm RPV} & = & -[C_{2u}-2\lambda_x
]\Delta_{12k}({\tilde e}^k_R)-\Delta^\prime_{11k}({\tilde d}^k_R),\\
\Delta C_{2d}^{\rm RPV} & = & -[C_{2d}+2\lambda_x
]\Delta_{12k}({\tilde e}^k_R)-\Delta^\prime_{1k1}({\tilde q}^k_L),
\end{eqnarray}
where $\lambda_x$ is defined in Eq.~(\ref{eq:rpv-weak}).

\begin{figure}[ht]
\hspace{0.00in}
\begin{center}
\resizebox{8.cm}{!}{\includegraphics*[30,200][520,600]{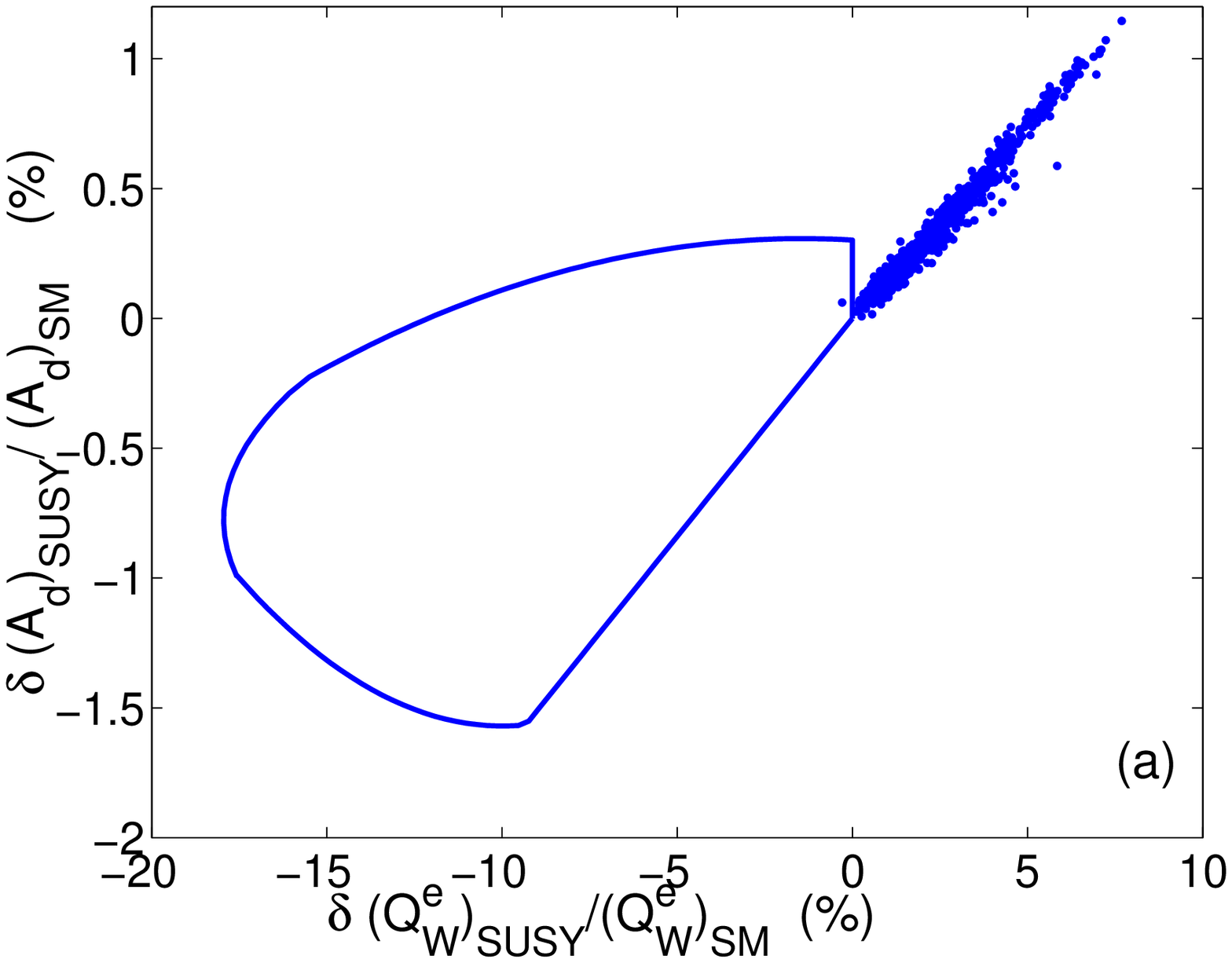}}
\resizebox{8.cm}{!}{\includegraphics*[30,200][520,600]{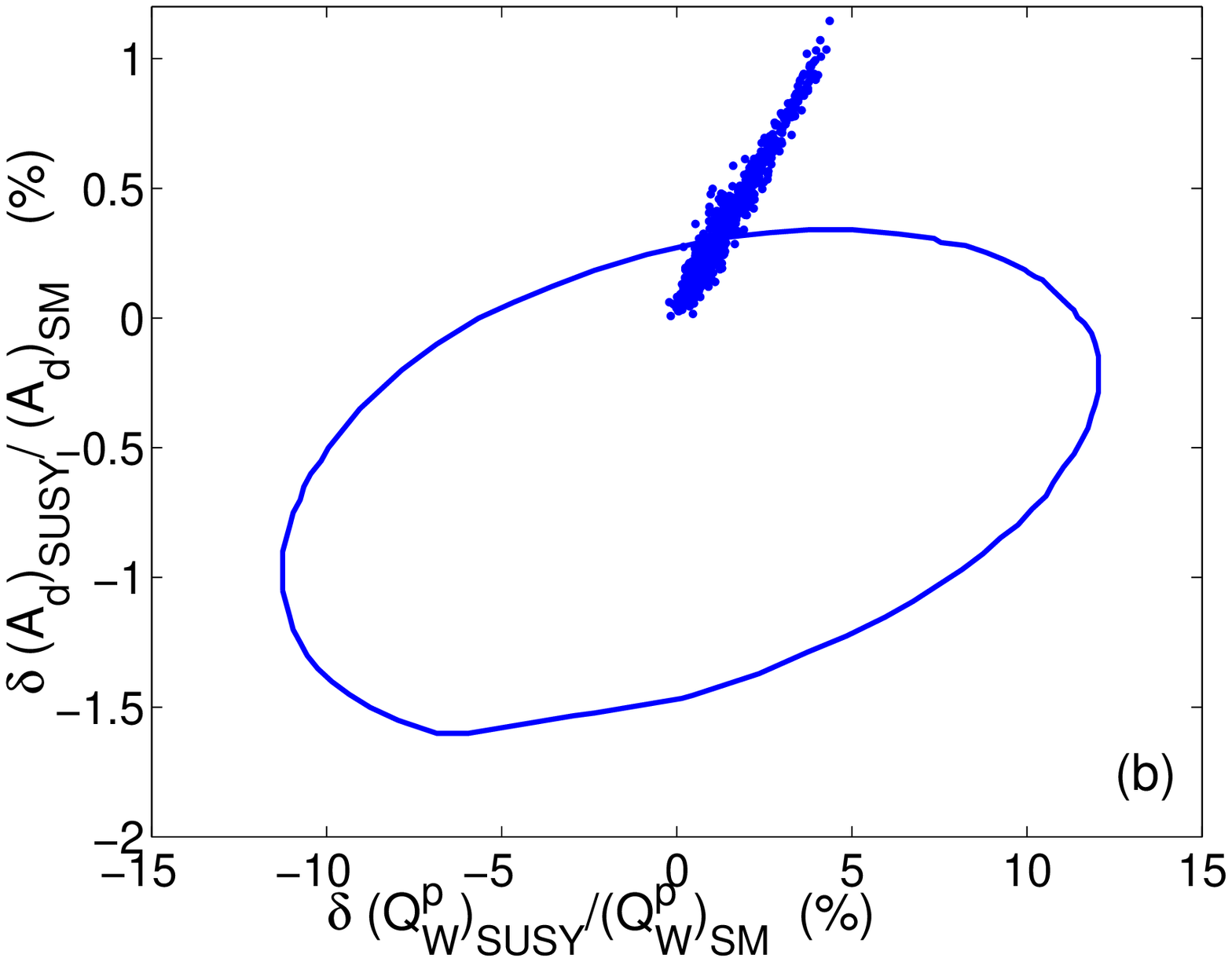}}
\caption{95 $\%$ CL allowed region for RPV contribution to 
$\alred(y=1, Q^2=3.7\ {\rm GeV}^2)$ 
vs. electron weak charge (a) and proton weak charge (b).  
The dots indicate the SUSY loop corrections. 
The figures are reprinted from  Ref.~\cite{eDDISsu-musolf} with                  
permission from Elsevier. }
\label{fig:qwe_qwp}
\end{center}
\end{figure}

In Fig.~\ref{fig:qwe_qwp}, we illustrate the sensitivity of $\alred$
to the effects of MSSM loop contributions and tree-level 
RPV effects \cite{eDDISsu-musolf}. We plot the relative shifts
in $\alred$ vs. those in $Q_W^e$ and $Q_W^p$. 
The interior of the truncated ellipse gives
the 95\% C.L. region from RPV effects 
allowed by other precision electroweak data.
A deviation of about 1\% could be expected from MSSM loop effects while
the maximum correction from RPV effects would be $-1.5\%$, 
corresponding to about 2$\sigma$ for the precision proposed
in Ref. \cite{eDDIS}.
The presence of RPV
effects would induce negative relative shifts in both $\alred$ and
$Q_W^e$, whereas the relative sign of the loop corrections is positive
in both cases.  A sizable positive
shift in $Q_W^p$ (up to $3\sigma$ for the proposed Qweak
measurement) due to RPV contributions could correspond to a tiny
effect on $\alred$ whereas a substantial negative shift in the proton
weak charge could also occur in tandem with a substantial negative
correction to $\alred$. On the other hand, even a result for $Q_W^p$
consistent with the SM would not rule out a sizable effect on
$\alred$.
The addition of an eD DIS measurement would provide a useful 
complement to the PV $ee$ and elastic $ep$ measurements, 
assuming it can be performed with $\sim$ 0.5\% precision or better.

\subsection{Atomic Parity Violation}
\label{sec:apv}
The effects of atomic parity violation (APV) can be measured either by observing the rotation of the 
polarization plane of linearly polarized light, or by measuring the rate for a Stark 
induced transition.  The most precise measurement of an APV effect has been 
performed by the Boulder group using the Stark interference method on a 
beam of Cesium atoms \cite{APV}.  The PV transition between
two atomic states can be expressed as
\begin{equation}
M(n^\prime P_{1/2} \rightarrow n S_{1/2} )
\sim \frac{G_{\mu}}{2\sqrt{2}}C_{SP}(Z)Q_W(Z,N)+\cdots,
\end{equation}
where the dots are the effects from finite nuclear size, nucleon substructure,
and nuclear spin-dependent term \cite{Erl-MJRM-Kur02,apvnuclear}.  
The  uncertainty associated with these effects is small, about 0.15\%.  
$Q_W(Z,N)$ is the weak charge for atom, which is a combination of 
the weak charge of the up and down quark:
\begin{equation}
Q_W(Z,N)=(2Z+N)Q_W^u+(Z+2N)Q_W^d\approx Z(1-4 \sstw)-N \approx -N
\end{equation}
The coefficient $C_{SP}(Z)$ parametrizes 
the contribution related to the atomic structure \cite{apvatomic}.
A precise measurement on the cesium transit dipole amplitude 
can be used to determine $C_{SP}$ with relatively small 
uncertainty \cite{apvC}.

The cesium weak charge extracted from the APV measurement is \cite{APV}
\beq
Q_W^{\rm Cs}({\rm exp.})=-72.69 \pm 0.48,
\eeq
with the combined experimental and theoretical uncertainty
to be about 0.6$\%$.  This is consistent with the SM 
prediction \cite{Erl-MJRM-Kur02}
\beq
Q_W^{\rm Cs}({\rm SM})=-73.16 \pm 0.13.
\eeq

The correction to 
the atom weak charge can be written as
a sum of the weak charge of the up and down type quarks:
\beq
\delta Q_W(Z,N) = (2Z+N)\delta Q_W^u + (2N+Z)\delta Q_W^d.
\eeq
Using the results of the MSSM corrections to the up and down quark 
weak charge as described in Sec.~\ref{sec:pves}, 
we can obtain the contribution of SUSY loop
corrections to the weak charge of heavy nuclei probed with APV.
Since the  sign
of $\delta Q_W^f/Q_W^f$ due to superpartner loops is nearly always the
same, and since $Q_W^u>0$ and $Q_W^d<0$ in the SM, a strong
cancellation between $\delta Q_W^u$ and $\delta Q_W^d$ occurs in heavy
nuclei. This cancellation implies that the magnitude of superpartner loop contributions to $\delta
Q_W(Z,N)/Q_W(Z,N)$ is generally less than about 0.2\% for cesium and
is equally likely to have either sign. Since the presently quoted
uncertainty for the cesium nuclear weak charge is about 0.6\%
\cite{sushov02}, cesium APV does not substantially constrain the SUSY
parameter space.  Equally as important, the present agreement of
$Q_W^{\rm Cs}$ with the SM prediction does not preclude significant
shifts in $Q_W^{e,p}$ arising from SUSY.  The situation is rather
different, for example, in the $E_6~Z^\prime$ scenario, where sizable
shifts in $Q_W^{e,p}$ would also imply observable deviations of
$Q_W^{\rm Cs}$ from the SM prediction.

There are several ongoing atomic parity violation experiments following
the cesium APV results.  A more precise cesium APV measurement is 
underway \cite{apvparis}. The Seattle group is measuring the APV effects using
${\rm Ba}^+$ ions, which is able to achieve the same level of precision
as for cesium \cite{apvseattle}.  
The measurement of APV effects along an isotope chain 
would eliminate the large theory uncertainties from  atomic structure.  
Efforts are underway at Berkeley \cite{apvberkeley} to measure
APV along Yb isotope chain.

A measurement of the Helium weak charge via $0^+\rightarrow 0^+$ transition 
is under investigation at JLab, which could be used to cross check
the Cesium APV experiments \cite{QWHe}.  
Furthermore, weak charge data from JLab
on both Hydrogen and Helium would have the
advantage of correlated errors which would cancel in the ratio
$Q_W^p/Q_W^{\rm He}$, potentially yielding significantly tighter
constraints on new physics.  A preliminary study showed that a 1\% measurement
of the Helium weak charge can constrain RPV SUSY at nearly the same
level as a 4\% proton weak charge experiment \cite{QWHe, suprivate}.

\subsection{Neutrino-Nuclei Deep Inelastic Scattering}
\label{sec:nutev}
Neutrino  scattering experiments have played a key role in elucidating the
structure of the SM. Recently, the NuTeV collaboration has performed a
precise determination of the ratio $\rnu$
($\rnubar$) of neutral current and charged current deep-inelastic
$\nu_\mu$ ($\bar\nu_\mu$)-nucleus cross sections \cite{NuTeV}, which can be 
expressed as:
\begin{equation}
\label{eq:rnudef}
R_{\nu ({\bar\nu})} = \frac{\sigma(\nu (\bar\nu)N \rightarrow \nu X)}
{\sigma(\nu (\bar\nu) N \rightarrow l^{-(+)} X)}=
(g_L^{\rm eff})^2 + r^{(-1)} (g_R^{\rm eff})^2\ \ \ ,
\end{equation}
where $r=\sigma^{CC}_{{\bar\nu} N}/\sigma^{CC}_{\nu N}$ and
$(g_{L,R}^{\rm eff})^2$ are effective hadronic couplings (defined below).
Comparing
the SM predictions \cite{pdg} for $(g_{L,R}^{\rm eff})^2$ with the values
obtained by the NuTeV
Collaboration yields deviations:
$\delta R_{\nu({\bar\nu})}=
R_{\nu({\bar\nu})}^{\rm exp}-R_{\nu({\bar\nu})}^{\rm SM}$
\begin{equation}
\delta R_{\nu}=-0.0033 \pm 0.0007, \ \ \
\delta R_{\bar{\nu}}=-0.0019 \pm 0.0016.
\label{eq:deltaRnu}
\end{equation}

Within the SM, these results may be interpreted as a test of the
scale-dependence of the $\sinhat$ since the $(g_{L,R}^{\rm eff})^2$ depend on the
weak mixing angle. The
results from the NuTeV measurement
imply a $+3\sigma$ deviation at $Q \sim 3$ GeV.
This interpretation of the NuTeV results has been the subject of considerable
debate. Unaccounted for effects, such as
NLO QCD corrections \cite{nutevqcd, davidson}, Electroweak radiative
corrections \cite{nutevew, davidson}, 
strange sea asymmetry \cite{nutevstrange}, isospin 
violation \cite{nuteviso}, nuclear shadowing \cite{nutevshadow},
other nuclear effects such as neutron excess in the target
and nuclear cross section effects \cite{nutevnuclear},
electron neutrino content in the NuTeV beam \cite{nutevnue}
have
been proposed as possible remedies for the
anomaly. Alternatively, one may consider physics beyond the 
SM \cite{davidson, kur-rm-su-nutev, nutevnew}.  
In this review, we focus on the 
MSSM effects on the neutrino-nuclei scattering \cite{kur-rm-su-nutev}.

For momentum transfers $q^\mu$ satisfying $|q^2| \ll \mzs$, the neutrino-quark
interactions can be represented
with sufficient accuracy by an effective four fermion Lagrangian:
\begin{eqnarray}
{\cal L}_{\nu q}^{NC}  & = & -\frac{G_\mu\hat \rho_{NC}}{\sqrt{2}}
{\bar\nu}_\mu\gamma^\lambda (1-\gamma_5) \nu_\mu
\sum_q {\bar q}\gamma_\lambda [\epsilon_L^q (1-\gamma_5)+\epsilon_R^q 
(1+\gamma_5)]q\ , \\
{\cal L}_{\nu q}^{CC} & = & -\frac{G_\mu\hat\rho_{CC}}{\sqrt{2}}
{\bar\mu}\gamma^\lambda (1-\gamma_5) \nu_\mu
{\bar u}\gamma_\lambda(1-\gamma_5) d + {\rm h.c.} \ ,
\end{eqnarray}
where 
\begin{eqnarray}
\epsilon_L^q & = & I_L^3 - Q_q\hat\kappa_\nu\sinhat + \hat\lambda_L^q\ , \\
\epsilon_R^q & = & -Q_q\hat\kappa_\nu\sinhat +\hat\lambda_R^q \ .
\end{eqnarray}
The parameters $\hat\rho^{CC}$  and
$\hat\lambda_{L,R}^q$ are similar to those quantities that are defined 
in Sec.~\ref{sec:renorm}.  The relevant expressions can be found 
in Ref.~\cite{kur-rm-su-nutev}\footnote{In that reference a subscript \lq\lq $\nu N$" on $\hat\rho_{NC,\, CC}$ was included.}.  

The NC to CC cross section ratios $\rnu$ and $\rnubar$ can be expressed in
terms of the above parameters via the effective couplings $(g_{L,R}^{\rm
eff})^2$
appearing in Eq. (\ref{eq:rnudef}) in
a straightforward way:
\begin{equation}
\label{eq:glrdef}
(g_{L,R}^{\rm eff})^2 = \left(\frac{{\hat M}_Z^2}{{\hat M}_W^2}\right)^2\left(
\frac{{\hat M}_W^2-Q^2}{{\hat M}_Z^2-Q^2}\right)^2
\left(\frac{\hat\rho_{NC}}{\hat\rho_{CC}}\right)^2\sum_q\ 
(\epsilon_{L,R}^q)^2 \ \ \ .
\end{equation}
The SM values for these quantities are \cite{pdg} $(g_L^{\rm
eff})^2=0.3042$
and $(g_R^{\rm eff})^2 = 0.0301$ while the NuTeV results imply
$(g_L^{\rm eff})^2=0.3005\pm 0.0014$ and $(g_R^{\rm eff})^2=0.0310\pm 0.0011$.

The MSSM loop contributions to $R_{\nu}$ and $R_{\bar\nu}$ are   
highly correlated, in the range of about 
$0-1.5\times 10^{-3}$ \cite{kur-rm-su-nutev}.
The sign of the SUSY loop corrections is
nearly always positive, in contrast to the sign of the NuTeV anomaly.
There is one corner of the parameter space which admits a negative
loop contribution. This scenario involves gluino loops, whose effect can become
large and negative when  the first generation
up-type squark and down-type squarks are nearly degenerate  and
left-right mixing is close to maximal.
Although the gluino
contribution could be as much as few$\times 10^{-3}$ in magnitude, 
equal and large left-right mixing for both up- and down-type squarks
is inconsistent with the other precision electroweak
inputs, such as the $M_W$ and charged current universality \cite{kurylov02}, and color neutrality of the vacuum (unless the  $H^0$, $A^0$, and $H^\pm$ become super heavy).
In addition, the negative contribution from gluino to 
$\rnu$ and $\rnubar$ could not account for the apparent deviation of 
$\sstw$ from the SM prediction implied by the NuTeV analysis when a 
Paschos-Wolfenstein type relation is used.

When $R$-parity is not conserved, 
one obtains the effective Lagrangian for neutrino-quark scattering
via tree-level contribution:
\begin{eqnarray}
\label{eq:rpveffective}
{\cal L}_\sst{RPV}^\sst{EFF} &=&
-\Delta^{\prime}_{2k1}(\tilde{d}^k_L){\bar d}_R\gamma^\mu d_R
{\bar\nu}_{\mu L}\gamma_\mu \nu_{\mu L} +
\Delta^{\prime}_{21k}(\tilde d^k_R){\bar d}_L\gamma^\mu
d_L {\bar\nu}_{\mu L}\gamma_\mu \nu_{\mu L} \nonumber\\
&&- \Delta^{\prime}_{21k}(\tilde d^k_R)\left[{\bar
u}_L\gamma^\mu d_L
{\bar\mu}_{ L}\gamma_\mu \nu_{\mu L}+{\rm h.c.}\right]\ \ \ .
\end{eqnarray}
The corresponding shifts in $R_{\nu(\bar\nu)}$ are
\begin{eqnarray}
\label{eq:rnurpv}
\delta R_{\nu (\bar\nu)}&=&\lambda_x
[-\frac{4}{3} \epsilon_L^u + \frac{2}{3}\epsilon_L^d ]
[1+r^{(-1)} ]\Delta_{12k}(\tilde{e}_R^k)
-2[R_{\nu (\bar\nu)}^{\rm SM}+\epsilon_L^d]
\Delta_{21k}^{\prime}(\tilde{d}_{R}^k)
+2r^{(-1)}\epsilon_R^d 
\Delta_{2k1}^{\prime}(\tilde{d}_{L}^k)\nonumber \\
&\approx& -0.25 [1+r^{(-1)} ]\Delta_{12k}(\tilde{e}_R^k) -
2[R_{\nu (\bar\nu)}^{\rm SM}-0.43]
\Delta_{21k}^{\prime}(\tilde{d}_{R}^k)+ 1.6 r^{(-1)}
\Delta_{2k1}^{\prime}(\tilde{d}_{L}^k).
\end{eqnarray}

As we discuss in Section \ref{sec:susy}, $\Delta_{12k}({\tilde e}^k_R)$ and
$\Delta^{\prime}_{21k}({\tilde d}^k_R)$ are constrained by other
precision electroweak data, while $\Delta^{\prime}_{2k1}({\tilde
d}^k_L)$ is relatively unconstrained. In Eq.~(\ref{eq:rnurpv}), the
coefficients of $\Delta^{\prime}_{21k}({\tilde d}^k_R)$ and
$\Delta^{\prime}_{2k1}({\tilde d}^k_L)$ are positive, while the
coefficient of $\Delta_{12k}({\tilde e}^k_R)$ is negative. Since the
$\Delta_{ijk}$ are non-negative, we would require sizable value of
$\Delta_{12k}({\tilde e}^k_R)$ and rather small values of
$\Delta^{\prime}_{21k}({\tilde d}^k_R)$ and
$\Delta^{\prime}_{2k1}({\tilde d}^k_L)$  to account for the negative
shifts in $\rnu$ and $\rnubar$ implied by the NuTeV result. The
present constraints on $\Delta_{12k}({\tilde e}^k_R)$ from other
precision electroweak observables, as listed in
Table~\ref{tab:rpv-constrain}, however, are fairly stringent.  The
possible effects on $\rnu$ and $\rnubar$ from RPV interactions are by
and large positive. While small negative corrections are also
possible, they are numerically too small to be interesting
\cite{kur-rm-su-nutev}.

In short, the MSSM -- with or
without R-parity conservation -- is likely not responsible for the NuTeV
anomaly. The  culprit,
apparently, is to be found elsewhere.

Finally, we note that another proposal to measure the weak mixing angle at 
$Q^2 =4 \times 10^{-6} \ {\rm GeV}^{2}$
with a reactor-based experiment via $\bar{\nu}_e e^-$
elastic scattering  has been proposed in \cite{nureactor}. 
The estimated error on $\sinhat$ is about 1\%, comparable
to APV and NuTeV results, but with substantially
different systematic contributions.  Such measurement could explore 
the electroweak corrections, though the possible implications for SUSY have not been analyzed.

\section{Flavor, CP, Neutrinos, and Cosmology}
\label{sec:cpv}

The issues of flavor and CP symmetries are generally challenging for SUSY phenomenology. In the case of the MSSM with R-parity conservation, the structure of the soft SUSY-breaking Langrangian allows for a variety of flavor changing neutral current (FCNC) processes that must be suppressed in order to be consistent with experiment. By itself, the general structure of the soft Lagrangian does not provide for this suppression, so models of SUSY-breaking mediation must be constructed that provide it in a natural way. Similarly, ${\cal L}_{\rm soft}$ contains a host of new CP-violating phases beyond the phase of the Standard Model CKM phase that accounts for CP-violation (CPV) in the neutral kaon and B-meson systems. If these phases are ${\cal O}(1)$ and if the soft masses are on the order of a TeV, the associated CPV interactions can give rise to permanent electric dipole moments (EDMs) of the electron, neutron, and neutral atoms that are up to two orders of magnitude larger than experimental EDM limits. Short of any fortuitous cancellations between various CPV effects, there is no {\em a priori} reason to expect large suppressions of these phases as needed for consistency with experiment. The corresponding \lq\lq SUSY CP problem" again provides a challenge to model builders. 

Both the SUSY flavor and CP problems have been reviewed extensively elsewhere, and we refer the reader to excellent recent discussions (see, {\em e.g.}, Refs.~\cite{Chung:2003fi,Masiero:2003fy,Nir:2002gu, Buras:1999tb,Masiero:1997bv,Dimopoulos:1995ju}). Here, we focus on aspects of these issues most relevant to the current experimental efforts in the low-energy sector as well as on recent theoretical developments pertaining to their broader implications for particle physics and cosmology. After reviewing general features of SUSY flavor physics and CPV, we concentrate on three areas of interest: (a) lepton flavor violation and the corresponding implications for neutrino physics; (b) EDM searches and their theoretical interpretation; and (c) implications for SUSY baryogenesis and dark matter.

\subsection{General Considerations}

\noindent {\em Flavor}

\vskip 0.25in

Within the Standard Model, the GIM mechanism provides an elegant explanation for the suppression of FCNCs among quarks. In the limit of degenerate quarks, a sum over all intermediate quark states in loop contributions to FCNC processes yields a vanishing result due to the unitarity of the CKM matrix. The natural scale for FCNC effects  -- such as $K^0$-${\bar K}^0$ mixing and $b\to s\gamma$ -- is thus governed by differences in the squares of quark Yukawa couplings  (and products of off diagonal elements of the CKM matrix). The presence of these factors provides a natural way to understand the observed suppression of FCNCs.

In general, superpartner loop contributions to FCNCs can upset the GIM suppression mechanism. For example, the difference of scalar quark masses need not be small compared to the weak scale, so that the corresponding loop contributions can be enhanced relative to those arising from quark loops. Similarly, flavor mixing among squarks need not be suppressed since there exists no {\em a priori} reason to expect flavor non-diagonal terms in the squark mass matrix to be small compared to the diagonal terms. Thus, studies of FCNC semileptonic or hadronic weak interactions can provide important constraints on the flavor structure of ${\cal L}_{\rm soft}$. 

It has become conventional to characterize these SUSY flavor-violating effects by assuming that they are small enough to be described by single insertions of the relevant flavor-violating soft mass parameter. In this \lq\lq mass insertion" approximation, one may consider the parameter\cite{Chung:2003fi} 
\be
\label{eq:massinsert}
\left(\delta_{AB}\right)_{ij} = \frac{\left(M_{AB}^2\right)_{ij}}{\left[\left(M_{AA}^2\right)_{ii}\left(M_{BB}^2\right)_{jj}\right]^{1/2}}
\ee
where $A$, $B$ denote $L$ or $R$ and $i$ and $j$ are flavor indices. Note that flavor can be violated separately  among the left- and right-handed fermion superpartners as well as in mass terms that mix them after electroweak symmetry breaking, {\em viz}
\be
\left( M_{LR}^2\right)_{ij} = \frac{v_d}{\sqrt{2}}\left[- \mu\, \left(Y_d\right)_{ij}\tan\beta+\left(a_d\right)_{ij}\right]
\ee
for down-type squarks. While the term containing the Yukawa matrix can be diagonalized by performing the same rotation on $L$- and $R$-squarks that diagonalize the quark mass matrix, the term containing the triscalar coupling $(a_d)_{ij}$ will generally remain flavor non-diagonal after this rotation. 

Although the mass insertion approximation may not always accurately reflect the scale of flavor-violation in a given scenario, it provides a useful framework for comparing the constraints on SUSY flavor violation obtained from different experiments. A summary of present limits on the $(\delta_{AB})_{ij}$ for various processes can be found in Ref.~\cite{Chung:2003fi}. As emphasized by the authors of that work, there does not exist a sufficient set of experimental observables to completely determine the flavor-violating parameters that enter the 
$(\delta_{AB})_{ij}$ even within the MSSM. Consequently, one must use the experimental limits as input for model building. To this end, several broad approaches have been pursued. Among the most popular are:

\begin{itemize}

\item[(i)] Universality, which assumes that the soft terms are flavor diagonal and universal. For example, one may take \cite{susy}
\be
\label{eq:univ1}
\mbold{M}^2_f=\tilde{m}^2\,\mbold{1}
\ee
for $f=Q, U, D, L, E$ and
\be
\label{eq:univ2}
\mbold{a}_f=A_f\,\mbold{Y}_f
\ee
for $f=U,D,E$.  Assuming that Eqs.~(\ref{eq:univ1}-\ref{eq:univ2}) hold at  the SUSY-breaking scale, RG evolution to the electroweak scale will induce corrections to the relations (\ref{eq:univ1}-\ref{eq:univ2}) at the electroweak scale. To the extent that the RG evolution is dominated by Yukawa interactions, one would expect the soft parameters at the electroweak scale to have an expansion in the Yukawa matrices(see, {\em e.g.}, Ref.~\cite{Isidori:2006qy} and references therein):
\bea
\nonumber
\mbold{M}^2_Q & = & {\tilde m}^2\left[{\tilde a}_1 \mbold{1}+{\tilde b}_1 \mbold{Y}_u^\dag\mbold{Y}_u +{\tilde b}_2 \mbold{Y}_d^\dag \mbold{Y}_d
+{\tilde b}_3\left(\mbold{Y}^\dag_d \mbold{Y}_d \mbold{Y}^\dag_u \mbold{Y}_u
+\mbold{Y}^\dag_u \mbold{Y}_u \mbold{Y}^\dag_d \mbold{Y}_d\right)\right]\\
\label{eq:mfv}
\mbold{M}^2_U & = & {\tilde m}^2\left[{\tilde a}_2\mbold{1}+{\tilde b}_4 \mbold{Y}_u \mbold{Y}^\dag_u\right]\\
\nonumber
\mbold{M}^2_D & = & {\tilde m}^2\left[{\tilde a}_3\mbold{1}+{\tilde b}_5 \mbold{Y}_d \mbold{Y}^\dag_d \right]
\eea
with similar expressions for the triscalar couplings $\mbold{A}_{U,D}$ to third order in the Yukawa matrices. Equations~(\ref{eq:mfv}) illustrate an alternative approach known as \lq\lq minimal flavor violation" (MFV) in which all of the flavor violation in the soft sector is dictated solely by the structure of the Yukawa interactions. 

\item[(ii)] Alignment, a scenario in which the soft interactions can be diagonalized by the same rotations that diagonalize the SM Yukawa interactions. 

\end{itemize}

Both the universality and alignment approaches build  a \lq\lq super GIM mechanism" into the soft SUSY-breaking Lagrangian and protect one against the appearance of large FCNC effects. Specific models for SUSY-breaking mediation may or may not lead to either universality or alignment (for a discussion, see {\em e.g.}, Ref.~\cite{Randall:1998te}), and neither of these approaches may ultimately be correct. Nevertheless, they provide a useful starting point for the phenomenology of low-energy precision tests of SUSY in the flavor sector.

\vskip 0.25in

\noindent{\em Lepton Flavor and Number}

\vskip 0.2in

In lepton sector, total lepton number (LN)  is an exact, accidental symmetry of the SM, while lepton flavor violation (LFV) involving charged leptons is highly suppressed by the scale of neutrino mass. Neither feature generally carries over to SUSY. In addition to the long-standing scrutiny of the flavor structure of ${\cal L}_{\rm soft}$, there has been considerable recent interest in the possibility of total lepton number violation (LNV) in SUSY models. 
In the MSSM, LNV arises when the requirement of $P_R$ conservation is relaxed, allowing for the existence of the $\lambda$, $\lambda^\prime$, and $\mu^\prime$ terms in the superpotential of Eq.~(\ref{eq:RPVL}). As discussed earlier, such interactions may generate tree-level corrections to SM CC and NC interactions, some of which provide rather stringent constraints on the associated coupling to mass ratios. The interactions in $W_{\Delta L=1}$ may also give rise to a non-zero rate for neutrinoless double $\beta$-decay ($0\nu\beta\beta$)\cite{Faessler:1997db,Faessler:1996ph} -- a $\Delta L=2$ process -- as well as to LN  conserving but LFV processes such as $\mu\to e\gamma$, $\mu\to e$ conversion, and $\mu\to 3e$ (for a recent discussion, see, {\em e.g.}, Ref.~\cite{deGouvea:2000cf} and references therein). For TeV scale superpartner masses, the interactions in $W_{\Delta L=1}$ may have important consequences for the interpretation of these LNV and LFV processes as we discuss below.  In addition, the existence of $\Delta L=1$ SUSY interactions imply the existence of a radiatively-induced Majorana mass term for the neutrino\cite{Schechter:1981bd}, while extensions of the MSSM that explicitly allow for RH neutrino superfields have taken on renewed interest (the literature on the topic is vast; for representive discussions, see, {\em e.g.}, Refs.~\cite{Mohapatra:2005nc,Dong:2006vk, Kang:2004ix,Frank:2002hk,Hisano:2001qz,Hisano:1995cp}). In SUSY models that explicitly contain tree-level Majorana masses, one may also find a possibility of relatively low-scale leptogenesis. Below, we review some of these recent developments involving LNV and the neutrino sector in SUSY. 

\vskip 0.25in

\vskip 0.25in 

\noindent{\em CP}

The plethora of new CPV phases that arise in ${\cal L}_{\rm soft}$ leads to similar complications for phenomenology as in the case of flavor. There simply do not exist a sufficient number of experimentally accessible CPV observables to independently constrain all of the phases, and one is generally forced to adopt simplifying, model assumptions. The situation is simplest in the Higgs and gauge sectors, where all CPV phases may be rotated into the following relative phases between the $\mu$ parameter and the gaugino mass parameter(see, {\em e.g.}, Ref.~\cite{Pospelov:2005pr}):
\be
 {\rm Arg}\, \left(\mu M_i b^\ast \right)\qquad \qquad {\rm Arg}\, \left(M_i M_j^\ast\right)
\ee
where $i,j$ run over the three gauge groups of the MSSM (leading to a total of three independent phases in this sector). The analysis of CPV in this sector is often further simplified by assuming a common gaugino mass parameter at high scales, thereby reducing the number of independent phases to one. When discussing this special case, we refer to the common relative phase of the $\mu$-parameter and gaugino mass parameters as $\phi_\mu$.

In the most general situation, the parameters in the scalar sector of ${\cal L}_{\rm soft}$ allow for an additional 37 independent CPV phases (for a discussion of parameter counting, see, {\em e.g.}, Ref.~\cite{Dimopoulos:1995ju}). As with the Higgsino-gaugino sector, the number of independent phases can be reduced by adopting a version of the flavor universality or MFV scenarios. In the latter case, for example, most of the CPV phases can be absorbed by sfermion field redefinitions, leaving only the parameters in $\mbold{A}_U$ as complex. Alternatively, assuming flavor diagonality for the scalar mass matrices $\mbold{M}^2_{Q,U,D}$ but a general set of triscalar couplings, one has the additional phases\cite{Pospelov:2005pr}
\be
{\rm Arg}\, \left(A_f M_i^\ast\right) \qquad \qquad {\rm Arg}\, \left(A_f A_{f^\prime}^\ast\right)\ \ \ .
\ee
A third, commonly employed scenario is that of a common triscalar coupling and a single CPV phase for the gaugino masses at high scales, leaving only one additional phase $\phi_A$. In what follows, we will consider this minimal scenario wherein only two phases -- $\phi_\mu$ and $\phi_A$ -- need be considered. 

The presence of non-zero RPV terms in the superpotential can introduce a large number of additional CPV phases. We will note review the implications of these additional sources of CPV in this review and refer the reader to recent literature (see, {\em e.g.}, Ref.~\cite{Faessler:2006at}).

\vskip 0.25in

\noindent {\em Connections with Cosmology}

\vskip 0.2in

While the search for CPV beyond that of the SM is interesting in its own right, there exists additional motivation from cosmological considerations. In particular, an explanation of the small (but anthropically relevant) baryonic component of the energy density of the universe points to the need for CPV beyond that of the SM. Indeed, as first observed by Sakharov nearly four decades ago\cite{Sakharov:1967dj}, arriving at a particle physics-based accounting for the baryon asymmetry of the universe (BAU) requires three ingredients in the particle physics of the early universe: (a) violation of baryon number ($B$); (b) violation of both C and CP symmetry; and (c) a departure from thermal equilibrium, assuming that CPT is an exact symmetry. In principle,  SM contains all three ingredients.  Baryon number  violation arises through anomalous \lq\lq sphaleron" processes that cause transitions between different electroweak vacua having different Chern-Simons number and, therefore, different total $B+L$. At temperatures above the electroweak scale, these transitions are mediated by the excitation of gauge field configurations called sphalerons. At lower temperatures, the probability of sphaleron excitations is Boltzmann suppressed, and transitions between different vacua can only occur via exponentially suppressed tunneling. The SM also contains electroweak CP violation as well as C violating interactions (the gauge-boson couplings to axial vector currents). Finally, a departure from thermal equilibrium occurs as the universe cools through the electroweak temperature and the gauge symmetry of the SM is spontaneously broken. The strength of the CPV effects are strongly suppressed by the light quark Yukawa couplings and Jarlskog invariant associated with the CKM matrix\cite{Shaposhnikov:1987tw,Farrar:1993sp,Farrar:1993hn}, while the LEP II lower bound on mass of the SM Higgs boson precludes a strongly first order electroweak phase transition as needed to prevent washout of the BAU (see, {\em e.g.}, Ref.~\cite{Balazs:2005tu}). 

A variety of particle physics scenarios have been proposed that attempt to circumvent these SM shortcomings in explaining the BAU via baryogenesis at different cosmic epochs. At present, have no conclusive evidence favoring either early time/high scale scenarios such as leptogenesis, or relatively late time, electroweak scale baryogenesis. From the standpoint of phenomenology, consideration of the latter is particularly attractive, since a combination of CPV searches, precision electroweak measurements, and collider studies can highly constrain and possibly even rule out electroweak baryogenesis (EWB). In addition, it is interesting to consider the possibility that new physics at the electroweak scale may provide both an explanation of the BAU and viable candidate for cold dark matter (DM). In this case, DM considerations can provide additional constraints on EWB. 

The study of SUSY DM remains an active, on-going field that has been reviewed extenstively elsewhere (see, {\em e.g.}, Refs.~\cite{Bertone:2004pz,Jungman:1995df}). Below, we focus on the viability of SUSY EWB, taking into account recent field theoretical developments  and  the phenomenology of EDM searches, precision electroweak measurements, collider studies, and DM considerations. In this respect, we provide an up-date to the extensive reviews of baryogenesis provided by Trodden and Riotto\cite{Riotto:1999yt} and Dine and Thomas\cite{Dine:2003ax}. 

\vskip 0.25in

\subsection{Lepton Flavor and Number Violation}

The best-known probe of LFV is to search for the SM-forbidden decay $\mu\to e\gamma$. The current best limit on the branching ratio is\cite{Brooks:1999pu}
\be
B_{\mu\to e\gamma} \equiv \frac{\Gamma(\mu^+\to e^+\gamma)}{\Gamma(\mu^+\to e^+\nu{\bar\nu})}
< 1.2 \times 10^{-11}\qquad\qquad {\rm 90\% C.L.}
\ee
obtained by the MEGA collaboration. A similarly interesting bound on the rate for $\mu\to e$ converstion in gold nuclei has been obtained by the SINDRUM collaboration\cite{Wintz:rp}:
\be
B^{\rm Au}_{\mu\to e} \equiv \frac{\Gamma[\mu^-+A(N,Z)\to e^-+A(N,Z)]}{\Gamma[\mu^-+A(N,Z) \to \nu
A(Z-1,N+1)]} < 8\times 10^{-13}\qquad {\rm 90\% C.L.} \ \ \ .
\ee
In addition, stringent limits have been obtained for other LFV rates: $1.0\times 10^{-12}$ for $B_{\mu^+\to e^+ e^- e^+}$\cite{Bellgardt:1987du}; $4.3\times 10^{-12}$ for $B_{\mu\to e}^{\rm Ti}$\cite{Dohmen:1993mp}; and $4.6\times 10^{-11}$ for 
$B_{\mu\to e}^{\rm Pb}$\cite{Honecker:1996zf}. A new experiment is being performed by the MEG collaboration at PSI that hopes to reach a sensitivity of $\sim 5\times 10^{-14}$ for $B_{\mu\to e\gamma}$\cite{Yashima:2000qz}, while until recently there had been an effort by the MECO collaboration to probe $B^{\rm Al}_{\mu\to e}$ at the level of $5\times10^{-17}$ using the AGS at Brookhaven.
Although that experiment has now been derailed by the U.S. funding agencies, efforts are underway to pursue an experiment at an alternate site, possibly using a future muon strorage ring at JPARC in Japan.

The prospective implications of these experiments for supersymmetric  lepton flavor structure has been analyzed by several authors (for a discussion within non-SUSY models, see, {\em e.g.}, Refs.~\cite{Cirigliano:2004mv,Cirigliano:2004tc,Atre:2005eb} and references therein). A comprehensive analysis in a minimally-extended MSSM that includes RH neutrino supermultiplets and neutrino mass generation {\em via} the see-saw mechanism has been carried out in Ref.~\cite{Hisano:1995cp}. While the RH neutrino sector decouples from low-energy LFV observables due to the large RH neutrino mass ($M_R\sim 10^{12}$ GeV), the LFV effects of this sector can be communicated to the charged slepton mass matrices and triscalar couplings in ${\cal L}_{\rm soft}$ through RG running from high scales. The authors obtain general expressions for the rates for $\mu\to e\gamma$ and $\mu\to e$ conversion in terms of the slepton and light sneutrino rotation matrices, $Z_L^{Ij}$ and $Z_\nu^{Ij}$, introduced earlier without relying on the mass insertion approximation. One has
\be
\label{eq:meg1}
B_{\mu\to e\gamma} = {48\pi^3 \alpha}\, \left(\left\vert {\tilde A}_2^L\right\vert^2+ \left\vert {\tilde A}_2^R\right\vert^2 \right)
\ee
where ${\tilde A}_2^{L,R}$ are the dipole amplitudes appearing in the amplitude for $\mu\to e\gamma^{(\ast)}$:
\begin{eqnarray}
\nonumber
{\cal M}_{\mu\to e\gamma^{(\ast)}} &=& e G_\mu\, \varepsilon^{\alpha\, \ast}\, {\bar \mu}(p-q)\Bigl[ \left(q^2\gamma_\alpha -\dslash{q} q_\alpha\right) \left({\tilde A}_1^R P_R +{\tilde A}_1^L P_L\right) \\
\label{eq:meg2}
&&
+ i m_\mu \sigma_{\alpha\beta} q^\beta \left({\tilde A}_2^R P_R +{\tilde A}_2^L P_L\right)\Bigr] \, e(p)
\end{eqnarray}
where, in contrast to Ref.~\cite{Hisano:1995cp} we have normalized the amplitudes to $G_\mu$. 

\begin{figure}
\resizebox{4 in}{!}{
\includegraphics*[60,520][300,640]{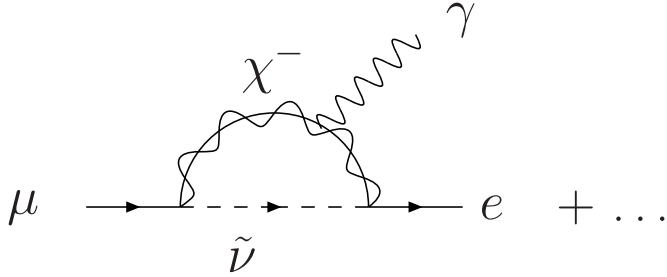}}
\caption{Contributions to the LFV amplitudes entering $\mu\to e\gamma^{(\ast)}$ .}
\label{fig:mueg}
\end{figure}

Contributions to these amplitudes are generated by the diagrams of Fig. \ref{fig:mueg}. The chargino loop contribution to ${\tilde A}_2^R$, for example, gives 
\bea
\label{eq:meg3}
{\tilde A}_2^{R\ (\chi^\pm)} &=& -\left(\frac{1}{4\sqrt{2}\pi^2}\right) \, \sum_{ j,k}\, \left(\frac{M_W^2}{m_{\tilde\nu_j}^2}\right)\, Z_\nu^{1j} Z_\nu^{2j\, \ast}\, \biggl[ \vert U_{k1}\vert^2\, Z_\nu^{1j} Z_\nu^{2j\, \ast}\ F_1(x_{jk})\\
\nonumber
&& \qquad -U_{k1}^\ast\, V_{k2}\, \left(\frac{m_{\chi_k}}{m_\mu}\right)F_1(x_{jk})\biggr]
\eea
where $x_{jk}=m^2_{\chi_k}/m_{\tilde\nu_j}^2$,  and the $F_{1,2}(x)$ are loop functions defined in Ref.~\cite{Hisano:1995cp}.
Analogous expressions for the other chargino and neutralino loop contributions to the ${\tilde A}_2^{R,L}$ are given in Ref.~\cite{Hisano:1995cp}. Assuming no cancellations among the various contributions, one may use Eqs.~(\ref{eq:meg1}-\ref{eq:meg3}) and experimental limits on $B_{\mu\to e\gamma}$ to derive bounds on the LFV parameters $Z_\nu^{1j} Z_\nu^{2j\, \ast}$ {\em etc.}. Note that these LFV couplings also arise in the non $(V-A)\times (V-A)$ SUSY box graph contributions to the muon decay parameter, $g^S_{RR}$ as in Eq.~(\ref{eq:grrloop}). The bounds on these quantities are sufficiently stringent that only the L-R mixing terms can make appreciable contributions to $g^S_{RR}$. In the case of $\mu\to e$ conversion in nuclei, the \lq\lq penguin" amplitudes proportional to ${\tilde A}_1^{L,\, R}$ contribute to the four fermion ${\bar e}\mu {\bar q}q$ conversion operators, as do LFV $Z^0$-exchange amplitudes and box graphs. As we discuss below, the ${\tilde A}_1^{L,R}$ may give the dominant contributions  in the presence of $P_R$ non-conservation. 

An older analysis using the mass insertion approximation has been performed by the authors of Refs.~\cite{Gabbiani:1996hi,Gabbiani:1988rb}. For the special case in which the amplitude is dominated by photino loops one has
\be
\label{eq:meg4}
B_{\mu\to e\gamma} =\frac{24\alpha\sin^2\theta_W}{\pi}\, \left\{ \left\vert M_3(x)\, \left(\delta_{LL}^\ell\right)_{21} +\frac{M_{\tilde\gamma}}{m_{\tilde\ell}}\, M_1(x)\, \left(\delta_{LR}^\ell\right)_{21}
\right\vert^2+\, L\leftrightarrow R\right\}
\ee
where $m_{\tilde\ell}$ is an average slepton mass, $x=M_{\tilde\gamma}^2/m_{\tilde\ell}^2$, and the $M_i(x)$ are defined in Ref.~\cite{Gabbiani:1996hi}. The limits on the parameters $(\delta_{AB}^\ell)_{21}$ vary with $x$. At $x=1.0$ one has 
\be
\left\vert\left(\delta_{LL}^\ell\right)_{21}\right\vert \leq1.9\times 10^{-3}\qquad \qquad
\left\vert\left(\delta_{LR}^\ell\right)_{21}\right\vert\leq 4.3\times 10^{-7}
\ee
for $m_{\tilde\ell}=100$ GeV. These authors did not analyze the $\mu\to e$ conversion process, so no limits on the $(\delta_{AB}^\ell)_{21}$ are available for this process.

In some scenarios, the mass insertion approximation may not provide a realistic guide to the magnitude of LFV effects. A well-known illustration occurs in SUSY GUT models\cite{Barbieri:1994pv,Barbieri:1995tw}, wherein contributions from the large, third generation up-type Yukawa coupling to the RG running of the sfermion and triscalar couplings from the Planck scale to the GUT scale lead to splittings among these parameters. After implementing symmetry-breaking at the GUT scale and evolving these parameters to the weak scale one has for an SU(5) GUT
\be
{\bf m_L^2}=m_{\tilde{L}}^2 \mbold{1}\quad {\bf m_{\bar{e}}^2}=m_{\tilde{e}}^2-\mbold{I}_G\quad
{\bf a_e}=\left[A_e\mbold{1}-\frac{1}{3}\mbold{I}_G^\prime\right]\mbold{y}_e
\ee
where $\mbold{I}_G$ and $\mbold{I}_G^\prime$ are contributions arising from the Yukawa induced running from $M_P$ to $M_G$. It is convenient to rotate the sfermion fields by the same transformations that diagonalize the Yukawa matrices, 
\be
{\tilde e}_R^\prime = \mbold{V}_e\, {\tilde e}_R \qquad \qquad {\tilde L}^\prime = \mbold{U}_e^\dag L
\ee
so that the term in ${\bf M_{LR}^2}$ containing $\mbold{I}_G^\prime$ becomes 
\be
-\frac{1}{3} {\tilde e}_R^{\prime \dag} \mbold{I}_G^\prime \mbold{V}_e^\ast \mbold{M}_e
{\tilde e}_L^{\prime} +{\rm h.c.}
\ee
where $\mbold{M}_e$ is the charged lepton mass matrix. For an SO(10) GUT, the LH slepton mass matrix ${\bf m_L^2}$ picks up an additional contribution from running above $M_G$, while the induced triscalar contribution becomes
\be
-\frac{5}{14} {\tilde e}_R^{\prime \dag} \mbold{I}_G^\prime \mbold{V}_e^\ast \mbold{M}_e \mbold{V}_e^\dag {\tilde e}_L^\prime +{\rm h.c.}
\ee
In general, the numerical impact on $B_{\mu\to e\gamma} $ of these LFV couplings is not well reproduced using the mass approximation. Importantly, the expected magnitude of this LFV observable is expected to be well within the reach of the MEG experiment for superpartner masses on the order of a few hundred GeV. For sufficiently large $m_{\tilde e_R}$, however, $B_{\mu\to e\gamma} $ scales as $1/m_{\tilde e_R}^4$, so a null result for the MEG experiment would imply TeV-scale superpartner masses in this GUT scenario. 

The foregoing analyses assume conservation of $P_R$, so that the potentially significant LFV effects at the weak scale are not accompanied by corresponding LNV. If $P_R$ is not conserved, however, then the terms in the superpotential $W_{\Delta L=1}$ can lead to observable low-scale LNV as well as LFV. An analyses of this possibility of low-scale LFV that is accompanied by low-scale LNV has recently been carried out in Ref.~\cite{Cirigliano:2004tc} using an effective operator approach, following on the earlier analysis of Raidal and Santamaria\cite{Raidal:1997hq}. The leading LFV and LNV operators have dimension six and nine, respectively, and appear in the effective Lagrangians valid below the electroweak scale\footnote{Consequently, the operators do not generally respect the SU(2)$_L\times$U(1)$_Y$ symmetry of the SM.}.
\begin{eqnarray}
{\cal L}_{\rm LFV} &=& \sum_i\frac{c_i}{\Lambda^2} {\cal O}_i^{(6)} +\cdots \\
{\cal L}_{\rm LNV} &=& \sum_i\frac{\tilde c_i}{\Lambda^5} {\cal O}_i^{(9)} +\cdots 
\end{eqnarray}
where the $+\cdots$ indicate terms containing higher dimension operators and where 
\begin{eqnarray}
{\cal O}^{(6)}_{\sigma L} & = & {\bar \ell}_{iL} \sigma_{\mu\nu} i D\!\!\!\!/ \, \ell_{jL} F^{\mu\nu} +{\rm h.c.}
\\
{\cal O}^{(6)}_{\ell L} & = & {\bar \ell}_{iL} \ell^c_{jL} {\bar\ell}^c_{kL} \ell_{mL}\\
{\cal O}^{(6)}_{\ell q} & = & {\bar \ell}_{iL} \Gamma_\ell \ell_{j} {\bar q_L}\Gamma_q q
\end{eqnarray}
with $\Gamma_{\ell,q}$ is a shorthand for all possible $\gamma$-matrix insertions and where the corresponding operators with RH fields have not been explicitly included\footnote{In the semileptonic operator ${\cal O}^{(6)}_{\ell q}$ we have not included chirality labels on all the fields since scalar and tensor interactions flip chirality while vector and axial vector interactions preserve it.}. In the case of ${\cal L}_{\rm LNV}$ one has operators of the general form that will contribute to $0\nu\beta\beta$
\be
{\cal O}^{(9)} = {\bar q} \Gamma_1 q\ {\bar q}\Gamma_2q\ {\bar e}\Gamma_3 e^c +{\rm h.c.}
\ee
where a complete list of operators of this form can be found in Ref.~\cite{Prezeau:2003xn}. All searches for $0\nu\beta\beta$-decay involve $0^+\to 0^+$ transitions, and for these cases the dominate operator is \cite{Faessler:1997db,Prezeau:2003xn}
\be
{\cal O}^{(9)++}_{+}=\left[
 {\bar q}_R\tau^+ q_L {\bar q}_R \tau^+ q_L + {\bar q}_L\tau^+ q_R {\bar q}_L \tau^+ q_R\right] {\bar e}(1+\gamma_5) e^c
\ee
with coefficient in the case of $P_R$ non-conservation given by
\be
\frac{\tilde c}{\Lambda^5} = \frac{8\pi\alpha_s}{9}\frac{|\lambda_{111}^\prime|^2}{m_{\tilde q}^4 m_{\tilde g}}\, + \cdots
\ee
where $m_{\tilde q}$ is an average squark mass, $m_{\tilde g}$ is the gluino mass, and the $+\cdots$ indicate contributions proportional to the semiweak coupling $\alpha_2$. 

The chiral structure of ${\cal O}^{(9)++}_{+}$ implies that it can contribute to an effective $\pi\pi ee$ operator that contains no derivatives\cite{Prezeau:2003xn}, thereby leading to an enhanced, long-range pion-exchange contribution to the $0\nu\beta\beta$-decay rate\cite{Faessler:1997db}. From experimental upper limits on this rate and taking into account this long-range contribution, the authors of Ref.~\cite{Faessler:1997db} obtained the constraint
\be
\label{eq:0nurpv}
\lambda^\prime_{111} \leq 2\times 10^{-4} \left(\frac{m_{\tilde q}}{100\, {\rm GeV}}\right)^2\, \left(\frac{m_{\tilde g}}{100\, {\rm GeV}}\right)^{1/2}\ \ \ .
\ee

The coefficients of the LFV operators can similarly be expressed in terms of the $\lambda$ and $\lambda^\prime$ couplings:
\begin{eqnarray}
\nonumber
\frac{c_\sigma}{\Lambda^2} &\sim & \frac{\lambda\lambda^\ast}{m_{\tilde\ell}^2}\, , 
\frac{\lambda^\prime\lambda^{\prime\ast}}{m_{\tilde q}^2} \\
\frac{c_\ell}{\Lambda^2} &\sim& \frac{\lambda_{i11}\lambda_{i21}^\ast}{m_{\tilde\nu_i}^2},\quad 
\frac{\lambda_{i11}^\ast\lambda_{i12}}{m_{\tilde\nu_i}^2}\\
\nonumber
\frac{c_{\ell q}}{\Lambda^2} &\sim& \frac{\lambda^{\prime\ast}_{11i}\lambda_{21i}^{\prime}}{m_{\tilde d_i}^2},\quad 
\frac{\lambda_{1i1}^{\prime\, \ast} \lambda_{2i1}^\prime}{m_{\tilde u_i}^2}
\ \ \ ,
\end{eqnarray}
where the various combinations of the $\lambda$ and $\lambda^\prime$ entering $c_\sigma$ are given in Ref.~\cite{deGouvea:2000cf}. Limits on various combinations of the $P_R$-violating couplings are given in Tables II and III of that work. From the present limits on $B_{\mu\to e\gamma}$, one has for example
\begin{eqnarray}
\label{eq:megrpv}
|\lambda_{131}\lambda_{231}| \leq 2.3 \times 10^{-4} \left(\frac{m_{\tilde \ell}}{100\ {\rm GeV}}\right)^2\\
|\lambda_{111}^\prime\lambda_{211}^\prime| \leq 7.6 \times 10^{-5} \left(\frac{m_{\tilde q}}{100\ {\rm GeV}}\right)^2
\end{eqnarray}
while from $B^{\rm Au}_{\mu\to e}$ one obtains
\begin{eqnarray}
\label{eq:merpv}
|\lambda_{131}\lambda_{231}| \leq 1.1 \times 10^{-5} \left(\frac{m_{\tilde \ell}}{100\ {\rm GeV}}\right)^2\\
|\lambda_{111}^\prime\lambda_{211}^\prime| \leq 6.0 \times 10^{-7} \left(\frac{m_{\tilde q}}{100\ {\rm GeV}}\right)^2
\end{eqnarray}
Analogous limits on other combination of the couplings can be found in Ref.~\cite{deGouvea:2000cf}.

The results in Eqs.~(\ref{eq:0nurpv}-\ref{eq:merpv}) lead to several observations. First, for superpartner masses of ${\cal O}(100)$ GeV, the bounds on $\lambda^\prime_{111}$ obtained from $0\nu\beta\beta$ are stronger than those derived from the LFV observables, though the gap between them shrinks as the superpartner masses are increased. Second, the LFV $\mu\to e$ conversion limits are more stringent than those derived from $\mu\to e\gamma$, even though $B^{\rm Au}_{\mu\to e}$ contains an extra factor of $e^2$ suppression compared to $B_{\mu\to e\gamma}$. The reason is that the penguin operators generated by the terms proportional to  ${\tilde A}_1^{L,R}$ in Eq.~(\ref{eq:meg2}) contain large logarithmic enhancements while the dipole operators do not, and these large logs overcome the nominal $4\pi\alpha$ suppression of the conversion process. In addition, tree-level exchange of superpartners can occur in the presence of $P_R$ non-conservation, leading to the presence of four-fermion operators not directly related to the LFV electromagnetic amplitudes of Eq.~(\ref{eq:meg2}). This situation differs from that of the $P_R$-conserving GUT scenario of Refs.~\cite{Barbieri:1994pv,Barbieri:1995tw}, wherein the magnetic amplitudes are dominant and the naive expectations for the relative magnitudes of conversion and $\mu\to e\gamma$ branching ratios obtains. 

\begin{figure}
\resizebox{6 in}{!}{
\includegraphics*[50,500][580,660]{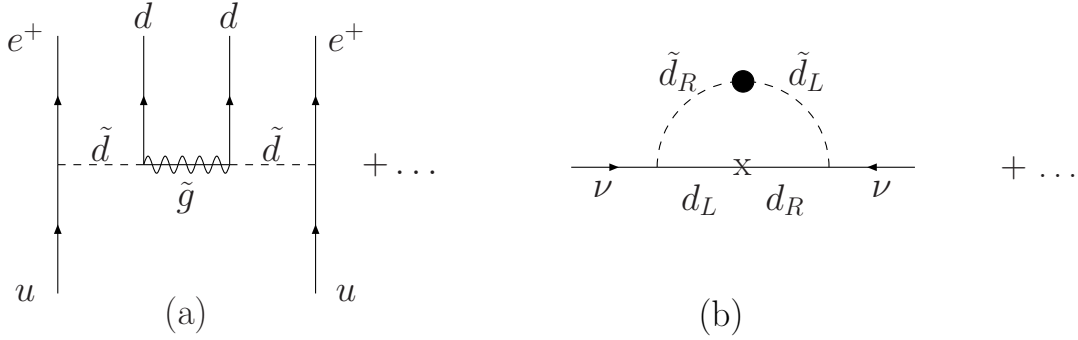}}
\caption{Contributions from RPV interactions to (a) $0\nu\beta\beta$-decay    
and (b) neutrino mass. Figure (a) gives representative contribution from        
semileptonic trilinear RPV interactions parameterized by the coupling           
$\lambda^\prime_{111}$; the \lq\lq $+\cdots$ indicate contributions             
involving neutralino exchange. Figure (b) shows analogous neutrino mass         
contribution from semileptonic trilinear RPV; the cross and shaded circle       
denote fermion mass and sfermion L-R mixing insertions, respectively. In        
both cases generation indices are suppressed. }
\end{figure}

This observations can have important consequences for the interpretation of $0\nu\beta\beta$ in terms of light Majorana neutrino exchange. Indeed, 
For $\lsim{\cal O}$(TeV) masses in ${\cal L}_{\rm soft}$, the contribution from ${\tilde c}{\cal O}^{(9)++}_{+}/\Lambda^5$ to the rate for $0\nu\beta\beta$ can be comparable to the contribution from the exchange of a light Majorana neutrino. Denoting the amplitudes for the former and latter as $A_H$ and $A_L$, respectively, one has\cite{Cirigliano:2004tc}
\be
\label{eq:heavylight}
\frac{A_H}{A_L}\sim \frac{M_W^2 {\bar k}^2}{\Lambda^5 m_{\beta\beta}}
\ee
where $m_{\beta\beta}$ is the effective mass associated with the exchange of the light Majorana neutrino and ${\bar k}\sim 50$ MeV is its typical virtuality. For $\Lambda\sim 1$ TeV and $m_{\beta\beta}\sim 0.1-0.5$ eV, the ratio in Eq.~(\ref{eq:heavylight}) can be order unity. In general, one would like to know whether or not there exist important heavy particle contributions when attempting to extract robust bounds on $m_{\beta\beta}$ from the $0\nu\beta\beta$ rate limits. To this end, it was observed in Ref.~\cite{Cirigliano:2004tc} that the presence of important heavy particle (superpartner), $P_R$-violating exchange contributions is also accompanied by logarithmically-enhanced contributions to the conversion rate. Should the results of future studies find that the relative sizes of the LFV branching ratios are in accord with the naive expectations, one would conclude that there are no large $P_R$-violating contributions to the $0\nu\beta\beta$-decay process. A similar conclusion holds for other, non-SUSY scenarios, pointing to the usefulness of LFV studies as a diagnostic for the presence of low scale LNV.

\subsection{R Parity Violation and Neutrino Mass}

One may incorporate massive neutrinos in SUSY models by supersymmetrizing any of the SM extensions that incorporate non-vanishing $m_\nu$. As discussed above, the authors of Ref.~\cite{Hisano:1995cp} carried out such an analysis in an extension of the MSSM that contains right-handed neutrino superfields $N_j^C$  that are singlets under the MSSM gauge groups. The small masses of the light neutrinos arise via the see saw mechanism with a right-handed neutrino mass parameter $M_R\sim10^{12}$ GeV.  Except for the effects of the resulting non-zero light neutrino Majorana mass, the effects of the fields in $N_j^C$ decouple from low-energy observables. Variations on this class of SUSY models with massive neutrinos can be found  
in the literature (see Refs.~~\cite{Mohapatra:2005nc,Dong:2006vk, Kang:2004ix,Frank:2002hk,Hisano:2001qz,Hisano:1995cp} and references therein). 

An alternate mechanism for generating neutrino mass involves the  LNV interactions in $W_{\Delta L=1}$ (for a recent analysis and survey of the literature, see {\em e.g.}, Ref.~\cite{Grossman:2003gq}). Tree-level Majorana masses can arise from the bilinear terms that lead to mixing between neutrinos and neutral Higgsinos [see Eq.~(\ref{eq:RPVL})]. Specifically, one has the contribution to the light neutrino mass matrix\cite{Grossman:2003gq}
\be
\label{eq:neutrino1}
\left[ m_\nu\right]_{ij}^{(\mu\mu)} \sim \mu_i^\prime \mu_j^\prime \, \frac{\cos^2\beta}{\tilde m}
\ee
where $\tilde m$ is a characteristic soft mass parameter. In the absence of fine-tuning between the contribution in Eq.~(\ref{eq:neutrino1}) and other tree-level neutrino mass terms, one obtains the following rough upper bound on the scale of the $P_R$-violating bilinear coupling
\be
\label{eq:neutrino2}
\left\vert \frac{\mu_i^\prime}{\tilde m}\right\vert \lsim \frac{3\times 10^{-6}}{\cos\beta}\left(\frac{m_\nu}{1\, {\rm eV}}\right)^{1/2} \, \left(\frac{100\, {\rm GeV}}{\tilde m}\right)^{1/2}.
\ee
The neutrino mass implications for the magnitude of the dimensionful $P_R$ violating SUSY couplings are thus quite severe. 

The triscalar couplings in $W_{\Delta L=1}$ may also induce contributions to $m_\nu$ through one-loop radiative corrections. Such contributions have been studied extensively, and we give only a flavor of these analyses here (for an extensive list of the literature, see Ref. [4] of Ref.~\cite{Grossman:2003gq}). Roughly speaking, one finds
\begin{eqnarray}
\label{eq:neutrino3}
\left[ m_\nu\right]_{ij}^{(\lambda\lambda)} & = & \frac{1}{32\pi^2}\, \sum_{\ell, k}\, \lambda_{i\ell k}\lambda_{jkl} m_{L_\ell} \left(\delta^L_{LR}\right)_{kk}\ {\bar\xi}^L_{kk}\\
\nonumber
\left[ m_\nu\right]_{ij}^{(\lambda^\prime\lambda^\prime )} & = & \frac{N_C}{32\pi^2}\, \sum_{\ell, k}\, \lambda_{i\ell k}^\prime \lambda_{jkl}^\prime m_{d_\ell} \left(\delta^d_{LR}\right)_{kk}\ {\bar\xi}^d_{kk}
\end{eqnarray}
where
\be
{\bar\xi}^f_{kk} = \frac{
\sqrt{(M_L^2)^f_{kk} (M_R^2)^f_{kk}}}{ (M_L^2)^f_{kk} +(M_R^2)^f_{kk} }
\ee
is a number typically of ${\cal O}(1)$ and where we have assumed that $(M_L^2)^f_{kk} \sim(M_R^2)^f_{kk} >> (M_{LR}^2)^f_{kk} $. From Eqs.~(\ref{eq:neutrino3}) we may derive neutrino mass naturalness bounds on products of the triscalar couplings for a given value of the flavor diagonal LR mixing parameters, $(\delta^f_{LR})_{kk}$. Assuming that $(M_{LR}^2)^f\propto m_f$, so that 
$(\delta^f_{LR})_{kk}\sim m_f/{\tilde m}$, we obtain the most restrictive bounds for third generation fermions ($\ell=3$):
\begin{eqnarray}
\label{eq:neutrino4}
\lambda^\prime_{i3k}\lambda^\prime_{jk3}\lsim 4\times 10^{-7}\,  \left(\frac{m_\nu}{1\, {\rm eV}}\right)\left(\frac{\tilde m}{100\, {\rm GeV}}\right)\\
\lambda_{i3k}\lambda_{jk3}\lsim 4\times 10^{-5}\,  \left(\frac{m_\nu}{1\, {\rm eV}}\right)\left(\frac{\tilde m}{100\, {\rm GeV}}\right)
\end{eqnarray}
Note that these limits are comparable to those obtained from LFV and LNV observables for ${\tilde m}\sim 100$ GeV, though the dependence on the soft masses differ in the various cases.

\subsection{EDM Searches: Implications for SUSY}

Recent advances in experimental techniques have put the field of EDM searches on the brink of a revolution. The present limits on the EDMs of the electron\cite{Regan:ta}, neutron\cite{baker06}, and mercury atom\cite{Romalis:2000mg} -- shown in Table \ref{tab:edm} -- are already remarkably stringent. The pursuit of EDMs that began with the pioneering studies of the neutron by Purcell and Ramsey in the 1950's \cite{Purcell:1950,Smith:ht} is poised to enter a new era with prospective improvements in sensitivity of up to four orders of magnitude. New efforts are underway that aim to push the sensitivity of the EDM searches for the electron\cite{DeMille:2000,Kawall:2003ga,Liu:2004,hunter05}, neutron\cite{Mischke:ac,Aleksandrov:2002}, and neutral atoms\cite{Romalis:2001,Romalis:2004,Holt:2004}, as well as for the muon and deuteron\cite{Semertzidis:2003iq}.  These prospective experimental advances, as well as related theoretical issues and developments, have been discussed in several recent reviews\cite{Erler:2004cx,Pospelov:2005pr,Fortson:fi,Ginges:2003qt} as well as the text by Lamoreaux and Khriplovich\cite{Khriplovich:ga}, and we refer the reader to those publications for an extensive survey of the literature.

While the potential improvements shown in Table \ref{tab:edm} would not provide access to EDMs associated with the CPV phase of the CKM matrix\cite{Shabalin:rs,Shabalin:sg,Bernreuther:1990jx}, they could allow one to observe an EDM associated with SUSY CPV. As noted above, present EDM limits already preclude ${\cal O}(1)$ CPV phases and TeV scale superpartner masses in the MSSM, and the future measurements will make these constraints even more stringent.

\begin{table}
\begin{center}
\begin{minipage}[t]{16.5 cm}
\caption[]{ Present and prospective EDM limits. Expectations based on SM (CKM)
CP violation are also shown.}
\label{tab:edm}
\vspace*{4pt}
\end{minipage}
\begin{tabular}{|c|r|l|r|c|}
\hline
&&&&\\[-8pt]
System & Present Limit ($e$-cm)& Group & Future Sensitivity & Standard Model 
(CKM) \\
\hline
&&&&\\[-8pt]
$e^-$ & $1.6\times 10^{-27}$  (90\%~CL)  & Berkeley & &  $<10^{-38}$  \\
$e^-$ &  &   Yale (PbO) & $\sim 10^{-29}$ & \\
$e^-$ &    & Indiana/Yale & $\sim 10^{-30}$ & \\
$e^-$ &    & Amherst & $\sim 10^{-30}$ & \\
$e^-$ & & Sussex (YbF) & $\sim 10^{-29} $ & \\
&&&&\\
\hline
&&&&\\
$\mu$ & $ 9.3\times 10^{-19}$ (90\%~CL)  & CERN & &$<10^{-36}$ \\
$\mu$ & & BNL & $\sim 10^{-24}$ & \\
&&&&\\
\hline
&&&&\\
$n$ & $2.9\times 10^{-26}$ (90\%~CL)  & ILL & $1.5\times 10^{-26}$ & 
$1.4\times 10^{-33} - 1.6\times 10^{-31} $ \\
$n$ &  & PSI &  $7\times 10^{-28}$  &\\
$n$ &  & SNS &  $2\times 10^{-28}$ & \\
$n$ &  & ILL &  $2\times 10^{-28}$ & \\
&&&&\\
\hline
&&&&\\
$^{199}$Hg & $2.1 \times 10^{-27}$ (95\%~CL) & Seattle & $5\times 10^{-28}$& 
$\lsim 10^{-33}$ \\
$^{225}$Ra & & Argonne & $10^{-28}$ & \\
$^{129}$Xe & & Princeton & $10^{-31}$ & $\lsim 10^{-34}$ \\
D & & BNL  & $\sim 10^{-27}$ & \\
$^{223}$Rn & & TRIUMF  & $\sim 10^{-28}$ & \\
[-8pt] &&&&\\
\hline
\end{tabular}
\end{center}
\end{table}     

The theoretical interpretation of the present and prospective EDM searches in terms of the parameters of ${\cal L}_{\rm soft}$ requires a careful delineation of a variety of effects. The most straightforward analysis occurs at the level of operators involving the SM fermion and gauge boson fields. In the strong sector of the SM, the lowest-dimension, gauge invariant operator that can generate an EDM is the QCD $\theta$-term:
\be
{\cal L}_{(4)}^{\rm CPV} = \frac{\alpha_s \bar\theta}{8\pi} {\rm Tr}\left( G_{\mu\nu}{\tilde G}^{\mu\nu}\right)
\ee
where $G_{\mu\nu}$ is the SU(3)$_C$ field strength tensor and $\tilde G^{\mu\nu} = (1/2)\epsilon^{\alpha\beta\mu\nu} G_{\alpha\beta}$. The present limits from the EDM of the neutron, $d_n$, lead to the most stringent bounds on ${\bar\theta}$ \cite{baker06}:
\be
\label{eq:thetabar}
|{\bar\theta}| < (1.2\pm 0.6) \times 10^{-10}\ \ \  (90\% \ {\rm C.L.})\ \ \ ,
\ee
where the $(1.2\pm 0.6)$ prefactor is obtained using the QCD sum rule result of  Ref.~\cite{Pospelov:2005pr} and includes the theoretical error quoted in that work.

CP-violation in the electroweak sector arises from the complex phase in the CKM matrix that enters the renormalizable interactions of quarks with $W^\pm$ gauge bosons. The magnitude of its effects is governed by the Jarlskog invariant~\cite{Jarlskog:1985ht}, 
\be
\label{eq:jarlskog}
   J = \cos\theta_1\cos\theta_2\cos\theta_3\sin\theta_1^2\sin\theta_2
   \sin\theta_3\sin\delta = (2.88\pm 0.33)\times 10^{-5},
\ee
with the $\theta_i$ and $\delta$ being the three angles and complex phase in the CKM matrix. The EDMs generated by SM electroweak CP violation arise at multi-loop level\cite{Shabalin:rs,Shabalin:sg,Gavela:1981sm,Khriplovich:1981ca,He:1989xj} and are proportional to $J$, suppressing their effects to the levels indicated in Table \ref{tab:edm} (for a discussion of the corresponding SM effects in atoms and nuclei, see, {\em e.g.}, Refs.~\cite{Haxton:dq,Flambaum:1984fb,Donoghue:dd,Schiff:1963,Engel:2003rz,Engel:1999np,Dzuba:2002kg,Dmitriev:2003sc,Khriplovich:1999qr}). 

The effects of supersymmetric CPV arise through loop corrections to operators involving SM fields. For example, one-loop SUSY corrections to quark propagators can generate complex phases in the quark mass matrix, and through redefinitions of the quark fields, these phases can be absorbed into ${\bar\theta}$. In the absence of a Peccei-Quinn (PQ) symmetry that allows one to absorb these contributions into the axion field and maintain a vanishing ${\bar\theta}$ prior to symmetry-breaking, the experimental limits on ${\bar \theta}$ given in Eq.~(\ref{eq:thetabar}) lead to tight constraints on CPV in the ${\rm SU(3)}_C$ sector of the MSSM\cite{Pospelov:2005pr}. Generally speaking, however, phenomenological analyses of supersymmetric CPV implicitly assume such a PQ mechanism, leading one to consider SUSY contributions to higher dimension operators.

The lowest dimension non-renormalizable, gauge-invariant CPV operators arise at dimension six:
\bea
\label{eq:L6}
{\cal L}_{(6)}^{\rm CPV} & = & \frac{i\, g_1 d_u^B}{\Lambda^2} {\bar Q} \sigma_{\mu\nu}\gamma_5 B^{\mu\nu} H_u U + \frac{i\, g_1 d_d^B}{\Lambda^2} {\bar Q} \sigma_{\mu\nu} \gamma_5 B^{\mu\nu} H_d D \\
\nonumber
& +&  \frac{i\, g_2 d_u^W}{\Lambda^2} {\bar Q} \sigma_{\mu\nu} \gamma_5 \tau^A W^{\mu\nu\, A} H_u U + \frac{i\, g_2 d_d^W}{\Lambda^2} {\bar Q} \sigma_{\mu\nu}\gamma_5 \tau^A W^{\mu\nu\, A} H_d D\\
\nonumber
& +&  \frac{i\, g_3 d_u^G}{\Lambda^2} {\bar Q} \sigma_{\mu\nu}\gamma_5 \lambda^A G^{\mu\nu\, A} H_u U + \frac{i\, g_3 d_d^G}{\Lambda^2} {\bar Q} \sigma_{\mu\nu} \gamma_5\lambda^A G^{\mu\nu\, A} H_d D\\
\nonumber
&+& \frac{w}{\Lambda^2} {\rm Tr}\left(G^{\mu\nu} G_{\nu\alpha} {\tilde G}^{\alpha}_\mu\right)\\
\nonumber
&+& \frac{1}{\Lambda^2} {\rm Tr}\left(G^{\mu\nu} {\tilde G_{\mu\nu}}\right)\, \left[w_u H^\dag_u H_u+w_d H^\dag_d H_d + w_{ud} \left(H_u^\dag \epsilon H_d+{\rm h.c.}\right)\right] \\
\nonumber
& +&  \frac{1}{\Lambda^2} {\rm Tr}\left(W^{\mu\nu} {\tilde W_{\mu\nu}}\right)\, \left[c_u H^\dag_u H_u+c_d H^\dag_d H_d + c_{ud} \left(H_u^\dag \epsilon H_d+{\rm h.c.}\right)\right] \\
\nonumber
&+& \frac{1}{\Lambda^2} {\rm Tr}\left(B^{\mu\nu} {\tilde B_{\mu\nu}}\right)\, \left[b_u H^\dag_u H_u+b_d H^\dag_d H_d + b_{ud} \left(H_u^\dag \epsilon H_d+{\rm h.c.}\right)\right] \\
\nonumber
&+& \frac{1}{\Lambda^2} {\rm Tr}\left(W^{\mu\nu\, a} {\tilde B_{\mu\nu}}\right)\, \left[a_u H^\dag_u\tau^a H_u+a_d H^\dag_d\tau^a H_d + a_{ud} \left(H_u^\dag \tau^a \epsilon H_d+{\rm h.c.}\right)\right] \\
\nonumber
& +& \sum_{ab} \frac{C_{abcd}}{\Lambda^2} \epsilon_{ij} {\bar Q}_i^a d^c {\bar Q}_j^b i\gamma_5 u^d +\cdots
\eea
where the $+\cdots$ indicate gauge invariant operators involving lepton fields. Here, we have chosen to normalize the operators in terms of a new physics scale $\Lambda$ taken to be greater than the scale of electroweak symmetry breaking. After electroweak symmetry-breaking, the terms containing the SU(2)$_L$ and U(1)$_Y$ field strength tensors give rise to the electric dipole moments of the elementary fermions:
\be
{\cal L}_{EDM} = -\frac{i\, d_u^\gamma}{2\Lambda} {\bar U}_L \sigma_{\mu\nu} F^{\mu\nu} U_R- \frac{i\, d_d^\gamma}{2\Lambda} {\bar D}_L \sigma_{\mu\nu} F^{\mu\nu} D_R-\frac{i\, d_\ell^\gamma}{2\Lambda} {\bar \ell}_L \sigma_{\mu\nu} F^{\mu\nu} \ell_R
\ee
where 
\bea
d_u^\gamma & =& -\frac{\sqrt{2}\, v_u\left(c_W\, d_U^B+s_W\, d_U^W\right)}{\Lambda}\\
d_d^\gamma & =& -\frac{\sqrt{2}\, v_d\left(c_W\, d_d^B+s_W\, d_U^W\right)}{\Lambda}\\
d_\ell^\gamma  & =& -\frac{\sqrt{2}\, v_d\left(c_W\, d_\ell^B+s_W\, d_\ell^W\right)}{\Lambda} \ \ \ 
\eea
where $c_W\equiv\cos\theta_W$ and $s_W=\sin\theta_W$. 

The terms coupling $G^{\mu\nu}$ to quarks are the chromoelectric dipole moment operators; the term containing three powers of $G^{\mu\nu}$ is the CPV Weinberg three gluon operator; and the four fermion operator involves products of SU(2)$_L$ doublet and singlet fields (with SU(2)$_L$ indices $i,j$) of flavors $a, \ldots, d$. As discussed in Ref.~\cite{Manohar:2006gz}, the terms containing products of $H^\dag_u H_u$ and $G {\tilde G}$ {\em etc.}, become topological when the Higgs fields are replaced by their vevs, and these effects cannot be analyzed in perturbation theory. They do, however, contribute to the operator ${\cal L}_{(4)}^{\rm CPV}$ and amount to a shift in the value of ${\bar\theta}$ that is constrained by $d_n$. The parts of the $H^\dag_u H_u G {\tilde G}$ operators containing physical scalar degrees of freedom contribute to a renormalization of ${\cal L}_{(4)}^{\rm CPV}$ and, thus, to EDMs. Naive dimensional analysis suggests that values of the operator coefficients $w_u\sim \alpha_s/4\pi$ would lead to EDMs that are consistent with present limits for the mass scale $\Lambda\gsim 1$ TeV.  A comprehensive study of these operators, however, has not been performed at present. 

\begin{figure}
\resizebox{6 in}{!}{
\includegraphics*[60,480][550,660]{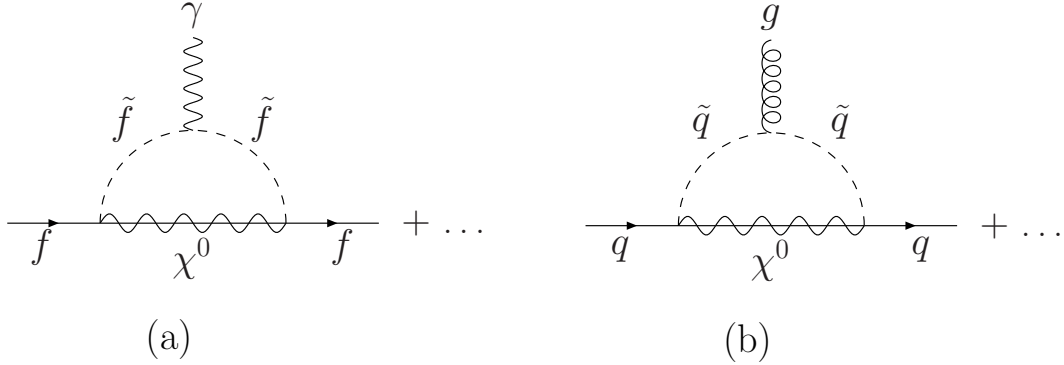}}
\caption{Representative one loop supersymmetric contributions to elementary     
fermion (a) electric dipole moment                                              
and (b) chromoelectric dipole moment (quarks only).}
\label{fig:dfoneloop}
\end{figure}

\begin{figure}
\resizebox{5 in}{!}{
\includegraphics*[0,0][680,240]{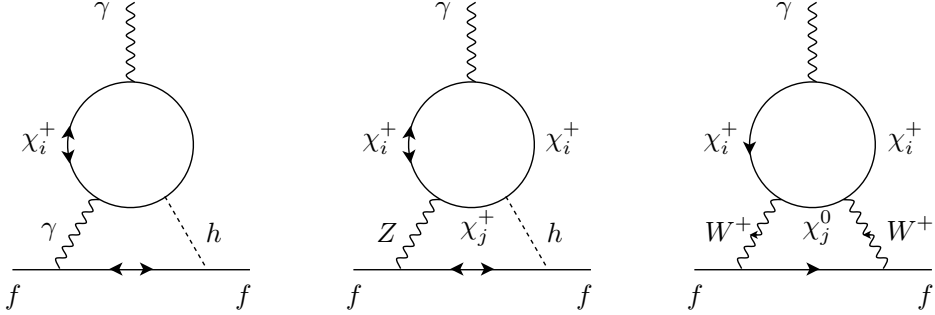}}
\caption{Two-loop contributions to elementary fermion EDMs.  The figures are 
reprinted from Ref.~\cite{Giudice:2005rz} with                  
permission from Elsevier.}
\label{fig:twoloopedm}
\end{figure}

One-loop SUSY contributions to the operator coefficients in Eq. (\ref{eq:L6}) have been computed in Refs.~\cite{Bernreuther:1990jx,Ibrahim:1997gj,Falk:1999tm} (for older one-loop computations, see the literature cited in these studies), while two-loop contributions to the elementary fermion EDMs $d_f^\gamma$ have recently been analyzed in Refs.~\cite{Giudice:2005rz,Chang:2005ac,Chang:2002ex,Pilaftsis:2002fe}. Illustrative contributions are show in Figs. \ref{fig:dfoneloop} and \ref{fig:twoloopedm}.
Approximate results for the one-loop quark and lepton EDMs and quark chromo-EDMs have been given in Ref.~\cite{Pospelov:2005pr}   for a simplified scenario in which only two CP-violating phases contribute: $\phi_\mu$, a common relative phase between the electroweak gaugino masses and the $\mu$ parameter, and $\phi_A$, a common phase associated with the triscalar couplings:
\bea
\frac{d_e^\gamma}{e\kappa_e} & = & -\frac{g_1^2}{12}\sin\phi_A+\left(\frac{5g_2^2}{24}+\frac{g_1^2}{24}\right)\sin\phi_\mu\tan\beta\\
\frac{d_q^\gamma}{e_q\kappa_q} & = & \frac{2 g_3^2}{9}\left(\sin\phi_\mu[\tan\beta]^{\pm 1} +\sin\phi_A\right)
+{\cal O}(g_2^2,g_1^2)\\
\frac{d_q^G}{\kappa_q} & = & \frac{5 g_3^3}{18}\left(\sin\phi_\mu[\tan\beta]^{\pm 1} +\sin\phi_A\right)
+{\cal O}(g_2^2,g_1^2)
\eea
where $e_q$ is the quark charge, 
\be
\kappa_f = \frac{m_f}{16\pi^2\, {\tilde m}}
\ee
and where for simplicity we take $\Lambda={\tilde m}$, a common soft mass, and where the upper (lower) sign corresponds to negatively (positively) charged quarks. 
These results, together with two-loop contributions to the Weinberg three gluon operator coefficient $w$, can be used to compute the EDMs of charged leptons, the neutron, and neutral atoms. In the case of hadrons and atoms, one must contend with a variety of theoretical issues involving non-perturbative QCD, nuclear structure, and atomic structure theory. Extensive, recent reviews  of these issues can be found in Refs.~\cite{Erler:2004cx,Pospelov:2005pr,Fortson:fi,Ginges:2003qt,Khriplovich:ga},  so we do not reproduce those discussions here. Instead, we illustrate the sensitivity of these EDMs to the CPV phases. Considering first the electron, we scale the SUSY mass scale ${\tilde m}$ to 100 GeV and use
$g_1^2 = 4\pi\alpha\tan/\cos^2\theta_W$ and $g_2^2 = 4\pi\alpha/\sin^2\theta_W$ to obtain
\be
\label{eq:oneloopest}
\frac{d_e^\gamma}{\Lambda} \approx 5\times 10^{-25}\, \left(\frac{100\, {\rm GeV}}{{\tilde m}}\right)^2 \left[ \tan\beta\sin\phi_\mu - 0.05\sin\phi_A\right]\ e-{\rm cm}\ \ \ .
\ee
In this case, we see that for $\tan\beta \gsim 1$, $d_e^\gamma$ is overwhelmingly sensitive to the relative phase of the $\mu$ parameter and electroweak gaugino soft masses and that the present experimental bounds in Table \ref{tab:edm} imply
\be
\label{eq:deoneloopest}
\sin\phi_\mu \lsim 3\times 10^{-3} \, \left(\frac{{\tilde m}}{100\, {\rm GeV}}\right)^2\, \cot\beta\ \ \ .
\ee
In short, only for ${\tilde m}\gsim $ a few TeV can one accomodate an ${\cal O}(1)$ phase and remain consistent with present limits.

\begin{figure}
\includegraphics*[width=3 in]{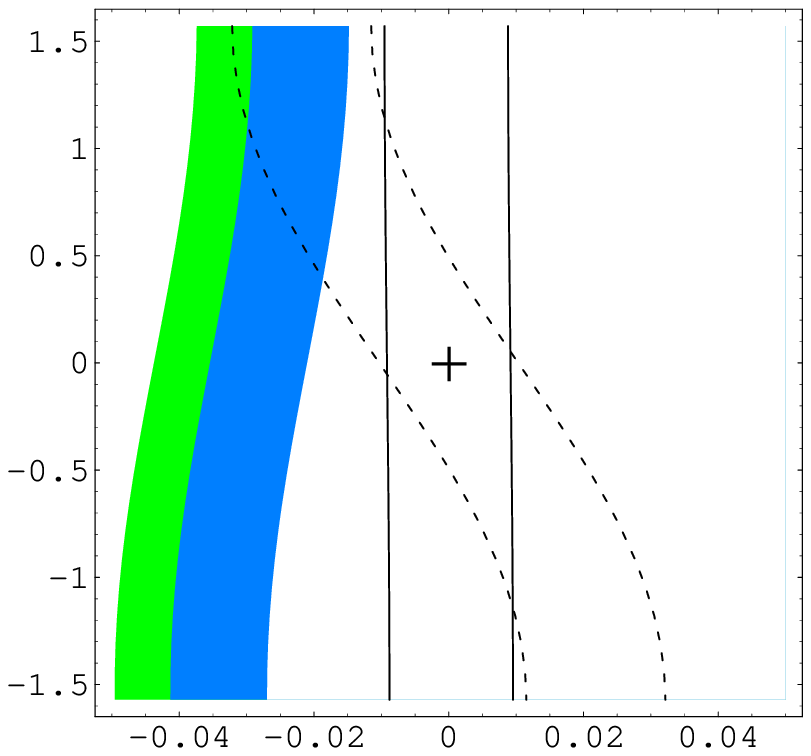}
\includegraphics*[width=3 in]{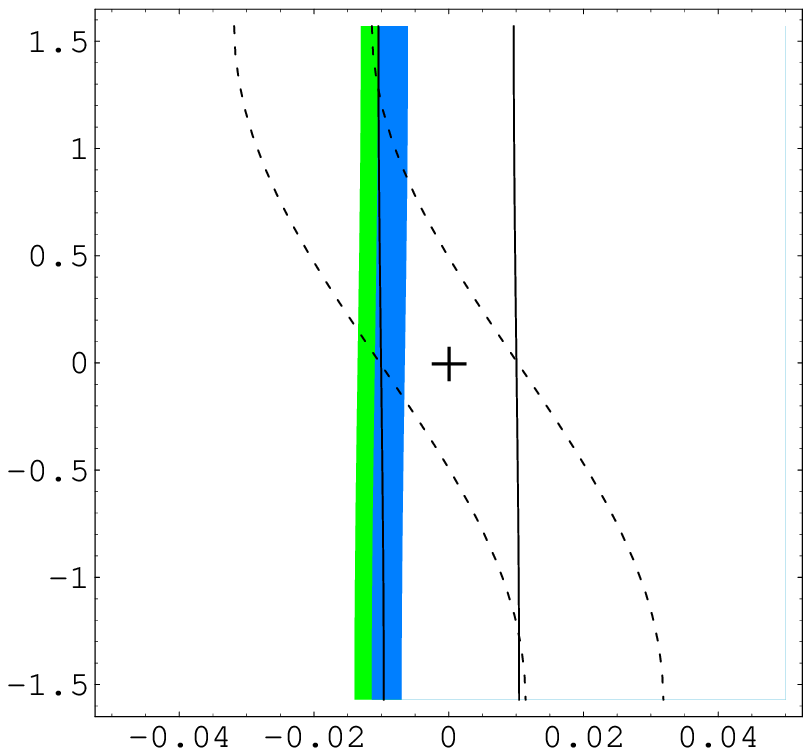}
\caption{
One loop constraints on CPV phases $\phi_\mu$ (horizontal axis) and    
$\phi_A$ (vertical axis) from present 95\% C.L. limits on the EDMs of            
the electron (solid lines), neutron (dashed lines) and the baryon asymmetry     
(colored bands). Constraints consistent with the WMAP value for the baryon      
asymmetry are given by the blue band and those obtained from BBN are given      
by the blue + green bands. Left panel corresponds to choosing $(|\mu|,          
M_2)=(250, 200)$ GeV (non-resonant baryogenesis) and the right panel            
corresponds to $(200, 200)$ GeV (resonant baryogenesis). In obtaining both      
figures, a common sfermion mass of 1 TeV was used.  These figures are courtesy of C.~Lee. }
\label{fig:e-n-oneloop}
\end{figure}

In Fig. \ref{fig:e-n-oneloop}, we illustrate the complementary sensitivity of various EDMs to the SUSY phases using illustrative results for the electron and neutron. Here, we have assumed a common sfermion mass 
$m_{\tilde f} = $ 1 TeV for the first two generations  and show results for two different scenarios for the values of $|\mu|$ and $M_2$. The widths of the bands correspond solely to the experimental error and contains no  theoretical uncertainty associated with the non-perturbative methods used to obtain $d_n$. 
The shaded band indicates the regions required to produce the baryon asymmetry during the supersymmetric electroweak phase transition. The band associated with the $^{199}$Hg EDM limit is similar to that for $d_n$ (see, {\rm e.g.},  \cite{Pospelov:2005pr}). The bands in Fig. \ref{fig:e-n-oneloop} illustrate the general feature that various EDMs display complementary dependences on SUSY CP-violating phases and that results from a variety of EDM searches are needed to obtain meaningful constraints in a given scenario. Moreover, the scale of the allowed phases is quite small: $\sim {\rm few}\, \times 10^{-2}$. This small scale does not appear to be {\em a priori} natural, and leads to the SUSY CP problem indicated earlier. 

As discussed in Ref.~\cite{Pospelov:2005pr} two-loop \lq\lq Barr-Zee" contributions to the EDMs of elementary fermions may become important for TeV scale sfermions in the limit of large $\tan\beta$. For the electron, these contributions are given approximately by
\be
\label{eq:debarrzee}
\frac{d_e^\gamma}{e\kappa_e} \approx \frac{\alpha\,  y_t^2}{9\pi}\, \ln\left(\frac{\tilde m^2}{m_A^2}\right)\, \tan\beta\, \sin(\phi_\mu+\phi_A)\ \ \ ,
\ee
where $Y_t$ is the top Yukawa coupling and $m_A$ is the mass of the CP-odd Higgs. The dependence on these parameters arises from the coupling of the CP-odd Higgs to scalar top quarks in the two-loop amplitude. Comparing with the dominant one-loop contribution we observe that  -- apart from the logarithmic factor -- the two-loop Barr-Zee contribution is suppressed relative to the one-loop EDM by a little more than the $1/16\pi^2$ loop factor. Since the dependence on $\tan\beta$ is similar in both cases,  the two-loop Barr-Zee contribution will be comparable with the one-loop contribution only if $|\phi_\mu| << |\phi_A| $ and a large splitting between the masses of the scalar fermions and CP-odd Higgs generate logarithmic enhancements of the two-loop amplitude. 

The tight one-loop limits on the SUSY CP violating phases may be relaxed for sufficiently heavy sfermion masses, as in \lq\lq split supersymmetry" scenarios. In this case, the dominant contributions arise from the two-loop graphs of Fig. \ref{fig:twoloopedm}. Omitting the Barr-Zee contribution and  letting 
\be
d_f^\gamma({\rm 2\ loop}) = d_f^\gamma(\gamma h) + d_f^\gamma(Zh) + d_f^\gamma(WW)
\ee
corresponding to the three different graphs of Fig. \ref{fig:twoloopedm}, one has, for example, \cite{Giudice:2005rz,Chang:2005ac,Pilaftsis:2002fe}
\be
\label{eq:dftwoloop}
\frac{d_f^\gamma(\gamma h)}{\Lambda^2} = 
\frac{e Q_f \alpha^2}{4\sqrt{2}\pi^2 s_W^2}\, {\rm Im}\left(D_{ii}^R\right) \frac{m_f M_{\chi^+_i}}{M_W m_{h^0}^2}\, F_{\gamma H}(r_{iH}^+)
\ee
where $r^+_{iH} = (m_{\chi^+_i}/m_{h^0})^2$ and $D_{ii}^R$ involves combinations of the chargino diagonalization matrices and Higgs-Higgsino-Gaugino couplings. In the simplified scenario discussed above in which the electroweak gaugino mass parameters have a common phase, ${\rm Im}(D_{ii}^R)\propto \sin\phi_\mu$. Analogous expressions can be obtained for the other contributions in Fig. \ref{fig:twoloopedm}. However, the literature does not agree on the results for these graphs. Numerically, the authors of Ref.~\cite{Giudice:2005rz} obtain in the heavy chargino limit 
\bea
d_e^\gamma(Zh) & \approx & 0.05 d_e^\gamma(\gamma h) \\
d_e^\gamma(WW) & \approx & -0.3 d_e^\gamma(\gamma h) \\
d_n^\gamma(Zh) & \approx &  d_e^\gamma(\gamma h) \\
d_n^\gamma(WW) & \approx & -0.7 d_e^\gamma(\gamma h) 
\eea
where the results for $d_n$ include QCD evolution from the electroweak scale to the hadronic scale.
Note that in the case of the electron, the $Zh$ contribution is suppressed by the small vector coupling of the $Z$-boson to the electron. Moreover, the authors find substantial cancellations between the $WW$ and $\gamma h$ contributions to $d_n$.  In contrast, the authors of Ref.~\cite{Chang:2005ac} find an opposite sign for the $WW$ contribution and no cancellation. Since the relative importance of this disagreement is less severe for $d_e$ we concentrate on the electron EDM below. The sensitivity of the two-loop EDM to the CPV phases is given approximately by
\be
\label{eq:twoloopest}
\frac{d_e^\gamma(\gamma h)}{\Lambda} \approx 2\times 10^{-27} \left(\frac{m_{\chi^\pm}}{m_{h^0}}\right)\, \left(\frac{100\, {\rm GeV}}{m_{h^0}}\right) \left({\rm Im} D_{ii}^R\right) F_{\gamma H}(r_{iH}^+)\ \ e-{\rm cm}\ \ \ .
\ee
For $F_{\gamma H}(r_{iH}^+)\sim {\cal O}(1)$ and ${\rm Im} D_{ii}^R\sim\sin\phi_\mu$ the two-loop EDM is roughly 300 times less sensitive to $\sin\phi_\mu$ than the one-loop contribution. Thus, 
the present $d_e$ limits may accommodate ${\cal O}(1)$ phases for sfermion masses of ${\cal O}(10\, {\rm TeV})$. For $\sin\phi_\mu=0.5$, for example, the one-loop contributions become suppressed relative to the two-loop effects for $m_{\tilde f}\gsim 4-5$ TeV and for $200\, {\rm GeV}\lsim \mu, M_2 \lsim 1$ TeV\cite{Cirigliano:2006dg}.  We discuss the corresponding implications for SUSY baryogenesis and dark matter below.

While the foregoing discussion relies largely on flavor-diagonal CPV, elementary fermion EDMs may also provide information on the flavor structure of ${\cal L}_{\rm soft}$. For example, one-loop gluino contributions to the d-quark EDM can be enhanced by a factor of $m_b/m_d$ in the presence of flavor non-diagonal CPV\cite{Pospelov:2005pr}
\be
\label{eq:dquarkedm}
d_d^\gamma = e\, Q_d\, \delta^d_{131}\, \left(\frac{m_b}{{\tilde m}^2}\right)\, \frac{\alpha_s\tan\beta}{45\pi}
\ee
where
\be
\delta^d_{131}\, = {\rm Arg}\left[\left(\delta^d_{LL}\right)_{13}\, \left(\delta^d_{LR}\right)_{33}\, \left(\delta^d_{RR}\right)_{31}\right]
\ee
and where the $\left(\delta^d_{AB}\right)_{ij}$ have been normalized to an average sfermion mass-squared rather than to the denominator appearing in Eq.~(\ref{eq:massinsert}). The quantity $\delta^d_{131}$ also enters the imaginary part of the one-loop quark mass renormalization, and so will be constrained to the $10^{-9}$ level in the absence of a PQ symmetry. Imposing the latter and taking the present EDM limits one finds bounds on the $\delta^f_{131}$ ranging from $10^{-3}$ to $10^{-6}$ for ${\tilde m}=1$ TeV.

Apart from the Weinberg operator proportional to ${\rm Tr}\, GG{\tilde G}$, the other dimension six operators appearing in Eq.~(\ref{eq:L6}) have received less scrutiny than the the dipole operators. It has been observed in Ref.~\cite{Lebedev:2002ne} that the EDMs of the neutron and neutral atoms may receive important contributions from the four fermion operators in the limit of large $\tan\beta$, as these contributions grow with $\tan^3\beta$. 

\subsection{SUSY Baryogenesis and Dark Matter}

Explaining the origin of the matter density of the universe is an ongoing task that lies at the interface of particle physics, nuclear physics, and cosmology. As is well known, the smallest -- but most anthropically relevant -- component of the energy of the universe consists of baryonic matter, while the next largest component is the non-baryonic cold dark matter (DM). In principle, SUSY could provide a particle physics basis for explaining how both of these matter components came to be and take on their observed abundance. The literature pertaining to both supersymmetric dark matter and baryogenesis is vast, and comprehensive reviews have appeared over the past decade\cite{Jungman:1995df,Bertone:2004pz,Riotto:1999yt,Dine:2003ax}. Here, we review recent theoretical developments and the corresponding implications for SUSY CP-violation.

The baryon asymmetry of the universe (BAU) can be characterized by the ratio of the baryon density to the entropy density of photons:
\be
\label{eq:ewb1}
Y_B\equiv \frac{n_B}{s} = 
\biggl\{
\begin{array}{cc}
(7.3\pm 2.5)\times 10^{-11}, & \text{BBN \cite{Eidelman:2004wy}}\\
(9.2\pm 1.1)\times 10^{-11}, & \text{WMAP \cite{Spergel:2003cb}}
\end{array}
\ee
where the first value (BBN) is obtained from observed light element abundances and standard Big Bang Nucleosynthesis and the second value is obtained from the cosmic microwave background as probed by the WMAP collaboration.

Despite the presence of all three Sakharov  ingredients discussed above, it was shown by Shaposhnikov\cite{Shaposhnikov:1987tw} that they are not sufficiently effective within the SM to account for the observed BAU. In order to prevent ``washout" of any baryon number created at the electroweak temperature, the electroweak phase transition (EPWT) has to be strongly first order. The strength of the EWPT depends on the parameters of the Higgs potential, $V_H$,  that also govern the mass of the Higgs boson. In the SM, the mass of the Higgs must be below 45 GeV to allow for a strong first order EWPT, so the present LEP 2 direct search limit $m_h > 114.4$ GeV precludes this possibility. 
Roughly speaking, this bound arises from a competition between thermal contributions to $V_H$ and those generated by parameters in the Lagrangian at $T=0$:
\bea
\label{eq:vh1}
V_H(T,\phi) & = & V_{\rm eff}(\phi)+V_{\rm th}(T,\phi)\\
V_{\rm eff}(\phi) & = & \frac{1}{2} m^2\phi^\dag \phi+\frac{\lambda}{4}(\phi^\dag\phi)^2+\cdots
\eea
where $V_{\rm eff}(\phi)$ is the zero temperature one-loop effective potential and $V_{\rm th}(T,\phi)$
 gives thermal contributions to the potential.  One may then use the high temperature expansion to determine the shape of $V_H(T,\phi)$ as a function of $T$, leading to (see, {\em e.g.}, Ref.~\cite{Quiros:1999jp})
 \be
 \label{eq:veff}
 V_H(T,\phi) = D(T^2-T_0^2) \phi^2 - ET \phi^3 +\frac{\lambda(T)}{4}\phi^4
 \ee
 where $T_0$ is the temperature below which the quadratic term becomes negative and where $D$, $E$, $\lambda(T)$ and $T_0$ are functions of the scalar, vector boson,  and fermion masses, $v$, and $T$. The EWPT is characterized by two additional temperatures in addition to $T_0$:
\bea
\label{eq:ewpt1}
T_1^2 & =&  \frac{8\lambda D T_0^2}{8\lambda D - 9E^2}\\
\nonumber
T_C^2 & = & \frac{\lambda D T_0^2}{\lambda D-E^2}\ \ \ .
\eea
For $T< T_1$, a potential barrier forms between the phases of broken and unbroken electroweak symmetry, associated with minima of the potential at $\phi^0=0$ and $\phi^0=v(T)/\sqrt{2}\equiv \phi_m$. The formation of a barrier is 
accompanied by the onset of tunneling between the two phases and the formation of bubbles of broken electroweak symmetry. For $T<T_C$, one has $V_H(\phi_m,T) < V_H(\phi=0,T)$ while for $T<T_0 < T_C$, the potential at $\phi=0$ is a local maximum. 

In order to avoid washout of the baryon asymmetry during the EWPT, the energy associated with the sphaleron configurations, $E_{\rm sph}$, must be sufficiently large as to suppress the associated transition rate between different vacua. The sphaleron energy is also tied to the value of $v(T)$ since the gauge and Higgs fields are coupled through the associated equations of motion. From numerical studies, one obtains
\be
\label{eq:ewpt2}
\frac{E_{\rm sph}}{T_C} = \frac{4\pi}{g}\, \frac{v(T_C)}{T_C}\, B(\lambda/g^2) \gsim 40
\ee
where $B(\lambda/g^2)$ is a constant of ${\cal O}(1)$ that depends gently on $\lambda/g^2$ and where the last inequality is imposed in order to ensure a value of $Y_B$ no smaller than given by Eq.~(\ref{eq:ewb1}). The latter requirement, thus, leads to
\be
\label{eq:ewpt3}
\frac{v(T_C)}{T_C}\gsim 1\ \ \ .
\ee
The value of $v(T_C)$ is, in turn, determined by minimization condition $V^\prime(\phi,T_C)=0$ that yields
\be
\label{eq:ewpt4}
\frac{v(T_C)}{T_C} = \frac{2 E}{\lambda(T_C)} = 4E\, \left( \frac{v_0^2}{m_h^2}\right)+\cdots \ \ \ ,
\ee
where the the \lq\lq$+\cdots$" indicate small corrections arising from the logarithmic $T$-dependence of $\lambda$. From Eqs.~(\ref{eq:ewpt3},\ref{eq:ewpt4}) one has
\be
\label{eq:ewpt5}
4 E\, \left( \frac{v_0^2}{m_h^2}\right)\gsim 1
 \ee
as the condition on the parameters in $V_H(\phi,T)$ that must be satisfied in order to obtain a strong first order EWPT that prevents washout of the baryon asymmetry. 

 In the SM, the cubic coupling is given by
 \be
 \label{eq:ewpt6}
 E = \frac{2 m_W^3+m_Z^3}{4\pi v^3} \approx 0.01
 \ee
implying that $m_h$ must be lighter than about 45 GeV in order to satisfy the condition (\ref{eq:ewpt5}). As this bound is well below the present LEP II lower bound, the electroweak baryogenesis in the SM is clearly ruled out experimentally. 
 
In the MSSM, the bound on the lightest CP-even Higgs mass can be relaxed either through additional scalar contributions to $V_H(\phi,T)$ that increase the value of $E$ or choices of the other parameters that allow for a lighter Higgs boson (see the discussion in Section \ref{sec:higgs}). The former possibility is realized in the presence of a light, right-handed stop that couples strongly to the Higgs fields, contributes to $E$ at the one-loop level, and enhances $E$ by roughly an order of magnitude :
\bea
\label{eq:ewpt7}
E_{MSSM} & \approx &  E_{SM}+\frac{y_t^3\, \sin^3\beta\, \left(1-{\bar A}_t^2/M^2_{Q_3}\right)^{3/2}}{4\sqrt{2}\pi} \approx 9\, E_{SM}\\
\nonumber
{\bar A}_t & = & A_t-\mu\cot\beta \ \ \ .
\eea
In principle, light left-handed stops could also generate large enhancements, but precision electroweak data rule out the existence of such a light ${\tilde t}_L$. The feasibility of the light ${\tilde t}_R$-induced enhancement of $E$ also depends on the avoidance of color- and charge-breaking minima, leading to conditions on the soft parameter $M^2_{U_3}$. From the analysis of this requirement in Ref.~\cite{Carena:1996wj}, one finds that RH stops lighter than about 130 GeV are disfavored. Assuming these conditions are satisfied, the enhanced value of $E_{MSSM}$ increases the upper bound on the lightest Higgs from Eq.~(\ref{eq:ewpt5}) to $\sim 120$ GeV. 

Going beyond the MSSM, one may strengthen the EWPT by introducing singlet Higgs supermultiplets $S$ via the superpotential\cite{Gunion:1984yn,Pietroni:1992in,Davies:1996qn,huber96}
\be
\label{eq:nmssm1}
W_{\rm singlet}=\left(\mu+\alpha S\right)\, H_1 H_2 + \beta S + \frac{\kappa}{3} S^3
\ee
and the corresponding soft Lagrangian that contains triscalar couplings
\be
\label{eq:nmssm2}
{\cal L}_{\rm singlet} = -\left(\alpha A_\alpha H_1\epsilon H_2 S+{\rm h.c.}\right)-\left(\frac{1}{3}\kappa A_\kappa S^3+{\rm h.c.}\right)+\cdots \ \ \ ,
\ee
where $\epsilon$ is the antisymmetric SU(2) tensor, and $A_\alpha$ and $A_\kappa$ are the soft SUSY breaking parameters.
Writing the vevs of the neutral fields as 
\bea
\nonumber
{\rm Re}\langle H_1^0\rangle & = & \frac{\phi}{\sqrt{2}}\cos\gamma\cos\beta\\
\label{eq:nmssm3}
{\rm Re}\langle H_2^0\rangle & = & \frac{\phi}{\sqrt{2}}\cos\gamma\sin\beta\\
\nonumber
{\rm Re} \langle S\rangle  & = & \frac{\phi}{\sqrt{2}}\sin\gamma
\eea
the corresponding singlet contribution to the cubic term is given by
\be
\label{eq:nmssm4}
E_{\rm singlet} = \frac{2\sqrt{2}\, \sin\gamma}{T}\, \left(\alpha A_\alpha \cos^2\gamma\sin 2\beta +\frac{2}{3}\kappa A_\kappa \sin^2\gamma\right)\ \ \ .
\ee
Supersymmetric models with singlet Higgs are generically referred to as \lq\lq next-to-minimal" and are motivated largely by a desire to generate the $\mu$ parameter as the vev of the singlet field\footnote{In writing Eq.~(\ref{eq:nmssm1}), we follow the conventions of Ref.~\cite{Davies:1996qn}, where a possible quadratic term in  eliminated by a constant shift of the field $S\to S+c$. In this case, an explicit $\mu$ parameter appears.} . Models of this type have received considerable attention recently, as in the studies of Refs.~\cite{nonminimal}.

The feasibility of choosing the soft parameters to obtain a sufficiently larger value of $E_{\rm singlet}$ while respecting the experimental lower bounds on the lightest CP-even Higgs were initially studied in Ref.~\cite{Pietroni:1992in}. It  was shown in that there exist considerable regions of the singlet parameter space leading to a sufficiently strong first order EWPT. The more stringent LEP II lower bounds on the mass of the Higgs reduce this available parameter space, but there remain considerable regions that admit $v(T_C)/T_C \gsim 1$. 
 
In addition to these EWPT considerations, the size of the CP-violating asymmetries generated by particle physics interactions at the phase boundary must also be studied. In the SM, these asymmetries are highly suppressed by the Jarlskog invariant (\ref{eq:jarlskog}) and by 
\be
\label{eq:ewb2}
(y_t^2-y_c^2)(y_t^2-y_u^2)(y_c^2-y_u^2)(y_b^2-y_s^2)(y_b^2-y_d^2)(y_s^2-y_d^2)\approx 4\times 10^{-17}
\ee
since CP-violating effects should vanish in the limit that any two quarks become degenerate. 
Farrar and Shaposhnikov\cite{Farrar:1993sp,Farrar:1993hn} subsequently argued, however, that the relevant CP-violating asymmetry depends solely on the difference between the probabilities for reflection and transmission of $s$ and $d$-quark currents and the phase transition boundary, so that $Y_B$ is proportional to $y_s-y_d$ rather than the combination in Eq.~(\ref{eq:ewb2}). Assuming that a non-SM mechanism generates a strong first order EWPT, the resulting expectation for the BAU in the SM is much closer to the observed value than in Shaposhnikov's original work. 

As with the EWPT, the presence of supersymmetric interactions at the phase boundary may also lead to larger CP-violating asymmetries, as there exist a plethora of CPV interactions that are not Jarlskog suppressed. The computation of these asymmetries was initiated by the authors of Refs.~\cite{Cohen:1994ss,Joyce:1994zn} using conventional transport methods and was followed up by a number of subsequent studies in the MSSM\cite{Huet:1995sh,mssmewb}. The results generally indicated that SUSY CPV phases of $\lsim {\cal O}(1)$ would be needed to obtain the observed values of $Y_B$\cite{Huet:1995sh}. In the late 1990's, however, Riotto pointed out -- using more sophisticated non-equilibrium field theory methods -- that memory effects in the EWPT plasma could resonantly enhance the CP-violating sources needed for  successful EWB with smaller CP-violating phases\cite{Riotto:1998zb} for appropriately tuned values of the MSSM parameters. Subsequent detailed analyses of these resonant enhancements were performed in Refs.~\cite{carena97,resonantewb}. These resonant enhancements allow for successful EWB in the MSSM with significantly smaller CPV phases than implied by earlier work, thereby allowing for consistency with the corresponding EDM bounds on these phases.

The CPV sources appear in the transport equations for Higgs and quark supermultiplet densities that  govern the projection of chiral charge at the phase boundary.
In the case of the Higgs supermultiplet current density, $H_\mu$, for example, one has 
\be
\label{eq:Heq}
\partial^\mu H_\mu = - \Gamma_H\frac{H}{k_H}
-\Gamma_Y\biggl(\frac{Q}{k_Q} - \frac{T}{k_T} + \frac{H}{k_H}\biggr) -
{\tilde\Gamma}_Y\biggl(\frac{B}{k_B} - \frac{Q}{k_Q} +
\frac{H}{k_H}\biggr)+ \bar\Gamma_h\frac{h}{k_h} + S_{\widetilde
H}^{\CPV} \ \ \ .  
\ee 
$Q$ and
($B$,$T$) are the number densities of particles in the third
generation left- and right-handed quark supermultiplets, respectively;
the $k_{H,h,Q,T,B}$ are statistical weights; $S_{\widetilde H}^{\CPV}$
is a CP-violating source; and $\Gamma_H$, $\Gamma_Y$,
${\tilde\Gamma}_Y$, and $\bar\Gamma_h$ are transport coefficients. The terms proportional to $\Gamma_H$ and ${\bar\Gamma}_h$ cause any non-zero Higgs supermultiplet asymmetry to relax to zero, as favored by minimization of the free energy. The terms containing $\Gamma_Y$ and 
${\tilde\Gamma}_Y$ favor the transfer of the Higgs asymmetry into the baryon sector and are, thus, essential for the generation of a non-vanishing $Y_B$ from CPV in the Higgs sector. The net baryon asymmetry depends on a detailed competition between the effects of the CPV sources and the CP-conserving relaxation and Higgs-to-baryon transfer rates.

Analogous equations obtain for the quark supermultiplet densities. Carena {\em et al} observed that the enhancements of the sources $S_{{\widetilde H},T,Q}^{\CPV}$ occur when LH and RH scalar top quarks or Higgsinos and gauginos are nearly degenerate, leading to resonant scattering from the spacetime varying Higgs vevs\cite{carena97}. As noted above, however, the requirements of a strong first order EWPT and of precision electroweak data preclude the occurrence of such degeneracies in the scalar top sector, implying that resonant supersymmetric EWB may only occur via gauginos and Higgsinos. 

These developments were followed up by studies in 
Refs.~\cite{Konstandin:2004gy,Konstandin:2003dx,Konstandin:2005cd}, who found somewhat smaller resonance effects from the sources, and by the work of Refs.~\cite{Lee:2004we,Cirigliano:2006wh}, in which the relaxation and Higgs-to-baryon transfer coefficients terms in the quantum transport equations were computed using the same non-equilibrium methods. In particular, the authors of the latter work observed that the various terms in Eq.~(\ref{eq:Heq}) (and the analogous quark supermultiplet equations) could be derived by expanding the Greens functions and self-energies entering the non-equilibtrium Schwinger-Dyson equations in ratios of physical scales present during the EWPT.
It was also pointed  that resonance effects could also enhance  Higgsino and chiral charge relaxation coefficients ($\Gamma_H$, {\em etc}.) , thereby mitigating the effect of enhanced CP violating sources. In addition, the lowest order contributions to the Higgs-Quark transfer coefficient, $\Gamma_Y$, had been neglected in earlier analyses. The net impact of these refinements implies that the viability of supersymmetric EWB in the MSSM is a quantitative question, depending in detail on the parameters of the theory and requirement input from EDMs, present and future collider studies, and precision electroweak data.

\begin{figure}
\resizebox{3 in}{!}{
\includegraphics{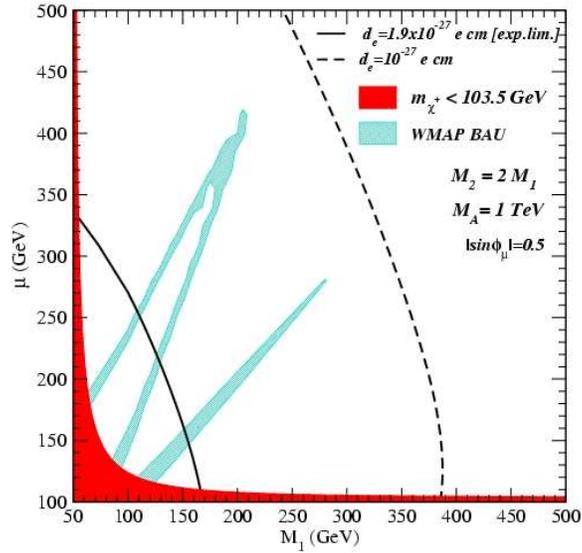}}
\caption{Constraints on MSSM parameters from resonant electroweak baryogenesis, two-loop electric dipole moment of the electron, and LEP II .  This figure is courtesy of S. Profumo.}
\label{fig:edm-dm-ewb}
\end{figure}

Although  formal theoretical issues in EWB remain to be addressed, it is instructive to consider the phenomenological implications of the recent studies for EDM searches. Recent analyses have appeared in Refs.~\cite{Balazs:2004ae,Cirigliano:2006dg,Huber:2006ri,YaserAyazi:2006zw}. 
Illustrative results are given in Fig.~\ref{fig:edm-dm-ewb} , where we show  the regions of the $\mu$-$M_1$ parameter space that correspond to resonant MSSM EWB (light blue bands), assuming GUT relation between the gaugino masses. The funnel-like structure corresponds to the resonance conditions: $\mu\sim M_1$ or $\mu\sim M_2$. The red region is excluded by LEP 2. The semicircular bands indicate the exclusion region derived from the present (dark blue) and prospective (black) electron EDM measurements for $\sin\phi_\mu\sim 0.5$ using the two-loop computations of Refs.~\cite{Giudice:2005rz,Chang:2005ac}. An analogous set of plots for various values of $\tan\beta$ and $m_A$ are given in Ref.~\cite{Cirigliano:2006dg} for AMSB type gaugino mass relations.  

The study of Ref.~\cite{Cirigliano:2006dg} indicates that the viability of resonant EWB in the MSSM implies $|d_e|\gsim 10^{-28}$ e-cm and gaugino/Higgsino mass parameters $\lsim 1$ TeV. Although some portion of the remaining parameter space can be explored with LHC studies, full collider exploration of the needed MSSM parameters will await the ILC. In addition, future dark matter detection experiments may provide a complementary probe, as the character of the LSP is governed by the same parameters $\mu$ and $M_{1,2}$ that determine the viability of resonant EWB. Assuming that the relevant portions of the MSSM parameter space leads to the observed DM relic abundance, one could expect resonant EWB to be accompanied by enhanced production of, and detection rates for, high energy neutrinos produced by neutralino annihilation in the sun.   The absence of evidence for high energy solar neutrinos in Super Kamiokande data implies that a portion of the parameter space in Fig.~\ref{fig:edm-dm-ewb}   near the base of the EWB funnels would be excluded under such scenarios. Future neutrino telescopes and ton-sized direct detection experiments will considerably extend the reach of these DM probes of EWB. In most cases, however, obtaining the observed relic abundance requires modifications from standard cosmology, such as the presence of an additional energy-density that increases the Hubble parameter and allows for earlier decoupling of neutralinos.

\section{$Z$-pole Electroweak Precision Measurements}
\label{sec:zpole}
The electron-positron colliders SLC and the LEP have produced millions of $Z$ bosons
at $\sqrt{s}\simeq M_Z$.  The measurements of $Z$ lineshape, forward-backward 
asymmetries and  polarized asymmetries at $Z$-pole 
lead to a precise determination of the $Z$-boson pole mass, the total 
$Z$-width and the $Z$ couplings to fermions pairs. 
The $Z$-pole precision observables are also combined
with the results from other experiments, like CDF, D0, and low 
energy atomic parity violation (APV) and scattering measurements to confront 
theories such as the Standard Model or other new physics extensions.
The global fit to the electroweak precision observables are 
in excellent agreement with 
the Standard Model.   Any new physics beyond  the SM that could 
contribute to these precision observables is, therefore,  tightly constrained by 
the existing data.  
In the $R$-parity conserving MSSM, new corrections to the precision observables arise
from superparticle loops.  The SUSY contributions
could be significant when the superparticles are light.
Global analyses of precision observables in the 
framework of MSSM have appeared in the  
literature
\cite{cho, boer, erler, Heinemeyer:2004gx, ellis,DjouadiYK,AltarelliWX,Belanger:2004ag}, 
which will be briefly reviewed in this section.

\subsection{Precision Observables}
The $Z$ pole observables can be organized into several 
groups \cite{erler,EWG}:
\begin{itemize}
\item{9 lineshape observables}\\
A fit to the $Z$ lineshape and the leptonic forward-backward asymmetries 
determines the $Z$ boson mass $M_Z$, the total $Z$-width $\Gamma_Z$,
the hadronic cross section $\sigma_{\rm had}^0$, and, for each lepton flavor,
$l=e$, $\mu$, $\tau$, the ratio
$R_l=\Gamma_{\rm had}/\Gamma_{ll}$\footnote{The results are combined into one
$R_l$ if lepton flavor universality is assumed.},
and the pole asymmetry $A_{FB}^{0,l}$. Defining 
\beq
A_f=\frac{1-4 Q_f \sin^2\theta_{\rm eff}^f}{1-4 Q_f \sin^2\theta_{\rm eff}^f
+8 Q_f^2 \sin^4\theta_{\rm eff}^f},
\eeq
where $\sin^2\theta_{\rm eff}^f$ is the effective weak mixing angle for 
fermion $f$  at the $Z$-pole, we have 
\beq
A_{FB}^{0,f}=\frac{3}{4}A_e A_f.
\eeq

\item{3 further LEP asymmetries}\\
These include the $\tau$ polarization 
${\cal P}(\tau)=A_{\tau}$, its forward-backward asymmetry 
${\cal P}_{FB}(\tau)=A_e$, and the hadronic charge asymmetry,
$\langle Q_{FB} \rangle$, which is used as a determination of 
$\sin^2\theta_{eff}^e$.

\item{6 heavy flavor observables}\\
These include the ratios $R_q=\Gamma_{qq}/\Gamma_{\rm had}$,
the forward-backward pole asymmetries $A_{FB}^{0,q}$, 
and the left-right forward-backward asymmetries 
$A^{LR}_{FB}(q)=A_q$, each for $q=b,c$.

\item{3 further SLD asymmetries}\\
The precise measurements at SLD using the polarized electron beam
determine $A_e$, $A_\mu$ and $A_\tau$\footnote{The 
results are combined into one
$A_l$ if lepton flavor universality is assumed.}.
\end{itemize}

Besides the precision measurements performed at the $Z$-resonance,
the $W$ boson mass ($M_W$) and  width ($\Gamma_W$) 
have been measured to a relatively high precision
at both the Tevatron and the LEP 2.  These quantities are  usually included in the list of 
precision observables.

Low energy precision observables are sometimes included in  global fits.
These observables include
\begin{itemize}
\item{Two weak charge measurements from atomic parity violation:
$Q_W^{\rm Tl}$ and $Q_W^{\rm Cs}$.}
\item{Deep inelastic scattering experiments that yield
$\kappa$, which is a linear 
combination of effective 4-Fermi operator coefficients.}
\item{$\nu_\mu e$ scattering experiments that yield the leptonic 4-Fermi
operator coefficients $g_V^{\nu e}$ and $g_A^{\nu e}$.}
\item{NuTeV results on the neutrino-nucleus deep inelastic scattering.}
\item{The branching ratio of $B\rightarrow X_s \gamma$.}
\item{Muon anomalous magnetic moment $(g-2)_{\mu}$.}

\end{itemize}

\subsection{SM Global Fit}
The Standard Model contributions to the precision observables depend on 
the free parameters in the model. In the gauge and Higgs sectors one has five parameters: the three gauge couplings, the Higgs vev, and the physical Higgs boson mass ($m_h$).  From among these, one ordinarily chooses the fine structure constant ($\alpha$) and Fermi constant  ($G_\mu=1/\sqrt{2}v^2$) as independent inputs because they are known precisely from low-energy measurements. The remaining quantities in this sector that are relevant for $Z$-pole observables are the mass of the $Z$-boson ($M_Z$),  $m_h$, and the SU(3)$_C$ coupling ($\alpha_s$). In addition, $Z$-pole observables depend on the value of the running QED coupling $\alpha(M_Z)$. Since $\alpha(M_Z)/\alpha$ receives hadronic contributions associated with the five light quarks that are not calculable with the same precision as other  SM contributions, one treats these contributions, denoted as $\Delta \alpha_{\rm had}^{(5)}$, as an separate parameter to be obtained from the fits. The  top quark does not contribute to the running couplings at $\mu=M_Z$ and below because $m_t>M_Z$. However, one-loop radiative corrections depend strongly on $m_t$ so it is also treated as a fit parameter.

The $Z$-boson mass 
$M_Z$ has been determined at LEP to a high precision comparable to that of $G_{\mu}$.  
Therefore, $M_Z$ is sometime taken as fixed input instead of as a fitting parameter.  
The top quark mass $m_t$ has  been measured directly at CDF and D0.
Its value from the direct measurement is included in the global fit
as a constraint.  The strong coupling constant $\alpha_s$, can be determined from 
from non-$Z$ lineshape\cite{alphas}; 
and $\Delta \alpha_{\rm had}^{(5)}$,  is obtained 
from hadronic $\tau$ decay \cite{AlemanyTN}. Both determinations  
are included in the global fitting as extra constraints.
The lower bound on the Higgs mass from LEP Higgs searches is sometimes 
included as well.

A global fit to the precision observables with extra
constraints on $m_t$,  $\alpha_s$, $\Delta \alpha_{\rm had}^{(5)}$
and $m_h$ determines the fitted 
values of the input parameters.  The studies from LEP electroweak
working group \cite{EWG} included the latest results on the 
mass of the top quark from Tevatron: $m_t=171.4 \pm 2.1$ GeV, 
the width of the $W$ boson from the Tevatron and LEP-2: 
$\Gamma_W = 2.147\pm 0.060$ GeV,
and the mass of the $W$ from LEP-2: $M_W = 80.392 \pm 0.029$ GeV.
The program ZFITTER is used to calculate the SM 
predictions for those precision observables, including full
one-loop radiative corrections\footnote{ZFITTER uses on-shell renormalization.} and higher order QCD and electroweak 
corrections \cite{ZFITTER}.
The global fit to all the high $Q^2$ precision observables showed 
excellent agreement between the measurements and
the SM fitted values.  The discrepancies are usually 
less than 2 $\sigma$ for almost all the observables except for $A_{FB}^{b}$, 
where the deviation is about  3 $\sigma$.
The fitted Higgs mass
is in the range of $85_{-28}^{+39}$ GeV at 68\% C.L.  An upper limit of 
$m_h < 166$ GeV is obtained at 95\% C.L.  This limit increases to 199 GeV
when the LEP-2 direct Higgs search limit of 114 GeV is included in the fit.

\subsection{MSSM Contributions to the Precision Measurements}

The success of the SM global fit to the electroweak precision measurements
imposes strong constraints on any new physics extension beyond the Standard 
Model.  Supersymmetric models can always avoid such constraints
since supersymmetric corrections decouple in the ${\tilde m}\to\infty$ limit.  Thus,   supersymmetric  models 
look just like the Standard Model if the supersymmetric mass
scale is large enough.  As long as the Standard Model with a light 
Higgs boson provides a good fit to the data, supersymmetric models can as 
well.  On the other hand, large contributions from SUSY are possible when a
supersymmetric spectrum includes light superparticles. 
There have been numerous attempts to identify 
constraints on the MSSM parameters from precision 
observables, both in the general framework of 
MSSM \cite{cho, boer, AltarelliWX}, 
or in a specific SUSY breaking scenario, {\em e.g.}, 
mSUGRA\cite{boer, erler,ellis,Heinemeyer:2004gx,DjouadiYK, Belanger:2004ag}, 
or GMSB\cite{erler}.
The results from different group differ slightly, depending on the choices of 
the set of precision observables that are included in the fit,  the order of 
loop corrections, and the experimental values that are used.
We will review the general feature of SUSY contributions to the 
precision observables in this section, and leave the discussion of the 
global fit in a particular SUSY framework to section~\ref{sec:sugra_gmsb}.

When either the supersymmetric scalars (squarks and sleptons) or 
the supersymmetric fermions (charginos and neutralinos) are sufficiently 
heavy, the radiative corrections to  precision observables are 
dominated by the universal gauge boson propagator corrections,
or the oblique corrections $S$, $T$ and $U$ [see Eqs.~(\ref{eq:stu-sirlin})].  The authors of Ref.~\cite{cho}  systematically studied the  MSSM contributions to the oblique parameters from four different sectors:
squarks, sleptons, neutralinos/charginos, and the Higgs sector.  
They found that relatively light squarks/sleptons generally make the 
fit to the electroweak data worse than the SM fit.
The squark sector contributes essentially to the positive $T$ direction.
The slepton sector contributes negatively to $S$, but $T$ remains 
constant or slightly positive for large $\tan\beta$.  Both tend 
to be disfavored by data.
The contributions from 
light charginos and neutralinos make both 
$S$ and $T$ negative, which slightly improves the fit.  The best
fit is obtained when the lightest chargino
mass is near its experimental lower bound.  The 
contributions from the MSSM Higgs sector are found to be small 
when the light CP-even MSSM Higgs mass is taken to be the SM Higgs boson
mass.

When both the supersymmetric 
scalars and fermions are light, additional 
non-oblique  (vertex, external leg, and box graph) corrections to all the $Zff$ vertices
become important.  In addition to considering the non-oblique corrections to specific processes, one must also include the the MSSM contributions to the muon decay amplitude since we express the coupling ${\hat g}^2$ in terms of $G_\mu$ (see Section \ref{sec:renorm}).
Morevoer, for large $\tan\beta$, the bottom and tau Yukawa couplings are
large.  In this case, the MSSM Higgs boson loops can appreciably affect the 
$Zbb$, $Z\tau\tau$ and $Z\nu_{\tau}\nu_{\tau}$ vertices,
especially for small $m_A$.  As a result, the overall fit to the precision
observables for large $\tan\beta$  is worse than the SM.
On the other hand, the order $\alpha_s$ SUSY-QCD contributions to $Zqq$
vertex via gluino-squark-squark loop are found to be 
negligibly small when the mass for gluino and squarks are 
bigger than about 200 GeV.
The electroweak contributions to $Zqq$ vertices due to light
squarks and light neutralinos/charginos are 
insignificant when their masses are above the 
current direct search limit.  
When the masses of left-handed slepton and  neutralinos/charginos
are light, the $Zll$ vertices as well as the muon-decay
amplitude are affected significantly.  The fit is  improved 
slightly when the left-handed slepton mass 
is around 200$\sim$500 GeV \cite{cho}.
In contrast, the right-handed slepton
and squark masses are not constrained significantly, and hence 
smaller masses are still allowed.



In summary, for SUSY spectrum with light left-handed sleptons and 
chargino/neutralinos,  
the global fit in the MSSM has a lower $\chi^2$ value than in the SM.
Since the MSSM fit has fewer degrees of freedom than the SM fit\footnote{The increase in the number of parameters in the MSSM compared to the SM reduces the number of fitting degrees of freedom.}, 
the overall fit probability in the MSSM is very similar
to that in the SM.

\subsection{Global Analysis in mSUGRA and GMSB}
\label{sec:sugra_gmsb}
In a particular SUSY breaking scenario such as  mSUGRA or GMSB, the 
complete SUSY spectrum can be determined from the RGE running of only a 
few parameters from the high energy scale down to the weak scale.  
Due to the relatively small numbers of parameters in these scenarios,
a global fit to the precision observables 
can be used to exclude certain region of the parameter space if 
the fit is significantly worse than the SM.

Ref.~\cite{erler} studied the global fit of precision observables (including 
certain low energy measurements) in the framework of mSUGRA and 
GMSB.  It is shown that significant portions of the parameter spaces
of mSUGRA and GMSB are excluded.  Requiring 
$\chi^2_{\rm MSSM}-\chi^2_{\rm SM}< 3.84$, 
a lower limit on the mass of the light 
CP-even Higgs: $m_h \geq 78$ GeV can be obtained.  Also, the first
and second generation squark masses are constrained to be above 
280(325) GeV in the mSUGRA(GMSB) model.

The global fit in mSUGRA performed in Ref.~\cite{boer}
[including $(g-2)_{\mu}$ and $b\rightarrow s \gamma $]
showed that  at 95\% C.L., 
the value of $\tan\beta$ is constrained to be above 6.5,
while the value of the gaugino masses at the GUT scale has to be above
$\sim$ 220 GeV, which corresponds to a lower limit on the lightest
neutralino(chargino) of 95 (175) GeV.

Several analyses focusing 
on the cold dark matter relic density favorite region in mSUGRA have been 
performed \cite{ellis, Heinemeyer:2004gx, DjouadiYK,Belanger:2004ag}.
Ref.~\cite{ellis} includes all the high $Q^2$ 
precision observables, muon $(g-2)_{\mu}$ and 
$b\rightarrow s \gamma$.  The analysis shows a favored region for $\mu>0$
and small $m_0$ and $m_{1/2}$.
Ref.~\cite{Heinemeyer:2004gx} 
focuses on the SUSY contributions to the $W$ boson mass $M_W$,
the effective leptonic weak mixing angle 
$\sin^2\theta_{\rm eff}$, the anomalous magnetic moment of the 
muon $(g-2)_{\mu}$, and $b\rightarrow s \gamma$.  Higher order 
loop corrections are included and both theoretical and experimental 
errors are treated.
A fit to these precision quantities 
in mSUGRA shows 
a clear preference for relative small value of $m_{1/2}$, with
a best-fit value of about 300 GeV for $\tan\beta=10$.  
An upper bound of about 600 GeV on $m_{1/2}$ is obtained at 90\% C.L.. We note that only the fits of Refs.~\cite{Heinemeyer:2004gx,Belanger:2004ag} take into account the most recent $(g-2)_{\mu}$ results (for the final report of the Brookhaven E821 experiment, see Ref.~\cite{Bennett:2006fi}).

\section{The Experimental Limit on the MSSM Neutral Higgses}
\label{sec:higgs}

As emphasized in Section~\ref{sec:cpv}, the  Higgs           
potential plays a crucial role in determining the viability of electroweak      
baryogenesis. Both the shape of the potential and the mass of the lightest,     
Standard Model-like Higgs boson are important in this respect. Here, we         
review what is known about the neutral Higgs bosons in the MSSM and             
highlight the various assumptions associated with the corresponding Higgs       
mass limits. Importantly, some scenarios allow for weaker experimental lower    
bounds on $m_{h^0}$ than the bound for the mass of the SM Higgs that makes      
it too heavy to accommodate a strong first order EWPT in the SM.

There have  been extensive searches for Higgs bosons at the LEP. 
No signals have been found so far and a lower limit 
of 114.4 GeV has been set for the mass of the SM model Higgs boson at
95\% C.L. \cite{SMHiggssearch}.   In MSSM, two Higgs doublets need
to be introduced and there are five physical Higgses in the 
spectrum: three neutral ones and two charged ones.  When CP is conserved
in the Higgs sector, the three neutral Higgses are CP eigenstates: 
two CP-even ones $h^0$ and $H^0$ and one CP-odd one $A^0$.  However,
there is no reason to exclude the CP violation in the Higgs sector.
In particular, CP violation in MSSM could provide one of the ingredients 
to explain the observed matter-antimatter asymmetry in the 
Universe~\cite{Riotto:1999yt,Dine:2003ax}, as discussed above.  In the CP violating  scenario, the three neutral 
Higgs mass eigenstates $H_1$, $H_2$ and $H_3$ 
are a mixture of the CP-even and the CP-odd states. The Higgs production and 
decay might differ significantly from the CP conserved scenario.  Analysis
of the LEP Higgs searches have been performed in both 
scenarios \cite{LEPMSSMneutral, OPAL, ALEPH, DELPHI, L3}.  In this 
section, we will briefly review the Higgs searches and the 
exclusion bounds on the neutral Higgs masses and other parameters.

In the CP conserving scenario, the main production 
processes of $h^0$, $H^0$ and $A^0$ at the LEP are Higgsstraahlung 
$e^+e^-\rightarrow h^0 Z$ and $e^+e^-\rightarrow H^0Z$ (if kinematically possible) and 
pair production $e^+e^- \rightarrow h^0A^0$ and $e^+e^- \rightarrow H^0A^0$ (if kinematically possible).
These two processes are complementary since $\sigma_{h^0Z}$ is
governed by the $h^0ZZ$ coupling, which is proportional to $\sin(\beta-\alpha)$,
while $\sigma_{h^0A^0}$ is governed by the $Zh^0A^0$ coupling, which is 
proportional to $\cos(\beta-\alpha)$.  For the Heavy Higgs $H^0$, $\sin$ and $\cos$
are exchanged.

The light neutral Higgs $h^0$, whose mass is typically below 140 GeV 
\cite{Higgsmass},  decays dominantly into fermion pairs since its mass 
is below the threshold of $WW$ and $ZZ$.  Although, for particular 
choices of parameters, the fermionic decay may be strongly suppressed, 
which will be discussed below in the {\it large $\mu$}, 
{\it gluophobic} and {\it small $\alpha_{eff}$} benchmark models \cite{benchmark}.  For the 
CP-odd state $A^0$, it also dominantly decays into fermion pairs, since
its coupling to the gauge boson vanishes at tree level. For not too small 
$\tan\beta$, the Higgses decay into $b\bar{b}$ or $\tau^+\tau^-$, while
for $\tan\beta<1$, decays to $c\bar{c}$ might be important. 

In the CP violating scenario, the mass eigenstates $H_i$, ($i=1,2,3)$ are
a mixture of CP eigenstates.  Each of them can be produced by Higgsstrahlung
$e^+e^-\rightarrow H_iZ$ via the CP-even field components,
and also in pairs $e^+e^-\rightarrow H_i H_j (i\neq j)$.  The relative 
rates depend on the CP-even/odd mixing. Such mixing does not occur in the tree-level potential, but does appear at one-loop order. The degree of CP-mixing is proportional to the quantity
\beq
\frac{m_t^4}{v^2}\frac{{\rm Im}(\mu A)}{{\tilde m}^2}.
\label{eq:CPmixing}
\eeq
that arises from stop loops. 
Large CP violation in the Higgs sector is expected for small ${\tilde m}$
and large ${\rm Im}(\mu A)$. The CP violation effects are also very sensitive 
to the precise value of the top quark mass, which is known with a few
GeV experimental error.

The Higgs searches in the 
CP-violating scenario are, in general, more challenging than the CP-conserving
case.  The reason is that the production of the lightest Higgs $H_1$ is 
reduced due to its suppressed coupling to $Z$.  While the production of the
heavier Higgses are suppressed or forbidden due to kinematics.  The decay
of the Higgses in the CP-violating scenario is very similar to the 
CP-conserving case that discussed above.

Higgs searches have been performed at LEP up to the highest LEP energy of
209 GeV, carried out by the four LEP collaborations \cite{OPAL, ALEPH, DELPHI, L3}.
The searches include the Higgsstrahlung process and pair production process,
which are sensitive over the accessible MSSM parameters due to their 
complementarity.
For the Higgsstrahlung process, the principle signal topologies are Higgs
decays to fermion pairs $b\bar{b}$, $\tau^+\tau^-$ or flavor independent 
$q\bar{q}$, while the $Z$ decays into pair of jets ($q\bar{q}$),  
$\nu\bar\nu$ (associated with missing energy), or lepton pairs $e^+e^-$, $\mu^+\mu^-$, $\tau^+\tau^-$.  The 
reconstruction of the $Z$ mass offers a discrimination of the signal
over the background.  Searches including Higgs cascade decay 
$e^+e^-\rightarrow H_2 Z \rightarrow (H_1 H_1) Z$ have also been performed,
which might play an important role when this decay mode is open. 
For Higgs pair production process, $b$ pairs and $\tau$ pairs final states
have been studied when Higgs masses are above the $\tau^+\tau^-$ threshold.
When the $b\bar{b}$ decay mode of the Higgs is suppressed in certain
parameter spaces, flavor-independent searches are used as 
a supplementation or replacement.

The combined LEP data show no significant signal for Higgs boson 
production \cite{LEPMSSMneutral}.  Therefore, the search results are used to set an upper limit
on the Higgs production cross section, and they are interpreted
in a set of representative MSSM ``benchmark'' models \cite{benchmark}.

\begin{figure}
\includegraphics*[width=3 in]{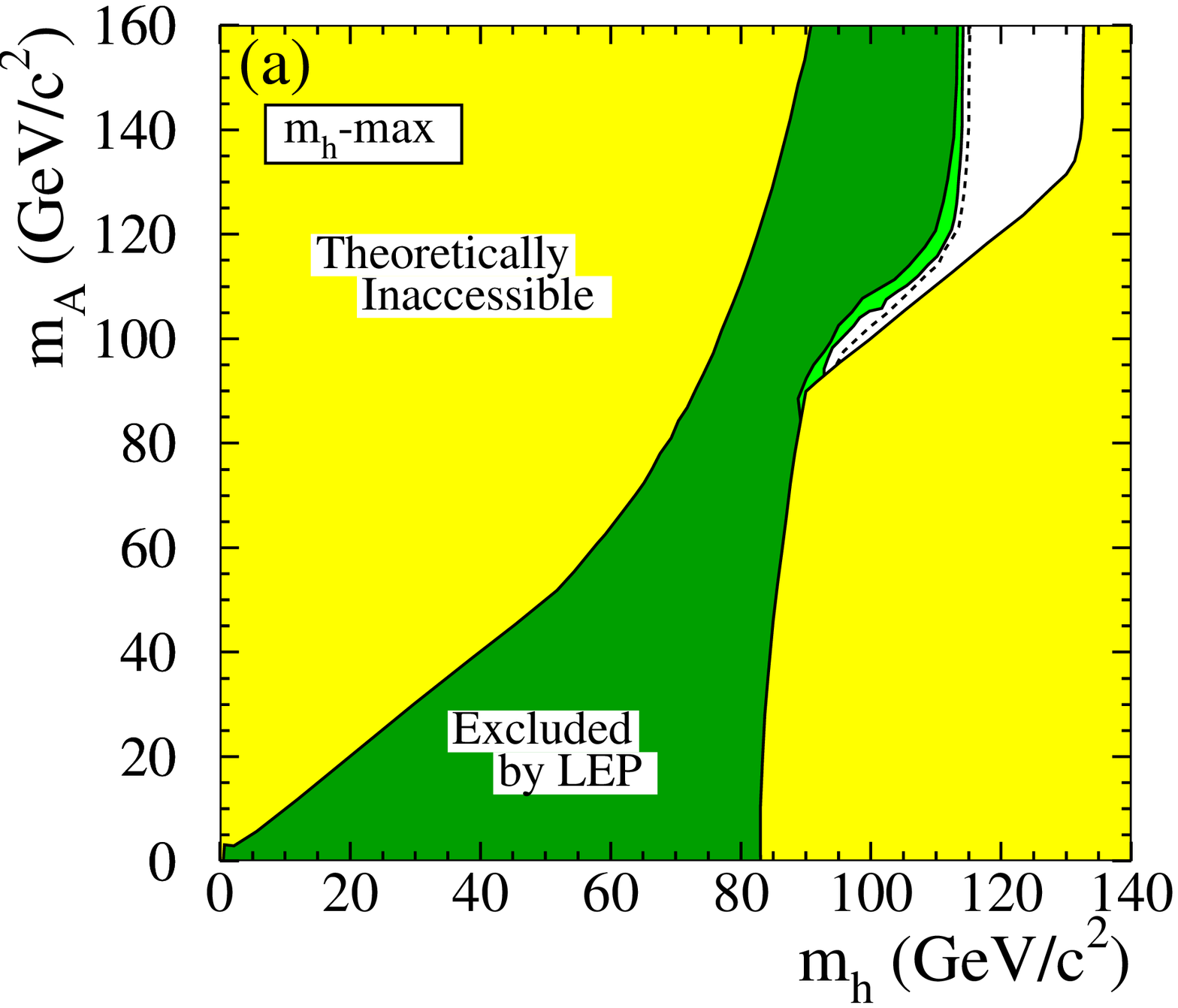}
\includegraphics*[width=3 in]{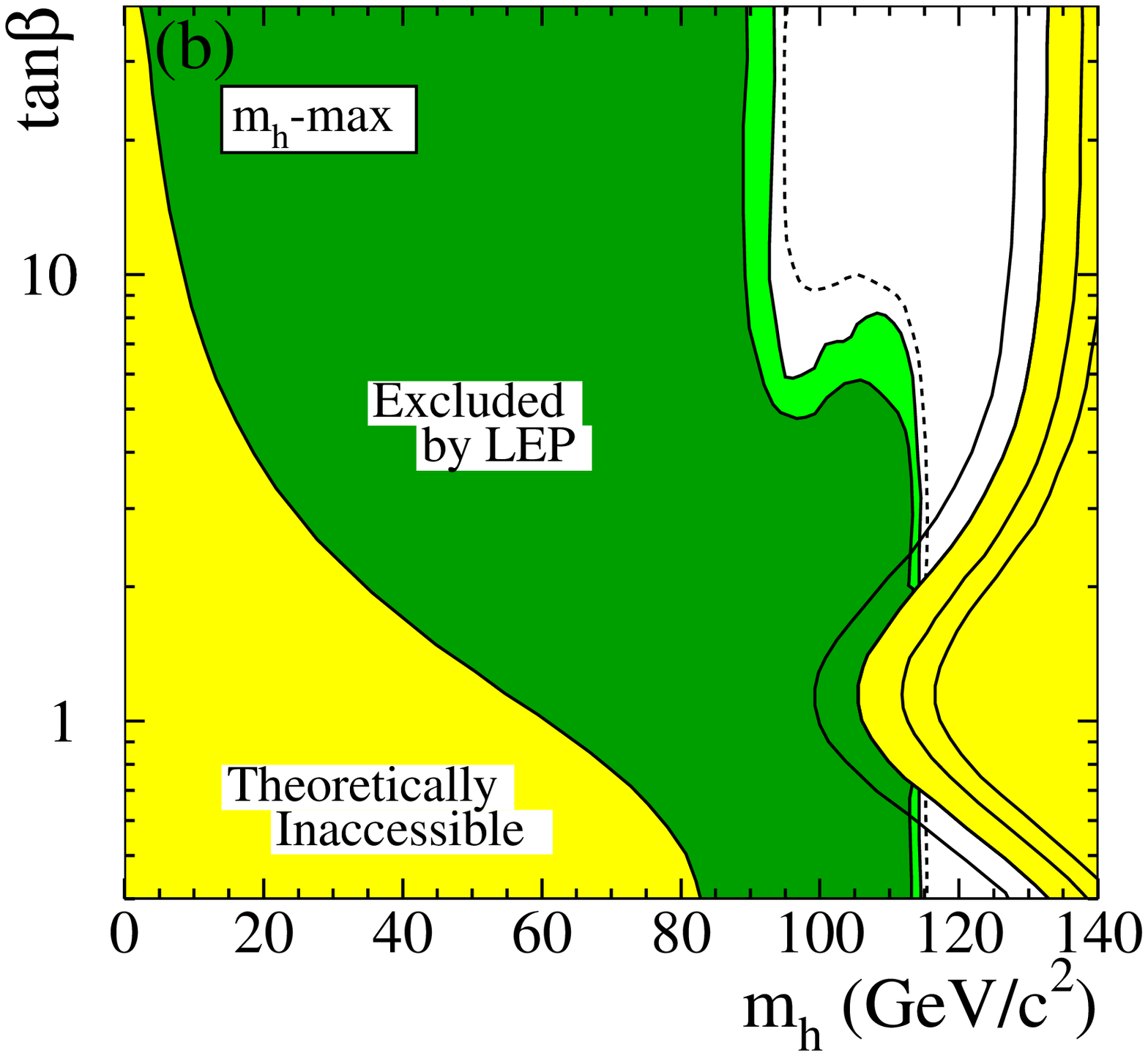}
\caption{The MSSM exclusions, at 95\% C.L. (light-green) and 99.7\% C.L. (dark-green),
for the  {\it $m_h$-max} benchmark scenario, with $m_t$=174.3 GeV.  The theoretically 
inaccessible regions are shown in yellow.  The dashed lines indicate the boundaries of the 
regions expected to be excluded on the basis of the Monte Carlo simulations with no
signal.  In the right plot, the upper edge of the parameter space is indicated for various
top quark masses; from left to right: $m_t=$ 169.3, 174.3, 179.3 and 183.0 GeV.
The figures are reproduced from Ref.~\cite{LEPMSSMneutral} with kind permission of            
Springer Science and Business Media.}
\label{fig:mhmax}
\end{figure}

For the CP-conserving scenario, the {\it $m_h$-max} benchmark \cite{benchmark} occurs  when 
the stop mixing parameter is set to a large value,
$X_t=A-\mu \cot\beta=2 M_{\rm SUSY}$.  This model is designed to maximize
the theoretical upper bound on $m_h^0$ for a 
given $\tan\beta$, thereby providing the largest parameter space 
and therefore the most conservative exclusion limits among all
the CP-conserving scenarios  studied.  
Fig.~\ref{fig:mhmax} \cite{LEPMSSMneutral} 
shows the excluded region in $(m_h^0,m_A^0)$ (left plot)
and in $(m_h^0, \tan\beta)$ (right plot).
For $\tan\beta<5$, the 95\% C.L. 
exclusion bound on the Higgs mass is about 114 GeV, provided by the 
Higgsstrahlung process, while for higher values of $\tan\beta$, 
pair production process dominates and the bounds is about 93 GeV
for both $m_h^0$ and $m_A^0$.  A certain region of $\tan\beta$ between 
0.5 and 3 is also excluded, which, however, depends on the 
precise value of the top quark mass.  The excluded region gets smaller 
for larger $m_t$, and no limit can be set for $\tan\beta$ for $m_t>183$ GeV.

For the {\it no-mixing} benchmark \cite{benchmark}, the stop mixing parameter $X_t$ is set to 
zero, thereby minimizing the stop contribution to the Higgs mass.  
The theoretical bounds of the parameter space are more restrictive than
in the {\it $m_h$-max} case, although the experimental bounds are similar. 
It is worth mentioning that a small domain at $m_h^0 \approx 80$ GeV,
$m_A^0 < 3$ GeV and $\tan\beta < 0.7$ is still allowed.  This domain is not 
covered by the current searches since the branching ratio of 
$h^0\rightarrow b \bar{b}$ is suppressed while $A^0\rightarrow \tau^+\tau^-$ is
not kinematically allowed. 

In the {\it large-$\mu$} scenario \cite{benchmark}, the experimental detection 
is {\it a priori} challenging due to the suppressed decay of 
$h^0\rightarrow b\bar{b}$, 
$\tau^+\tau^-$.   The dominant decay modes are 
$h^0\rightarrow c\bar{c}$, $gg$ and $W^+W^-$.   The flavor- and 
decay-mode-independent searches are used instead, which 
exclude almost all of the accessible MSSM parameter space. 

The {\it gluphobic} scenario \cite{benchmark} is constructed so that Higgs-gluon-gluon
coupling is suppressed, leading to a reduced Higgs production by gluon fusion
at the LHC.  The {\it small $\alpha_{eff}$} scenario refers to the case when
$h^0\rightarrow b \bar{b}$ and $\tau^+\tau^-$ are suppressed, since the 
corresponding couplings are proportional to $\alpha_{eff}$.  
Note that $\alpha_{eff}$ is the effective mixing angle of the neutral CP-even Higgs 
sector  (defined in Eq.~(\ref{eq:alpha}) in Sec.~\ref{sec:susy})
including radiative corrections.
Both scenarios 
were invented to test situations that might be problematic at LHC.  In both case 
large parameter spaces are excluded by the LEP searches.

\begin{figure}
\includegraphics*[width=3 in]{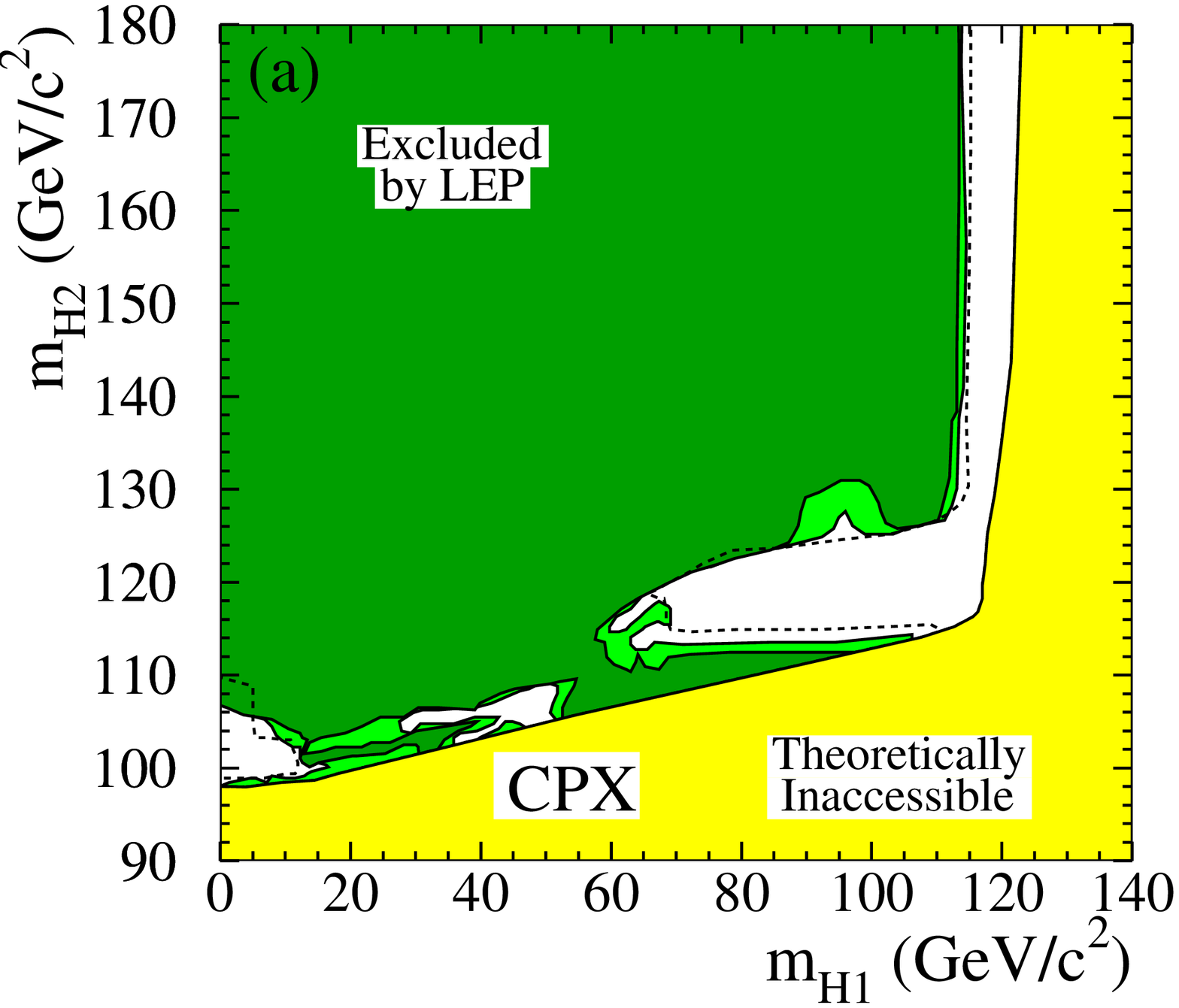}
\includegraphics*[width=3 in]{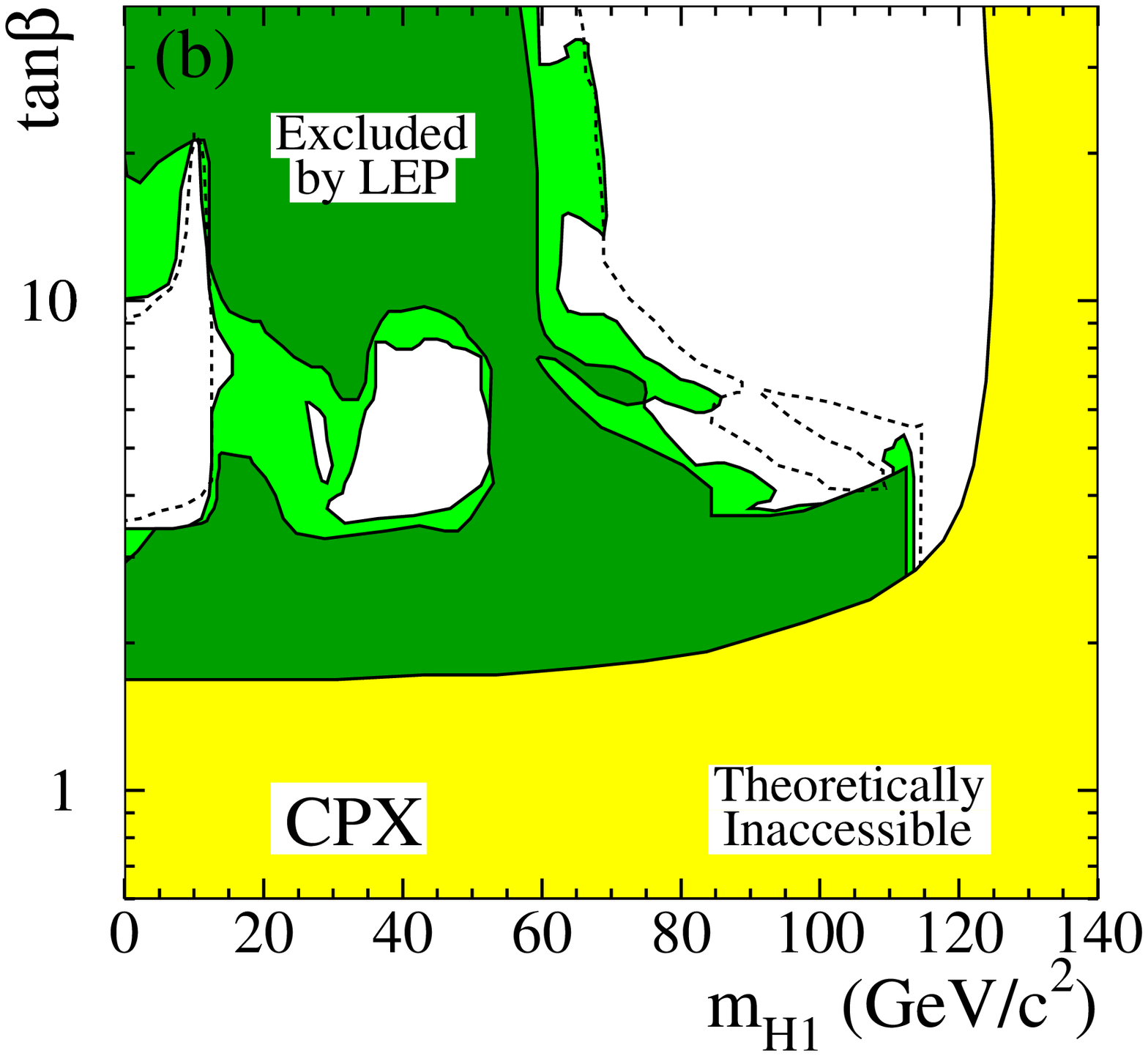}
\caption{Exclusions, at 95\% C.L. (light-green) and 99.7\% C.L. (dark-green),
for the  CP-violating scenario, with $m_t$=174.3 GeV.  The theoretically 
inaccessible regions are shown in yellow.  The dashed lines indicate the boundaries of the 
regions expected to be excluded, at the 95\% C.L., 
 on the basis of the Monte Carlo simulations with no
signal.  
The figures are reproduced from Ref.~\cite{LEPMSSMneutral} with kind permission of            
Springer Science and Business Media.}
\label{fig:mhcp}
\end{figure}

The parameters of the CP-violating benchmark \cite{CPXbenchmark} have been chosen to maximize
the difference with respect to the CP-conserving scenario: 
${\tilde m}=500$ GeV, $\mu=2000$ GeV and ${\rm arg}(\mu A)=90^0$.
Fig.~\ref{fig:mhcp} \cite{LEPMSSMneutral} shows the excluded region in $(m_{H_1},m_{H_2})$ (left plot)
and in $(m_{H_1}, \tan\beta)$ (right plot).
For large $m_{H_2}$, the $H_1$ is almost completely CP-even and the 
95\% C.L. exclusion bound on the $H_1$ mass is about 113 GeV.  For lighter 
mass of $m_{H_2}<130$ GeV, $H_1$ has a large CP-odd mixture, leading to 
unexcluded domain.   For $\tan\beta$ between 
about 3.5 and 10, the exclusion is particularly weak.
Nonetheless, the region of $m_{H_1}<114$ GeV and $\tan\beta<3.0$ are 
excluded by the data.
Furthermore,  
at 95\% C.L. $\tan\beta<2.6$ is excluded for all values of Higgs masses.

The exclusion region for the Higgs mass, 
however, depends strongly on the top quark mass. 
For the limits discussed above, $m_t=174.3$ GeV was used.  The exclusion
power is reduced for larger top quark mass, especially in the
region of $\tan\beta$ between 4 and 10.  The bound on $\tan\beta$ quoted
above, however, is barely sensitive to the precise value of $\tan\beta$.
The exclusion region also depends on the CP-violating phase,  ${\rm arg}(\mu A)$,
the $\mu$ parameter, and ${\tilde m}$.  The exclusion region is somewhat 
larger for values deviate away from the benchmark parameters. 

\section{Conclusions and Outlook}
\label{sec:conclude}

Precision measurements of electroweak observables played an important role in developing and testing the Standard Model, and they will undoubtedly be a crucial tool in determining the larger framework in which the SM lies. If that framework includes low-energy supersymmetry, then one would expect a rich array of effects to be discernible in precision measurements carried out at both high and low-energies. In this review, we have concentrated on the low-energy domain, where the \lq\lq precision frontier" will lie at least until the era of a future $e^+e^-$ linear collider. We hope to have demonstrated that through precise measurements of both SM observables  as well as those forbidden or highly-suppressed in the SM, studies of these low-energy observables will offer important information about SUSY that can complement what we may learn from the Large Hadron Collider. 

We also hope to have illustrated the opportunities and challenges in this field. Experimentally, recent advances have made the prospects for carrying out $\sim 0.1\%$ measurements of SM electroweak observables -- as needed to probe SUSY -- quite realistic, and a number of efforts are underway with such precision as a goal. Recent theoretical developments have also made it possible to interpret measurements at this level in terms of SUSY, as a number of strong interaction uncertainties have been circumvented or reduced. In both cases, going beyond the  $\sim 0.1\%$ precision level for SM observables represents the next horizon, one that both experimentalists and theorists are beginning to approach. At the same time, the prospects for performing measurements of rare and forbidden observables, such as electric dipole moments and lepton flavor violating effects, have improved dramatically, with several orders of magnitude increases in sensitivity now within reach. As we hope to have shown, the \lq\lq physics reach" of such experiments can match and even exceed what will be achievable at both the LHC and a linear collider, assuring that they will remain important avenues of study well into the future collider era. 

Finally, we emphasize that though we have concentrated here on the Minimal Supersymmetric Standard Model, it is by no means assured that the MSSM (together with the conventional assumptions about SUSY-breaking mediation) is the right low-energy manifestation of SUSY. There remains a   wealth of possible variations and a correspondingly rich field of low-energy, precision electroweak phenomenology to be explored. This opportunity, and the attendant experimental and theoretical challenges, will surely keep particle, nuclear, and atomic physicists busy for many years.

\begin{acknowledgments}

We would like to thank following people for discussions and interactions 
while finishing this paper:     
J. Erler, P. Langacker, W. Marciano,  V. Cirigliano, D. Hertzog, D. Bryman,     
J. Hardy, D. Pocanic, G. Savard, P. Reimer, X. Zheng, P. Souder, K. Kumar,     
R. Holt, Z. Lu, B. Filippone, G. Greene, M. Pospelov, M. Wise, D. Mack, C.      
Lee (also for assistance  with figures), S. Profumo (also for assistance  with figures), S. Martin,     
S. Tulin, P. Vogel, R. McKeown, C. Wagner, M. Carena, P. Herczeg, W.-F. Chang, K. Kumar, R. Carlini.       
M.J.  R.-M. thanks the Institute for Nuclear Theory, U. Pennsylvania, U.       
Arizona, and Los Alamos National Laboratory for hospitality during the          
completion of this work. 
S. Su thanks   California Institute of Technology for hospitality during the          
completion of this work. 
MRM  is supported under 
Department of Energy Contract \# DE-DE-FG02-05ER41361 and NSF Award              
PHY-0555674.  SS is 
supported under U.S. Department of Energy
contract \# DE-FG02-04ER-41298.
\end{acknowledgments}

\end{document}